\documentclass[12pt]{iopart}

\usepackage{graphicx}
\usepackage{xcolor}
\usepackage{hyperref}
\usepackage{sidecap}

\newcommand{\he}[1]{{\color{purple}#1}}

\begin{document}

\title[The exploration of hot and dense nuclear matter]{The exploration of hot and dense nuclear matter:\\ 
Introduction to relativistic heavy-ion physics}

\date{\today}

\author{Hannah Elfner$^{1,2,3,4}$ and Berndt M\"uller$^5$}
\address{$^1$GSI Helmholtzzentrum f\"{u}r Schwerionenforschung, Planckstr. 1, 64291 Darmstadt, Germany}
\address{$^2$Institut f\"{u}r Theoretische Physik, Goethe Universit\"{a}t, Max-von-Laue-Strasse 1, 60438 Frankfurt am Main, Germany}
\address{$^3$Frankfurt Institute for Advanced Studies, Ruth-Moufang-Strasse 1, 60438 Frankfurt am Main, Germany}
\address{$^4$Helmholtz Research Academy Hesse for FAIR (HFHF), GSI Helmholtz Center, Campus Frankfurt, Max-von-Laue-Stra{\ss}e 12, 60438 Frankfurt am Main, Germany}
\address{$^5$Department of Physics, Duke University, Durham, NC 27708, USA}

\begin{abstract}
This article summarizes our present knowledge about nuclear matter at the highest energy densities and its formation in relativistic heavy ion collisions. We review what is known about the structure and properties of the quark-gluon plasma and survey the observables that are used to glean information about it from experimental data. 
\end{abstract}

\maketitle

\section{Introduction}
\label{sec:intro}

{\em This Section gives an overview of the goals of exploring nuclear matter at high energy density and outlines the recent and ongoing program with relativistic heavy ions.}

What are the properties of the matter that permeated our universe \cite{Schwarz:2003du} during the first roughly $20~\mu$s of its existence when its temperature exceeded $2\times 10^{12}$ K? Theoretical considerations tell us that it was a quark-gluon plasma, i.~e.\ matter in which the quarks were not confined into color singlet objects, summarily called hadrons. Numerical simulations of quantum chromodynamics on a Euclidean space-time lattice (lattice-QCD) have established the presence of a rapid, but smooth cross-over transition in the properties of QCD matter in the temperature range 140 MeV $< T <$ 170 MeV from a phase that is well described as a gas of hadrons and resonances to a phase, called quark-gluon plasma, in which the color force between quarks is screened and the spontaneously broken chiral symmetry of the QCD vacuum is restored \cite{Aoki:2006br,Aoki:2006we}.  The near coincidence of these two transitions (quark deconfinement and chiral symmetry restoration) is due to their mutual reinforcement: Absent the action of a confining color force, single quarks are easier to excite than whole hadrons, which suppresses the quark condensate and, in turn, enhances the density of quarks available to screen the color force. The connection is effectively modeled by the Polyakov--Nambu--Jona-Lasinio (PNJL) model \cite{Fukushima:2003fw,Rossner:2007ik}.

The experimental exploration of nearly baryon number-free hot QCD matter commenced in the year 2000 with the start of operations of the Relativistic Heavy Ion Collider (RHIC). In the past two decades RHIC has produced and studied hot QCD matter in collisions of complex nuclei ranging from $^{16}$O to $^{238}$U, and it has performed baseline measurements in $p+p$ and $p(d,^3{\rm He})+{\rm Au}$ collisions. Hot QCD matter with a significant baryon excess was studied in fixed target experiments at the CERN--SPS and the BNL--AGS, and it has become a focus also at RHIC with the exploratory and high-statistics beam energy scans (BES-I and BES-II). At the lower energy end, SIS-18 at GSI is running and data is collected for very dense but moderate temperature systems. The conditions reached in low energy heavy-ion collisions are similar to the ones in neutron star mergers and provide complementary insights into nuclear matter at extreme densities. 

\begin{figure}[ht]
\centerline{\includegraphics[width=0.60\linewidth]{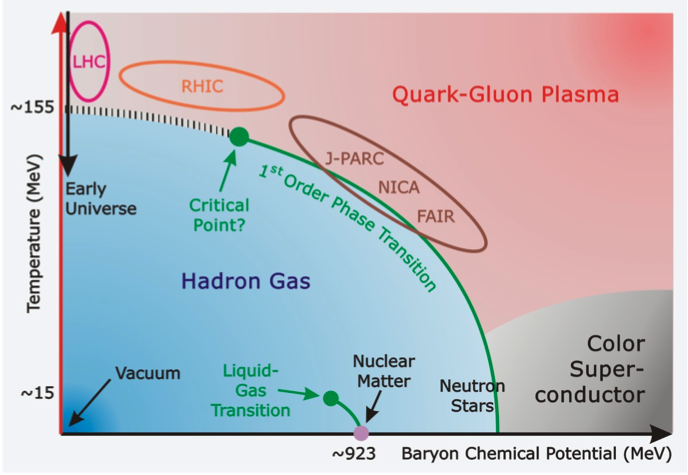}}
\caption{Sketch of the phase diagram of strongly interacting matter as a function of net baryon chemical potential and temperature.  [From A. Steidl, Frankfurt University]}
\label{fig:phase_diagram}
\end{figure}

Fig. \ref{fig:phase_diagram} shows a sketch of the phase diagram of strongly interacting matter. Putting firm constraints on the properties of matter is one of the major goals of heavy-ion research. The quark-gluon plasma phase is situated at high temperatures and/or densities with the cross-over transition to the hadron gas along the temperature axis. This is where the early universe has evolved and high energy heavy-ion experiments explore this net baryon number free regime. At finite densities and low temperatures the liquid-gas transition associated to bound nuclear matter is indicated as well as cold neutron stars at very high densities potentially including a color super conducting phase. At finite net baryon densities a first order phase transition with a critical endpoint is expected, that is explored by heavy-ion experiments at intermediate and low beam energies. 

The results obtained during the first five years of the RHIC experiments have been summarized in ``white papers'' \cite{Arsene:2004fa,Adcox:2004mh,Back:2004je,Adams:2005dq} published by the four experimental collaborations (BRAHMS, PHENIX, PHOBOS, and STAR).  The results provided compelling evidence that a novel form of thermal hot QCD matter was produced in nuclear collisions at RHIC, which clearly differs from hadronic gas and has the properties of a nearly inviscid liquid \cite{Muller:2006ee}.  Since a low viscosity implies strong coupling among the constituents of the medium, a natural interpretation of the RHIC data is provided by the hypothesis that the QCD matter produced in the experiments is a strongly coupled quark-gluon plasma \cite{Bannur:1998nq,Gyulassy:2004zy}.

The understanding of the dynamical properties of the quark-gluon plasma flowing from the RHIC program was extensively tested and expanded at much higher beam energies at the Large Hadron Collider (LHC) starting in 2010. Since the LHC beam energy is more than a factor of 10 higher than that of RHIC, the main question was whether the quark gluon plasma still exhibits the features of a strongly-coupled gauge liquid at these higher energies. This question was answered in the affirmative by the first reports from the three large experimental collaborations (ALICE \cite{Schukraft:2011cz}, ATLAS \cite{Steinberg:2011qq}, and CMS \cite{Wyslouch:2011zz}). All three collaborations have since published a wealth of data that all confirm this conclusion (see \cite{Muller:2012zq} for a first comprehensive assessment). A more narrowly focused review of bulk properties of the QGP based on results from RHIC and LHC can be found in \cite{Braun-Munzinger:2015hba}. A high-level summary of established insights and open questions is presented in \cite{Busza:2018rrf}.

As will be reviewed below, various properties of the baryon number-free quark-gluon plasma can now be rigorously and reliably calculated using lattice-QCD simulations, in particular, its equation of state. However, definitive calculations are still limited to thermodynamic quantities that can be formulated in terms of static observables. Many properties of hot QCD matter that are of great phenomenological interest, such as its transport coefficients or its excitation spectrum, are still inaccessible to rigorous lattice simulations. Also, the properties of QCD matter with a substantial baryon excess cannot be calculated using presently known techniques. For these properties we still have to rely on tenuous extrapolations of results obtained by means of thermal perturbation theory, on predictions from effective models of QCD, or on answers gleaned from rigorous calculations for strongly coupled gauge theories akin to, but different from, QCD that have known holographic duals. The results of such calculations must be regarded as qualitative, but they can provide useful guidance for the experimental program.

This review is structured as follows. Section~\ref{sec:obs} contains an overview of the connection between experimental observables and fundamental quantities characterizing the quark-gluon plasma. Section~\ref{sec:theory} covers the foundations of theoretical predictions for the properties of the quark-gluon plasma, based on lattice QCD, thermal perturbation theory, and holographic methods for strongly coupled gauge theories that serve as models for QCD. The basic phenomenological approaches to making this connection that are well established on a quantitative level are described in Section~\ref{sec:pheno}. In Section~\ref{sec:results}, we review the experimental results from RHIC and LHC for the observables introduced in Section~\ref{sec:obs} and discuss their interpretation. The review concludes in Section~\ref{sec:opps} with an overview of future opportunities for experimental and theoretical investigations of hot QCD matter. 

{\it How to read this article:} If the reader is mainly interested in getting an overview of the current state of experimental results, it would be advisable to jump directly to Section~\ref{sec:results} and then circle back to Sections~\ref{sec:obs},~\ref{sec:theory}, and~\ref{sec:pheno} depending on the nature of one's further interest. Alternatively, one might start with Section~\ref{sec:obs} if the primary interest is in getting an overview of the main experimental observables in relativistic heavy ion collisions, for example, if someone wants to prepare for the first attendance of a Quark Matter conference and then proceed to either Section~\ref{sec:pheno} or Section~\ref{sec:theory}. Those most interested in the theoretical motivation and foundation of the field may want to start with Section~\ref{sec:theory} and then proceed to Section~\ref{sec:pheno} and the other sections. We assume that any reader will have at least a cursory familiarity with the elements of QCD and its most important phenomenological aspects (asymptotic freedom, quark confinement, hadron spectrum).

Our review does not aim for completeness in citations. The published literature on the quark-gluon plasma and the phenomenology of relativistic heavy ion collisions is just too vast. An excellent source of foundational references is Kapusta {\it et al.}'s annotated reprint collection \cite{Rafelski:2003zz}. Concise presentations of many of the underlying concepts and phenomenological aspects can be found in the textbooks by Yagi {\it et al.} \cite{Yagi:2005yb} and by Letessier and Rafelski \cite{Letessier:2002gp}. Apart from specific citations of quoted experimental or theoretical results, we have mainly cited targeted reviews and publications that provide an overview of a specific area. 

We hope that this citation strategy will be of value to the readers at whom our review is primarily aimed: Graduate students and postdocs, scientists from other areas of nuclear and particle physics, who want to familiarize themselves with the status of this field of physics because they are considering contributing to it, and all those scientists who would like to  get a concise overview of the questions, methods and results of this area of physics. We refrain on purpose from historical accounts of events and rather report the current status of understanding of hot and dense QCD matter as explored in heavy ion collisions.

\section{Observables}
\label{sec:obs}

{\em This Section describes basic quantities characterizing high-energy density QCD matter and their relation to experimental observables. This Section contains very few references, since the details are discussed in Section \ref{sec:results} where references, mainly to review articles, are provided.}

\subsection{Overview}
\label{sec:obs-summary}

The approaches described in the previous Section allow theorists to relate various fundamental properties of the quark-gluon plasma to specific experimental observables:

\begin{enumerate}

\item
The equation of state, expressed as the dependence of pressure $p$ on energy density $\varepsilon$ and net baryon density $n$, of the quark-gluon plasma is reflected in the final particle spectra, collective flow properties, and the propagation of initial-state density fluctuations into the final state. The speed of sound, $c_s = \sqrt{\partial p/\partial\varepsilon}$, determines how the pressure gradients propagate during the expansion of the system. 

\item
The viscosities $\eta$ (shear) and $\zeta$ (bulk), which may be expressed as autocorrelation functions of the components of the stress tensor $T_{ij}$ \cite{Czajka:2017bod},
\begin{eqnarray}
\eta = \frac{1}{T} \int d^4x\, \langle T_{xy}(x) T_{xy}(0) \rangle ,
\\
\zeta = \frac{1}{9T} \sum_{i,j} \int d^4x\, \langle T_{ii}(x) T_{jj}(0) \rangle .
\end{eqnarray}
influence the collective transverse flow pattern, quantified by Fourier decomposition of the azimuthal distribution of particles in momentum space. The anisotropy of the transverse flow is especially sensitive to the dimensionless ratio $\eta/s$, where $s$ denotes the entropy density.

\item
Momentum transport coefficients, including the so-called jet quenching parameter $\hat{q}$, are expressed as correlation functions of the color force experienced by a fast parton in the quark-gluon plasma:
\begin{equation}
\hat{q} \propto  \int dx^- \langle {\rm Tr}\,[U_{0,x}F^{+i}(x) U_{x,0} F_{i}^{+}(0)] \rangle .
\label{eq:q-hat}
\end{equation}
Here $F$ denotes the field strength tensor and $U$ are so-called gauge links that ensure the gauge invariance of $\hat{q}$. The minus index in $x^-$ indicates that the correlation is sampled along the light cone. Together with two other coefficients, $\hat{e}$ and $\hat{e}_2$, $\hat{q}$ determines the rate of energy loss of a fast parton traversing the quark-gluon plasma. 
Experimentally, these coefficients can be determined by measuring the suppression of hadrons emitted with high momentum transverse to the beam direction.

\item
The screening length $\lambda_{\rm D}$ of the color force between a heavy quark-antiquark pair in the quark-gluon plasma, which is encoded in the correlator of the static gauge potential between two points separated by a spatial distance $x$:
\begin{equation}
\lambda_{\rm D}^{-1} = - \lim_{|x|\to\infty}\frac{\langle {\rm Tr}\,
[U_{0,x}A^{0}(x) U_{x,0} A^{0}(0)] \rangle}{|x|} ,
\end{equation}
controls the ability of a pair of heavy quarks to form a bound state. When $\lambda_{\rm D}$ is less than the size of the bound state, the state dissolves in the medium. Experimentally, this phenomenon is expected to be revealed by a strong suppression of charmonium ($c\overline{c}$) and Upsilon ($b\overline{b}$) states. 

\item
The electromagnetic current-current correlation function
\begin{equation}
C^{\mu\nu}_{\rm em}(q) = \int d^4x\, e^{iq\cdot x} \langle j^\mu(x) j_\nu(0) \rangle
\end{equation}
encodes the response of the quark-gluon plasma (or hadronic medium) to electromagnetic fields. It is directly only sensitive to quarks, but indirectly also to gluons, because quarks can be off-shell due to their interactions with thermal gluons. Observables carrying information about $C^{\mu\nu}(q)_{\rm em}$ are photons, sensitive to $q^2=0$, and dileptons, sensitive to time-like $q^2>4m^2$ with $m$ being the lepton mass.

\end{enumerate}

These well defined quantities summarize the main properties of the hot and dense QCD matter. Their linkage to experimental observables lays the ground for robust theoretical and experimental research programs in heavy ion physics. In making this linkage it is important to keep in mind that only the momentum-space properties of the matter is accessible to experiments, because the fireball is too small and short-lived to allow direct spatial of temporal resolution. It is possible to infer some information about the space- and time-dependence of the collision from two-particle correlations, which will be discussed in Section \ref{sec:coll_observables} under the topic ``HBT Femtoscopic Radii'', but even these indirect measurements are limited to last stage of the collision and require detailed modeling of the reaction.

From a practical viewpoint, on the experimental side, it is important that precision measurements of the relevant observables are made over a wide kinematic range so that theoretical models of the collision can be constrained. On the theoretical side, precise and reliable calculations of the transport coefficients under given conditions are needed in combination with realistic simulations of the dynamical evolution that connect the matter properties to the observables.

The following Subsections introduce the basic concepts underpinning the most commonly considered observables and point out their generally accepted connections to the properties of hot and dense QCD matter.

\subsection{Single-Particle Observables}

Single particle observables are all the observables that can be measured on the
basis of individual particle properties. 
\\

\noindent {\em Particle Yields and Multiplicities} 

The most basic analysis in a heavy ion collision is to count the numbers of particles of a certain species that are produced. One can either restrict the measurement of the yield to a region in momentum space around mid-rapidity or, if detector acceptance allows, extrapolate the number to the full multiplicity including the whole phase-space ($4\pi$ yields). By fitting the ratios of particle multiplicities to a grand canonical thermal distribution one can infer information about the conditions at the time of chemical freeze-out, after which the particle yields do not change anymore (see Section \ref{sec:chem_fo}).
\\

\noindent {\em Rapidity Distributions}

Rapidity distributions $dN/dy$ are sensitive to the initial energy deposition and thus to the stopping power of the colliding nuclei. Extreme limits of the charged particle rapidity distribution are associated with Landau (Gaussian shape) or Bjorken (flat shape) hydrodynamic behavior. The Landau limit corresponds to full stopping followed by explosive expansion. The Bjorken model corresponds to complete transparency for the valence quarks resulting in longitudinal boost invariance -- this approximation is only strictly valid in the infinite collision energy limit.
\\

\noindent {\em Transverse Momentum Spectra}

Counting the particles in bins of transverse momentum or transverse mass conveys information about the kinetic decoupling temperature and the flow profile  
\begin{equation}
\frac{1}{m_T} \frac{dN}{dm_T} \propto \exp^{-m_T/T_{\rm eff}}.
\end{equation}
By fitting the particle spectra with a thermal distribution one can extract an effective ``temperature'', which is often called the {\em slope parameter}. If the emitted hadrons originate from a locally equilibrated and collectively flowing momentum distribution, the slope parameter represents a blue-shifted temperature. The outward radial flow that the fireball develops during the dynamical evolution generally results in a flatter shape of the spectra of higher mass particles, e.~g.\ baryons, at low transverse momentum (see Section \ref{sec:trans_expansion}. 
\\

\noindent {\em Nuclear Modification Factor $R_{AA}$}

The nuclear modification factor $R_{AA}$ is the ratio of the transverse momentum spectrum in a heavy ion collision (AA) normalized by the transverse momentum spectrum
in the corresponding number of independent nucleon-nucleon collisions (pp)
\begin{equation}
R_{AA}=\frac{{d^2N/dp_T dy_{AA}}}{\langle N_{coll} \rangle {d^2N/dp_T dy_{pp}}}.
\label{eq:RAA}
\end{equation}
If there are final-state interactions with the medium that degrade the transverse momentum of energetic partons, $R_{AA}$ gets smaller than one, which explains the alternative name of this observable: nuclear suppression factor. This phenomenon is referred to as jet quenching. For light-heavy collision systems, such as $p+A$ or $d+A$, the $R_{p(d)A}$ is sensitive to initial-state effects that can be attributed to the behavior of cold nuclear matter relative to the reference of proton-proton collisions. For heavy-heavy collision systems the nuclear modification factor is mainly sensitive to final-state effects. The first determination of the energy loss parameter $\hat{q}$ is based on measurements of the nuclear modification factor (see Section \ref{sec:jet_quenching}). $R_{AA}$ is also used to quantify the suppression of heavy quarkonium production in heavy ion collisions.
\\

\noindent {\em Photon Spectra}

Direct photon spectra reflect the cumulative emission of thermal and prompt photons from all stages of a heavy ion reaction. The definition of a ``direct photon'' is that it does not originate from a particle decay, such as $\pi^0\rightarrow\gamma\gamma$. In the initial non-equilibrium evolution high-$p_T$ photons are produced in hard processes. Thermal emission from the quark-gluon plasma and the subsequent hadron gas phase are added during later collision stages. Since the mean free path of photons is larger than the fireball size, once photons are produced they end up in the detector and, different from strongly interacting probes, are minimally distorted by rescattering processes. By detailed dynamical modeling and matching of the various contributions to the photon spectrum, the goal is to extract information about the initial temperature and the lifetime of the QGP fireball from this observable. Photon spectra are also sensitive to the equation of state, since the emission depend strongly on the lifetime of the system that is substantially longer when a first order phase transition takes place (see Section \ref{sec:em_probes}).  
\\

\noindent {\em Dilepton Spectra}

Another electromagnetic probe are lepton pairs, commonly referred to as dileptons. These originate from virtual photons that decay into a $e^+ e^-$ or $\mu^+ \mu^-$ pair. The theoretical advantage is that leptons do not interact strongly and therefore can be measured from all stages of the heavy ion reaction. On the other hand, the cross-sections for leptonic pair production are orders of magnitude smaller than for hadronic emission. Since some hadron decays produce leptons, dilepton spectra constitute challenging measurements. Vector mesons, such as the $\rho$ and $J/\psi$ mesons, can decay into lepton pairs and show up as peaks in the dilepton invariant mass spectrum. By carefully analyzing the height and width of these peaks, one hopes to learn something about the influence of the medium on the resonance properties. Changes in the spectral function of the $\rho$ meson are associated with chiral symmetry restoration and deconfinement (see Section \ref{sec:em_probes}). 
\\

\noindent {\em Charmonium Suppression}

One telltale signature for the formation of a quark-gluon plasma is the suppression of charmonia ($J/\Psi$, $\Upsilon$, etc.). While number of heavy quark pairs created in initial hard interactions is deemed to be independent of the final-state nuclear medium, they are predicted to be unable to form bound meson states in a quark-gluon plasma due to color-screening. Lattice calculations for the heavy quark potential and the spectral function predict a sequential melting of charmonia states with increasing medium temperature. Therefore, the measurement of the ratio of charm spectra in nucleus-nucleus collisions divided by the spectrum expected from scaled proton-proton collisions is thought to serve as a direct way to extract the deconfinement temperature. This simplistic picture is complicated due to recombination of heavy quarks during hadronization and other effects (see Section \ref{sec:heavy_quarks}).  
\\

\noindent {\em Heavy Flavor}

The mass dependence of medium interactions can be probed by measuring single electron spectra that originate from the decay of open heavy flavor mesons (mesons containing an unpaired $c,b$ quark). Due to the dead-cone effect the expectation is that energetic charm and bottom quarks lose less energy in the quark-gluon plasma than light quarks. Measurements of the elliptic flow for open heavy flavor mesons provide insight into the question whether heavy quarks thermalize as well as light quarks (see Section \ref{sec:heavy_quarks}).

\subsection{Two- and Few-Particle Observables}

Two- and few-particle correlations are coincidence measurements of the probability of finding a particle with a certain property given that the event contains one particle with another (or the same) property. 
\\

\noindent {\em Photon-Hadron Correlations}

Photon-hadron correlations use a direct photon as the trigger particle for the correlation function. This has the advantage that the energy of the original parton in the hard process is constrained more tightly by the photon energy and, therefore, allows to access information about the fragmentation function. This observable is very clean for theoretical calculations, but poses a big challenge to experiments due to the contamination of the photon yield by decay photons (mainly from $\pi^0$) (see Section \ref{sec:jet_quenching}). 
\\

\noindent {\em Di-Hadron Correlations}

Hard di-hadron correlations are a sensitive tool for studying the path length dependence of jet energy loss. The trigger particle is a high-$p_T$ particle (usually with $p_T>5$ GeV/c), whereas the associate particle is chosen in a lower $p_T$ range. Di-hadron correlations are called ``hard'', if both $p_T$ ranges are 3 GeV/c and higher, so that a pQCD description is applicable. These correlations are often quantified as a distribution in the difference of the azimuthal angle between trigger and associated particle. One can then separate the near-side and the away-side structures, by an angle difference of roughly $180^\circ$ that originate from the back-to-back hard parton scattering. Given a near-side high-$p_T$ trigger particle the ratio of the yield of particles on the away-side in heavy ion collisions to the one in proton-proton collisions is called $I_{\rm AA}$ in analogy to $R_{\rm AA}$. Since $I_{AA}$ is a conditional yield, it potentially contains more detailed information about the energy loss, but also is subject to the so-called ``trigger bias'' (see Section \ref{sec:jet_quenching}). 
\\

\noindent {\em Triggered Correlations}

In addition to looking for correlations between two high-$p_T$ particles, jet medium interactions can be investigated by studying correlations between only one high-$p_T$ particle and an associated particle of lower $p_T$. The high-$p_T$ trigger particle preferentially emerges from the surface of the medium, having lost little energy on its way out, and therefore the backwards emitted parton experiences above average modification by the medium. Also, the medium itself might be modified by the traversing high energy parton resulting in characteristic excitation patterns like a Mach cone. To measure the effect of high $p_T$ partons on the medium, multi-particle correlations over various $p_T$ ranges and in longitudinal ($\Delta\eta$) and azimuthal ($\Delta\phi$) phase space have been studied.  
\\

\noindent {\em Untriggered Correlations} 

The last option for two- or three- particle angular correlations is to measure so-called untriggered correlations in minimum bias events, where one does not require one of the hadrons to have high transverse momentum. These $d^2N/d\Delta\phi\Delta\eta$ correlations have revealed interesting structure: There is an elongated enhancement  in pseudorapdity on the near side $\Delta\phi\approx 0$) that is often referred to as the ``ridge''. This structure is associated with higher Fourier coefficients in the azimuthal distribution with respect to the reaction plane (see below under collective flow) and will be discussed further in Section~\ref{sec:small_systems}.
\\

\noindent {\em Charge Asymmetry Correlations}

Another 3-particle measurement is the charge asymmetry with respect to the reaction plane. If there is an asymmetric production of up and down quarks in the initial state of a heavy ion reactions due to QCD vacuum fluctuations, these quarks could be separated by the magnetic field that is generated by the colliding nuclei. The chiral magnetic effect (see Section 5.13 for more details) is predicted to result in an asymmetry of the yield of charged particles with respect to the reaction plane that can be captured by observables of the type: 
\begin{equation}
\gamma^{\pm,\pm}=\langle \cos(\phi_\alpha^\pm + \phi_\beta^\pm - 2\Psi_{RP}) \rangle ,
\label{eq:gamma}
\end{equation}
where $\pm$ indicated the electric charge of the detected hadron. The observable (\ref{eq:gamma}) requires the detection of at least three particles, one of which is used to identify the reaction plane. A similar asymmetry of charged particle production with respect to the reaction plane can also be generated by a combination of balance functions and elliptic flow (see Section \ref{sec:chiral_magnetic}), which means that the observable has a large non-specific background.

\subsection{Multi-Particle and Collective Observables}
\label{sec:coll_observables} 

This section contains all of the observables that are based on multi-particle measurements and potentially reflect the collective behaviour of the system. 
\\

\noindent {\em Anisotropic Flow}

One of the most important observables in heavy ion reactions is collective flow of the particles. The Fourier coefficients $v_n$ of the azimuthal distribution of the final particles in momentum space are used to quantify anisotropic flow.
\begin{equation}
E \frac{d^3N}{dp^3} = \frac{1}{2\pi} \frac {d^2N}{p_t dp_t dy} +
\sum_{n=1}^\infty 2 v_n \cos[n(\phi-\Psi_n)] .
\label{eq:vn}
\end{equation}

The most prominent and well studied of these coefficients is $v_2$, commonly called elliptic flow, which arises due to the different pressure gradients in the transverse plane in non-central collisions. The initial almond shape in the coordinate space is translated by hydrodynamics into a momentum space anisotropy. As it is driven by pressure gradients, elliptic flow is highly sensitive to the shear viscosity and the equation of state of the expanding medium. Its collision energy dependence can, therefore, serve as a signature of the transition from hadronic matter to the quark-gluon plasma. Whereas $v_2$ can be measured by an event average, higher-order flow coefficients are non-zero require event-by-event measurements.They can be used to constrain the initial state profile of the fireball. The $v_n$ values can be determined by different analysis techniques like the event-plane method, the cumulant method, the flow vector method, or by means of Lee-Yang zeros (see Section \ref{sec:aniso_flow}). 
\\

\noindent {\em HBT Femtoscopic Radii}

Hanbury-Brown--Twiss (HBT) correlations of identical particles allow to infer the spatial volume, lifetime, and outward flow velocity of the fireball. The HBT correlations are caused by quantum (Bose or Fermi) interference between the two identical particles. The measurements are sometimes referred to as density interferometry. The measured correlation functions are usually fitted with a Gaussian that allows to extract three different radii
\begin{equation}
C({\bf q})=1+\lambda e^{-(q_o^2 R_{\rm out}^2+q_s^2 R_{\rm side}^2+ q_l^2 R_{\rm long}^2)}
\end{equation}
where $R_{\rm long}$ is aligned with the beam direction, $R_{\rm out}$ points along the center-of-momentum of the particle pair, and $R_{\rm side}$ is perpendicular to both. The HBT radii for different particle species contain information about the coherence region of the emission at kinetic freeze-out. The ratio of $R_{\rm out}/R_{\rm side}$ is indicative of the lifetime of the fireball and therefore sensitive to certain aspects of the equation of state, such as the nature of the QCD phase transition. Results and more details are discussed in Section \ref{sec:hbt}. 
\\

\noindent {\em Balance Functions}

Balance functions are a tool to study charge correlations in heavy ion reactions. They are similar to observables used to investigate hadronization in jets produced in $p\bar{p}$ or $e^+e^-$ collisions. The balance function describes the conditional probability that a particle in the momentum bin $p_1$ will be accompanied by a particle of opposite charge in the momentum bin $p_2$
\begin{eqnarray}
B_{ab}(p_2|p_1) &=&\frac{1}{2}\left(\rho(b,p_2|a,p_1) - \rho(b,p_2|b,p_1) \right. \nonumber \\
& & \qquad \left. + \rho(a,p_2|b,p_1) - \rho(a,p_2|a,p_1)\right) ,
\end{eqnarray}
where $\rho(b,p_2|a,p_1)$ is the conditional probability of observing a particle of type $b$ in bin $p_2$ given the existence of a particle of type $a$ in bin $p_1$. Balance functions can be defined for any conserved quantum number.
\\

\noindent {\em Reconstructed Jets}

Another observable sensitive to parton propagation in the quark-gluon plasma are fully reconstructed jets. For elementary collisions ($pp$, etc.) with normally only a single hard interaction the outgoing hard partons fragment into hadrons, and the energies of all fragment particles is measured in calorimeters. Applying sophisticated clustering algorithms the full jet can be reconstructed. Events containing a jet are identified online using a dedicated jet trigger. In the high-multiplicity environment of a heavy ion collision these measurements are much more challenging due to the underlying event background. Reconstructed jets allow comprehensive studies of jet energy loss and modifications of the internal jet structure (fragmentation functions, jet shape, etc.) by the comparison with predictions from jet shower Monte-Carlo algorithms tuned to jets in $pp$ collisions (see Section \ref{sec:jet_quenching}). 
\\

\noindent {\em Energy Flow and Track Functions}

Calorimeters measure energy flow carried by hadrons (hadronic calorimeter) or by photons and electrons (electromagnetic calorimeter) in a given direction. The most general realization of this concept is the energy flow operator \cite{Sveshnikov:1995vi}
\begin{equation}
    {\cal E}(\vec{n}) = \lim_{r\to\infty} \int_0^\infty dt\, r^2 n^i T^0_i(t,r\vec{n}) ,
\label{eq:Eflow}
\end{equation}
which measures energy flow into the direction $\vec{n}$. Observables of interest that provide a measure of the energy correlators of the form $\langle{\cal E}(\vec{n}_1) \cdots {\cal E}(\vec{n}_N)\rangle$, where one looks at the correlation of energy flow within a jet into different directional domains with an opening solid angle $\Delta R$ \cite{Komiske:2022enw}.

A related concept is that of track functions $T_{q\to h}(x)$, \cite{Chang:2013rca,Chang:2013iba}
which measure the probability that the total (light-cone) momentum fraction $x$ of a quark-initiated jet is carried by a certain type of hadron $h$. The track functions are similar to fragmentation functions, but instead of considering the momentum fraction of a single hadron, they sum over all hadrons of a given type within the jet. For a recent analysis of the QCD evolution of track function moments and a comparison with LHC data, see \cite{Li:2021zcf}.
\\

\noindent {\em Event-by-Event Fluctuations}

Event-by-event fluctuations of conserved charges, such as net electrical charge, net baryon number, or net strangeness, are a prominent signal for the phase transition to the quark-gluon plasma. Since quarks and gluons have different elementary units of the charges the predictions for event-by-event fluctuations are very different based on the assumption that a quark-gluon plasma or a hadron gas is present. The fluctuations of the mean transverse momentum can be regarded as the analogue to the temperature fluctuations in the cosmic microwave background, which are vestiges of quantum fluctuations in the initial state. The susceptibilities that are directly related to the fluctuations of conserved charges or their higher moments such as the skewness and the kurtosis are also calculable on the lattice and offer a direct connection between data and QCD predictions (see Section \ref{sec:fluctuations}).

\section{Theoretical Foundations}
\label{sec:theory}

{\em This Section reviews our present theoretical understanding of the structure and properties of the quark-gluon plasma, based on lattice-QCD simulations, thermal effective field theory, and exactly solvable strong coupling models.}

\subsection{Lattice QCD}

The quantitative {\it ab initio} calculation of the thermodynamic properties of QCD matter requires the model-independent evaluation of the functional integral that defines the quantum field theory. The only known rigorous method that allows to do this starts with the discretization of the QCD Lagrangian on a large space-time lattice and then evaluates the very high, but finite dimensional functional integral by Monte-Carlo methods. In order to study the properties of QCD at non-zero temperature, periodic boundary conditions (anti-periodic for quarks) are imposed on the imaginary time coordinate with period $\hbar/T$. The functional integral defining the quantum field theory is evaluated by means of Monte-Carlo sampling techniques that are used to generate a sufficient number (hundreds or thousands) of statistically independent, representative field configurations. Observables of interest are then evaluated by averaging over these stored configurations. The numerically most expensive part of the calculation is the integration over the fermion fields, which requires the evaluation of the determinant of a very large matrix. Much of the progress that has occurred over the past decade consists of finding improved ways to represent the QCD Lagrangian on a discrete lattice and speed up the evaluation of the fermion determinant. Other improvements in some simulations concern the implementation of manifest chiral symmetry by means of the domain wall or overlap fermion algorithms \cite{Kaplan:1992bt,Narayanan:1993sk}. 

Lattice QCD simulations have come a long way since the first calculations that demonstrated quark liberation in the high temperature phase of pure SU(2) lattice gauge theory \cite{Kuti:1980gh,McLerran:1981pb}. State-of-the-art lattice calculations include both, SU(3) gauge fields and dynamical $u$, $d$, and $s$ quarks (sometimes even $c$ quarks) with physical masses, employ improved lattice actions, extrapolate to the continuum limit (lattice distance $a \to 0$), and investigate the convergence of the results as a function of the number $N_t$ of Euclidean time slices. Fully converged lattice QCD simulations for baryon-symmetric QCD matter ($\mu_B=0$) have been made for the QCD equation of state and various other thermodynamic properties  \cite{Borsanyi:2010cj,Borsanyi:2013bia,HotQCD:2014kol}. 

Rigorous algorithms for the evaluation of the functional integral for thermal QCD are known only for $\mu_B=0$, because the integrand is not positive definite for non-zero real values of $\mu_B$. Nevertheless, attempts have been made to explore the QCD phase diagram away from the $\mu_B=0$ axis by means of lattice simulations using a number of indirect approaches. One method evaluates a certain number of derivatives of an observable with respect to $\mu_B$ at $\mu_B=0$ and then reconstructs the value at $\mu_B\neq 0$ from the truncated Taylor series \cite{Allton:2005gk}. Another method evaluates the functional integral for imaginary values of $\mu_B$, where the integrand is well behaved, and then attempts an analytic continuation to real values of $\mu_B$ \cite{deForcrand:2002ci}. A third method uses field configurations generated for $\mu_B=0$, but reweights them to simulate the $\mu_B$-dependence of the lattice action \cite{Fodor:2001pe}. The ``holy grail'' of such efforts is to locate a possible critical point in the QCD phase diagram. In spite of considerable progress in obtaining reliable results for $\mu_B/T < 2$, none of these approaches currently gives a positive indication for the possible location of such a critical point. Even the existence of a critical point in the $T-\mu_B$ plane, although extremely plausible on the basis of physical arguments and predicted by a multitude of QCD models, has not been established with absolute certainty \cite{deForcrand:2007rq}.

\subsubsection{Pseudocritical temperature.}

\begin{figure}[ht]
\centerline{\includegraphics[width=0.60\linewidth]{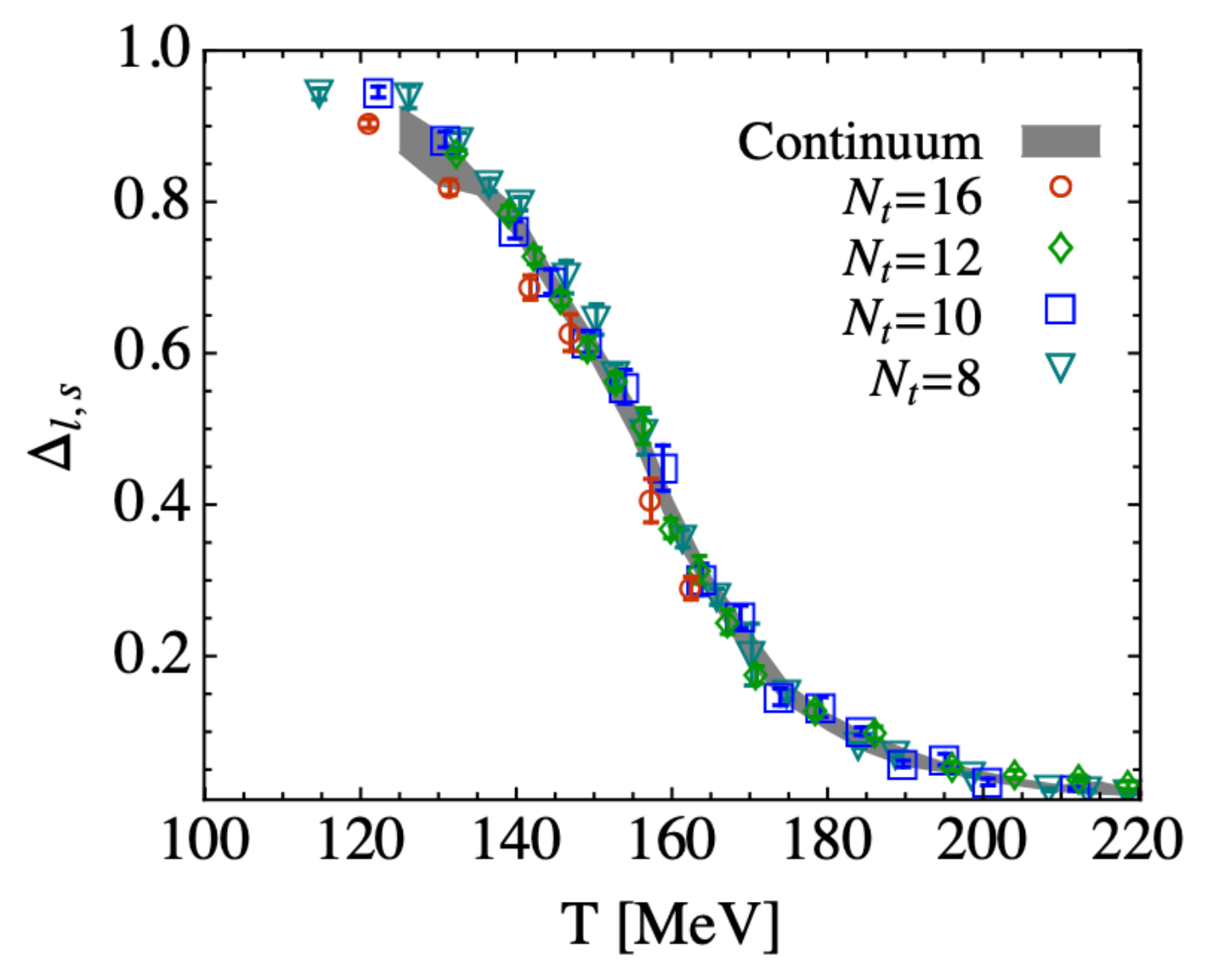}}
\caption{Temperature dependence of the renormalized light quark condensate $\Delta_{l,s}(T)$, divided by the value of the condensate in the vacuum, $\langle \overline\psi \psi \rangle_0$. The quark condensate, which becomes the order parameter of the chiral phase transition for massless quarks, serves as a measure of spontaneous chiral symmetry breaking. [From \cite{Borsanyi:2010bp}]}
\label{fig:psibarpsi}
\end{figure}

It has now been firmly established that the transition in baryon number-free ($\mu_B=0$) QCD matter is a smooth cross-over \cite{Aoki:2006we}. As a consequence, the value of the pseudo-critical transition temperature $T_c$ cannot be defined unambiguously. Usually, one identifies $T_c$ as the location of the peak or inflection point of a relevant thermodynamic quantity, but even this definition is ambiguous, because the value of $T_c$ depends upon the choice of this quantity. The transition temperature defined by the inflection point of the renormalized quark condensate \cite{Borsanyi:2010bp}
\begin{equation}
\Delta_{l,s} = \frac{\langle \overline\psi \psi \rangle_{l,T} -
\frac{m_l}{m_s}\langle \overline\psi \psi \rangle_{s,T}}
  {\langle \overline\psi \psi \rangle_{l,0} - \frac{m_l}{m_s}\langle
\overline\psi \psi \rangle_{s,0}}
\end{equation}
is $T_c^{(\chi)} \approx 158$ MeV \cite{Borsanyi:2020fev} with an uncertainty of less than 1 MeV (see Fig.~\ref{fig:psibarpsi}). 

\begin{figure}[ht]
\centering
\includegraphics[width=0.45\linewidth]{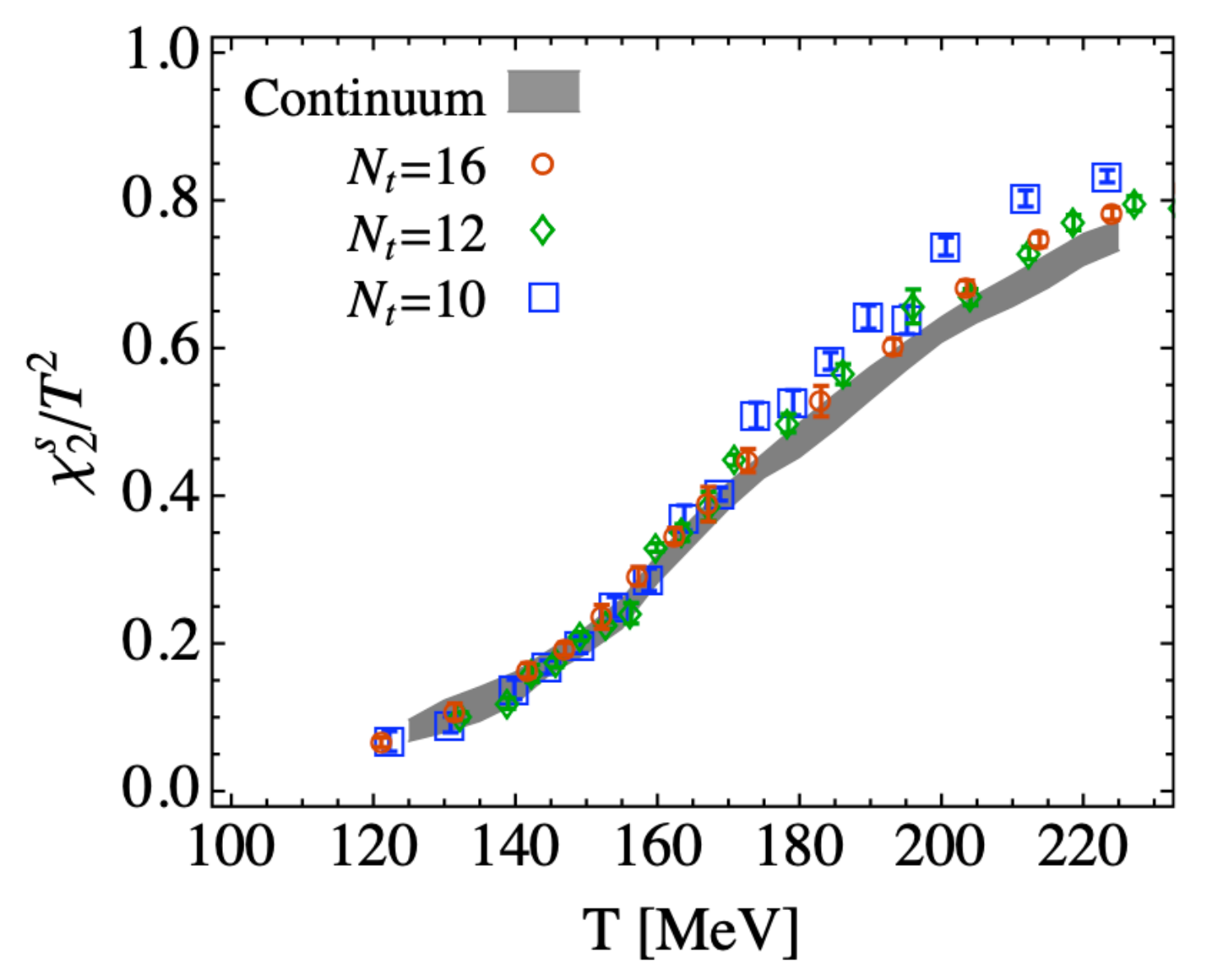}
\hspace{0.05\linewidth}
\includegraphics[width=0.45\linewidth]{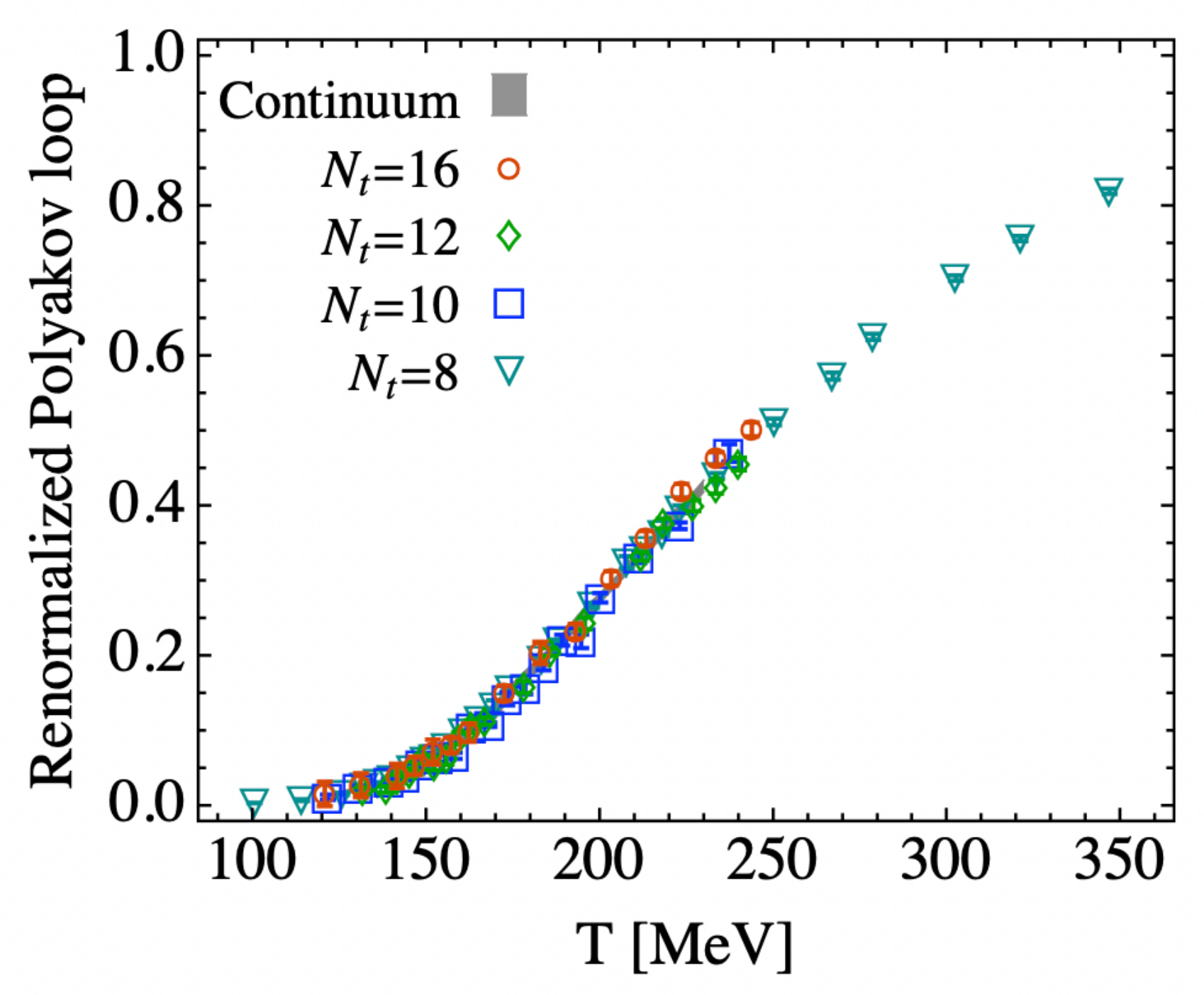}
\caption{Left panel: Temperature dependence of the strange quark number susceptibility $\chi_2^{(s)}/T^2$. Right panel: Temperature dependence of the expectation value of the renormalized Polyakov loop. Both quantities provide measures of quark liberation. [From \cite{Borsanyi:2010bp}]}
\label{fig:chis-Ploop}
\end{figure}

On the other hand, the transition region of quantities that are more direct measures of quark liberation, such as the expectation value of the Polyakov loop 
\begin{equation}
\langle L\rangle = \left\langle \frac{1}{N_c} {\rm Tr} \exp\left(ig\int_0^\beta
d\tau A^0(\tau,x)\right) \right\rangle
\end{equation} 
or the strange quark number susceptibility 
\begin{equation}
\frac{\chi_2^{(s)}(T)}{T^2} = \frac{1}{TV} \frac{\partial^2(\ln
Z)}{\partial\mu_s^2}
\end{equation}
is considerably broader. The expectation value of the Polyakov loop $\langle L\rangle = \exp(-E_Q/T)$ measures the interaction energy of an isolated heavy, static quark with the thermal bath of gluons. In the absence of light, dynamical quarks, it vanishes in the confined phase of QCD, because an isolated quark would have infinite energy. At high temperature, in the deconfined phase, the Polyakov loop approaches unity because the interaction energy is small compared with the temperature. This behavior is illustrated in Fig.~\ref{fig:chis-Ploop}, which shows the temperature dependence of the strange quark susceptibility (left panel) and of the renormalized Polyakov loop (right panel). Both observables yield values of the pseudocritical temperature, defined by the location of the inflection point, of $T_c{(L)} \approx T_c^{(s)} \approx 175$ MeV. (The higher value may be simply due to the greater width of the transition for these quantities.) The transition occurs over a range of about 20 MeV for the strange quark condensate and an even wider range for the Polyakov loop, which indicates that quark deconfinement is occurring gradually as the quark-gluon plasma is heated beyond $T_c$.

The transition is seen first, and over the narrowest interval, in the chiral quark condensate. This suggests that for the physical values of the parameters of QCD (the confinement scale $\Lambda_{\rm QCD}$ and the quark masses $m_u,m_d,m_s$) the transition is driven by the restoration of approximate chiral symmetry for the light quarks. The broad range of the transition seen in the strange quark susceptibility indicates that strange quark degrees of freedom thaw more gradually and at slightly higher temperature. The even more gradual transition of the renormalized Polyakov loop may be an indication that remnants of quark confinement in terms of correlations among quarks and antiquarks persist over a larger temperature range. This behavior is also evident in the QCD equation of state, which we discuss next.

\subsubsection{Equation of state.}

\begin{figure}[ht]
\centering
\includegraphics[width=0.95\linewidth]{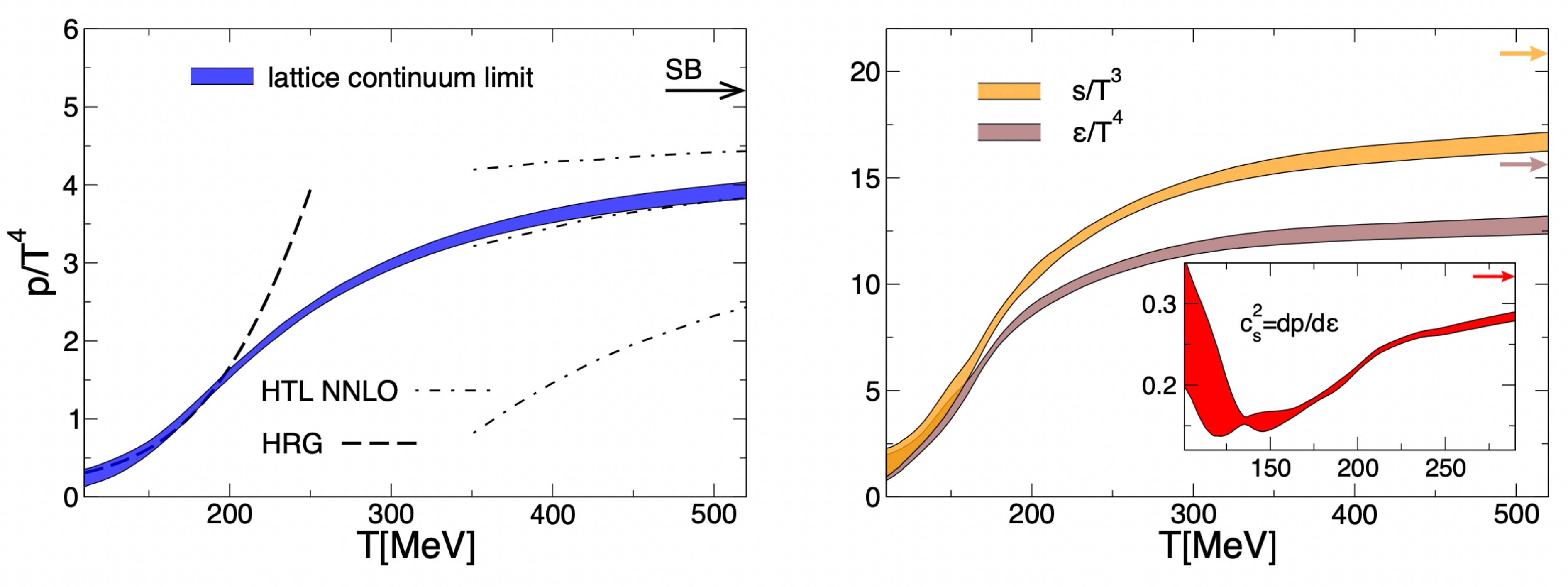}
\caption{Continuum extrapolated equation of state of (2+1)-flavor hot QCD matter at $\mu_B=0$. Left: Pressure $P(T)/T^4$; right panel: Energy density $\varepsilon(T)/T^4$ and entropy density $s(T)/T^3$. The insert on the right panel shows the speed of sound $c_s(T)^2$. The arrows indicate the Stefan-Boltzmann limit for a non-interacting gas of massless quanta. For an explanation of the theoretical curves labeled HRG and HTL NNLO, see text. [From \cite{Borsanyi:2013bia}]}
\label{fig:EOS_WB}
\end{figure}

The QCD equation of state at $\mu_B=0$ shows a rapid rise in the ratio of all thermodynamic quantities to the Stefan-Boltzmann limit for massless quanta over the temperature range 100 MeV $< T <$ 2,500 MeV, as shown in Fig.~\ref{fig:EOS_WB} \cite{Borsanyi:2010bp}. Perhaps the clearest manifestation of this behavior is in the effective number of degrees of freedom $d_{\rm eff}$ defined via the formula for the the entropy density $s(T)$ of a massless, noninteracting gas of particles:
\begin{equation}
s(T) = \frac{45}{2\pi^2}\, d_{\rm eff} T^3 .
\label{eq:deff}
\end{equation}
As seen in the right panel of Fig.~\ref{fig:EOS_WB}, the effective number of degrees of freedom in hot QCD matter rises steadily over this temperature range from $d_{\rm eff} \approx 8.5$ at $T=140$ MeV to $d_{\rm eff} \approx 38$ at $T=500$ MeV, which corresponds to 80\% of the Stefan-Boltzmann (SB) limit. The remaining deviation from the SB limit can be understood as the effect of perturbative color interactions in the quark-gluon plasma. 

Effective descriptions of the low-temperature and high-temperature domains are indicated in the left panel of Fig.~\ref{fig:EOS_WB}. At low temperatures, the hadron resonance gas (HRG) provides a rather good description, although in detail the equation of state indicates the existence of hadron resonances beyond those listed in the Particle Data Book \cite{Majumder:2010ik,Bazavov:2014xya}. The HRG gives a good description for most quantities up to $T \approx 150$ MeV. Some especially sensitive quantities may not only depend on the hadron mass spectrum, but also on the widths of resonances \cite{Andronic:2018qqt}. 

At high temperature ($T > 350$ MeV), hard-thermal loop resummed perturbation theory gives a good description of the equation of state and related observables. As indicated in the left panel of Fig.~\ref{fig:EOS_WB} the next-to-next-to leading order (NNLO) calculations in this scheme \cite{Andersen:2010wu,Andersen:2011sf} match the  lattice results for the pressure quite well, albeit with a large uncertainty deriving from the choice of the renormalization scale $\mu$. The three dash-dotted lines shown in the figure correspond to $\mu = \pi T$ (top), $2\pi T$ (center), $4\pi T$ (bottom); the central curve obviously matches best.

\begin{figure}[ht]
\centering
\includegraphics[width=0.95\linewidth]{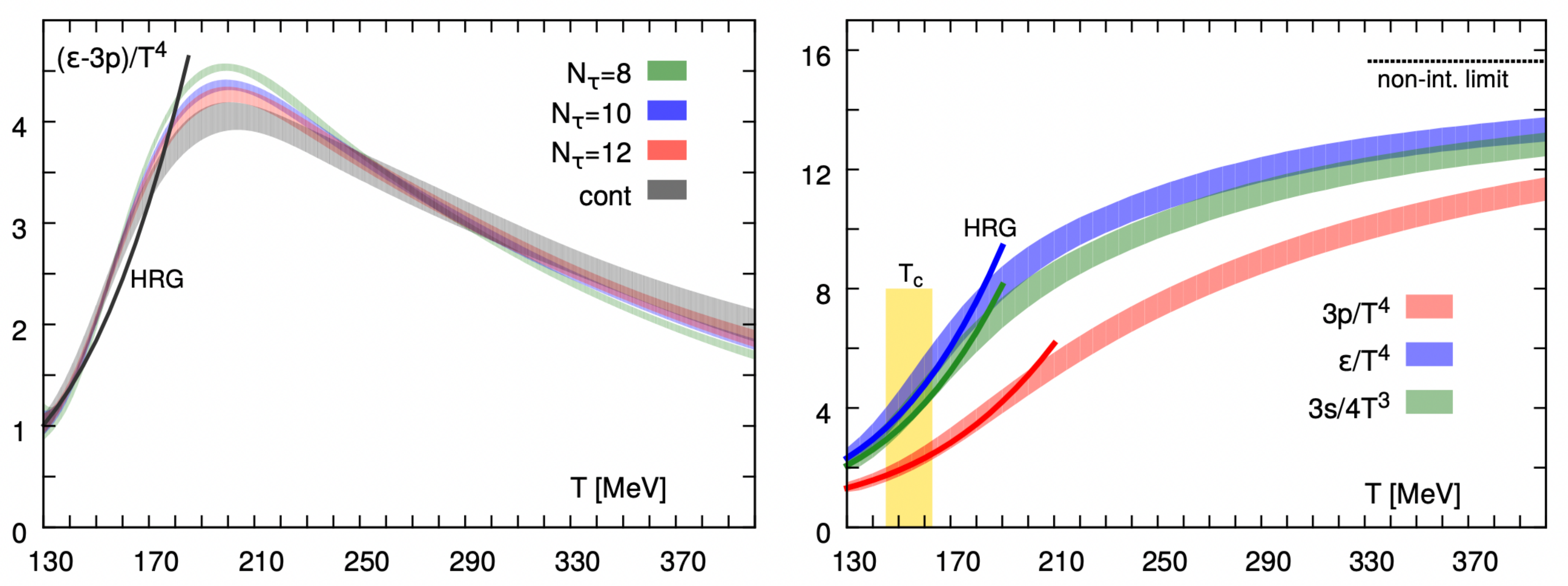}
\caption{Left panel: Continuum extrapolated scale anomaly $(\varepsilon-3P)/T^4$ for (2+1)-flavor hot QCD matter at $\mu_B=0$. Right panel: Energy density $\varepsilon(T)/T^4$, pressure $P(T)/T^4$, and entropy density $s(T)/T^3$. The results from the HotQCD Collaboration shown here agree with those of the Wuppertal-Budapest (WB) Collaboration shown in Fig.~\ref{fig:EOS_WB}. [From \cite{HotQCD:2014kol}]}
\label{fig:EOS_HotQCD}
\end{figure}

An especially interesting feature is the broad peak in the QCD scale anomaly, $I(t) = (\varepsilon-3P)/T^4$, shown in the left panel of Fig.~\ref{fig:EOS_HotQCD}, which measures the manifest violation of scale invariance in the QCD equation of state. The nonperturbative ``interaction measure'' $I(t)$ is distinct from the perturbative violation of scale invariance in the light quark and gauge sector of QCD that derives from the temperature dependence of the running coupling constant $\alpha_s(T)$. There have been multiple attempts to interpret this nonperturbative feature of hot QCD in terms of some quasi-particle picture \cite{Peshier:1999ww,Ivanov:2004gq}, but the strong coupling among the degrees of freedom has made it difficult to draw definite conclusions. The scale anomaly remains large throughout the entire temperature domain ($T\leq 400$ MeV) that is probed in relativistic heavy ion collisions. 

For $T<T_c$, the growth of $I(T)$ can be understood in terms of the hadron resonance gas model, which predicts that more massive states are being excited with increasing temperature. As the figure shows, the lattice result slightly exceeds the resonance gas model, which accounts for all experimentally known resonances, below 175 MeV. A possible explanation for this excess is the excitation of a large number of unknown resonances \cite{Majumder:2010ik}. Although the microscopic structure of these states is not known, many of them must contain internal gluon excitations, either as hybrid states (hadrons that contain both valence quarks and valence gluons) or as glueballs (hadrons composed solely of valence gluons). If this is so, the question what happens to the gluons when the quark-gluon plasma hadronizes, has a simple resolution: The thermally excited gluons become internal gluonic excitations of short-lived hadrons. As the hadron gas expands, these internal excitations will quickly decay into light hadrons, mostly pions, which absorb the additional entropy.

\subsection{Thermal Effective Field Theory}

Lattice gauge theory currently does not permit to perform reliable computations of real-time processes. For this reason, insight into transport processes in the hot QCD matter presently relies either on thermal perturbation theory or on QCD-inspired models of strongly coupled gauge theories with a holographic gravity dual. Holographic models will be discussed in the next subsection; here we focus on thermal perturbation theory. 

Thermal perturbation theory is usually formulated in terms of the hard thermal loop (HTL) effective theory \cite{Braaten:1989mz,Kapusta:2006pm,Lebellac:2000}.  The HTL formalism resums the leading thermal contributions to the gluon and quark self-energies into momentum-dependent effective masses. The effective theory relies on the separation of the scales $T \gg gT \gg g^2T$, which requires weak coupling (formally $g \ll 1$ corresponding $\alpha_s \ll 0.1$). QCD is not so weakly coupled at any physically relevant temperature, but it turns out that the power counting implied by the HTL scheme appears to work even when the scale ordering is not strictly realized, as it often is the case for effective field theories. 

Gauge invariance dictates that the thermal modifications of the propagators are balanced by vertex corrections that satisfy the Ward-Takahashi identities.  In the HTL effective theory the gauge field develops a collective longitudinal mode, the plasmon, with mass
\begin{equation}
m_{\rm pl}^2 = \frac{2N_c+N_f}{6} (gT)^2 + O(g^4T^2)
\end{equation}
for a plasmon at rest in the medium.  $m_{\rm pl}$ represents the additional mass scale that enters into the HTL effective Lagrangian and controls the thermal properties of the gauge theory. For example, the HTL formalism predicts that the static chromo-electric field around a heavy color charge in a QGP with $N_f$ light quark flavors is screened with a Debye mass $m_{\rm D} = \sqrt{3}m_{\rm pl} = gT\sqrt{1+N_f/6}$. Furthermore, all HTL gauge field modes are Landau damped in the space-like domain at the same scale.

An estimate of the domain of applicability of HTL perturbation theory can be obtained by calculating and comparing quantities that can be computed reliably on the lattice. An example is the QCD equation of state for $\mu_B=0$, which has been calculated up to three-loop (NNLO) order \cite{Andersen:2010wu,Andersen:2011sf} in the resummed HTL formalism and agrees well with the lattice results for $T > 350-400$ MeV (see Fig.~\ref{fig:EOS_WB}). It is also clear from these results that the HTL effective theory has very limited applicability in the temperature range $T < 2T_c$ that is most relevant to the heavy ion experiments.

Many properties of the quark-gluon plasma have been calculated in the HTL effective theory, mostly at leading order, but NLO results have increasingly become available. Examples include the shear viscosity $\eta$ \cite{Arnold:2000dr,Arnold:2003zc,Ghiglieri:2018dib}, the so-called jet quenching parameter $\hat{q}$ \cite{Caron-Huot:2008zna,Ghiglieri:2015zma,Ghiglieri:2015ala}, the rate of collisional energy loss of a fast parton $- dE/dx=\hat{e}$ \cite{Thoma:1991ea}, the heavy-quark diffusion constant \cite{Caron-Huot:2007rwy} and the rates of thermal photon \cite{Arnold:2001ms,Ghiglieri:2013gia} and lepton-pair \cite{Braaten:1990wp,Ghiglieri:2014kma,Ghiglieri:2015nba} production.

\subsection{Strongly Coupled Gauge Theories}

The discovery of the Anti-de Sitter/Conformal Field Theory (AdS/CFT) duality by Maldacena in 1997 fundamentally changed theorists' ability to perform exact calculations for strongly coupled gauge theories \cite{Maldacena:1997re,Aharony:1999ti,Casalderrey-Solana:2011dxg,Natsuume:2014sfa}. The AdS/CFT duality is based on the notion that conformal field theories in ($d+1$)-dimensional space-time can be mapped onto quantum gravity in ($d+2$)-dimensional Anti-de Sitter space (AdS$_{d+2}$), which has ($d+1$)-dimensional space-time as a boundary at infinity. The AdS space is known as the ``bulk''. Because the mapping is between theories defined on space-times with a different number of dimensions, one speaks of a {\it holographic} duality. 

The most ubiquitous case for relativistic heavy ion physics is the original duality between ${\cal N}=4$ super-Yang-Mills theory in ($3+1$) dimensions, which can be considered as a ``cousin'' of QCD, and superstring theory on the space AdS$_5\times{\rm S}_5$. In the limit of an infinite number of colors $N_c$, expressed as the limit of strong 't Hooft coupling ($\lambda = g^2 N_c \to \infty$), the holographic dual is reduced to (super-)gravity on AdS$_5$. While QCD is not conformally symmetric -- as we already mentioned, the violation of scale invariance is quite large in the temperature range of greatest interest to relativistic heavy ion physics -- there are a number of modifications of the original holographic model that incorporate scale symmetry breaking in a fundamental (see \cite{Gursoy:2007cb} for an overview) or phenomenological way (see \cite{Karch:2006pv,deTeramond:2008ht}). 

\begin{figure}[ht]
\centering
\includegraphics[width=0.50\linewidth]{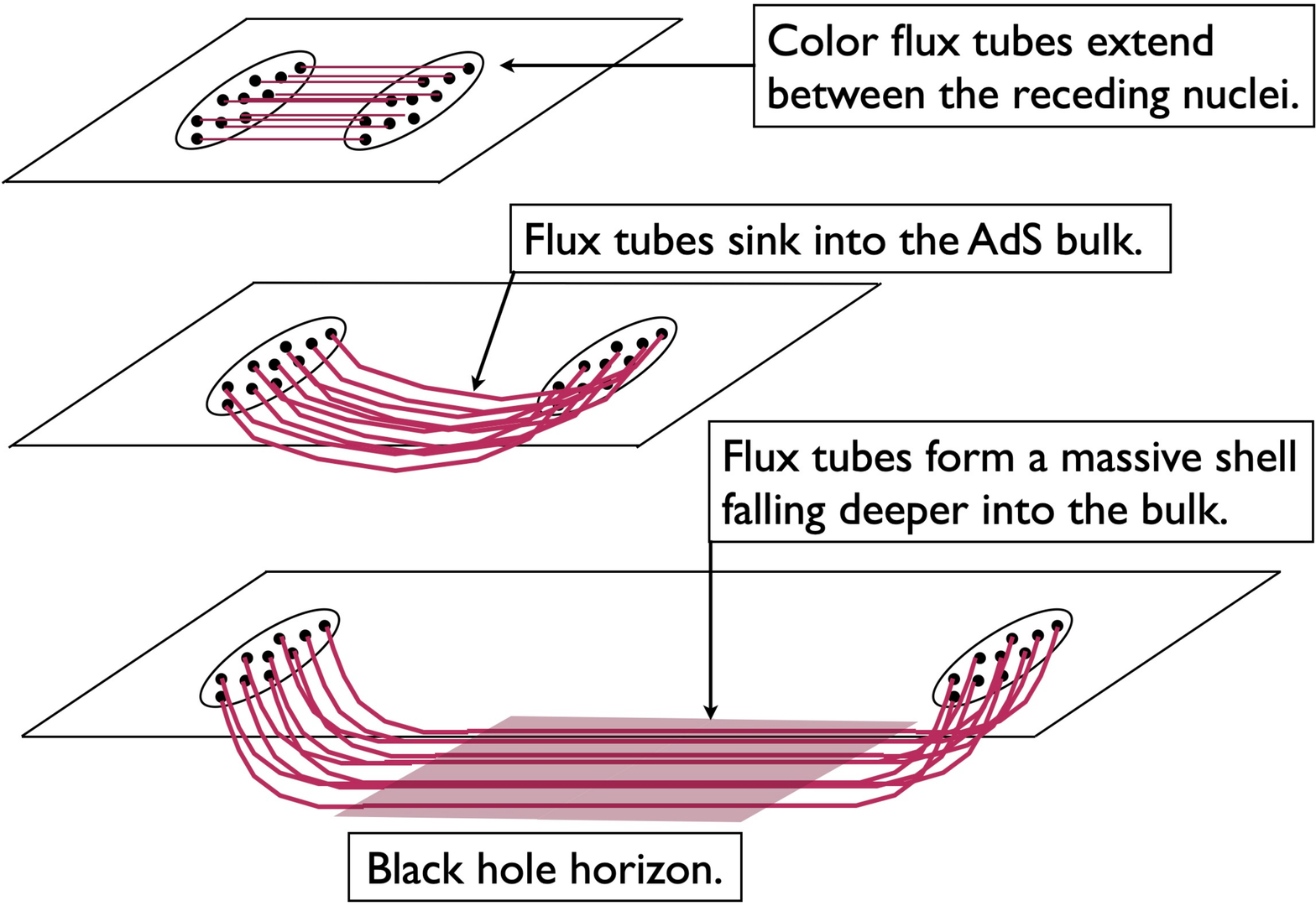}
\hspace{0.05\linewidth}
\includegraphics[width=0.40\linewidth]{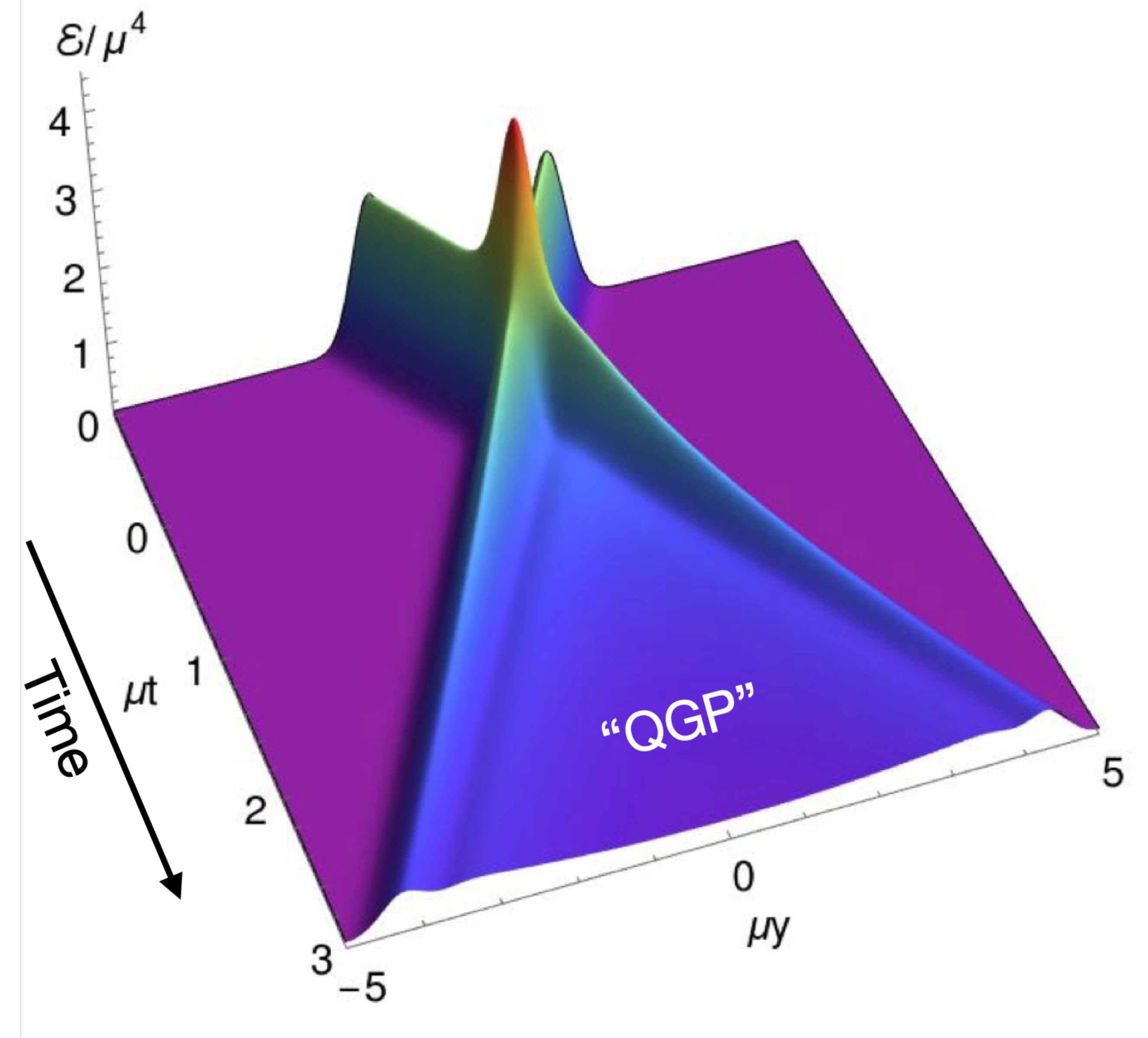}
\caption{Left panel: Schematic view of thermalization in a strong coupled gauge theory as formation of a black hole in the AdS bulk. Color flux tubes extending between the receding nuclei are idealized as a massive shell that falls into the depth of AdS space and eventually forms a black hole. Right panel: Collision of shock waves in the dual AdS gravity theory. The energy density extending between the receding shocks is the holographic dual of the hot gauge field plasma. [From \cite{Ecker:2016thn}]}
\label{fig:AdS/CFT}
\end{figure}

An especially attractive feature of AdS/CFT duality is that it enables rigorous calculations of dynamical processes at strong coupling, which is not possible in the framework of Euclidean lattice gauge theory \cite{Son:2002sd}. In particular, it is possible to study thermalization at strong coupling by mapping thermalization onto the process of formation of a black hole in the bulk \cite{Balasubramanian:2010ce,Balasubramanian:2011ur}, as illustrated in the left panel of Fig.~\ref{fig:AdS/CFT}. 

The dynamics of space-time near the event horizon of black holes has long been known to be governed by viscous relativistic hydrodynamics, a phenomenon known as the membrane paradigm \cite{Price:1986yy}. In the context of AdS space with a black hole, this translates into properties of viscous hydrodynamics in the gauge theory on the AdS boundary, some of which are deemed universal for any strongly coupled nonabelian gauge theory \cite{Policastro:2001yc,Kovtun:2004de}. The best known of these properties is the ratio of the shear viscosity to the entropy density, $\eta/s = 1/(4\pi)$ characteristic of a ``perfect'' fluid. 

Other experimentally relevant properties of strongly coupled gauge theories have also been calculated using holographic techniques, e.~g., the energy loss of heavy quarks \cite{Herzog:2006gh} and light quarks \cite{Liu:2006ug,Chesler:2008uy}. Quite generally, holographic approaches lend themselves most easily to the study of energy and momentum transport as these are directly mapped onto the dynamics of the gravitational field in the bulk. Modeling of other processes associated with quark flavor or spin is possible, but requires adding additional fields or geometric objects, such as D-branes, to the model thereby rendering the conclusions less universally valid.

The holographic approach allows to model the approach to hydrodynamics from arbitrary initial conditions without artificial approximations. Under a wide range of initial conditions, boost-invariant (Bjorken) hydrodynamics with $\eta/s = 1/(4\pi)$ emerges as asymptotic description of the dynamical evolution \cite{Janik:2005zt,Heller:2011ju,Liu:2006ug,Shuryak:2011aa}. The complete process of energy deposition, thermalization, and hydrodynamic expansion can be modeled in terms of collisions of two shock waves and can be rigorously solved using numerical techniques \cite{Chesler:2009cy,Chesler:2010bi,Chesler:2013lia}. By varying the height and width of the shock waves, the whole range from transparency to full stopping can be explored \cite{Casalderrey-Solana:2013aba}. Asymmetric collisions can also be modeled \cite{Muller:2020ziz}. By combining holographic modeling with other descriptions that are more appropriate for the late stages of a nuclear collision, a seamless end-to-end description of heavy ion collisions can be implemented \cite{vanderSchee:2013pia}.

\section{Phenomenological Approaches}
\label{sec:pheno}

{\em This Section describes the basics of effective approaches that are currently used to connect theoretical input from fundamental theories to observables.}

\subsection{Bulk Evolution}
\label{sec:bulk}

\begin{figure}[ht]
\centering
\includegraphics[width=0.80\linewidth]{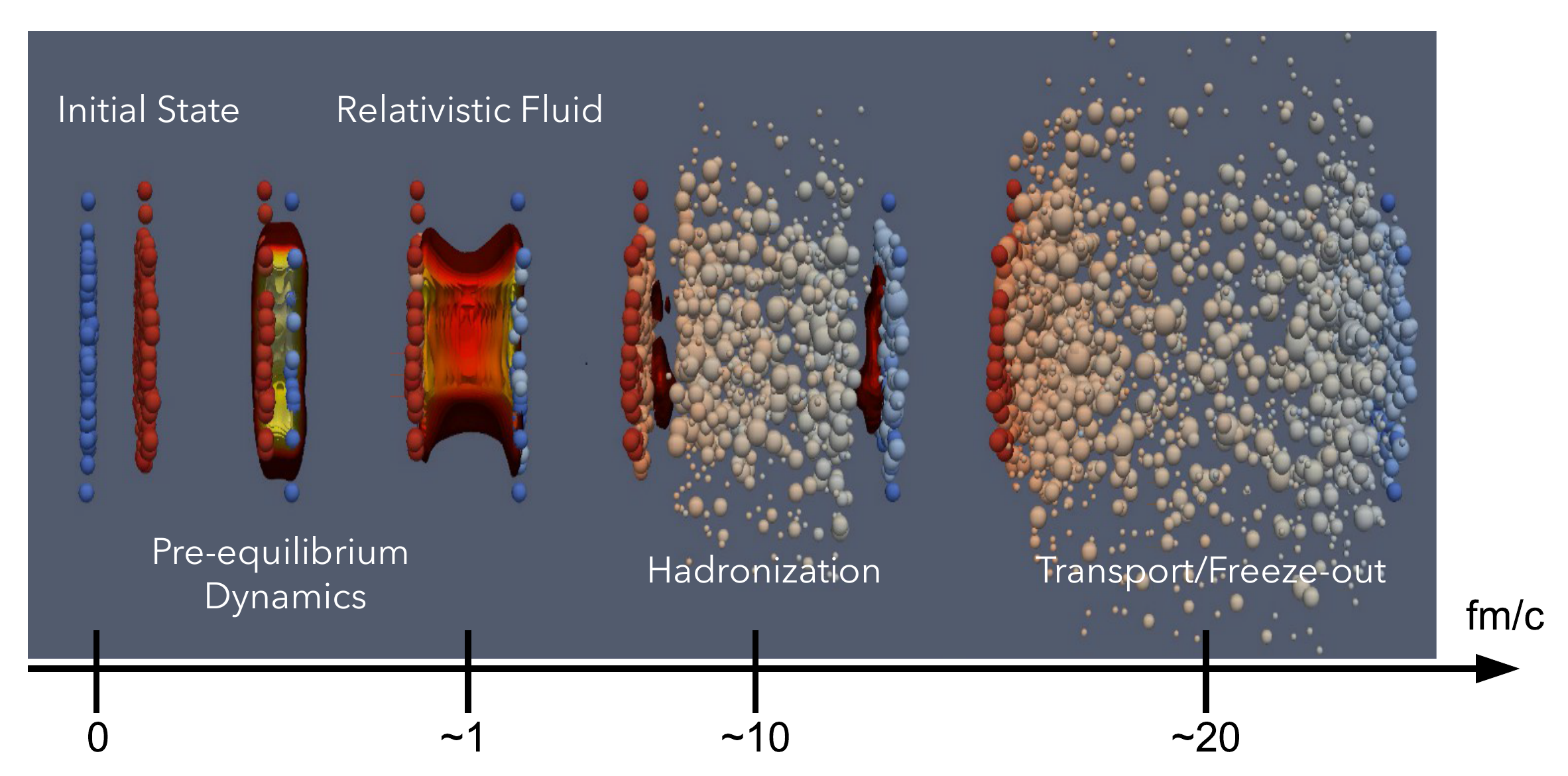}
\caption{Visualization of the different stages of heavy-ion collisions in a hybrid approach based on hydrodynamics and hadronic transport for the initial and final stages. [Adapted by G. Denicol from \url{madai.us}]}
\label{fig:evolution}
\end{figure}

The dynamical evolution of heavy-ion collisions can be divided into several stages (see Fig. \ref{fig:evolution}) that are described in detail in the following subsections. Initially, there are two Lorentz contracted nuclei approaching each other. Right after the impact there is a period characterized by non-equilibrium evolution. Eventually relativistic viscous fluid dynamics becomes applicable for the hot and dense stage. When the system dilutes enough, the fluid is converted to hadrons, which decay and interact with each other until the freeze-out. 

\subsubsection{Initial Conditions.}
\label{sec:ini_cond}

The initial conditions are an indispensable component of any dynamical description of heavy ion reactions. We have rather extensive knowledge of the structure of the colliding nuclei, but we lack a first-principle treatment on the basis of QCD of the evolution from the colliding nuclear ground states to the initial conditions that are used as input for hydrodynamics. 

The simplest picture for the initial geometry of the overlap region in the transverse plane is based on the Glauber model. While the original Glauber model provides a general formalism for multiple scattering, its application to relativistic heavy ion physics commonly makes use of the eikonal limit, where all particles are assumed to travel along straight line trajectories. (For a simplified introduction to the formalism, see \cite{Shukla:2001mb}; for an overview of its various uses in relativistic heavy ion physics, see \cite{Stock:2020blh}.) In this eikonalized Glauber model, the two nuclei are described by thickness functions $T_A(\vec{s})$, which represent Woods-Saxon density profiles denoted by $\rho_A$ that are integrated over the longitudinal direction: $\hat{T}_A(\vec{s})=\int \rho_A(\vec{s},z) dz$.  $\vec{s}$ is the transverse position with respect to the center-of mass of the nucleus $A$. The joint probability for the overlap of nucleus $A$ with nucleus $B$ colliding at an impact parameter $\vec{b}$ can be expressed as
\begin{equation}
\hat{T}_{AB} (\vec{b})=\int \hat{T}_A (\vec{s}) \hat{T}_B 
(\vec{s}-\vec{b}) d^2s .
\end{equation}

\begin{figure}[ht]
\centering
\includegraphics[width=0.60\linewidth]{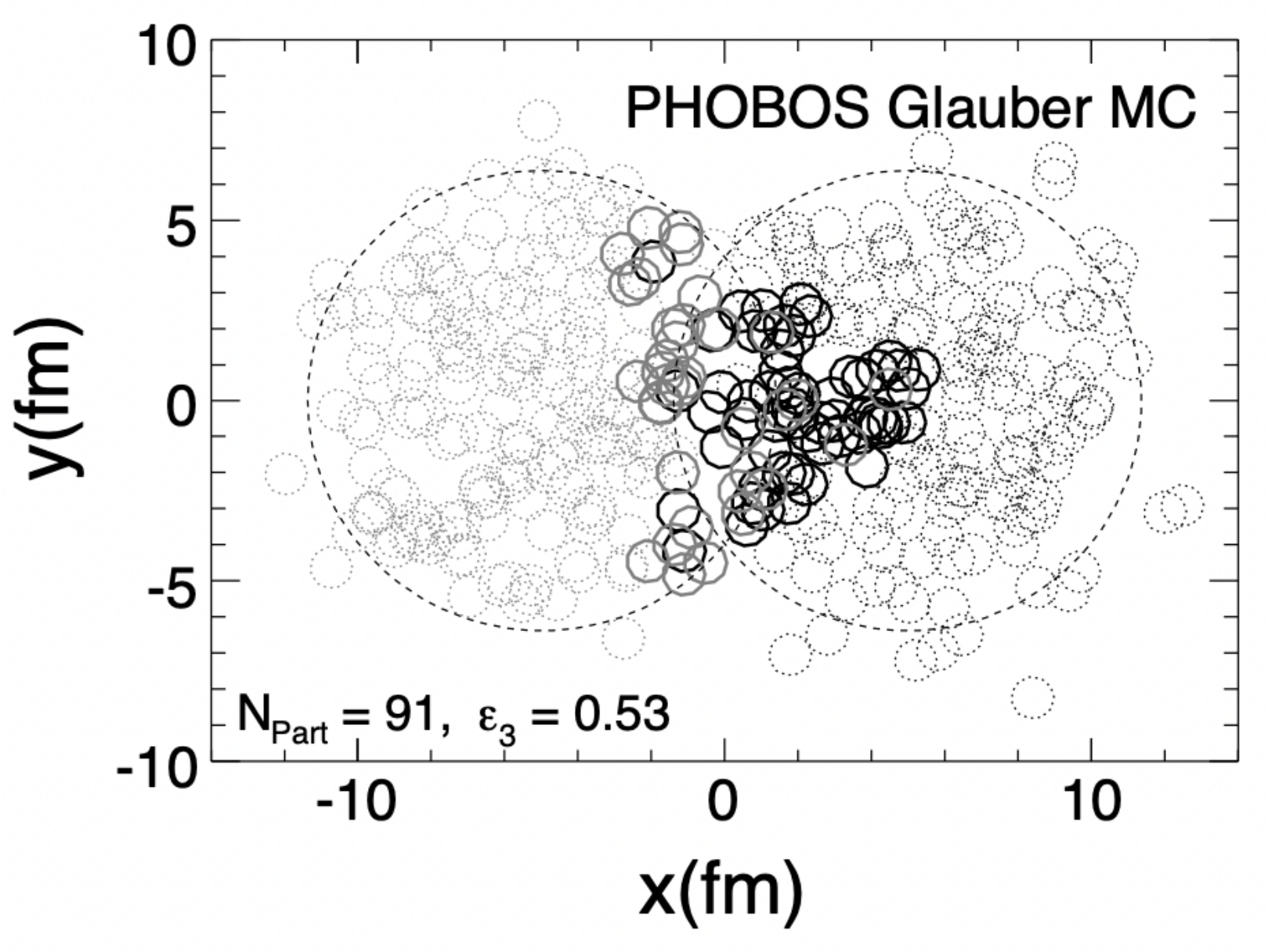}
\caption{Transverse distribution of participant nucleons (solid circles) and spectator nucleons (dotted circles) in a random Au+Au collision event at $\sqrt{s_{\rm NN}} = 200$ GeV. The large dashed circles outline the transverse positions of the two nuclei. The event was selected for its nearly triangular distribution of the participant nucleons. [From \cite{Alver:2010gr}]}
\label{fig:Glauber}
\end{figure}

Participant or ``wounded'' nucleons are defined as the nucleons that interact with at least one nucleon of the opposite nucleus, where the interaction is defined by the inelastic proton-proton cross section ($\sigma_{\rm NN}$) at a given energy. 
\begin{eqnarray}
N_{\rm part}&=&A \int \hat{T}_A (\vec{s}) \left\{ 1-\left[ 1 -\hat{T}_B(\vec{s}-\vec{b}) \sigma_{\rm NN} \right ]^B \right\} d^2s\\
&+& B \int   \hat{T}_B(\vec{s}-\vec{b})\left\{1-\left[ 1-\hat{T}_A (\vec{s}) \sigma_{\rm NN} \right ]^A \right\} d^2s
\end{eqnarray}
The nucleons outside the interaction region are called spectators and are not considered further. The number of binary nucleon-nucleon collisions ($N_{\rm coll}$) can be calculated as well. 
\begin{equation}
N_{\rm coll}=A B \hat{T}_{AB} (\vec{b}) \sigma_{\rm NN}
\end{equation}
In some heuristic models \cite{Kharzeev:2000ph} of the centrality dependence of the charged particle multiplicity binary collisions are assumed to contribute a fraction $\alpha < 1$ and while a fraction $(1-\alpha)$ is attributed  the participants:
\begin{equation}
N_{AA}(\vec{b})=\left[ \alpha N_{\rm coll} (\vec{b}) + 
\frac{1-\alpha}{2} N_{\rm part} (\vec{b}) \right] N_{pp}
\end{equation}
where $N_{pp}$ is the number of particles produced in a proton-proton collision at the same energy. The Glauber model is usually implemented as a Monte-Carlo process and therefore provides different initial condition profiles for individual events. Figure \ref{fig:Glauber} shows the distribution of participant nucleons (solid circles) and spectator nucleons (dotted circles) in a randomly chosen Au+Au collision event at top RHIC energy. Many improvements of this simplistic scheme have been studied over the years. These include treatments of nucleon-nucleon correlations in the initial state \cite{Blaizot:2014wba}, the finite extent and internal structure of nucleons, fluctuations in the energy deposition by individual nucleon-nucleon collision, and much more  (see \cite{dEnterria:2020dwq} for a recent review).

The {\it Color Glass Condensate} (CGC) model \cite{Gelis:2010nm} aims to describe the colliding nuclei directly in terms of their quark-gluon content and is thus limited to collisions at sufficiently high-energy. The CGC model assumes that energy deposition at midrapidity is dominated by the liberation of gluons from the colliding nuclei, and that the relevant gluon content of the nuclei can be described by semiclassical gauge fields. These fields are generated by sources (primarily valence quarks) that move at near-projectile rapidity and whose dynamics during the collision can be ignored because of time dilatation. In the simplest version of the CGC, the McLerran-Venugopalan model \cite{McLerran:1993ni,McLerran:1993ka}, the colliding nuclei are modeled as Gaussian distribution of color currents $J^{a\pm}(x) = \rho^a(\vec{x}_\perp)\delta(x^{\mp})n^{\pm}$ along the light-cone. The distribution of these color sources is given by
\begin{equation}
    W[\rho^a] = N \exp\left( - \frac{1}{2g^2 \mu^2}\int d^2x_\perp\, (\rho^a(\vec{x}_\perp))^2 \right) ,
\end{equation}
where $\mu$ is a scale setting parameter that is related to the saturation scale $Q_s \approx 0.6g^2\mu$ \cite{Lappi:2007ku}. The semiclassical color fields are obtained as solutions of the classical Yang-Mills equation $D^{ab}_\mu F^{a\mu\nu} = g J^{a\mu}$. 

\begin{figure}[ht]
\centering
\includegraphics[width=0.60\linewidth]{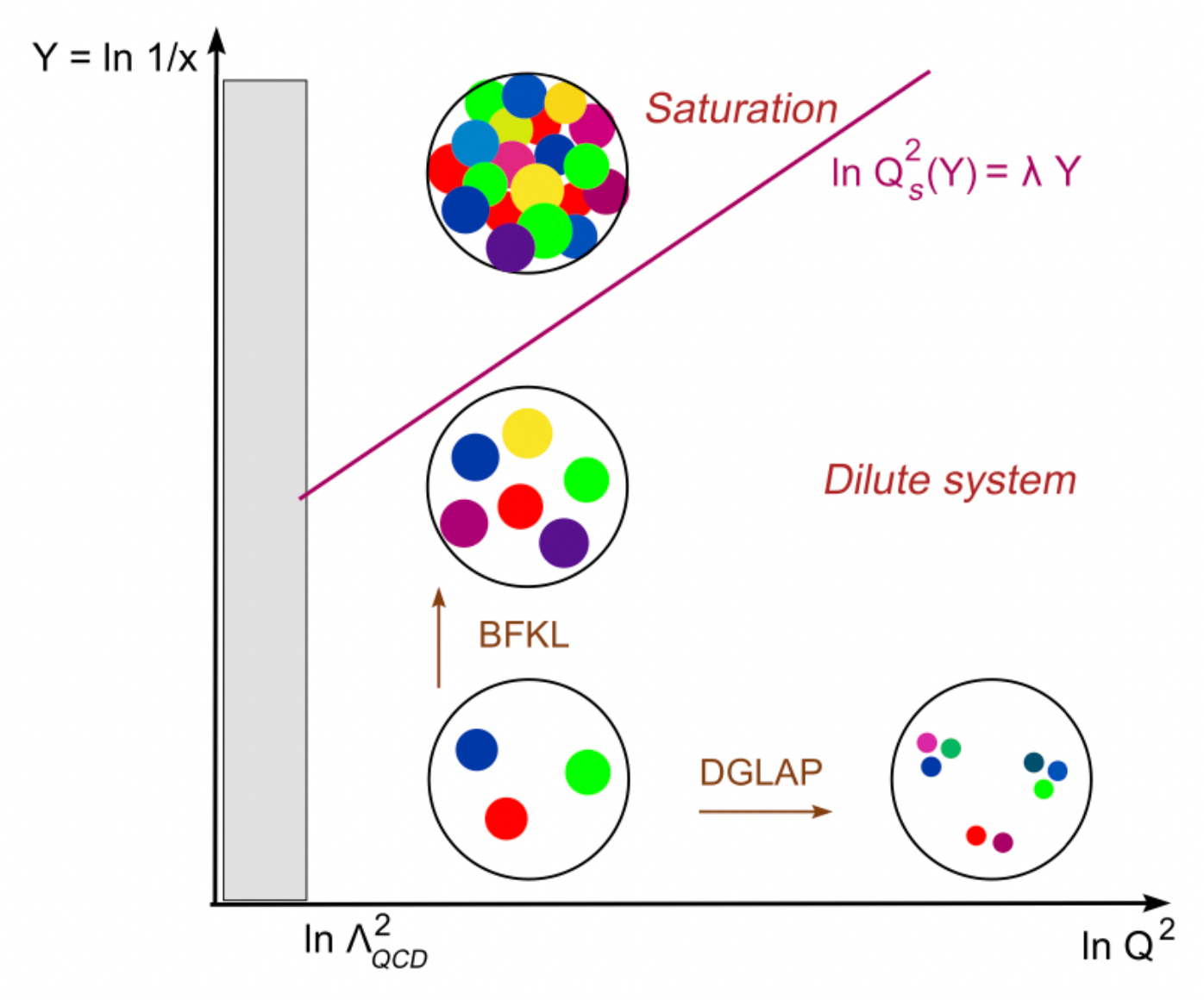}
\caption{Schematic view of different domains of the gluon distribution in nuclei. The CGC model applies above the $x$-dependent saturation scale $Q_s(x)$ shown by the straight line. [From \cite{Gelis:2010nm}]}
\label{fig:CGC}
\end{figure}

The classical approximation is only warranted when the occupation number of gauge field modes is of the order of $\alpha_s(Q_s)^{-1} \gg 1$. This condition requires that $Q_s \gg \Lambda_{\rm QCD}$. Because the saturation scale $Q_s$ scales as $Q_s^2 \sim A^{1/3} x^{-\lambda}$ with $\lambda \approx 0.3$, the condition can be fulfilled either for a large nuclear mass number $A \gg 1$ or for very small momentum fraction $x \ll 1$. Figure~\ref{fig:CGC} gives a graphical representation of this scaling. Small values of $x$ are only accessible at high collision energies, because the transverse momentum $p_T$ of a produced hard probe is given by the relation $p_T^2 = x_1 x_2 s_{rm NN}$, where $x_1$ and $x_2$ are the momentum fractions of the participating partons in the two colliding nuclei and $\sqrt{s_{\rm NN}}$ is the center-of-mass collision energy per nucleon pair.

The dependence of the color sources $\rho^a(x_\perp)$ on the momentum fraction $x$ is governed by renormalization group (RG) evolution. At a given value of $x_0$ all sources of color fields residing at $x>x_0$ contribute (valence quarks for all $x_0$ and gluons and sea quarks for small values of $x_0$). The RG evolution can either be expressed as a differential equation for the functional $W[\rho^a]$, the so-called JIMWLK equation \cite{Mueller:2001uk}, or as a nonlinear differential equation for the so-called dipole density $N_{\bf xy}$, the Balitsky-Kovchegov (BK) equation \cite{Rummukainen:2003ns}. Whereas the scaling with $A$ and $x$ can be calculated from geometry and QCD renormalization group arguments, the overall scale is nonperturbative. This means that the range of applicability of the CGC model is still under debate and awaits precise data from the future electron-ion collider (EIC). 

\begin{figure}[ht]
\centering
\includegraphics[width=0.80\linewidth]{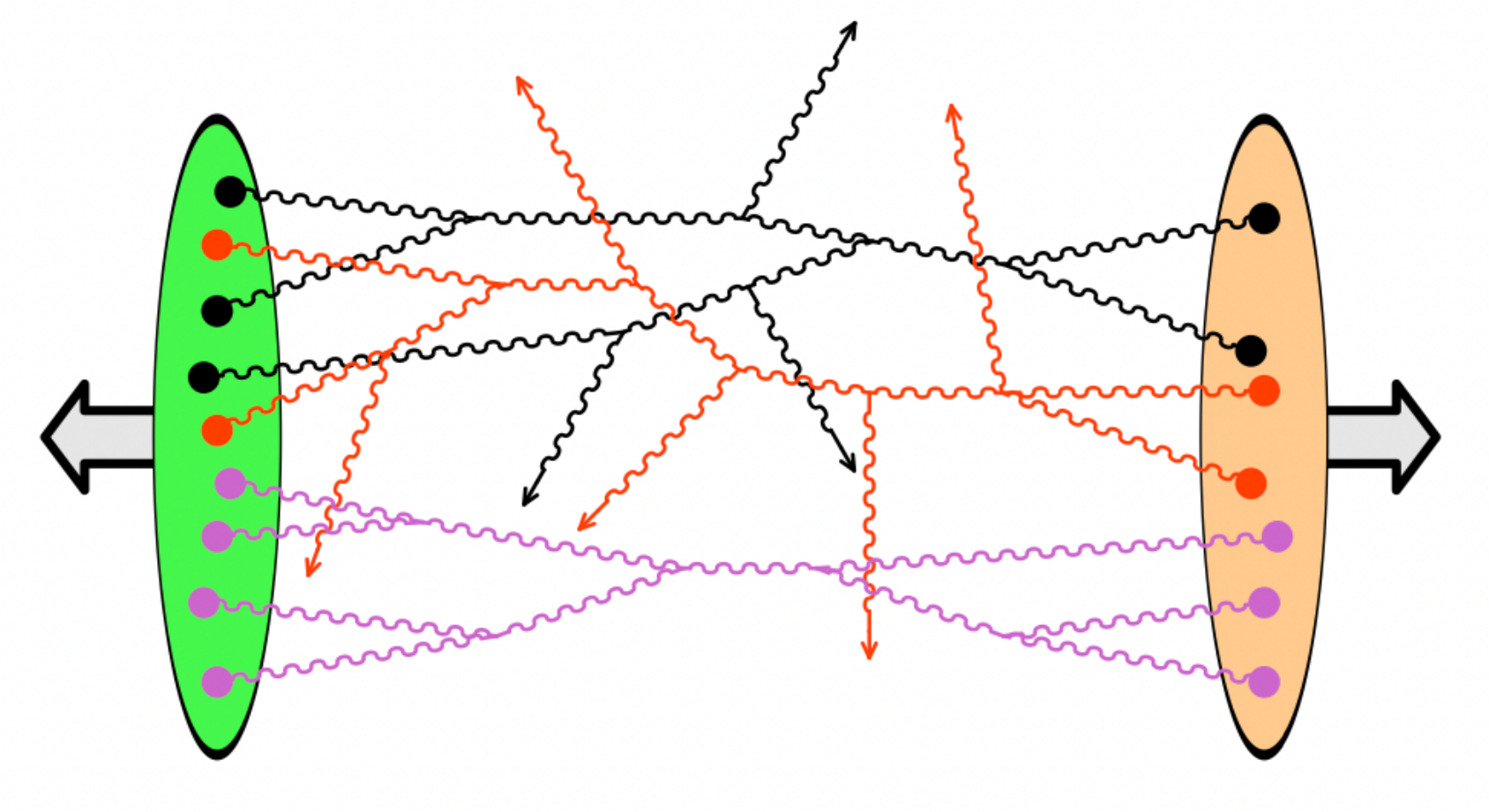}
\caption{The not yet thermalized distribution of gluons liberated by scattering from the CGC fields of colliding nuclei is called glasma.  [From \cite{Gelis:2010nm}]}
\label{fig:Glasma}
\end{figure}

The color fields generated in this way are virtual and remain an integral part of the nuclei until an interaction occurs.  During the collision of two heavy ion the CGC fields of the nuclei interact, and a large fraction of the virtual gluons are scattered on-shell, as sketched in Fig.~\ref{fig:Glasma}. The resulting nonequilibrium state is known as glasma. Because the CGC color fields are transversely polarized, the resulting energy-momentum tensor is far off equilibrium and cannot be directly used as an initial condition for the hydrodynamic evolution.

In order to describe individual heavy ion collision events, it is necessary to not only model the average energy deposition, but also the event-by-event fluctuations. There are two main sources of such initial-state fluctuations. One are the quantum fluctuations in the distribution of nucleons in the colliding nuclei or, if one instead looks at the collision at the parton level, in the distribution of valence quarks in the nuclei. The other source of fluctuations are the quantum fluctuations of particle production that are encoded in the S-matrix of a nucleon-nucleon or quark-quark collision. In the standard Glauber model, whether it is implemented at the nucleon or the valence quark lavel, the second type of fluctuations is commonly parametrized by a negative binomial distribution \cite{Praszalowicz:2011zza}
\begin{equation}
    F(t;k) = \frac{k^k}{\Gamma(k)} t^{k-1} e^{-kt} ,
\end{equation}
where the parameter $k$ (with $1 < k < 4$) depends on the energy. The probability distribution of the number of produced particles is given by
\begin{equation}
    P(n) = \int_0^\infty dt\, F(t;k) e^{-\bar{n}t} \frac{(\bar{n}t)^n}{n!} ,
\end{equation}
where $\bar{n}$ is the average number of produced particles.

For the CGC model, one can obtain fluctuating initial conditions using a formalism similar to the Glauber model, where the positions of nucleons are chosen randomly from the nuclear density distribution and the color sources are distributed within the individual nucleons. This implementation of the CGC model is known as impact parameter saturation (IP-sat) model \cite{Kowalski:2003hm}. When applied to nuclear collisions, the IP-sat model generates fluctuating initial conditions for the glasma (IP-glasma), which naturally result in a negative binomial distribution for the initial energy density \cite{Gelis:2009wh,Schenke:2012wb}.
\begin{figure}[ht]
\centering
\includegraphics[width=0.450\linewidth]{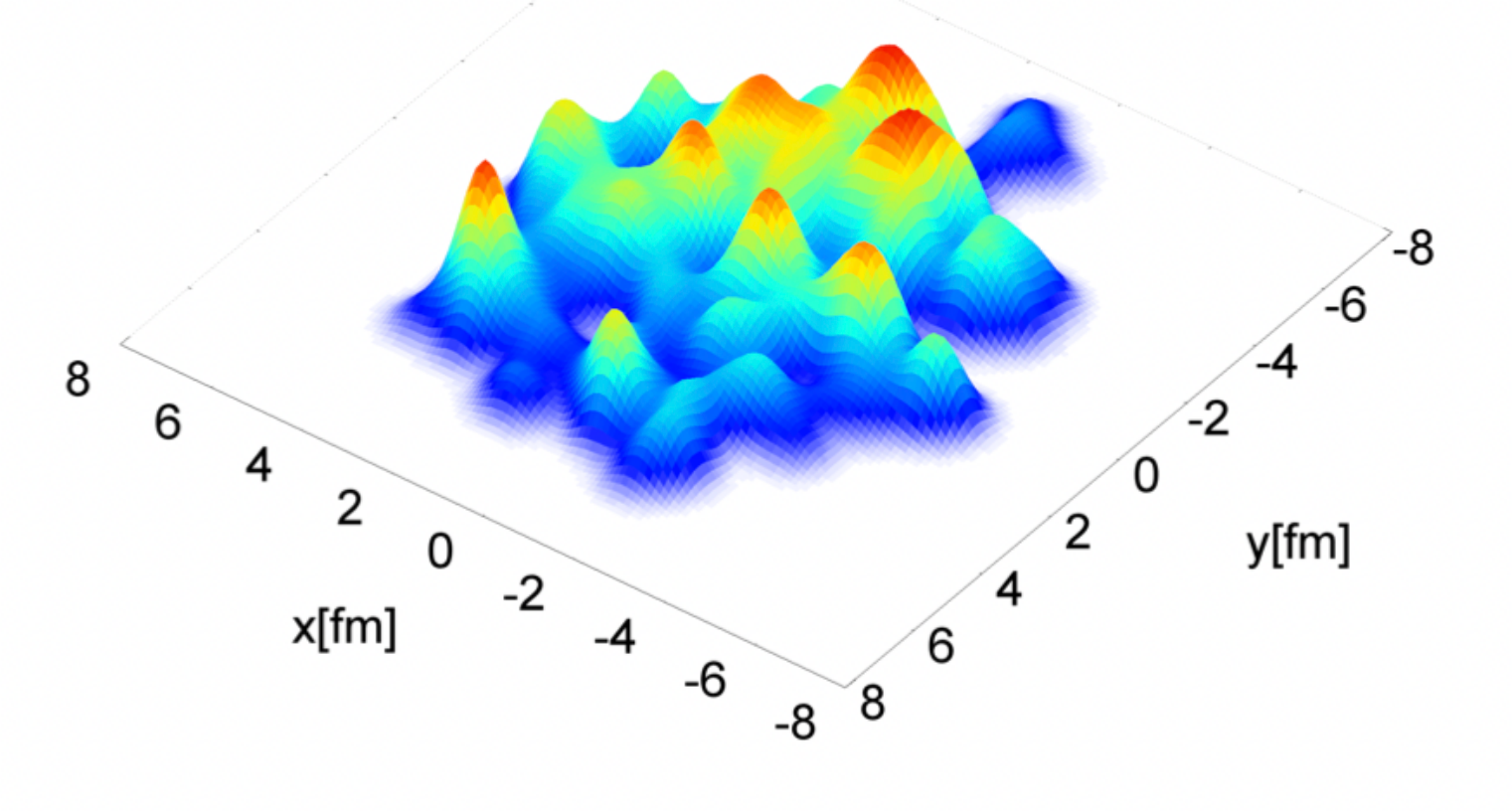}
\hspace{0.05\linewidth}
\includegraphics[width=0.450\linewidth]{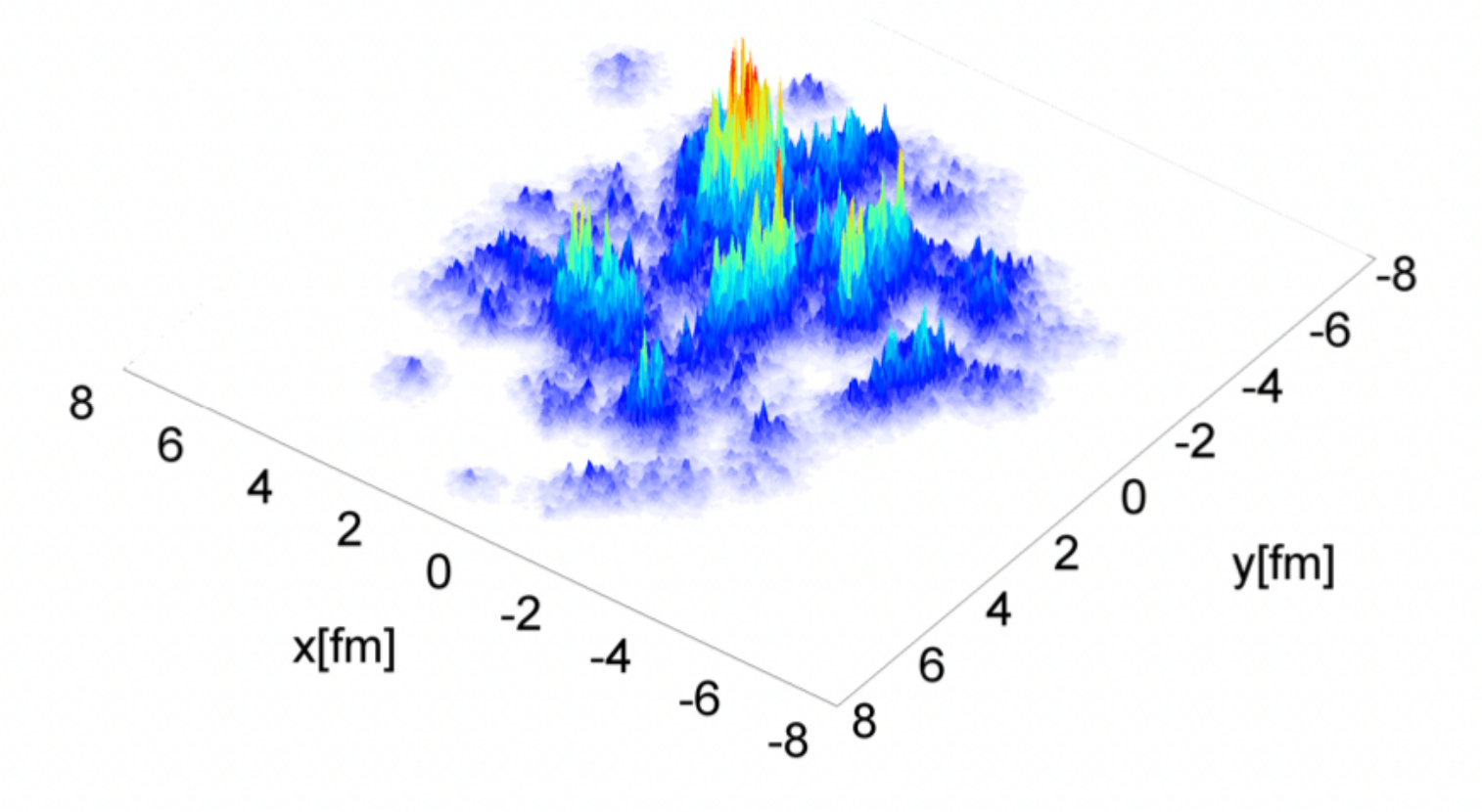}
\caption{Fluctuating initial energy density in the transverse $x-y$ plane for a single event of a Au+Au collision at top RHIC energy. Left panel: Nucleon-based Glauber model; right panel: IP-glasma model. In the nucleon-based Glauber model the fluctuations occur on the length scale of the nucleon radius ($\sim 0.8$ fm). In the IP-glasma model the fluctuations occur on the length scale of the inverse saturation scale $Q_s^{-1} \sim 0.1$ fm, and result in a much more spiky distribution. [From \cite{Schenke:2012wb}]}
\label{fig:fluct}
\end{figure}

The initial pattern of energy or entropy deposition in these models does not lend itself directly to be used as initial conditions for viscous hydrodynamics, because they do not correspond to an energy-momentum tensor near thermal equilibrium and the spatial gradients are too large in the case of event-by-event fluctuations. In the case of the standard Glauber model, this problem is commonly resolved by a short period of free-streaming evolution, which smoothes out the spatial gradients. Alternatively, one widely used version of a general Glauber-type initial condition generator, the TRENTO model \cite{Moreland:2014oya}, uses the entropy density to map the initial conditions generated by the model directly onto the initial conditions for viscous hydrodynamics.

For CGC initial conditions, the initial time evolution of the glasma is described by the nonlinear classical Yang-Mills equation; the resulting energy-momentum tensor is then mapped onto initial conditions for hydrodynamics \cite{Schenke:2012wb}. The microscopic thermalization process is thought to follow a scenario known as ``bottom-up'' thermalization \cite{Baier:2000sb}. In this scenario semi-hard gluons contained in the colliding nuclei radiate soft gluons which form a thermal bath that drains energy from the semi-hard gluons via elastic scattering and thereby thermalizes them. An effective kinetic description of this thermalization process has recently been developed and applied to the initial condition of hydrodynamics in relativistic heavy ion collisions \cite{Kurkela:2018vqr}. There are many additional details and possible refinements of this process, which are under investigation, but no generally accepted approach has emerged yet.

\subsubsection{The QGP Phase.}
\label{sec:qgp_phase}

Due to asymptotic freedom, the quark-gluon plasma was originally expected to behave as a weakly interacting gas with almost massless degrees of freedom, the quarks and gluons. When the first measurements of anisotropic flow in AuAu collisions at $\sqrt{s_{\rm NN}}=200$ GeV at RHIC were published, a surprisingly good agreement with predictions from ideal fluid dynamic calculations was observed \cite{Huovinen:2001cy, Kolb:2003dz} when the initial conditions were matched to fit the overall particle production yields and transverse momentum spectra. The elliptic flow is then a genuine prediction. 

Fluid dynamics is based on the equations for the conservation laws 
\begin{equation}
    \partial_\mu T^{\mu\nu}= 0, \quad \mbox{and} \quad \partial_\mu N_i^\mu = 0  
\end{equation}
of the energy-momentum tensor $T^{\mu\nu}$ and the current density $N_i^\mu$ of any conserved quantum number (e.g. $B,S,Q$). If the fluid is in local equilibrium, the energy-momentum tensor of an ideal fluid takes the form
\begin{equation}
T^{\mu\nu}= (\epsilon +P) u^\mu u^\nu +P g^{\mu\nu}    
\end{equation}
with $\epsilon$ being the energy density, $P$ the pressure and $u^\mu$ denoting the flow 4-velocity. $g^{\mu\nu}$ is the metric tensor and typically chosen as $(+,-,-,-)$ in special relativistic applications. The quantum number current density has the simple form $N^\mu = n u^\mu$, where $n$ is the density measured in the local rest frame.

In viscous hydrodynamics, the energy-momentum tensor is decomposed as 
\begin{equation}
    T^{\mu\nu}=\varepsilon u^\mu u^\nu - \Delta^{\mu\nu}(p+\Pi) + \pi^{\mu\nu} ,
\end{equation}
with $\Pi$ denoting the bulk viscous contribution and $\pi^{\mu\nu}$ the shear viscous tensor. First order viscous hydrodynamics is not causal, since it permits signal propagation that is faster than the speed of light. Therefore, second order viscous fluid dynamics must be applied with the following relaxation equations for shear and bulk viscosity including the coupling between the two
\begin{eqnarray}
    D\Pi&=&\frac{-\zeta\theta-\Pi}{\tau_\Pi}-\frac{\delta_{\Pi\Pi}}{\tau_\Pi}\Pi\theta+\frac{\lambda_{\Pi\pi}}{\tau_\Pi}\pi^{\mu\nu}\sigma_{\mu\nu}
    \\
    D\pi^{\langle\mu\nu\rangle} &=& \frac{2\eta\sigma^{\mu\nu}-\pi^{\mu\nu}}{\tau_\pi}-
   \frac{\delta_{\pi\pi}}{\tau_\pi}\pi^{\mu\nu}\theta + \frac{\phi_7}{\tau_\pi}\pi_\alpha^{\langle\mu}\pi^{\nu\rangle\alpha} \nonumber
   \\
   &-&\frac{\tau_{\pi\pi}}{\tau_\pi}\pi_\alpha^{\langle\mu}\sigma^{\nu\rangle\alpha} + \frac{\lambda_{\pi\Pi}}{\tau_\pi}\Pi\sigma^{\mu\nu}.
\end{eqnarray}
$\theta=\partial_\mu u^\mu$ is the expansion scalar, $\sigma^{\mu\nu}$ is the strain tensor, $\tau_\pi$ and $\tau_\Pi$ are the relaxation times for the shear and bulk viscous corrections and $\lambda_{\Pi\pi}$, $\lambda_{\pi\Pi}$, $\delta_{\pi\pi}$, $\tau_{\pi\pi}$ and $\phi_7$ are higher-order couplings, whose forms are taken from \cite{Denicol:2014vaa}.

There are two typical choices for the local rest frame, the Landau frame which moves with the energy density and the Eckart frame which moves with the conserved density, if one exists. More details on viscous relativistic hydrodynamics and its application to heavy ion physics can be found in  \cite{Ollitrault:2007du,Romatschke:2009im,Gale:2013da}.

Since the set of hydrodynamics equations listed above is insufficient to determine all independent components of the quantities of interest, one needs an additional equation as an input. This is the equation of state, which encodes the properties of the fluid under consideration. In the context of heavy ion physics this has the major benefit that hydrodynamics provides a controlled handle on the phase transition between a hadron gas and the quark-gluon plasma phase. One can study different equations of state to explore its influence on the dynamics of the system and various experimentally measurable observables. 

The simplest equation of state is that of a relativistic ideal gas composed of massless particles:
\begin{equation}
    P= \frac{\varepsilon}{3} . 
\end{equation}
At vanishing baryon chemical potential the equation of state is known from lattice QCD calculations and can be parametrized as originally done in \cite{Huovinen:2009yb}. Currently, there are many other possibilities including the extension to multiple conserved currents (see e.g. equations of state by the BEST collaboration \cite{Noronha-Hostler:2019ayj}, the BNL group \cite{Monnai:2019hkn} or from holography \cite{Grefa:2021qvt}). At finite density and vanishing temperatures there are constraints from the mass-radius relation of neutron stars. Also any equation of state should take the constraints based on stable nuclear matter into account. 

\begin{figure}[htb]
\flushright
\includegraphics[width=0.85\linewidth]{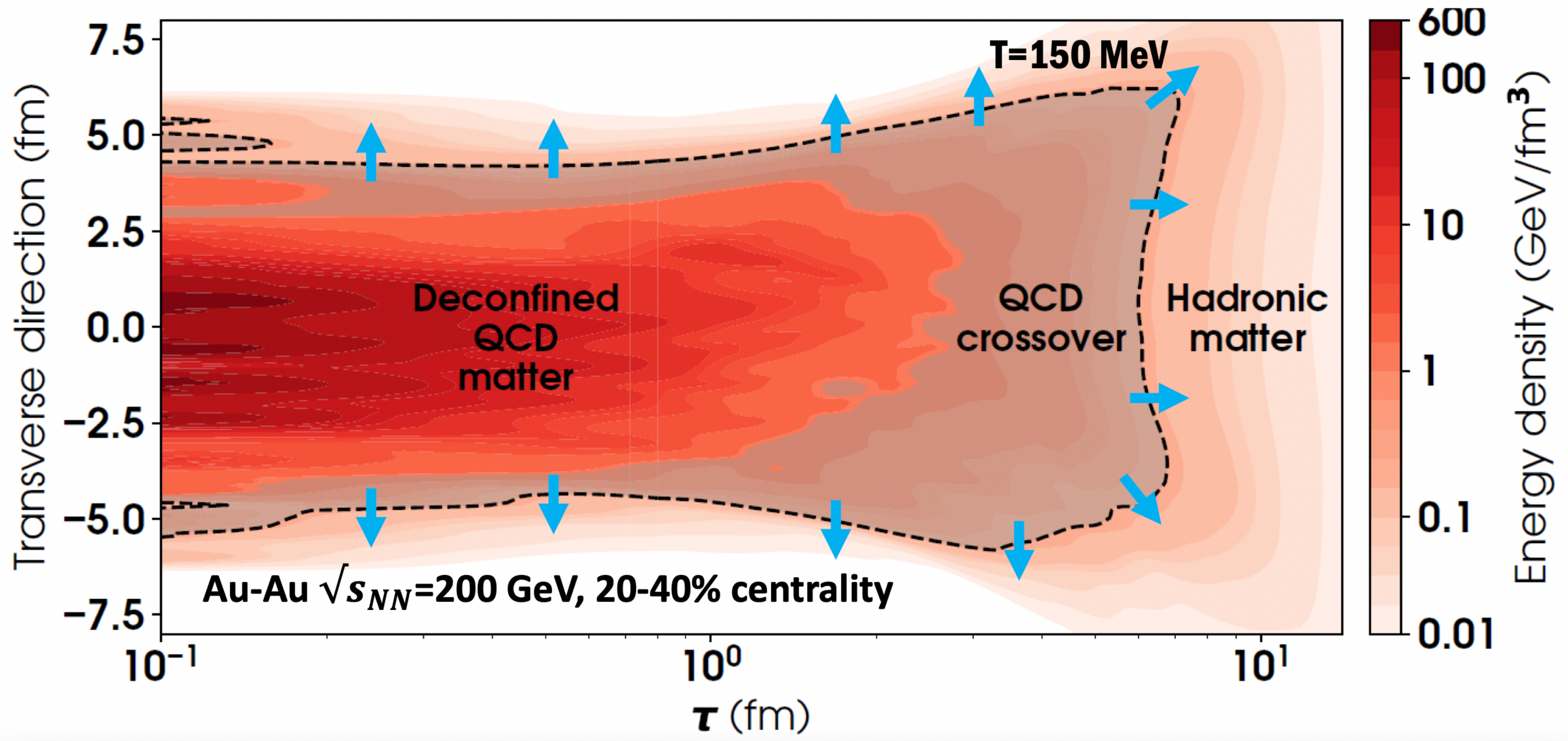}
\caption
{%
Contour plot of the evolution of the energy density in a midcentral Au+Au collision at the highest RHIC energy $\sqrt{s_{\rm NN}} = 200$ GeV. The horizontal axis shows the proper time $\tau$, the vertical axis shows one of the transverse coordinates. The black dashed line delineates the hadronization hypersurface $T = T_c = 150$ MeV \cite{Gale:2020dum}.} 
\label{fig:hydro}
\end{figure}

At high collision energies, hydrodynamics predicts the evolution of the fireball at midrapidity to be approximately boost invariant \cite{Bjorken:1982qr}. The hydrodynamical variables then become functions of the transverse coordinates $x_T$ and the local proper time $\tau$. The evolution of the fireball can be visualized by contour plots of the energy density, as shown in Fig.~\ref{fig:hydro} for a Au+Au collision at top RHIC energy. At lower beam energies the framework has to be extended to include finite net baryon charges and three spatial dimensions, since the assumption of a boost invariant longitudinal expansion breaks down. 

The second input into hydrodynamics calculations, which contains the information about the microscopic properties of the medium, are the transport coefficients. Main consideration has been given to the shear and bulk viscosity ($\eta$ and $\zeta$), but recently calculations have begun to include quantum number diffusion coefficients and even their cross-talk \cite{Fotakis:2019nbq}. The most interesting dimensionless quantities are $\eta/s$ and $\zeta/s$, where $s$ is the entropy density. It is extremely challenging to compute transport coefficients from fundamental lattice QCD, therefore one has to rely on models and approximations as well as limits for the high temperature--high density region (perturbative QCD) and the low temperature--low density region (hadron gas models). One of the major goals of the heavy ion program is to infer the values of the QCD transport coefficients from experimental data by detailed model-to-data comparison as explained in Section \ref{sec:results}.  

Fluid dynamics is only applicable when the system is sufficiently close to local equilibrium. Very shortly after the initial collision of the two nuclei, which originally only have longitudinal momentum, this condition is not satisfied. There is some transition time required for the transverse momenta of the medium constituents to become of similar magnitude as the longitudinal momenta. The initial state and associated thermalization process has been discussed in Section \ref{sec:ini_cond}. In the later stages of the reaction, the increasingly dilute medium will drop out of equilibrium, when the expansion rate is faster than the collision rate. This stage is the subject of the next Section \ref{sec:afterburner}. The boundary conditions for the differential equations that govern the fluid dynamic behaviour, are very important for a complete understanding of the dynamics. 

Conservation equations can be solved using a variety of algorithms. It is important to ensure that they also apply to relativistic hydrodynamics and event-by-event calculations with potentially spiky initial conditions without creating numerical artefacts. For this purpose a suite of tests of several scenarios, for which an analytic solution is known, was developed within the TEC-HQM collaboration \footnote{\url{https://wiki.bnl.gov/TECHQM/index.php/TECHQM_Main_Page} (see under {\it Documents})}. Several widely used implementations of viscous fluid dynamics for heavy ion collisions include MUSIC \cite{Schenke:2010nt}, vHLLE \cite{Karpenko:2013wva}, VISH2+1 \cite{Shen:2014vra} and CLVisc \cite{Pang:2018zzo}.

On the theoretical side, there are activities aimed at rooting relativistic viscous hydrodynamics in kinetic theory (see e.~g.~\cite{Denicol:2012cn}). Formalisms extending the applicability of fluid dynamics are also being developed. One example is anisotropic fluid dynamics that allows for larger differences between the longitudinal and transverse pressure in the system and therefore applies at earlier times than second-order viscous hydrodynamics \cite{McNelis:2018jho,Alqahtani:2017mhy}. Owing to the interest in the chiral magnetic effect and polarization observables, attention has been given to developments of fluid dynamic formulations with spin and or chiral currents (see e.g.~\cite{Weickgenannt:2022zxs, Bhadury:2020cop, Ammon:2020rvg}).

Fluid dynamics is not the only option to describe the hot and dense stage of heavy ion reactions, where the quark-gluon plasma is formed. Another possibility are microscopic transport codes that include partons as their degrees of freedom. Quarks and gluons are either treated as nearly massless particles interacting according to cross sections derived from perturbative QCD \cite{Xu:2004mz,Lin:2004en} or as dynamical quasiparticles \cite{Cassing:2009vt}. These microscopic non-equilibrium calculations have the advantage of providing detailed information about the complete phase-space over the entire dynamical evolution of the fireball at the expense of making more detailed assumptions that may be difficult to test.

\subsubsection{Hadron Transport and Freeze-Out.}
\label{sec:afterburner}

When the matter created in a heavy ion collision becomes more dilute again, the quark-gluon plasma hadronizes and the hadrons subsequently decouple from the system. The moment in time when hadron abundancies are fixed and inelastic interactions no longer occur is called chemical freeze-out, while the moment when also elastic scattering ceases is called kinetic freeze-out. The standard approach to the dynamical evolution at high beam energies involves hybrid models, based on (viscous) fluid dynamics as discussed in the previous Section \ref{sec:qgp_phase} for the hot and dense stage and Boltzmann-type hadronic transport equations for the late dilute stage of the reaction (see \cite{Petersen:2014yqa} for a review). 

The transition from partons to hadrons happens within the fluid dynamics calculation via a change of degrees of freedom in the equation of state. At some switching criterion, usually either the temperature or energy density, a three-dimensional hypersurface is constructed \cite{Huovinen:2012is} within the four-dimensional space-time that serves as the basis for the particlization process, where fluid quantities are mapped into particle properties \cite{Cooper:1974mv}. An example of such a freeze-out surface is shown as the dashed curve in Fig.~\ref{fig:hydro}.  The equation of state needs to be identical on both sides of the Cooper-Frye hypersurface and the transition happens according to 
\begin{equation}
     \frac{\mathrm{d}N_i}{d^3 p} = \frac{d_i}{(2\pi)^3} \  \int_\Sigma \left[ f_0(x, \vec{p}) + \delta f_\mathsf{shear}(x, \vec{p}) \ + \right. \\
     \left. \delta f_\mathsf{bulk}(x, \vec{p}) \right] \ \frac{p^\mu \ \mathrm{d} \Sigma_\mu}{E_{\vec{p}}} \qquad . 
\end{equation}
Here $\Sigma$ is the hypersurface, $f_0$ the equilibrium distribution function, and $\delta f$ denotes the viscous corrections to the distribution function. The specific form of the viscous corrections is one of the major uncertainties in the current hybrid approaches \cite{JETSCAPE:2020shq}. Once the particles are sampled according to their distributions on the hypersurface one can feed them into hadronic transport approaches like UrQMD \cite{Bass:1998ca,Bleicher:1999xi} or SMASH \cite{Weil:2016zrk}. 

\begin{figure}[htb]
\centering
\includegraphics[width=0.45\linewidth]{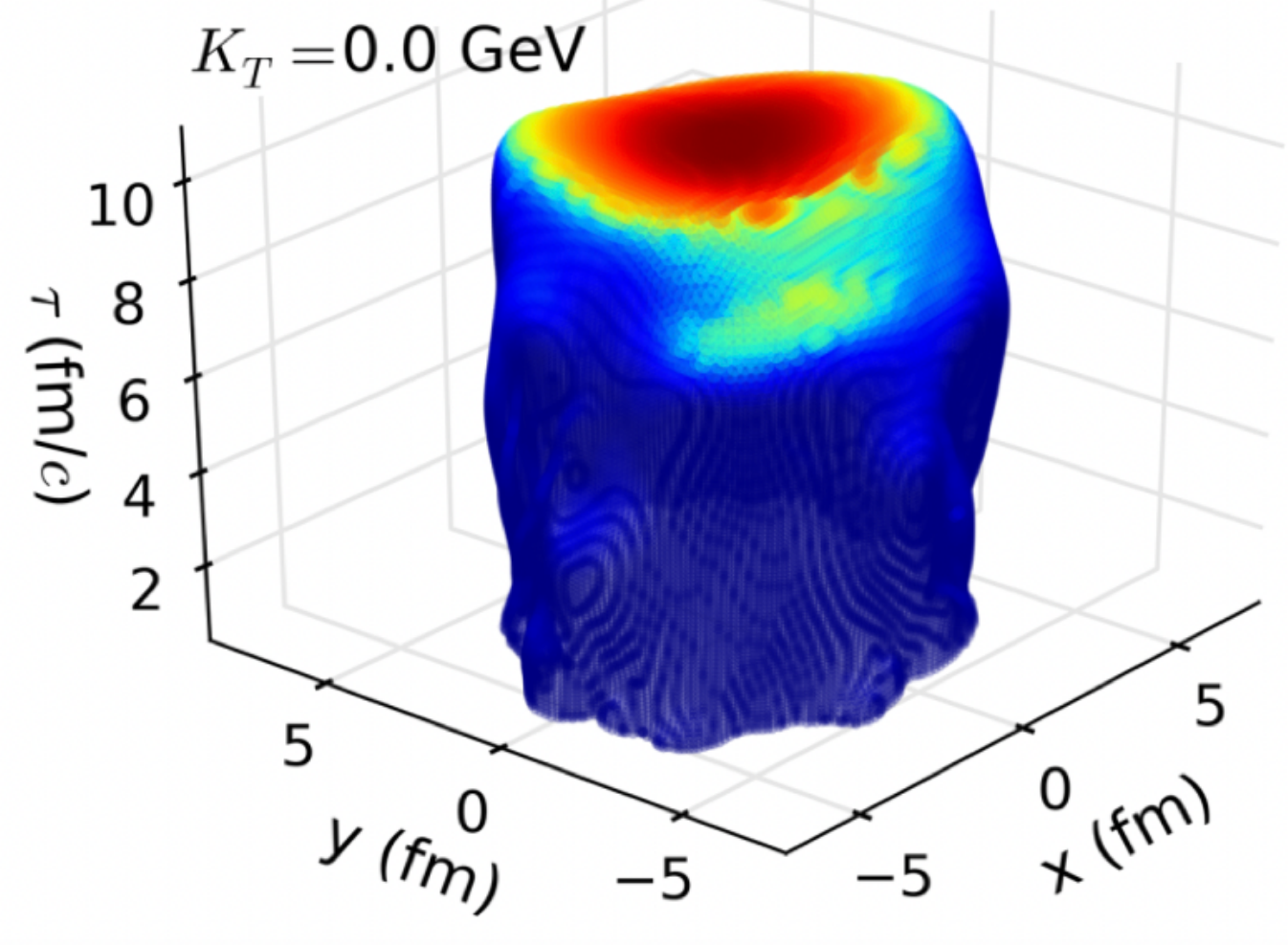}
\hspace{0.05\linewidth}
\includegraphics[width=0.45\linewidth]{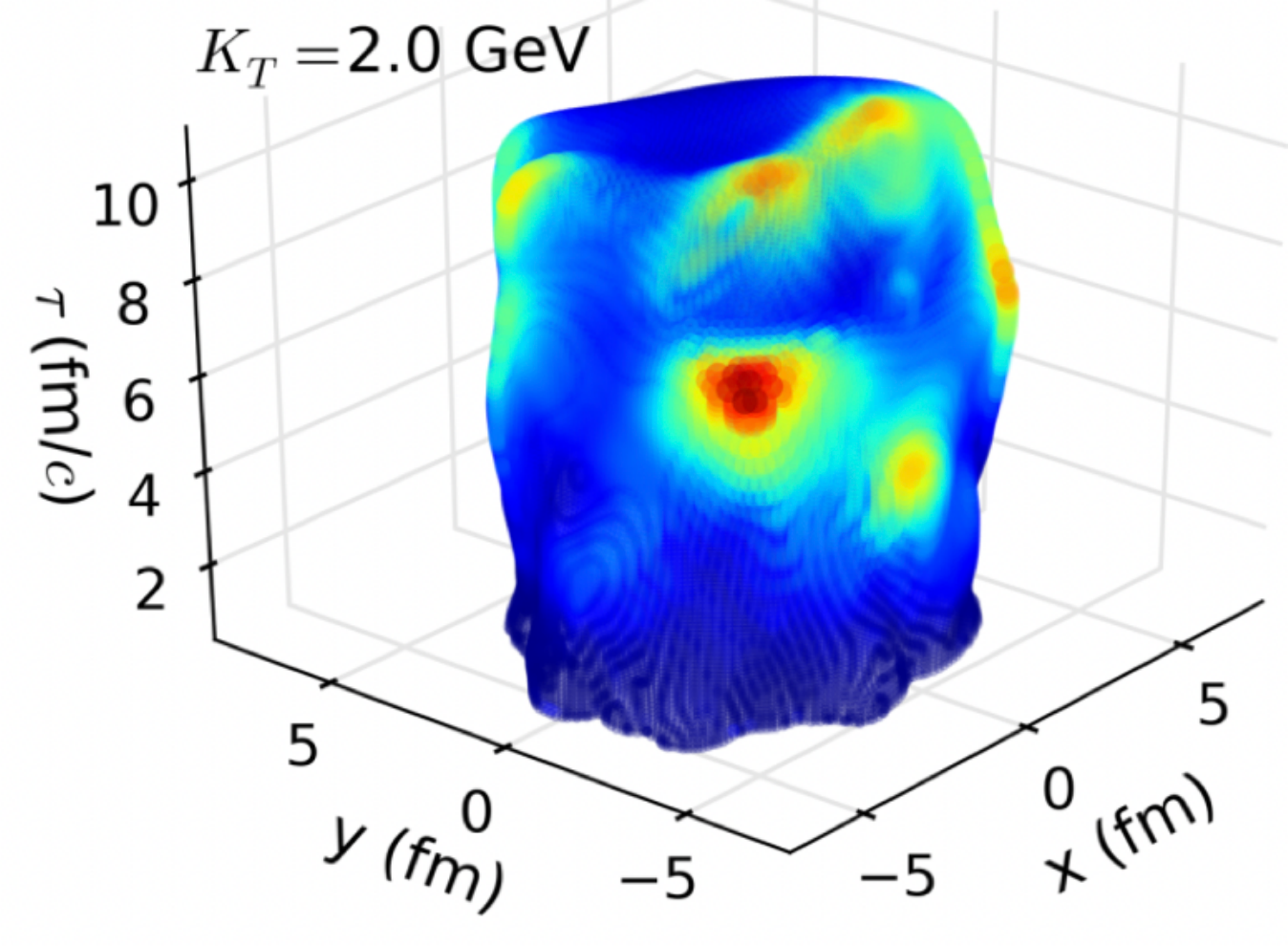}
\caption
{%
Intensity plots of the emission hypersurface for hadron pairs in Au+Au collisions at $\sqrt{s_{\rm NN}} = 200$ GeV. Left panel: Hadron pairs with total momentum $K_T = 0$, selectively weighting hadrons from bulk hadronization. Right panel: Hadron pairs with $K_T = 2$ GeV/c, selectively weighting hadrons from surface radiation. The emission intensity is highest in the dark red regions and lowest in the dark blue regions. [From \cite{Plumberg:2015eia}]} 
\label{fig:hadronization_surface}
\end{figure}

One generally assumes that the majority of particles escapes outward of this hypersurface but, in principle, there can be also inward directed contributions. These are particles that reenter the hydrodynamic evolution and are therefore very hard to deal with in practice \cite{Oliinychenko:2014tqa}. While all other parts of the evolution satisfy exact conservation laws event by event, in the sampling process they are only fulfilled for an ensemble of events. Therefore, very often many such samples and hadronic evolutions are calculated for each hydrodynamic event (oversampling method). To study correlation and fluctuation observables one might need a more careful consideration of conservation laws at the Cooper-Frye transition \cite{Oliinychenko:2019zfk}. For calculations of dilepton emission one needs to include the finite-width spectral functions of the resonances in the sampling process. For this purpose, it is necessary to match the degrees of freedom and their properties in the hadronic transport approach that is chosen for the final rescattering and decays. 

The advantage of the microscopic non-equilibrium treatment of the late hadronic stage is that the chemical and kinetic freeze-out are happening automatically, whenever the corresponding interactions cease for each particle species individually. An alternative option is to keep running hydrodynamics and introducing partial chemical equilibrium. In this approach one encounters a kinetic freeze-out hypersurface, an example of which is shown in Fig.~\ref{fig:hadronization_surface}. The explicit particle degrees of freedom have the added advantage that the outcome is very similar to what is measured in experiment. When using unified output formats, like OSCAR or the HepMC library \footnote{\url{https://ep-dep-sft.web.cern.ch/project/hepmc}}, the same analysis can then be applied to the simulation results and to the experimental data, which allows for more accurate comparisons. Recently, the RIVET library \footnote{\url{https://rivet.hepforge.org}} is also getting extended to include heavy ion observables, which could be an interesting option in the future. 

The same information on hadronic degrees of freedom and their interactions that one needs for final-stage hadron transport calculations in high-energy heavy ion collisions is employed for the full evolution in low energy heavy ion reactions. Hadronic transport approaches are based on the relativistic Boltzmann equation where, appropriate for low net baryon density, the mean-field term is omitted and only collisions are taken into account. The input for cross-sections and processes is taken from the Particle Data Book, experimental data for elementary reactions, or estimated using effective models. For low energy binary scatterings, the reactions are modeled via resonance excitation and decay, while at higher energies string excitation and fragmentation is used to model inelastic processes. Such microscopic approaches have the advantage that one has access to the complete phase-space information of all particles at all times, and they often serve as event generators for simulations of experimental results. 

\subsection{Hard and electromagnetic probes}
\label{sec:hard_probes}

Hard probes of hot and dense QCD matter are defined as those that are, in important parts, perturbatively calculable. This applies to strongly interacting (QCD) probes at high virtuality and to electromagnetic probes. In most instances, hard probes rely on input that is nonperturbative and must be derived from other experimental data. Parton distribution functions (PDFs) $f(x)$ and fragmentation functions $D(z)$ are salient examples. 

\begin{figure}[ht]
\flushright
\includegraphics[width=0.65\linewidth]{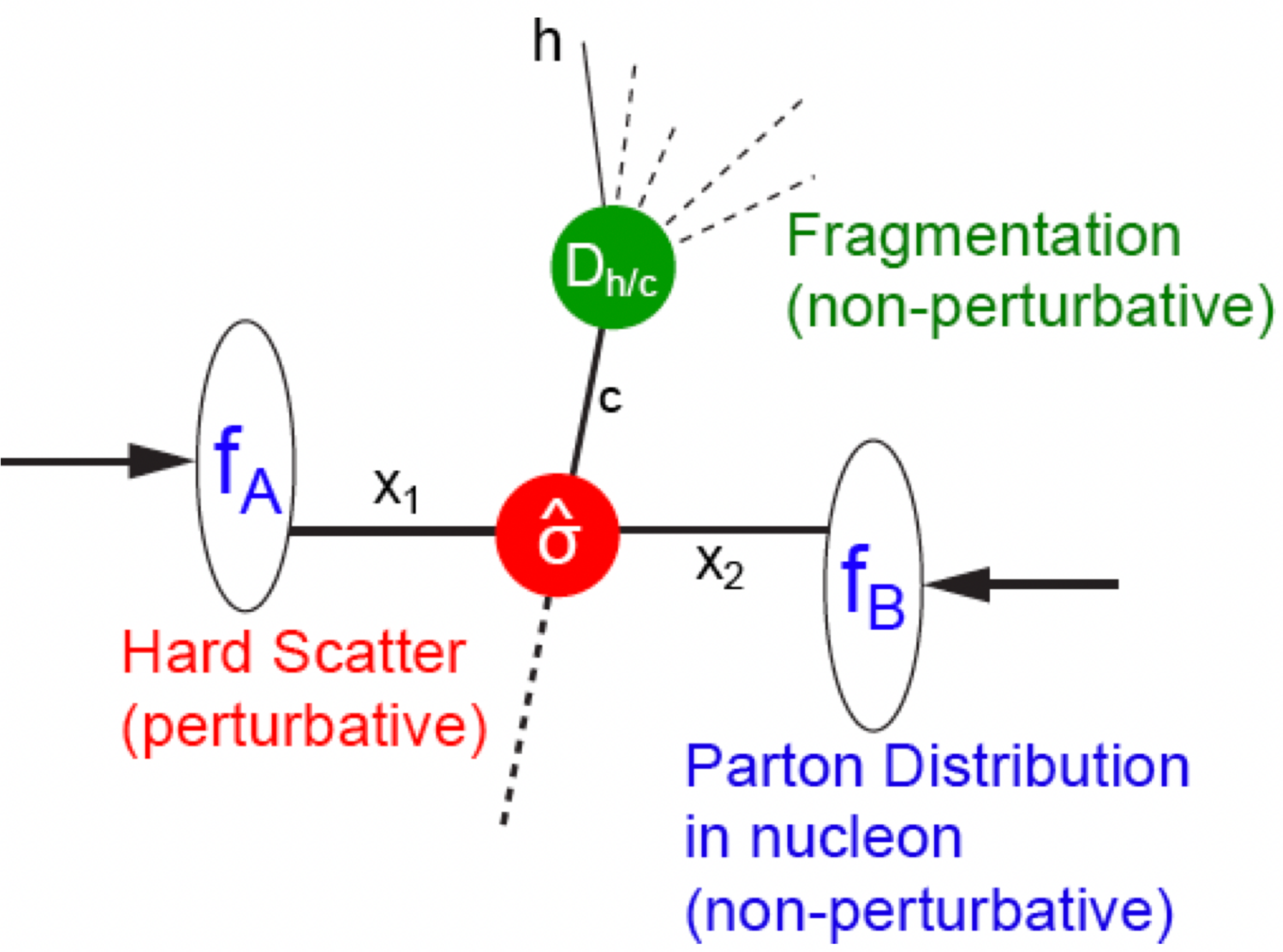}
\caption{Factorization of a hard scattering process into initial-state, hard scattering, and final state-state processes.  [From K.~Reygers (\url{https://www.physi.uni-heidelberg.de/~reygers/lectures/2017/qgp/qgp_ss17_08_jet_quenching.pdf})]}
\label{fig:factor}
\end{figure}

\begin{figure}[ht]
\flushright
\includegraphics[width=0.75\linewidth]{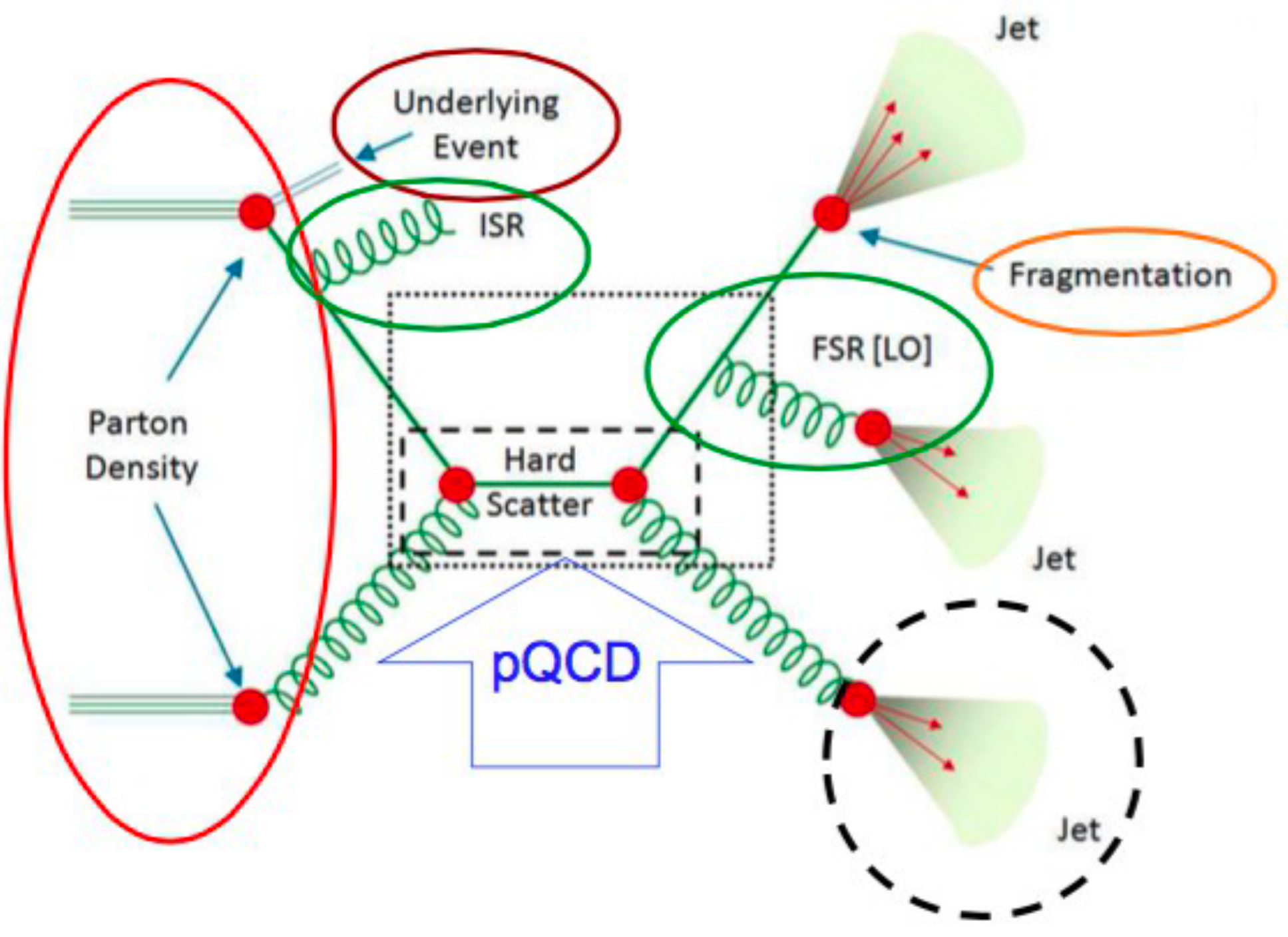}
\caption{Complexities hiding in the factorization of a hard scattering process into initial-state, hard scattering, and final state-state processes.  [From D.~S.~Cerci (\url{https://indico.cern.ch/event/614845/contributions/2728799/attachments/1529660/2398415/13_DSunarCerci.pdf})]}
\label{fig:factor_complex}
\end{figure}

Generally, hard probes in p+p or A+A collisions rely on the principle of factorization, illustrated in Fig.~\ref{fig:factor}: A scattering process is factorized into a nonperturbative initial-state matrix element, a perturbatively calculable hard scattering S-matrix element, and a nonperturbative final-state matrix element:
\begin{equation}
d\sigma = f_{\rm A}(x_1) \otimes f_B(x_2) \otimes d\hat{\sigma} \otimes D_{h/c}(z) .
\end{equation}
The process at the core of this scheme, the hard scattering cross section $d\hat{\sigma}$, is universal, but the initial- and final-state matrix elements can be modified by nuclear effects -- cold nuclear matter in the case of the initial state and hot QCD matter in the case of the final state.
The deceptively simple sketch in Fig.~\ref{fig:factor} hides many of the complexities that arise as a result of the nonperturbative interactions occurring in the initial and final states. This is illustrated in Fig.~\ref{fig:factor_complex}. Their sensitivity to the nuclear medium is what makes them of core interest to relativistic heavy ion physics. The challenge is to encapsulate these modifications in physics motivated descriptions that can be used to probe the properties of the nuclear medium. Examples are parton energy loss, diffusion coefficients, saturation scales, and so on.

We have already covered models of the initial-state in section \ref{sec:ini_cond}. We now discuss models for the final-state modifications. The output of the hard scattering process are two or more energetic partons, which then can interact with the medium created in the collision, often referred to as the underlying event. These interactions can result in energy or momentum loss \cite{Thoma:1991ea}, deflection \cite{Majumder:2009cf}, emission of radiation \cite{Guo:2000nz}, or transmutation of the parton into another particle \cite{Fries:2002kt}. A sequence of such interactions leads to modifications of the jet shower evolution, which manifests itself in various observables (see Section \ref{sec:jet_quenching}) and can be modeled using Monte-Carlo methods (see e.~g.~\cite{Zapp:2013vla,Putschke:2019yrg}).

The mechanisms causing an energetic parton to lose energy during its propagation through matter can be divided into two categories: elastic and inelastic. Elastic energy loss occurs in two-body scattering, where energy is transferred to the scattering partner. Inelastic energy loss occurs predominantly by radiation of a gluon following the interaction with a scattering center in the matter. At high energies, inelastic energy loss dominates; at low energies elastic energy loss dominates. The threshold at which inelastic energy loss begins to dominate increases with the mass of the energetic parton, because radiation from heavy particles is suppressed by the dead-cone effect \cite{Dokshitzer:1991fd}. 

Elastic energy loss is one aspect of momentum change in the medium caused by scattering. For heavy quarks with mass $M \gg T$ multiple scatterings are required to appreciably change the momentum. The process can then be considered as a diffusion process and described by a Langevin equation with a fluctuation constant $\kappa$ and drag coefficient $\eta$. In a thermal medium $\kappa$ and $\eta$ are related to the spatial diffusion constant $D$ and to each other by the Einstein relation
\begin{equation}
    \kappa = \frac{2T^2}{D} , \quad 
    \eta = \frac{T}{M D} = \frac{\kappa}{2 M T}
\end{equation}
The diffusion constant can be calculated in thermal perturbation theory \cite{Caron-Huot:2007rwy}, on the lattice \cite{Petreczky:2005nh}, or by holographic techniques \cite{Casalderrey-Solana:2006fio}. The Langevin framework can be easily simulated within a fluid dynamical description of the quark-gluon plasma. Rare scatterings with larger momentum transfer are not described by the Langevin approach. These can be treated in the framework of linearized Boltzmann transport in a thermalized, hydrodynamically flowing medium \cite{Ke:2018tsh}. Both descriptions can be merged into a unified framework that effectively describes all aspects of heavy quark transport \cite{Dai:2020rlu}.

Inelastic energy loss, i.~e.\ energy loss by gluon radiation, is the dominant mode of energy loss for light quarks and gluons. Medium induced radiation requires off-mass shell scattering of the energetic parton in the medium, followed by gluon emission. In a dense medium, multiple scattering events contribute coherently to a single radiation event. The resulting suppression of the radiation yield is known as Landau-Pomeranchuk-Migdal (LPM) effect. The scattering power of the medium (momentum transfer squared per unit length) is expressed in the transport coefficient $\hat{q}$ defined in (\ref{eq:q-hat}). Several different formalisms have been developed to describe the radiative energy loss by a quark or gluon. These correspond to different approximations or truncations of the multiple scattering process. For an overview and comparison of existing approaches, see \cite{JET:2013cls}. General overviews of energy loss and jet quenching can be found in \cite{Majumder:2010qh}. 

A fraction of the energy lost by an energetic parton traversing the quark-gluon plasma is deposited into the thermal medium. This applies to all energy lost by elastic collisions and to the soft component of the radiated energy which thermalizes. The response of the medium to the injection of energy and momentum along the trajectory of the energetic parton is treated by adding a source term to the hydrodynamic equations. The linear response of the medium can be understood in terms of sound propagation and results in the possible formation of a Mach cone \cite{Neufeld:2010tz}. A more microscopic approach treats the individual interactions within the evolution of a full jet as sources in a numerical hydrodynamics code \cite{Tachibana:2017syd}. For the effect of the medium response on jet observables, see \cite{KunnawalkamElayavalli:2017hxo}.

Electromagnetic (EM) probes, i.~e.\ photons and dileptons (often called real and virtual photons), are special in that they do not suffer from final state interactions. (They may, however, suffer from large backgrounds.) Their production can be related to the photon spectral function $\rho(\omega,\vec{p})$ in the medium \cite{Kapusta:2006pm}. In the vacuum, $\rho$ is a function of the invariant mass $M^2=\omega^2-p^2$ only; in a thermal medium, $\rho(\omega,M)$ is a function of, both, $M$ and the frequency measured in the fluid rest frame $\omega=-u_\mu p^\mu$. The rate of photon emission can be expressed as \cite{Floerchinger:2021xhb}
\begin{equation}
    \omega \frac{dR}{d^3p} = \frac{n_{\rm B}(\omega) \rho(\omega,0)}{(2\pi)^3}
\end{equation}
with $n_{\rm B}(\omega)=(e^{\omega/T}-1)^{-1}$. The dilepton emission rate is given by
\begin{equation}
    \frac{dR}{d^4p} = \alpha \frac{n_{\rm B}(\omega) \rho(\omega,M)}{12\pi^4 M^2}
    \left(1+\frac{2m^2}{M^2}\right)
    \sqrt{1-\frac{4m^2}{M^2}} \theta(M^2-4m^2) ,
\end{equation}
where $m$ denotes the lepton mass. Apart from the fact that the measured yields involve an integration over the space-time evolution of the emitting matter, EM probes thus provide direct information about the photon spectral function. A broad overview of electromagnetic probes of relativistic heavy ion collisions can be found in \cite{Gale:2003iz}.

\begin{figure}[ht]
\flushright
\includegraphics[width=0.95\linewidth]{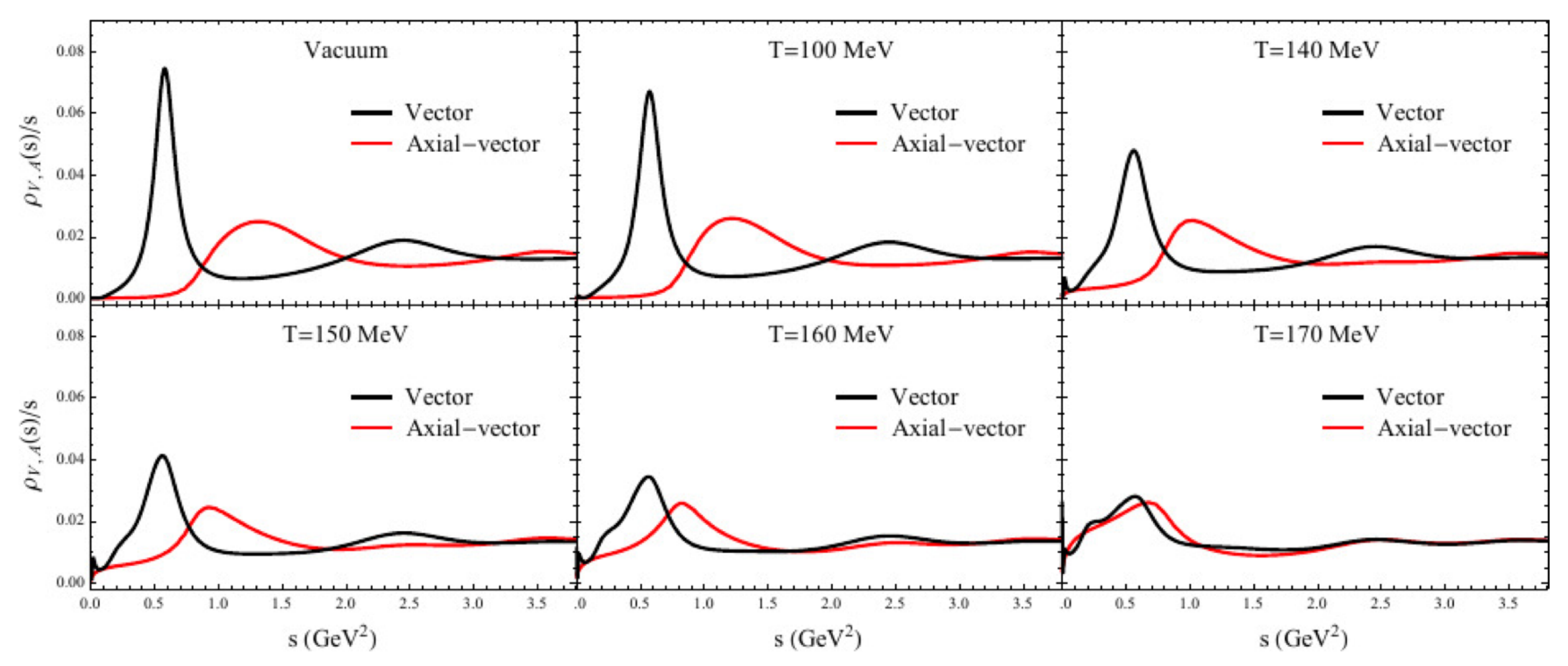}
\caption{Conjectured evolution of the vector (black curves) and axial vector (red curves) spectral functions with temperature $T$ for $\mu_B=0$ obtained by solution of QCD sum rules. [From \cite{Rapp:2016xzw}]}
\label{fig:spectral}
\end{figure}

The photon spectral function is predicted to change from one that is dominated by vector mesons at $T < T_c$ to a broad continuum corresponding to free quark-antiquark pair excitations at $T > T_c$ (see Fig.~\ref{fig:spectral}). It can be estimated using lattice QCD \cite{Ding:2016hua,Ghiglieri:2016tvj}, thermal perturbation theory \cite{Ghiglieri:2014kma}, or using QCD sum rules \cite{Kapusta:1993hq,Holt:2012wr}. In the sum rule approach, the transition of the photon spectral function is closely related to the phenomenon of chiral symmetry restoration. At low frequencies and long wavelengths, the spectral function is related to the electrical conductivity $\sigma$ of the QGP by
\begin{equation}
    \sigma 
    = \frac{1}{2} \lim_{\omega\to 0} \frac{1}{\omega}
    \left. \rho(\omega,p) \right|_{\omega^2=p^2}
    = \frac{1}{3} \lim_{\omega\to 0} \frac{1}{\omega}
    \left. \rho(\omega,p) \right|_{p=0} .
\end{equation}

It is a challenge for the future to combine the soft bulk evolution with the hard probes consistently. Many calculations us state-of-the-art background for the calculations of hard probes, but establishing the medium response and a fully coupled picture over the different kinematic regions is still lacking. 

\subsection{Hadronization Models}

Different kinematic regions of the particles produced in heavy ion collisions are governed by different regimes of quantum chromodynamics. The soft region below $\sim 3$ GeV is typically described by the bulk dynamic evolution (see Section \ref{sec:bulk}), while the hard region above $\sim 6$ GeV is covered by perturbative QCD (see Section \ref{sec:hard_probes}). Experimentally, it is only possible to detect final-state hadrons, while we would like to learn about the hot and dense quark-gluon plasma stage of the reaction. Therefore, the hadronization process plays an important role in our interpretation of the data from heavy ion reactions. 

\begin{figure}[ht]
\flushright
\includegraphics[width=0.75\linewidth]{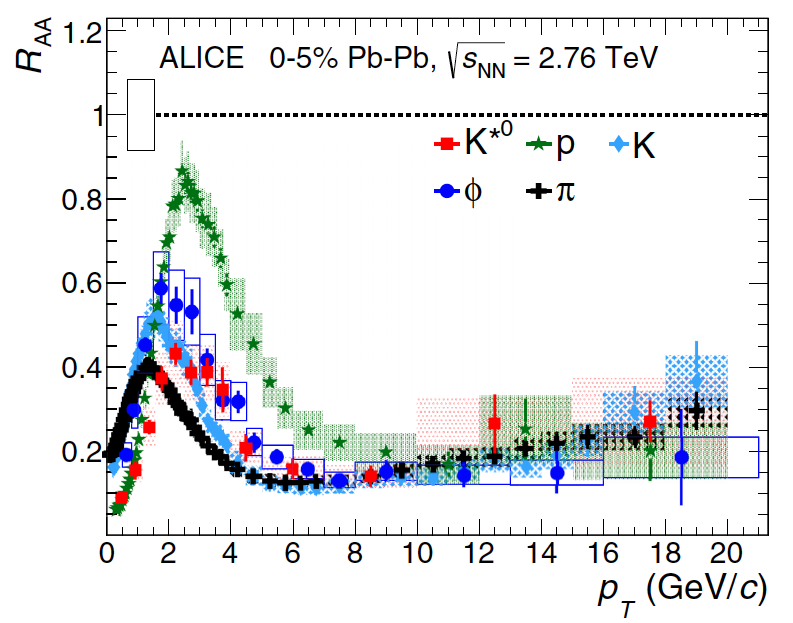}
\caption{Transverse momentum dependence of the nuclear modification factor of several particle species measured by the ALICE collaboration. [From \cite{ALICE:2017ban}] }
\label{fig:Raa_PID}
\end{figure}

Figure \ref{fig:Raa_PID} shows an example of the nuclear modification factor $R_{\rm AA}$ as a function of transverse momentum for several particle species from the ALICE collaboration in PbPb collisions at $\sqrt{s_{\rm NN}}=2.76$ TeV. This picture serves nicely as a basis to introduce the hadronization models that are applied in different regions. 

Below $\sim 2$ GeV the particle production is described by the viscous hydrodynamic evolution, and hadronization happens via the change of degrees of freedom in the equation of state. On the Cooper-Frye surface the hadronic fluid is converted to individual particles and those are then propagated by a hadronic cascade. This is one of the big advantages of hydrodynamics, that the transition from partonic to hadronic degrees of freedom is under nice theoretical control as long as local equilibrium can be assumed. In this region the nuclear modification factor shows the typical mass dependence expected from collective behaviour. For a more detailed discussion see Section \ref{sec:trans_expansion}.

Above a transverse momentum of $p_T \sim 8$ GeV one observes a rather universal behaviour, this is the region where hadronization is described by fragmentation of color flux tubes. The strings break apart when the leading quark-antiquark (or quark-diquark) pairs are pulled apart and the color field becomes strong enough to create particle-antiparticle pairs out of the vacuum. Subsequently, the newly produced partons connect with each other to form hadrons with well-defined quantum numbers. 

The intermediate region $p_T=2-8$ GeV is governed by the recombination and coalescence models for hadronization and the interaction of hard fragmentation and soft hadrons. Recombination means that the partons from the quark-gluon plasma are clustered into hadrons when they are close in phase space (see Section \ref{sec:ncq_scaling} for more details). In this region unusually large baryon-to-meson ratios can be observed in the nuclear modification factor. 

Understanding hadronization on a microscopic level poses a challenge for the future. It is important to advance the understanding on how hadrons emerge from the collective partonic state that is formed in ultra-relativistic heavy ion collisions. This involves understanding QCD on multiple different scales connecting the different regions outlined above in a smooth fashion.  

\section{Experimental Results}
\label{sec:results}

{\em This Section reviews selected experimental results from the experiments with relativistic heavy ion collisions at RHIC and LHC and their theoretical interpretation in terms of basic matter characteristics.}

\subsection{Location in the phase diagram} 
\label{sec:chem_fo}

The first question to be answered when studying heavy ion reactions is which temperature and density is reached in the collision. While full dynamical models as described in Section \ref{sec:pheno} provide (model-dependent) answers to this question, the possibility of defining a temperature for the system has more fundamental implications. To be able to assign a single temperature and net baryon chemical potential to the system created in such a highly dynamical process hints at the fact that an equilibrated plasma has formed. Another interpretation involves the idea that hadronization follows statistical behaviour and, therefore, the numbers of produced hadrons follow thermal expectations. The second interpretation is supported by the finding that even in particle production from $e^+ e^-$ collisions thermal particle yields are measured. \cite{Becattini:2008tx}

The basic idea to connect final yields at midrapidity or $4\pi$ multiplicities to a temperature relies on the formation of a fireball in global equilibrium, that emits all particles at the same instant in time. The thermal model actually has no notion of time evolution or spatial variations, therefore it is limited to a small set of observables, namely the particle abundances. The grand canonical partition function includes all mesons and baryons of the non-interacting hadron resonance gas. The fireball is treated as a grand canonical ensemble with Bose and Fermi statistics depending on the particle spin. The distribution function for particle species $i$ is given by 
\begin{equation}
\frac{dN_i}{dp^3 dx^3} = \frac{g_i}{(2\pi)^3} \, \frac{1}{\exp^{(E - \mu_i)/T} \pm \alpha_i} ,
\end{equation}
where $g_i$ is the degeneracy facter, $T$ is the temperature, $\mu_i = \mu_B B_i + \mu_S S_i$ the particle-specific chemical potential and $\alpha_i=\pm 1$ for fermions and bosons, respectively. Integrating over the entire phase space and assuming Boltzmann statistics -- a good approximation for all hadrons except pions when $|\mu_i|/T$ is small -- one obtains the following well-known expression for thermal particle production
\begin{equation}
N_i \;=\; \frac{g_i V  T^3}{2\pi^2}\, e^{\mu_i/T}\,\frac{m_i^2}{T^2}\,K_2\left(\frac{m_i}{T}\right)\, ,
\end{equation}
where $N_i$ is the number of particles of species $i$ produced in the fireball of volume $V$ at temperature $T$ and chemical potential $\mu_i$. $m_i$ denotes the mass of the particle and $K_2$ is the modified Bessel function of the second kind. The main quantity determining the particle yields is the mass of the particle species. One expects higher mass particles to be produced with smaller probabilities than lighter particles. To remove sensitivity to the volume, the usual strategy is to look at particle yield ratios that are assumed to be emitted from the same volume. By fitting ratios of several different particle species one can then obtain temperatures and chemical potentials. 

\begin{figure}[htb]
\includegraphics[width=0.5\linewidth]{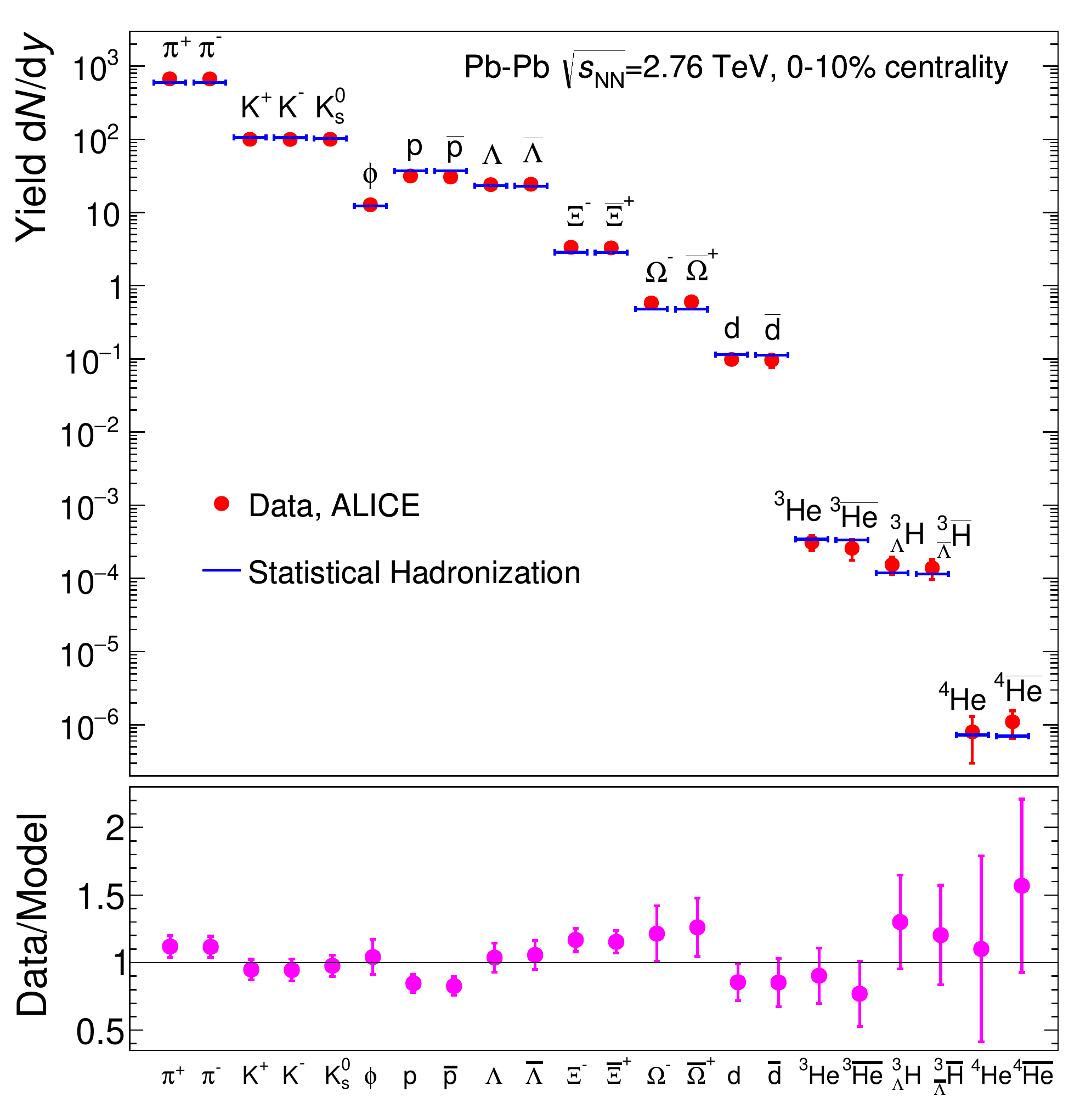}
\hspace{0.09\linewidth}
\includegraphics[width=0.45\linewidth]{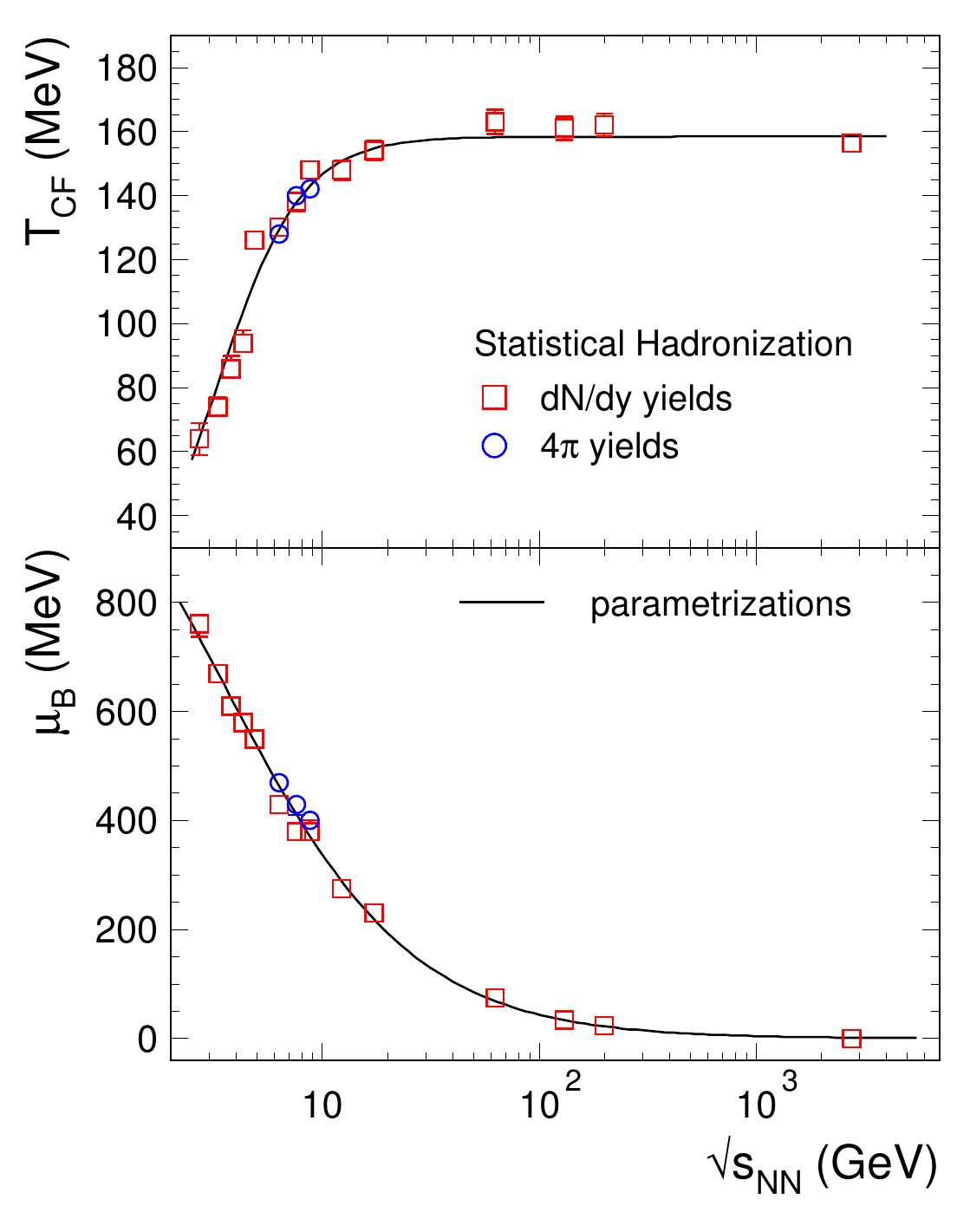}
\caption{Left: Hadron abundances ($dN/dy$ values at midrapidity) from the ALICE collaboration for central Pb+Pb collisions at $\sqrt{s_{\rm NN}}= 2.76$ TeV compared to statistical hadronization model analysis. Right: Energy dependence of chemical freeze-out parameters $T_{\rm CF}$ and $\mu_B$. The results are obtained from the statistical hadronization analysis of hadron yields (at midrapidity, $dN/dy$, and in full phase space, $4\pi$) for central collisions at different energies. 
[From \cite{Andronic:2017pug}]}
\label{fig:lhc_fits}
\end{figure}

It is impressive how well a vast amount of data on particle production in heavy ion collisions can be understood with such simple assumptions. One recent example is shown in Fig. \ref{fig:lhc_fits} (left) where particle ratios at midrapidity from Pb+Pb collisions at $\sqrt{s_{\rm NN}}=2.76$ TeV are fit within the statistical model. The bars denote the model fit and the symbols show experimental data from the ALICE collaboration. Even the light nuclei follow the trend of thermal production at a unique global temperature and chemical potential. A widely adopted interpretation is that these thermodynamic properties reflect the conditions prevailing at the chemical freeze-out, the moment at which the inelastic reactions cease and the abundances are frozen. (Elastic rescattering is still possible until the kinetic freeze-out that will be discussed in more detail in the next Section \ref{sec:trans_expansion}.) It is worth pointing out, however, that this interpretation is not logically consistent, because the model also describes the yields of particles that cannot be formed under the chemical freeze-out conditions, such as light nuclei. A physically credible interpretation of the observed yields of such states can be made either on the basis of the dynamical coalescence model \cite{Zhang:2018euf} or by invoking general properties of strongly coupled, highly excited quantum systems \cite{Muller:2017vnp}.

The right-hand part of Fig. \ref{fig:lhc_fits} shows the result of thermal fits to particle yields and multiplicities over a large range of beam energies. As expected, the temperature rises as a function of beam energy, while the net baryon chemical potential decreases. This can be understood by the stopping dynamics of the two nuclei passing through each other. At lower energy the nuclear passing time is longer and more baryons are stopped at midrapidity, while at high beam energies the nuclei fly through each other so fast that the baryon number remains localized at forward rapidities (transparency). The parametrizations shown in Fig. \ref{fig:lhc_fits} (right) are \cite{Andronic:2017pug}: 
\begin{equation}
    T_{\rm CF} = \frac{T_{\rm CF}^{\rm lim}} {1+\exp(2.60-(\ln \sqrt{s_{\rm NN}})/0.45)} ,
    \qquad
    \mu_B = \frac{a}{1+0.288\sqrt{s_{\rm NN}}} ,
\end{equation}
with $\sqrt{s_{\rm NN}}$ in GeV, $T_{\rm CF}^{\rm lim}=158.4$ MeV and $a=1307.5$ MeV.

\begin{figure}[ht]
\centerline{\includegraphics[width=0.5\linewidth]{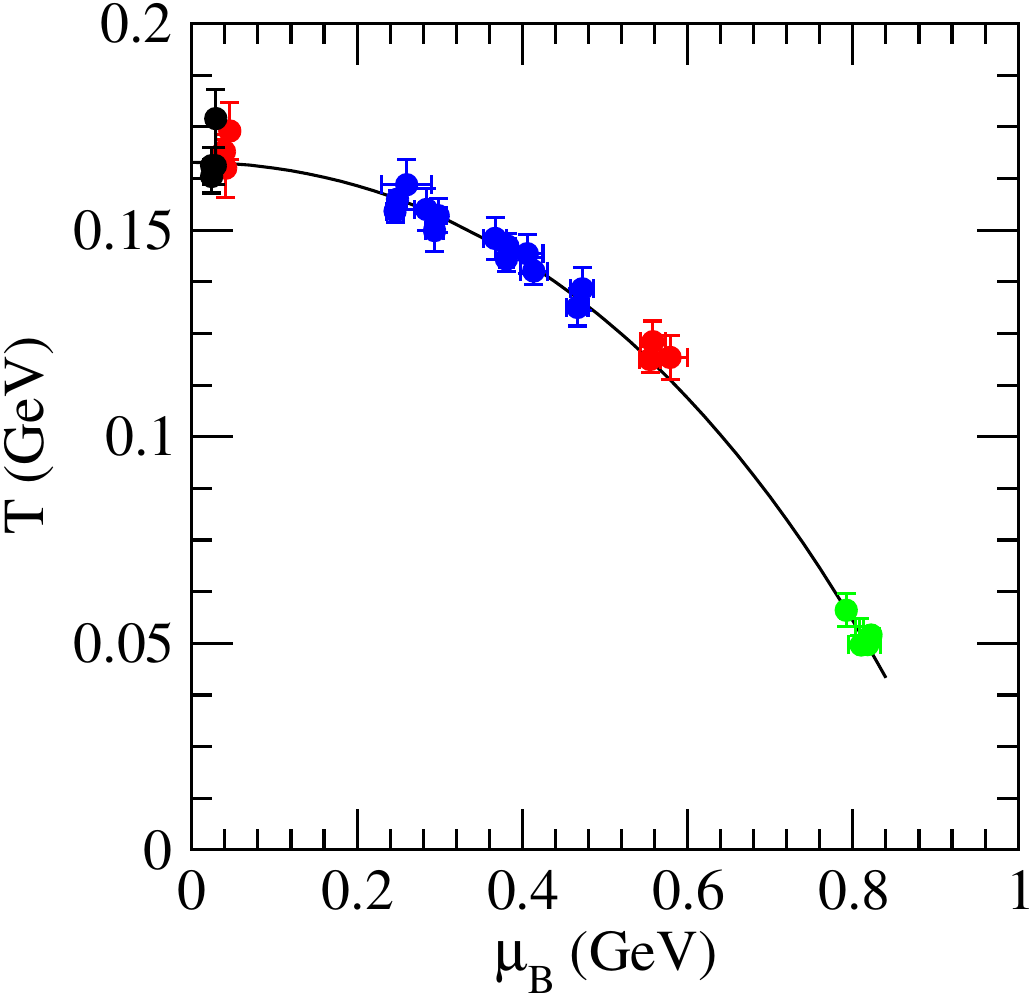}}
\caption{ Values of $\mu_B$ and $T$ for different energies. The solid line is a parameterization corresponding to
$T(\mu_B) \approx 0.17 - 0.13\mu_B^2 -0.06 \mu_B^4 $.[From
\cite{Cleymans:2005xv}]}
\label{fig:Tch_phase_diagram}
\end{figure}

When plotting the values for temperatures and net baryon chemical potential obtained in thermal model fits on the phase diagram one obtains a curve as specified in Fig. \ref{fig:Tch_phase_diagram}. Since those values correspond to a hadron resonance gas, the quark-gluon plasma formation can only happen beyond this line. These values define a lower bound of temperatures and densities that are reached in the corresponding heavy ion collisions, since the chemical freeze-out happens at the end of the evolution. The shape of the line is consistent with the hypothesis that the chemical freeze-out is associated with a constant energy per baryon in the system \cite{Cleymans:2005xv}. 

\begin{figure}[htb]
\includegraphics[width=0.5\linewidth]{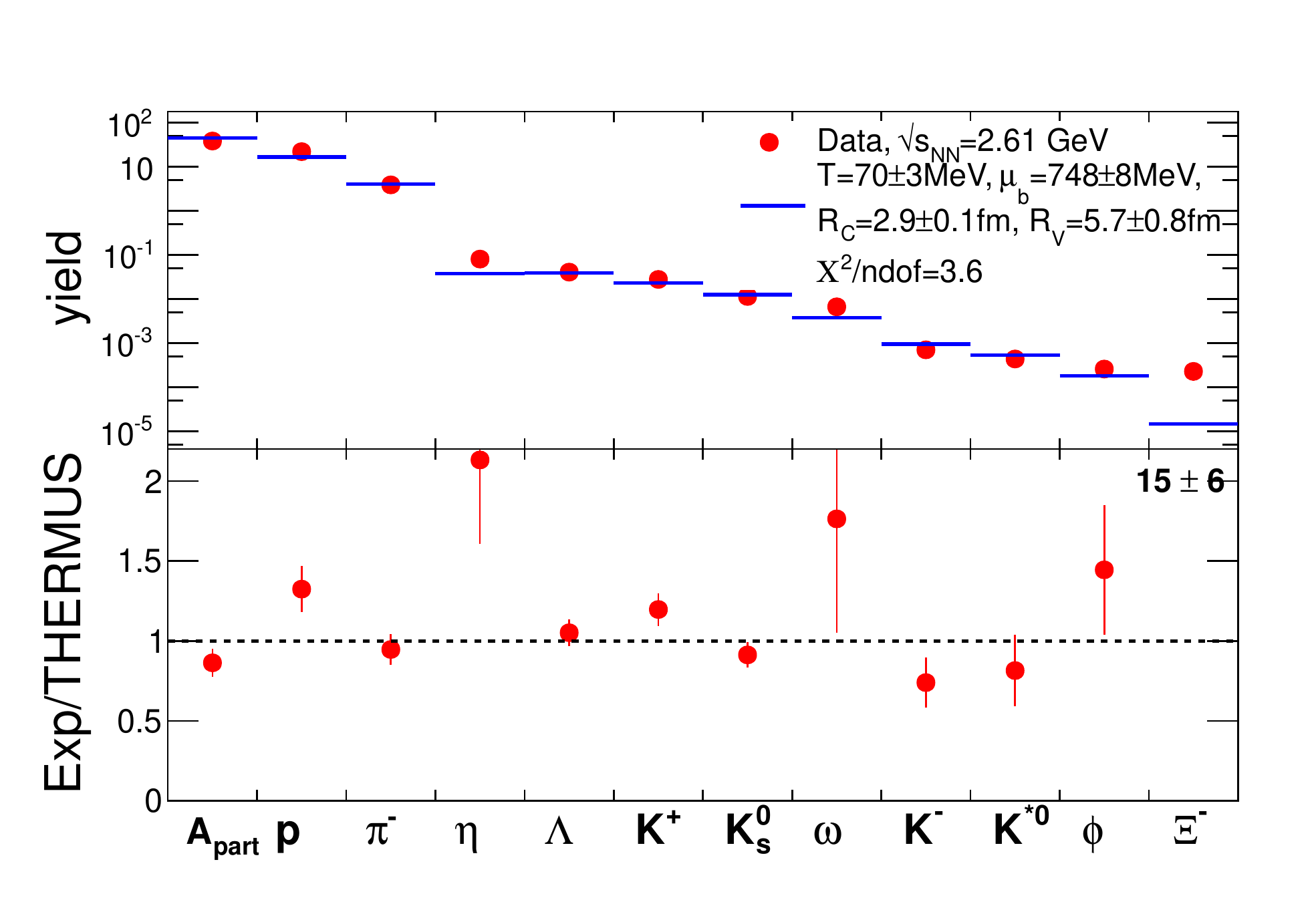}
\hspace{0.09\linewidth}
\includegraphics[width=0.45\linewidth]{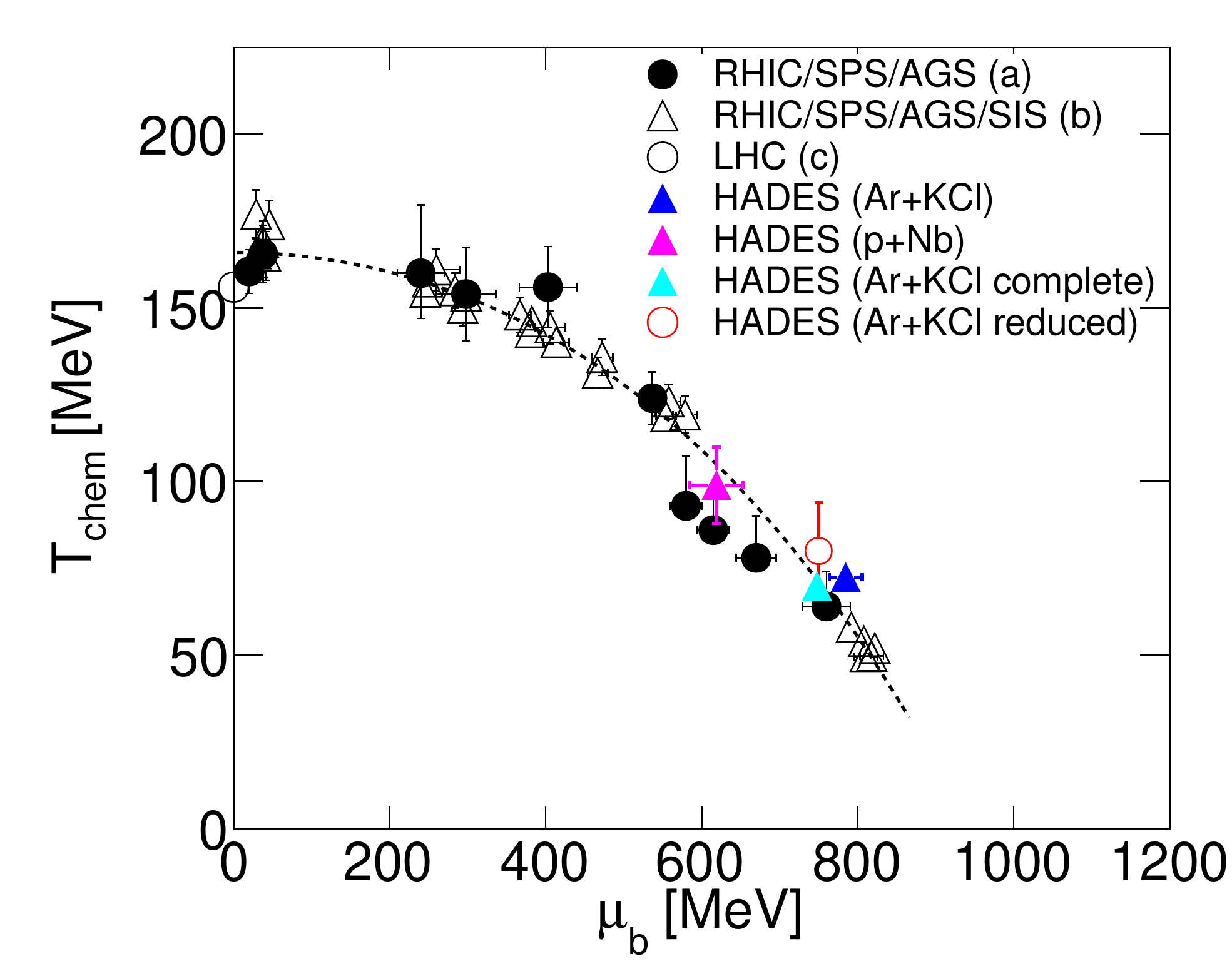}
\caption{Left: Hadron abundances ($dN/dy$ values at midrapidity) from the HADES collaboration for Ar+KCl collisions at $\sqrt{s_{\rm NN}}= 2.6$ GeV compared to THERMUS calculation. Right: Chemical freeze-out parameters from different systems at different beam energies in the $T-\mu_B$ plane. 
  [From
\cite{HADES:2015oef}]}
\label{fig:hades_fits}
\end{figure}

At lower beam energies and for smaller system sizes, the assumptions of a grand canonical ensemble may no longer be valid, especially for particles carrying a conserved quantum number such as strangeness. Since $s$ and $\overline{s}$ quarks must be produced in pairs, the conservation of net strangeness in the volume must be taken into consideration. Using a thermal model that includes these ideas, one can even obtain decent fits to particle ratios from low energy heavy ion reactions as shown in the left panel of Fig.~\ref{fig:hades_fits} by the HADES collaboration. While it is remarkable how well the extracted temperatures and net baryon chemical potentials fit into the world data (see right panel of Fig.~\ref{fig:hades_fits}), some discrepancies are also appearing. For example, the high yield of $\Xi$ baryons needs a different explanation. 

At high beam energies, there are some indications that strange particles have a higher chemical freeze-out temperature than hadrons consisting of light quarks only \cite{Bellwied:2018tkc}. This could be understood by the smaller cross-sections of strange hadrons with the surrounding matter. Those studies go beyond fitting yields by investigating the relevant susceptibilities through particle number fluctuations of various orders. There are a lot of caveats when comparing these observables directly to grand canonical lattice QCD calculations but, if properly taken into account, allow to extract valuable information on the thermodynamic properties of the system. 

If strangeness is found to be in thermal equilibrium with the rest of the produced particle species, it is an indication of quark-gluon plasma formation since the strangeness production is enhanced through partonic production channels compared to reactions at the hadron level \cite{Rafelski:1982pu}. There have been even claims that the peak in the kaon-to-pion ratio as a function of beam energy around $E_{\rm lab}= 40A$ GeV is a signature for the phase transition to the quark-gluon plasma \cite{NA49:2002pzu}. Also the ``step'' in the mean transverse momentum of kaons as function of beam energy has been attributed to the latent heat of the phase transition. However, in dynamical model calculations both phenomena can be attributed to more conventional effects, namely the transition from a baryon-dominated to a meson-dominated system, as well as to the softening of the equation of state due to the excitation of more active degrees of freedom associated with hadron resonances \cite{Steinheimer:2011mp,Petersen:2009mz}. 

Besides canonical suppression, there are other more advanced variants of the thermal model including, for example, van der Waals interactions \cite{Vovchenko:2019pjl} or excluded volume effects that can vary for different species \cite{Gorenstein:2007mw}. The limitations of the thermal model are that it is a static picture. It is not possible to address observables of more dynamical origin (e.g., anisotropic flow, HBT radii,..) within this approach. Also, resonances cannot be described since the rescattering of daughter particles restricts the ability of experimental collaborations to reconstruct all resonances produced at chemical freeze-out. At LHC energies, baryon annihilation during the late stage of the fireball expansion plays a significant role and affects the proton yields, which also leads to challenges in a fully thermal description \cite{Steinheimer:2012rd}. 

In addition to looking globally at particle production at midrapidity or integrated yields, one can investigate the variation of particle yields as function of rapidity. For example, the rapidity distribution of net protons gives valuable insight into the baryon stopping dynamics \cite{Blume:2005ru}. In the context of the present Section, it is important to realize that by studying more forward rapidities higher net baryon densities are accessible also at colliders like RHIC and LHC \cite{Anishetty:1980zp}. Of course, the high rapidity region is typically not equipped with as many detector elements capable of particle identification, but the high energies and high multiplicities might still provide an advantage over low beam energies, especially in fixed target experiments as they have been conducted in the STAR detector at RHIC and at the CERN-SPS. 

\subsection{Transverse Flow} 
\label{sec:trans_expansion}

The most basic dynamic observables related to transverse flow are transverse momentum distributions of (identified) particles. The particle spectra carry information about the radial expansion of the system in addition to thermal information. While the yields of particles are fixed at chemical freeze-out when inelastic reactions cease, the slopes of the transverse momentum spectra contain information about the kinetic freeze-out of the different particle species. Kinetic freeze-out is the moment when the particles no longer suffer elastic collisions and the momenta become fixed. Because elastic cross sections differ from particle to particle the effective temperatures extracted from spectra of different particles do not necessarily agree. Note that the decay of a resonance into the same (final) state  from which it was formed, such as the reaction $\pi N \rightarrow \Delta \rightarrow N\pi,$ is understood as a pseudo-elastic process, since it does not alter the chemical composition.

\begin{figure}[htb]
\centering
\includegraphics[width=0.32\linewidth]{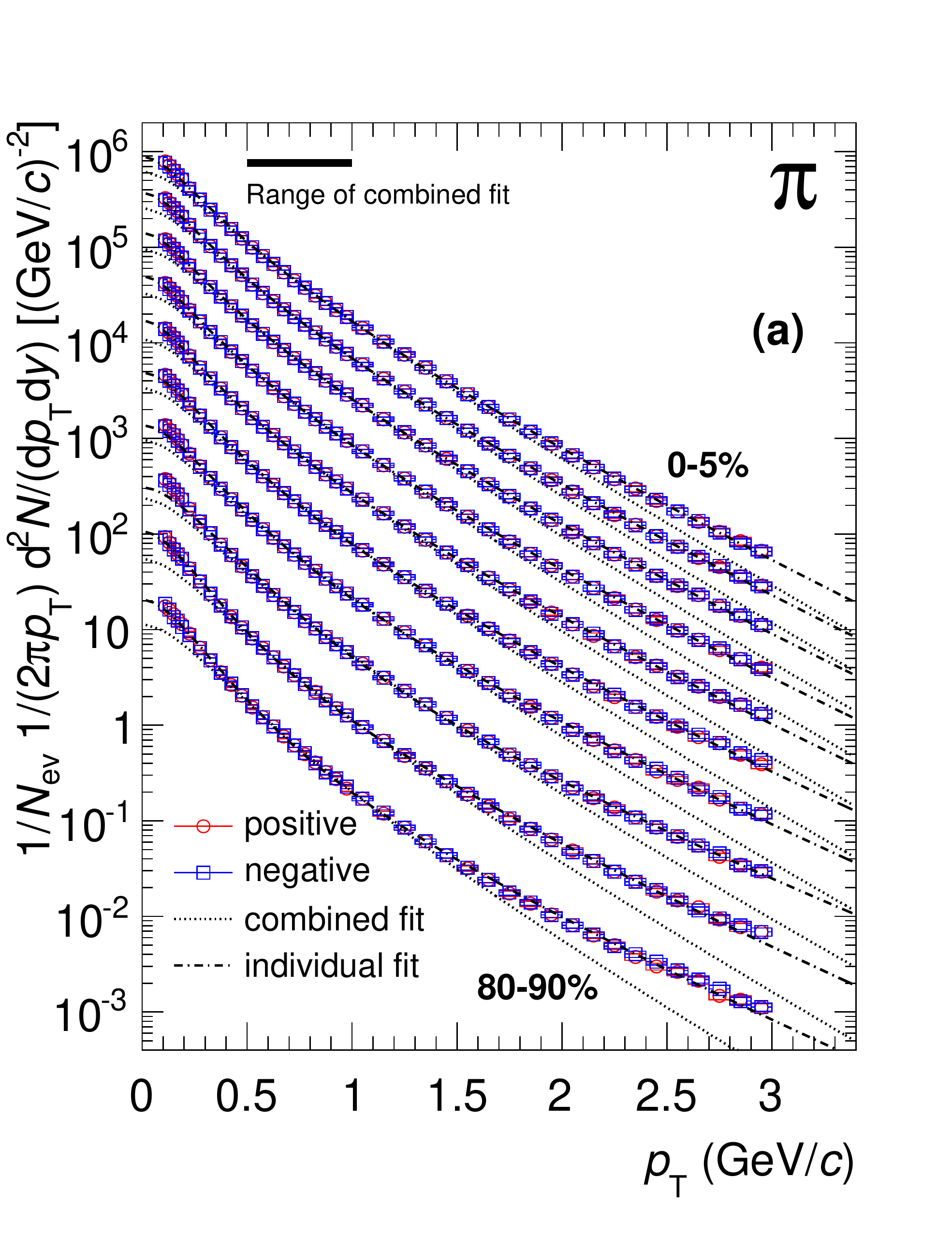}
\includegraphics[width=0.32\linewidth]{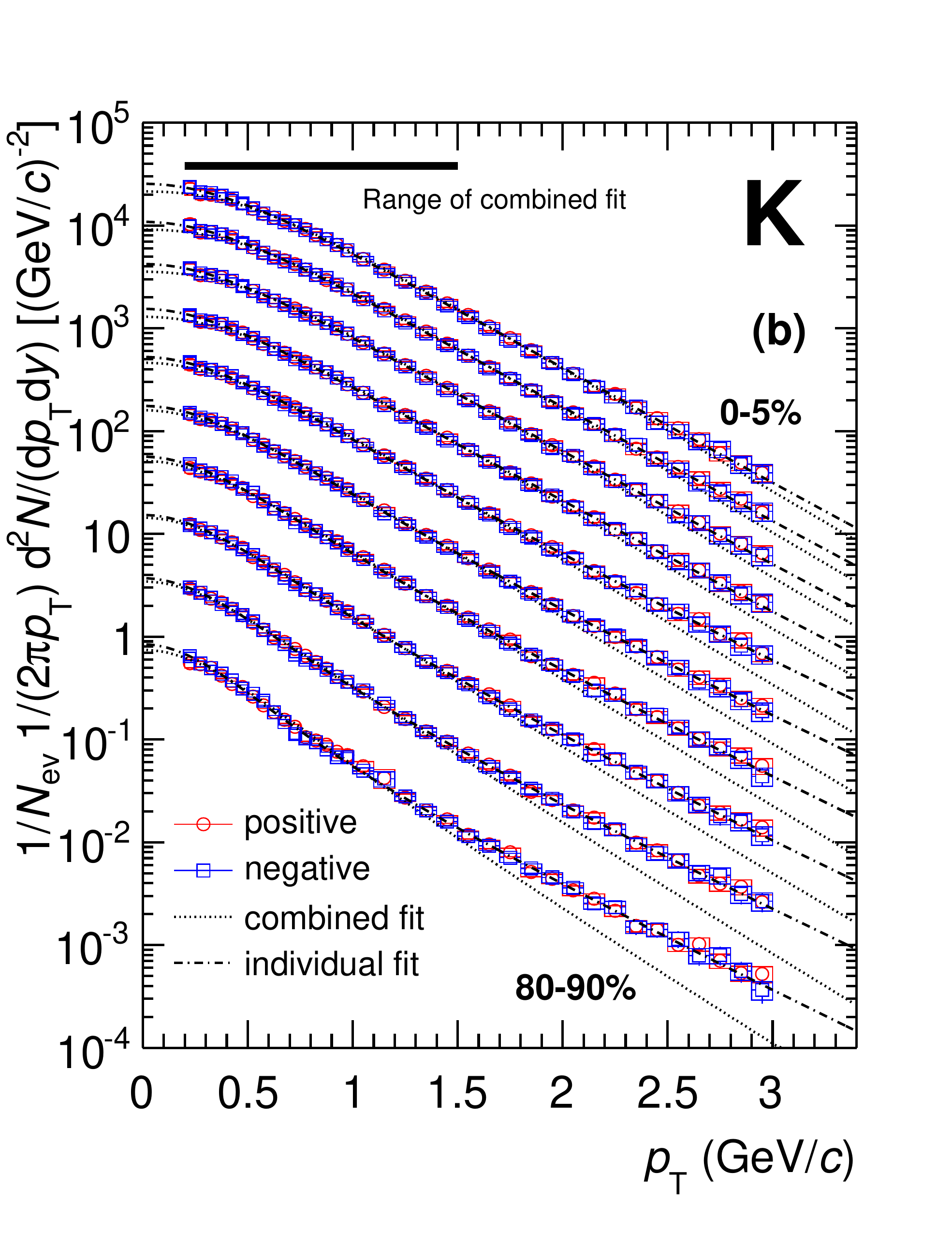}
\includegraphics[width=0.32\linewidth]{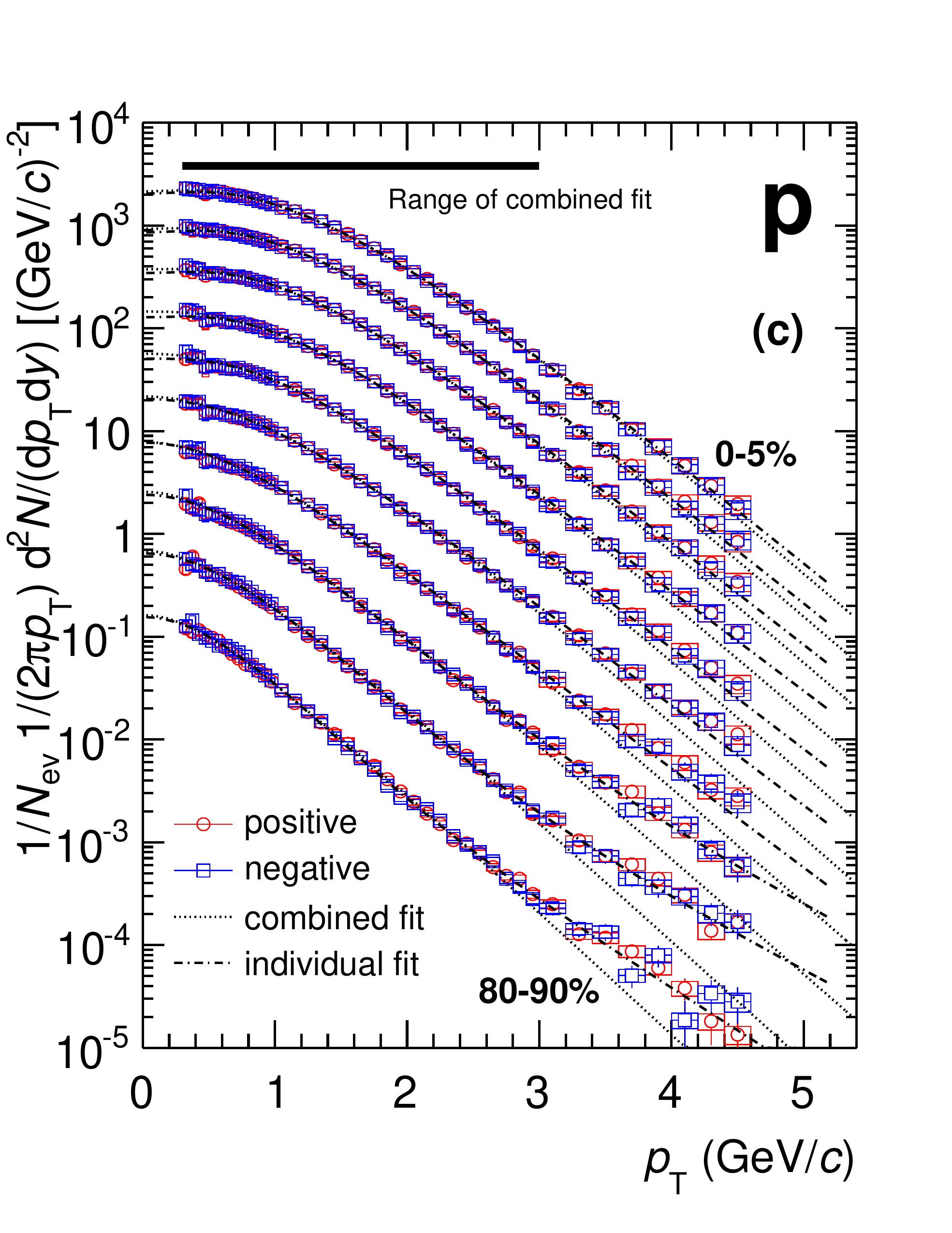}
\caption{Transverse momentum spectra of pions (a), kaons (b) and protons (c) in Pb+Pb collisions at $\sqrt{s_{\rm NN}}= 2.76$ GeV from the ALICE collaboration for different centralities (central ones are scaled). The dashed curves represent blast-wave fits to individual particles, while dotted curves indicate combined blast-wave fits.  
  [From
\cite{ALICE:2013mez}]}
\label{fig:alice_spectra}
\end{figure}

Employing the ideas introduced in the previous section \ref{sec:chem_fo} one can calculate the thermal expectation for particle spectra differentially along transverse momentum under the assumption of a Boltzmann distribution and obtains that 
\begin{equation}
  \frac{1}{p_T} \frac{\mathrm{d}N}{\mathrm{d}p_T} \propto e^{-p_T/T_{\rm eff}}.
\end{equation}
where $p_T$ is the transverse momentum, and $T_{\rm eff}$ is the slope of the distribution in a semi-logarithmic plot. In Fig. \ref{fig:alice_spectra} the transverse momentum spectra for pions, kaons and protons in heavy ion collisions at LHC energies are shown. The shoulder at low $p_T$ indicates a deviation from the purely thermal expectation that becomes more pronounced with increasing hadron mass, in particular, for protons. To obtain exponential distributions one can apply a variable transformation and fit transverse mass spectra $m_T^{-1}(dN/dm_T)$ instead of transverse momentum spectra. It is interesting that most of the information on particle production can be summarized by the integrated yield at midrapidity and the mean transverse mass. If a theoretical calculation reproduces both, the full spectra are typically described rather well. 

By measuring transverse momentum distributions in different centrality windows and rapidity bins, one can obtain the yields as a function of centrality and rapidity by integrating over these spectra. Since the acceptance of experimental detectors is often limited in the very low momentum region, one needs adequate fit functions. The particles with very low momenta are bent strongly in the magnetic fields and therefore do not reach the tracking detectors. To estimate systematic uncertainties, different functional forms are employed, e.g. pure exponential functions in $p_T$ or $m_T$, functions including quantum statistics or Tsallis-Levy distributions. 

\begin{figure}[htb]
\includegraphics[width=0.556\linewidth]{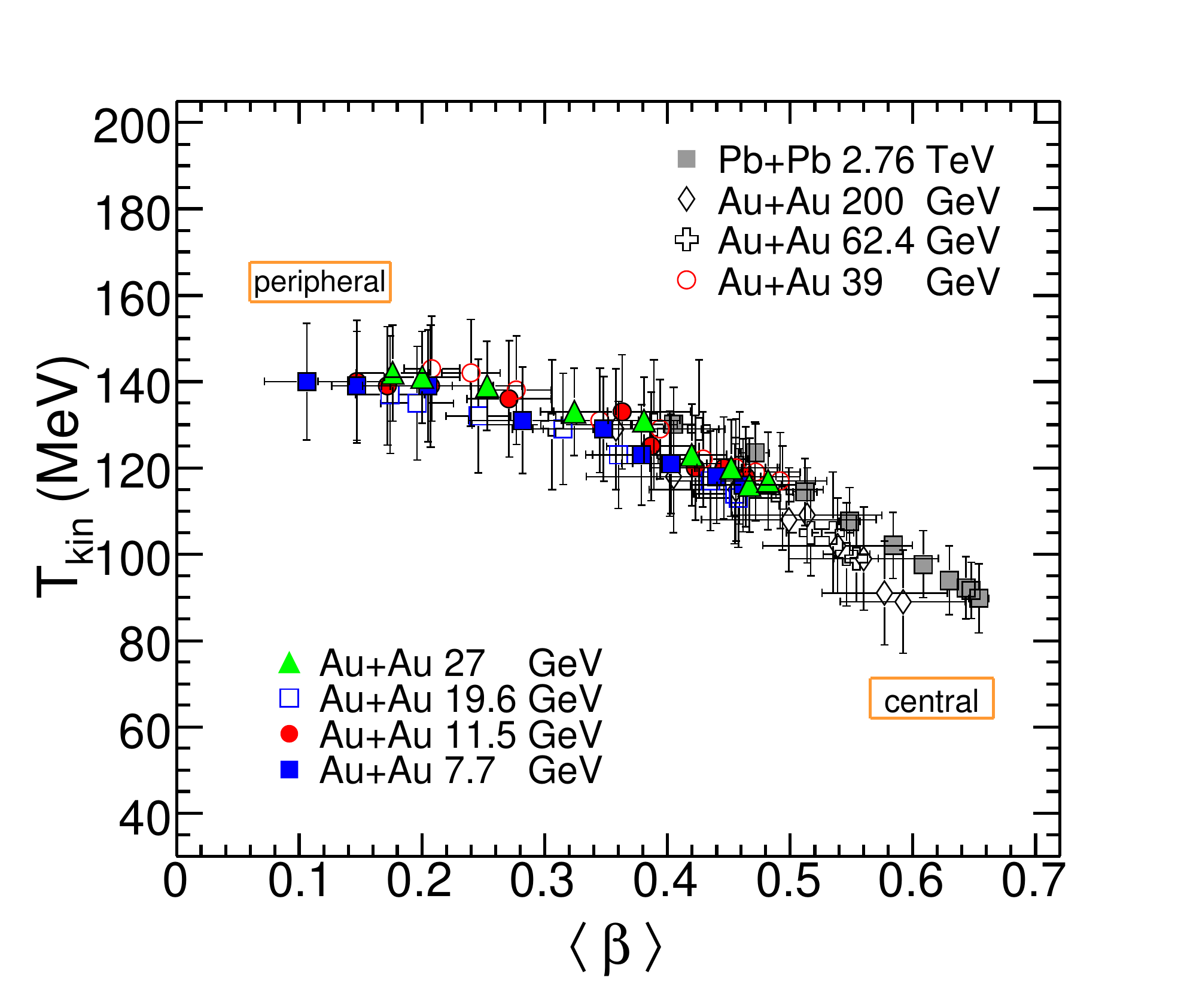}
\includegraphics[width=0.4\linewidth]{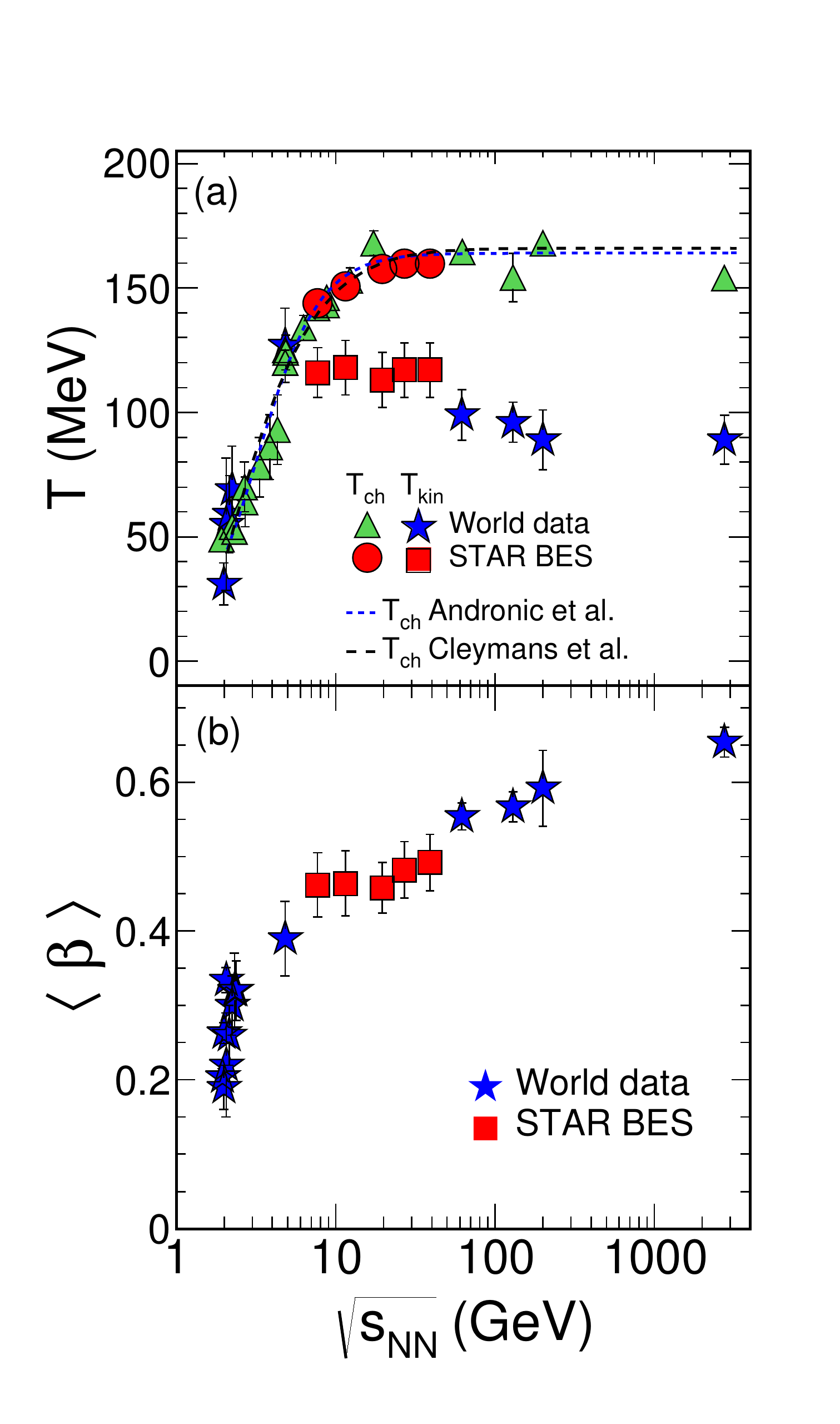}
\caption{Left: Results for kinetic freeze-out temperature and radial velocity from fits to transverse momentum spectra from the RHIC beam energy scan. Right: Kinetic freeze-out temperatures and radial velocities compared to the world data also on chemical freeze-out properties.  
  [From
\cite{STAR:2017sal}]}
\label{fig:star_fits}
\end{figure}

To extract kinetic freeze-out temperatures and average flow velocities from transverse momentum spectra of several particle species the blast-wave model is employed. The assumption here is that all particles are emitted from a boosted fireball that follows a common radial expansion profile. The different behaviour of the different species is then mainly attributed to the different masses that result in different mean transverse momenta even for the same underlying velocity profile. To obtain integrated yields and mean transverse momenta, the spectra shown in Fig. \ref{fig:alice_spectra} have been fitted individually for each species with a blast-wave function~\cite{Schnedermann:1993ws}:

\begin{equation}
  \frac{1}{p_T} \frac{\mathrm{d}N}{\mathrm{d}p_T} \propto \int_0^R r \mathrm{d}r\, m_{\rm T}\, I_0 \left( \frac{p_{\rm T}\sinh \rho}{T_{kin}} \right) K_1 \left( \frac{m_{\rm T}\cosh \rho}{T_{kin}} \right),
  \label{eq:blast-wave}
\end{equation}
where the velocity profile $\rho$ is described by
\begin{equation}
 \rho = \tanh^{-1} \beta_{\rm T}(r) = \tanh^{-1} \left( \beta_{\rm s}(r/R)^{n} \right) \; .
 \label{eq:rhoBWdefintion}
\end{equation}
Here, $m_T = \sqrt{p_T^2+m^2}$ is the transverse mass, $I_0$ and $K_1$ are modified Bessel functions, $r$ is the radial distance in the transverse plane, $R$ is the transverse radius of the fireball, $\beta_{\rm T}(r)$ is the transverse expansion velocity and $\beta_{\rm s}$ is the transverse expansion velocity at the surface.

Combining the spectra of different particle species in a joint fit makes it possible to extract kinetic freeze-out temperatures and average flow velocities for different beam energies and centralities as shown in Fig. \ref{fig:star_fits}. The results are rather similar at high RHIC and LHC energies. As a function of centrality one can see a trend that more central collisions result in lower temperatures and higher transverse velocities, while more peripheral collisions end up at higher temperatures and with less explosive radial expansion. The apparently counter-intuitive ordering of the deduced temperatures indicates that the fireball in central collisions produces more particles and is ``cooking'' longer, such that the particles decouple at lower temperatures when the system had more time to expand. One caveat is that the results are rather sensitive to the fit ranges, which have to be carefully chosen (see the black bars in Fig. \ref{fig:alice_spectra}). 

The world data shown in the right panel of Fig. \ref{fig:star_fits} reveal the expected behaviour of rising radial expansion velocities with increasing collision energy. The extracted temperatures show a more interesting pattern. At low beam energies the chemical freeze-out overlaps with the kinetic freeze-out, while above $\sqrt{s_{\rm NN}}\sim 8$ GeV the kinetic freeze-out temperatures are lower than the chemical ones and stay constant or even decrease at higher beam energies. One way to understand this follows the same argument as above for central collisions that the fireball has a longer lifetime at high beam energies and, therefore, the particles decouple later in the evolution. In particular, the cross section for pions and protons is large due to the $\Delta$-resonance, and these are the most abundant species at high beam energies.

One important consideration when comparing model calculations to experimental data is to ``match apples to apples''. For transverse mass spectra this seems straight forward, but there can be complications like feed-down corrections that have to be taken into account. In many models the $\Lambda$ hyperon is regarded as stable particle, because it is stable with respect to the strong interaction. In practice, some of the produced $\Lambda$ hyperons undergo a weak decay before they reach the detector, which is at a macroscopic distance from the collision vertex. Therefore, the reported number of protons differs by about 40 \% depending on whether the feed-down from $\Lambda$ hyperons is being considered or not. Similarly, the $\Sigma^0$ decays to almost 100\% into $\Lambda$ and needs to be added in model calculations before comparing to experimental data. 

\begin{figure}[htb]
\centering
\includegraphics[width=0.7\linewidth]{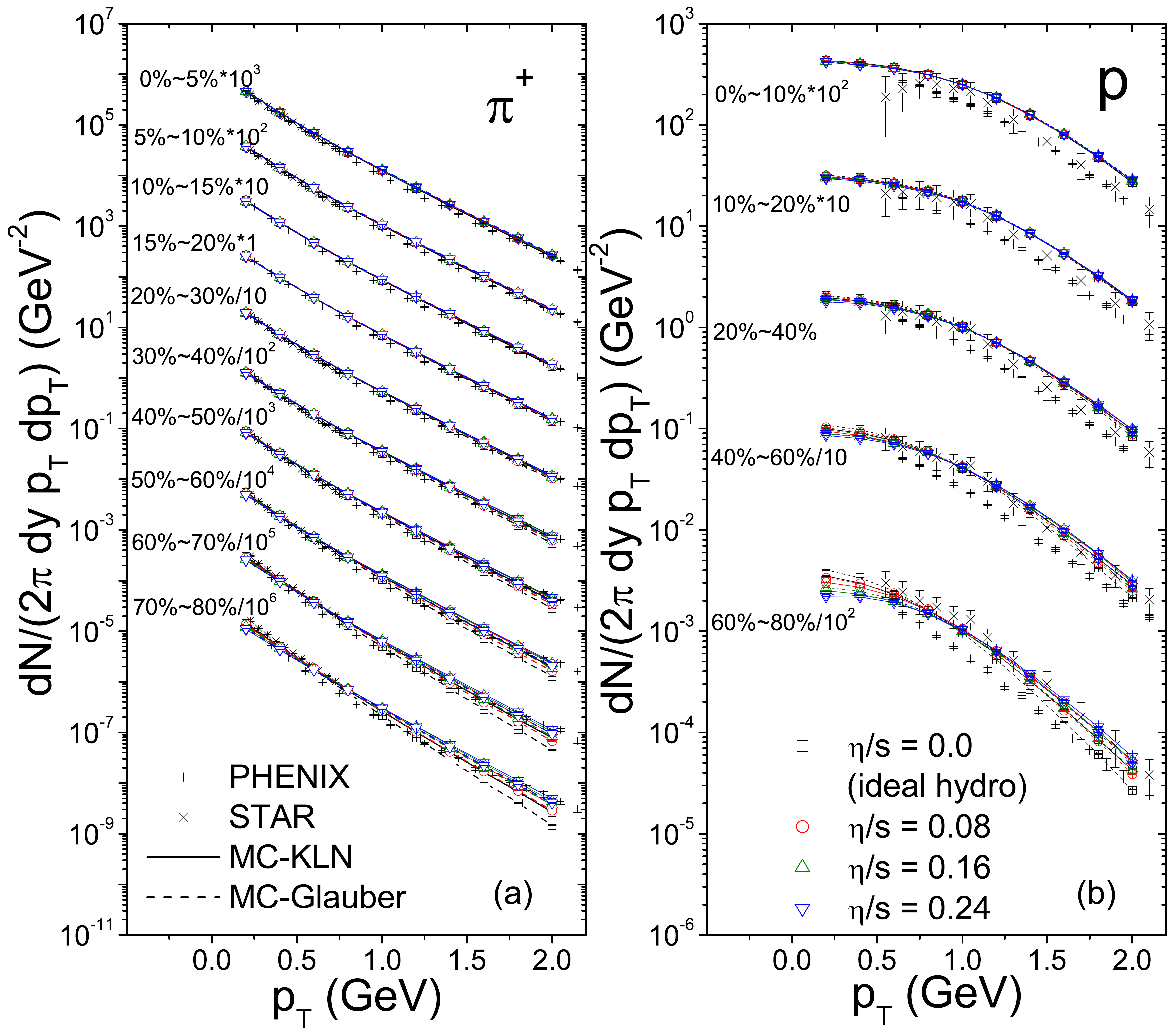}
\caption{Transverse momentum spectra of pions and protons for different centrality windows in Au+Au collisions at $\sqrt{s_{\rm NN}}= 200$ GeV compared to the VISHNU hybrid approach with different initial conditions and different values of shear viscosity over entropy density. 
  [From
\cite{Song:2011hk}]}
\label{fig:vishnu_spectra}
\end{figure}

Transverse momentum spectra and particle yields are also the basic observables that a hydrodynamic evolution needs to describe. In Fig. \ref{fig:vishnu_spectra} calculations within the VISHNU hybrid approach based on the VISH2+1 dimensional viscous hydrodynamic code and hadronic rescattering by UrQMD are compared to pion and proton spectra in Au+Au collisions at the highest RHIC energy. One can see that different levels of initial state fluctuations described by Monte-Carlo Glauber or a saturation based MC-KLN approach do not affect the spectra much. Also, varying the constant shear viscosity-over-entropy density ratio $\eta/s$ employed in the hydrodynamic evolution from 0 to 0.24 does not have a large effect. (See below for the dependence on bulk viscosity.) Of course, smaller variations might become visible when one compares results on a linear scale. In general, transverse momentum spectra can be used to gauge the basic hydrodynamic evolution parameters. The anisotropic flow discussed in the next Section \ref{sec:aniso_flow} has a higher sensitivity to the initial state fluctuations and to the properties of the quark-gluon plasma and its transport coefficients. 

\begin{figure}[htb]
\includegraphics[width=0.45\linewidth]{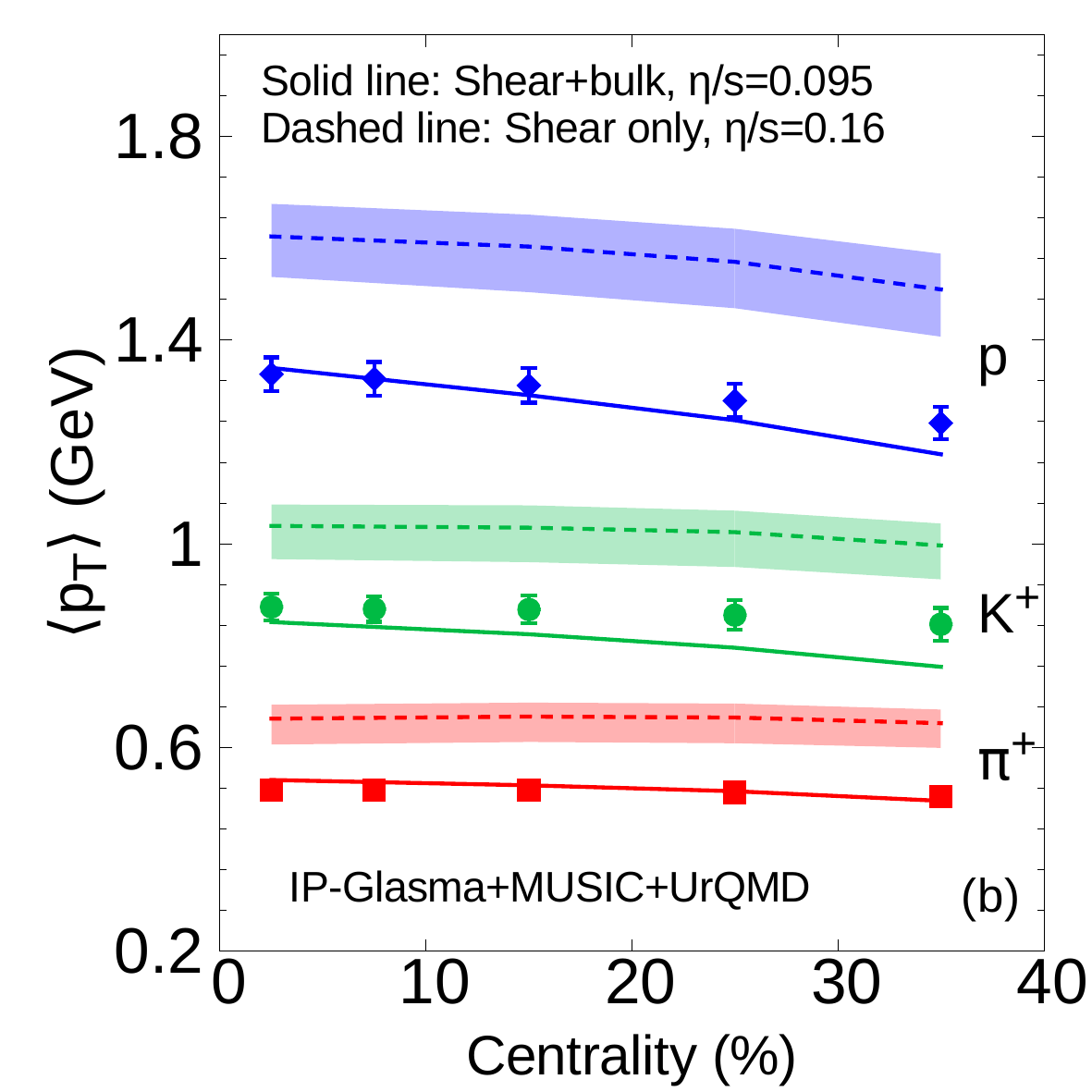}
\hspace{0.07\linewidth}
\includegraphics[width=0.48\linewidth]{figs/dNdpT_spc_allCent.pdf}
\caption{Left: Mean transverse momenta as a function of centrality in Pb+Pb collisions at $\sqrt{s_{\rm NN}}=2.76$ GeV, the bands around the dashed lines indicate the effect of changing $T_{\mathrm{switch}}$ within the MUSIC+UrQMD hybrid approach compared to experimental data. [From \cite{Ryu:2015vwa}] Right: Transverse momentum spectra for pions, kaons and protons in different centrality bins calculated within a hybrid approach compared to PHENIX measurements in AuAu collisions at $\sqrt{s_{\rm NN}}= 200$ GeV and the difference between the dashed and full lines indicates the effect of hadronic rescattering. 
  [From
\cite{Ryu:2017qzn}]}
\label{fig:hybrid_spectra}
\end{figure}

Fig. \ref{fig:hybrid_spectra} (left) shows the mean transverse momentum of pions, kaons and protons in heavy ion collisions at LHC energies, measured by the ALICE collaboration, in comparison with a hybrid calculation relying on IP-Glasma fluctuating initial conditions, the 3+1 dimensional viscous hydrodynamic implementation MUSIC and hadronic rescattering by UrQMD. As a function of centrality, the mean transverse momenta are found to be very sensitive to the inclusion of bulk viscosity in the evolution. A temperature dependent bulk viscosity with a peak around the transition temperature was employed in this study. While there is the caveat that this calculation still used an equation of state with a transition temperature around $T_c=190$ MeV, the qualitative differences are expected to remain also with an equation of state fitted to state-of-the-art lattice calculations (see e.g. the recent study of deuterons and their sensitivity to bulk viscosity in \cite{JETSCAPE:2022cob}). The right panel of Fig. \ref{fig:hybrid_spectra} shows the effect of hadronic rescattering. While pions and kaons are only mildly affected, the transverse momentum of protons increases by about 30\% during the late stage due to rescattering. This can be attributed to the large cross sections of nucleons with fast-moving pions; the phenomenon is sometimes referred to as ``pion wind''. 

\subsection{Geometry of the Fireball}
\label{sec:hbt}

\begin{figure}[htb]
\centering
\includegraphics[width=0.48\linewidth]{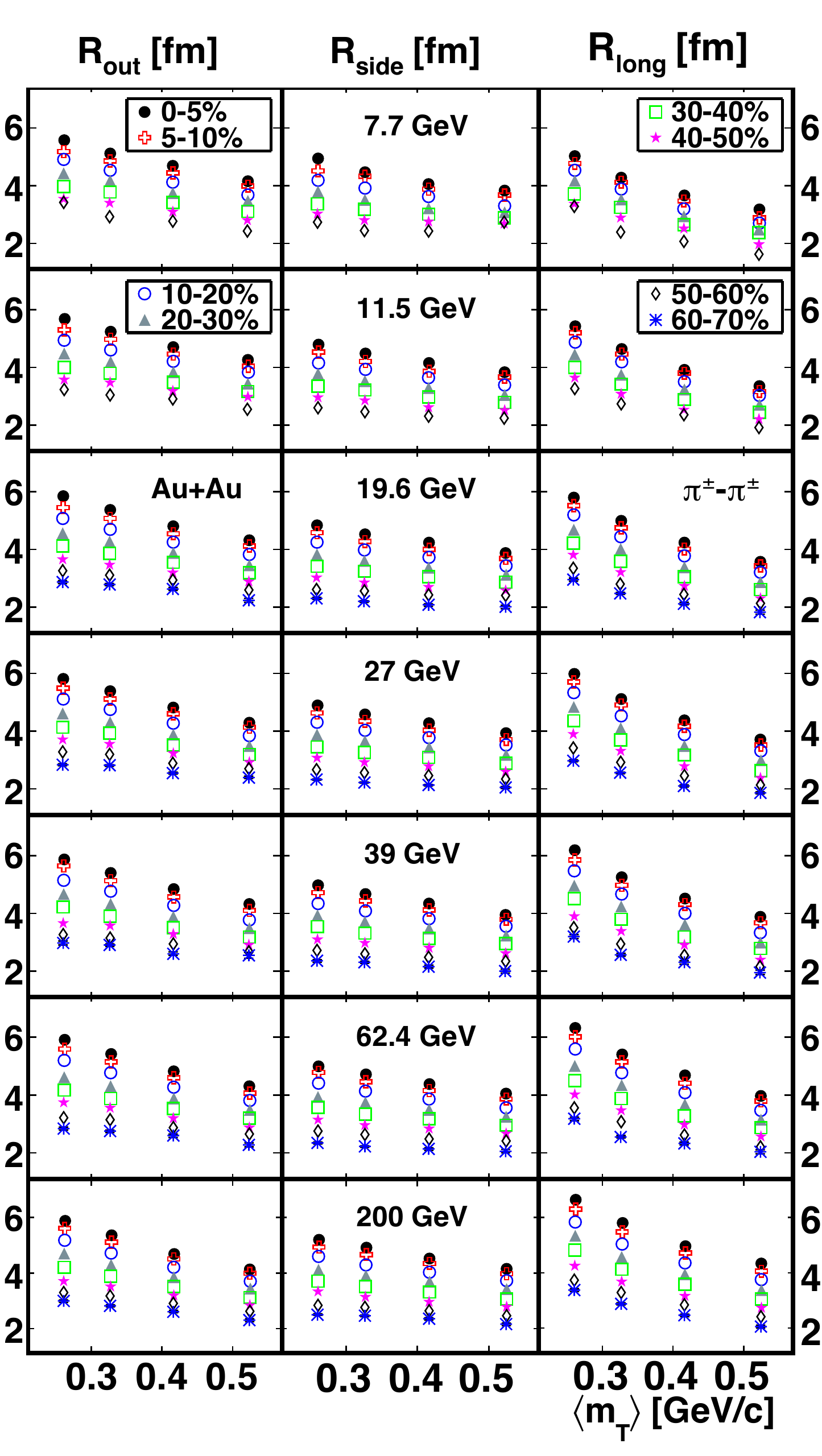}
\hspace{0.04\linewidth}
\includegraphics[width=0.42\linewidth]{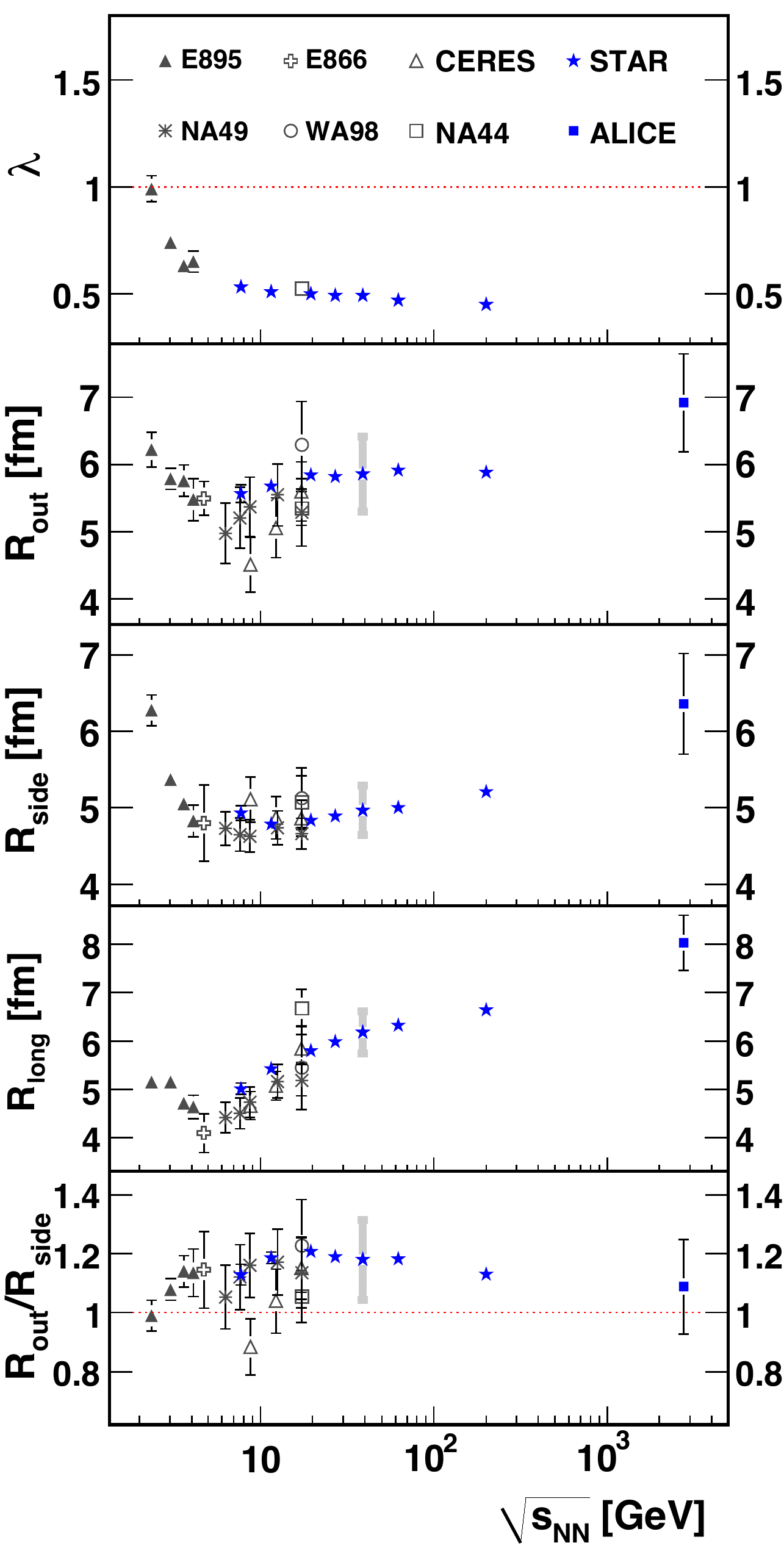}
\caption{Left: HBT radii as a function of transverse mass for different beam energies and centralities extracted from identical pion pairs. Right: Excitation function of $\lambda$ parameter, and the HBT radii including the ratio $R_{\rm o}/R_{\rm s}$ from identical pion pairs.  [From
\cite{STAR:2014shf}]}
\label{fig:hbt_radii}
\end{figure}

Measurements of correlation functions between particle pairs provide information about the geometry of the emission region in heavy-ion collisions. The inferred size of the so-called ``region of homogeneity'' reflects the kinetic freeze-out, since these measurements are performed on final state hadrons. This technique is originally applied in the size estimation of stars by Hanbury-Brown and Twiss (HBT). The shape of the correlation function is typically assumed to be Gaussian with the following parametrization: 
\begin{eqnarray}
C \left(\vec{q}\right)&=& \left(1-\lambda\right) + \lambda\, K_{\mathrm{Coul}}(q_{\mathrm{inv}}) 
\nonumber \\
&& \quad \times\exp\left(-q^{2}_{\rm o}R^2_{\rm o}-q^{2}_{\rm s}R^{2}_{\rm s}-q^{2}_{\rm l}R^{2}_{\rm l}-2q_{\rm o}q_{\rm s}R^{2}_{\rm os}-2q_{\rm o}q_{\rm l}R^{2}_{\rm ol}\right)
\end{eqnarray}
where a value of the chaoticity parameter $\lambda < 1$ accounts for dilution of the HBT signal due to particle correlations from resonance decays, and the different radii ($R_{\rm o}$, $R_{\rm s}$ and $R_{\rm l}$) refer to different directions. The out-direction ($R_{\rm o}$) points along the direction of the average momentum of the particle pair, the long-direction ($R_{\rm l}$) along the beam direction and the side-direction $R_{\rm s}$ denotes the third orthogonal direction. The contribution proportional to $\lambda$ results from quantum correlations due to quantum statistics. When the data are insufficient for a three-dimensional analysis, the correlation function is often assumed to be spherically symmetric, 
\begin{equation}
C \left(\vec{q}\right) = \left(1-\lambda\right) + \lambda\, K_{\mathrm{Coul}}(q_{\mathrm{inv}})\,  \exp\left(-|\vec{q}|^{2}R^2_{\rm inv}\right)
\end{equation}
with a single Gaussian source radius $R_{\rm inv}$.  A review on the topic can be found in \cite{Lisa:2005dd}.

Fig.~\ref{fig:hbt_radii} depicts the results from experimental measurements of HBT radii for identical pion pairs over a large beam energy range and for different centrality classes. While at lower beam energies the radii have almost the same size, at higher beam energies $R_{\rm l}$ is longer than the two transverse radii due to the dominating longitudinal expansion of the fireball. As a function of mean transverse mass of the particle pairs, there is a global decreasing trend. This can be easily understood due to the fact that faster particles are emitted earlier in the evolution and therefore from smaller source sizes than slower ones. 

Identical particle correlations have also bee measured for neutral and charged kaons and for protons \cite{ALICE:2015hvw}. The results are generally consistent with those obtained for pions, but slightly smaller source sizes have been measured for kaons perhaps pointing to their earlier kinetic freeze-out. Information about source sizes can also be gleaned from unlike charged particle correlations, where the  information resides in the factor $K_{\rm Coul}(q)$ that accounts for final-state Coulomb and other interactions between the emitted particles (see e.~g.~\cite{ALICE:2020mkb}). In the future, looking at HBT correlations for dileptons and photons can provide more information about the geometry of the system during the evolution since electromagnetic probes leave the fireball from all stages without disturbance from rescattering. 

\begin{figure}[htb]
\flushright
\vspace{-1.5cm}
\includegraphics[width=0.6\linewidth]{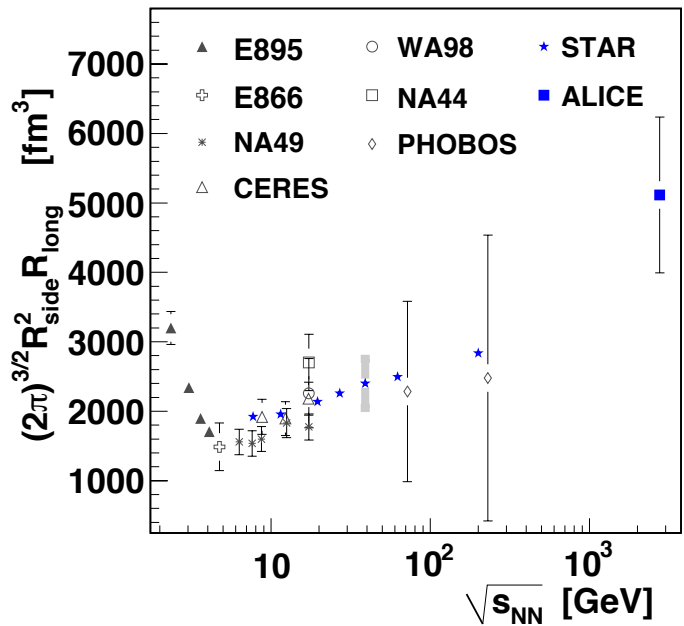}
\vspace{-1.5cm}
\caption{Volume of the emission region extracted from HBT radii of identical pions as a function of beam energy. [From \cite{STAR:2014shf}]}
\label{fig:hbt_volume}
\end{figure}

The excitation function of the radius parameters as well as the $\lambda$ parameter and the $R_{\rm o}/R_{\rm s}$ ratio is summarized in the right panel of Fig.~\ref{fig:hbt_radii}. The ratio of the two transverse radii is predicted to be sensitive to the lifetime of the system. Interestingly, this ratio stays rather constant over a large range of beam energies. The fraction of correlated particles $\lambda$ is higher at lower beam energies and saturates at higher energies to a lower constant value, which points to an increasing fraction of particles originating from longer-lived resonances in the sample. $R_{\rm out}$ and $R_{\rm side}$ stay more or less constant as a function of beam energy while $R_{\rm long}$ is rising significantly indicating a longer lifetime of the overall system evolution. 
By combining the transverse HBT radii one can infer a measure of the fireball volume at kinetic freeze-out as shown in Fig. \ref{fig:hbt_volume}. These volumes can be compared to the one required by the thermodynamic description in Section \ref{sec:chem_fo} and the qualitative behaviour as a function of beam energy is consistent. In general, the analysis of HBT radii and their comparison to hydrodynamic calculations allows to connect coordinate and momentum space descriptions that are otherwise hard to achieve, see e.~g.~\cite{Plumberg:2015eia}. The complexity of this connection was analyzed in \cite{Pratt:2008qv} where it was shown that multiple aspects of the dynamical evolution influence the size of the HBT radii and that it is non-trivial to get the bulk evolution in agreement with the correlation measurements. 

\begin{figure}[htb]
\flushright
\includegraphics[width=0.7\linewidth]{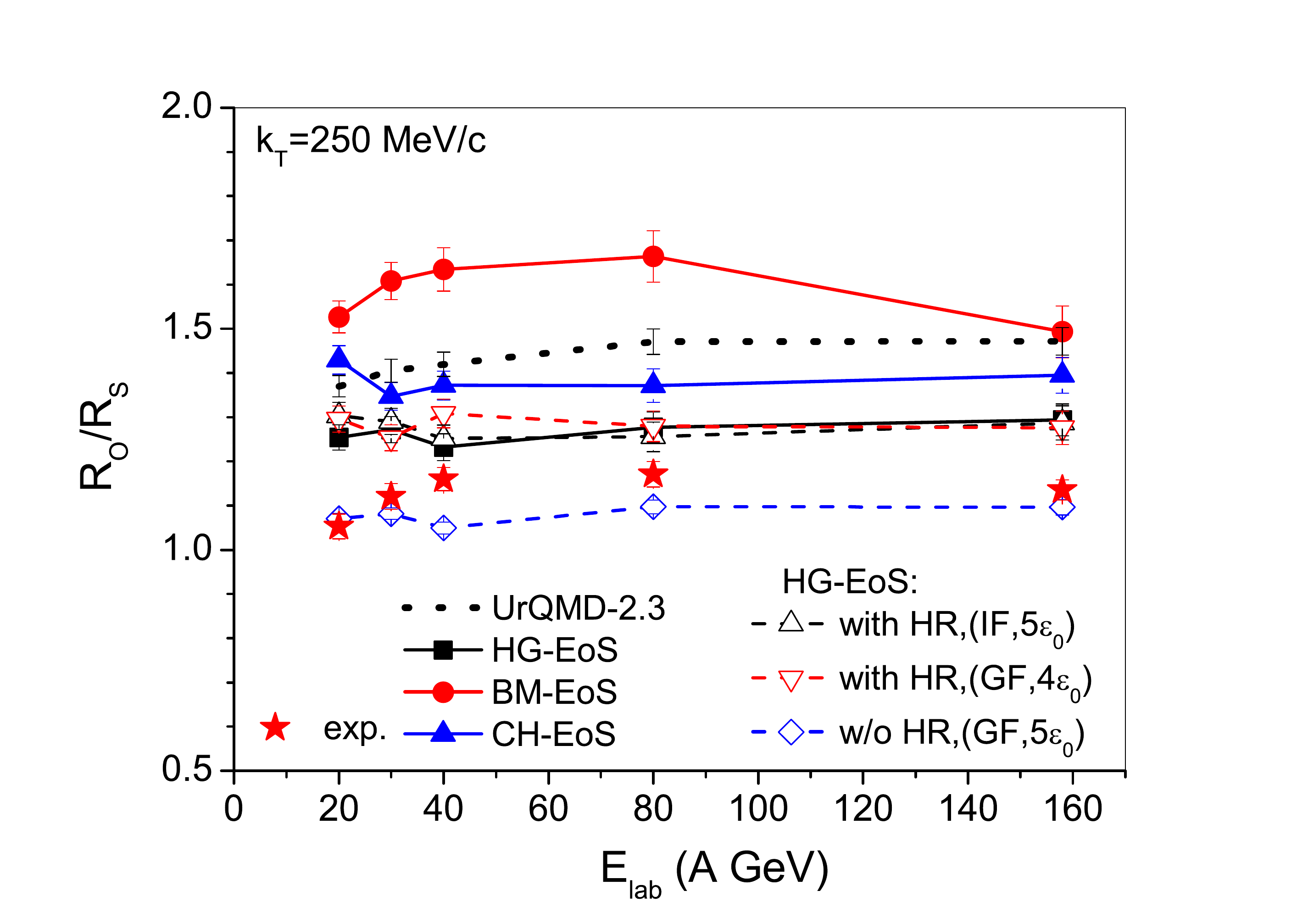}
\caption{$R_{\rm o}/R_{\rm s}$ ratio measured by NA49 as a function of beam energy compared to several calculations within a transport+hydrodynamics hybrid approach. [From \cite{Li:2008qm}]}
\label{fig:hbt_phase_transition}
\end{figure}

A large value of the $R_{\rm o}/R_{\rm s}$ ratio was proposed as one of the signatures of a first-order phase transition as a long duration of particle emission increases the outward directed correlation length $R_{\rm o}$. When the system undergoes a phase transition and the pressure is reduced significantly, the lifetime is expected to be significantly extended \cite{Rischke:1996em, Heinz:2002un}. In such a scenario one expects an increase in the $R_{\rm o}/R_{\rm s}$ ratio as a function of beam energy. Fig.~\ref{fig:hbt_phase_transition} shows a calculation within a hybrid approach based on UrQMD initial conditions, (3+1)-dimensional ideal hydrodynamic evolution, and final hadronic transport within UrQMD. Different switching energy density criteria as well as the equation of state influence the lifetime. The noticeable result is that the only curve that qualitatively follows the slight peak as a function of beam energy that the experimental data from the NA49 collaboration suggests is the one with a first-order phase transition. Detailed correlation observables have the potential to hint at a first-order phase transition between hadron gas and quark-gluon plasma although, when the dynamical evolution is modeled more realistically, the effect is much smaller than originally estimated.

\subsection{Anisotropic Flow and the ``Perfect'' Fluid}
\label{sec:aniso_flow}

\begin{figure}[htb]
\includegraphics[width=0.48\linewidth]{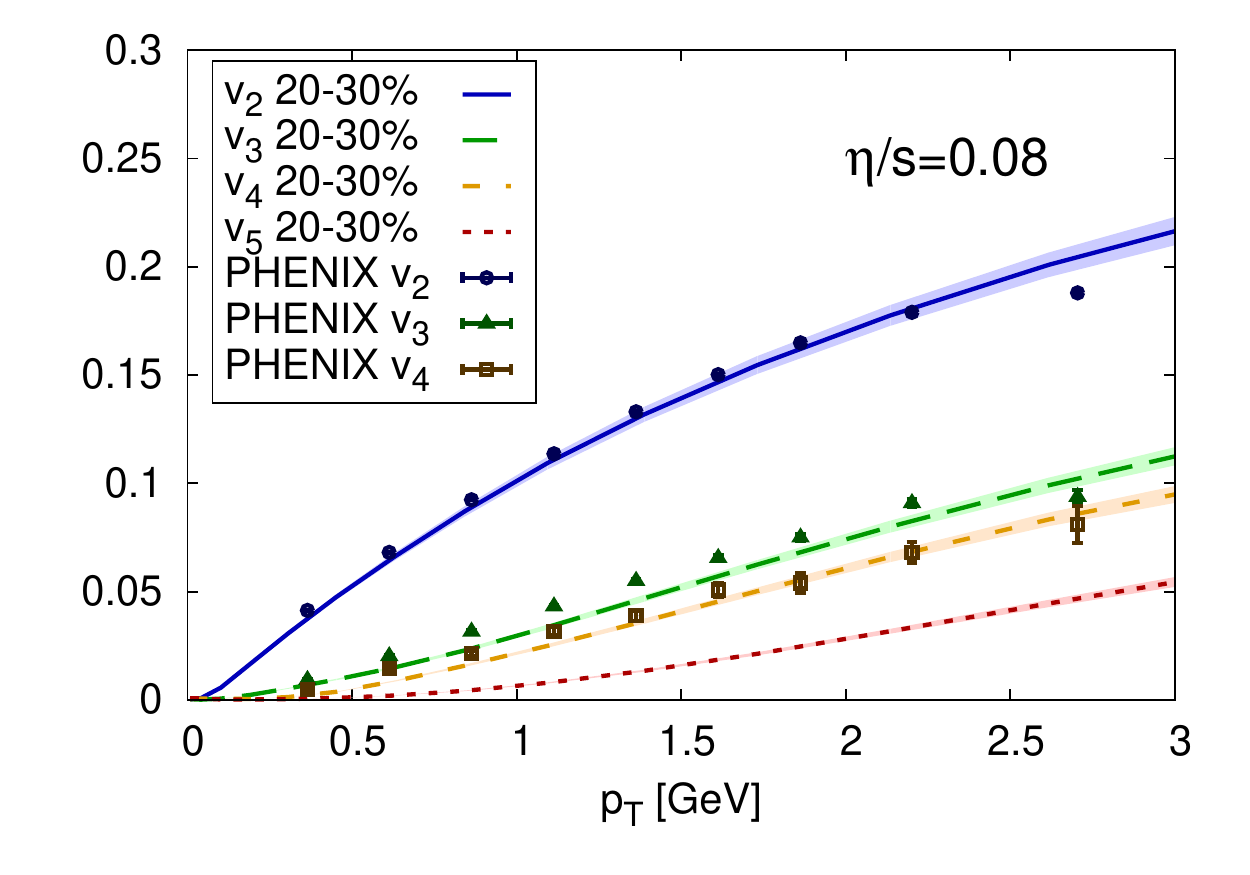}
\hspace{0.04\linewidth}
\includegraphics[width=0.48\linewidth]{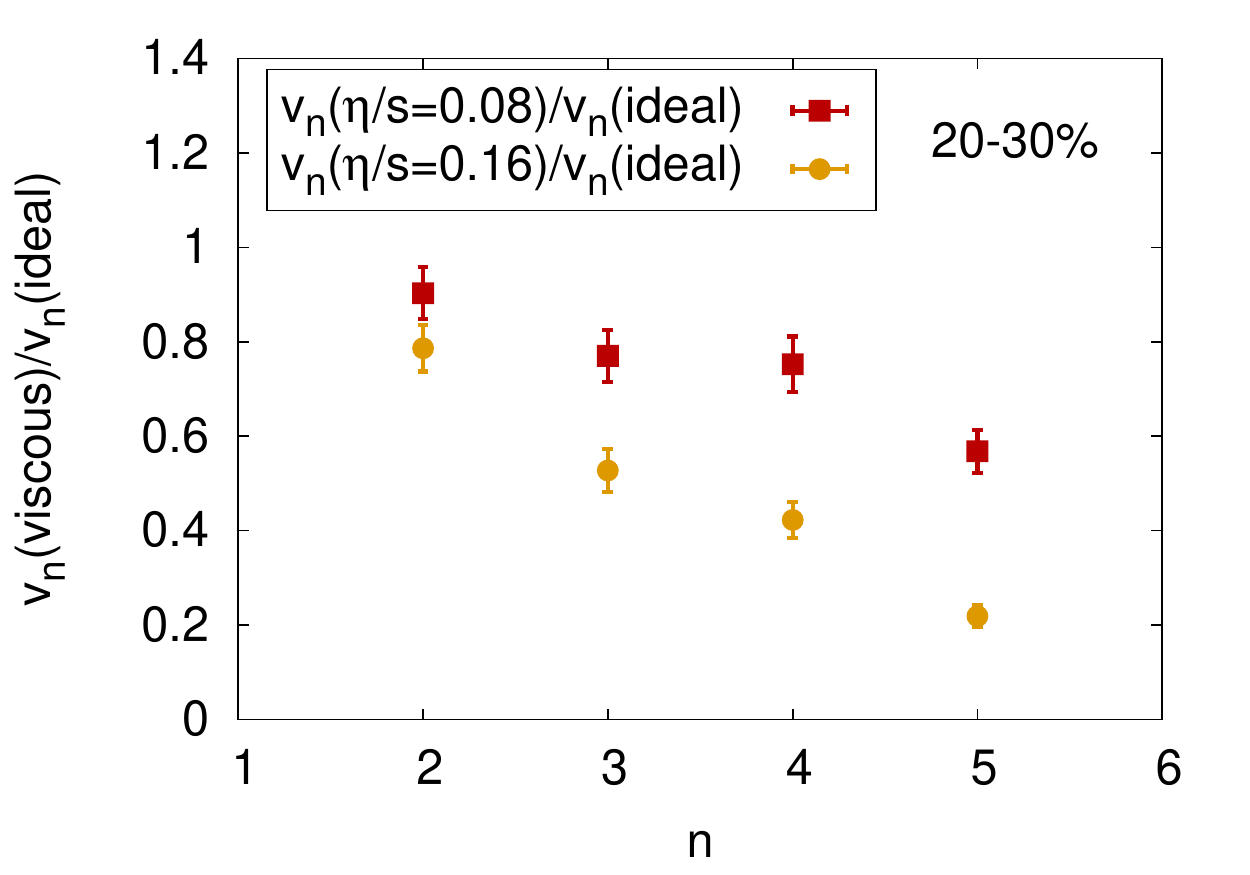}
\caption{Left: $v_n$ coefficients calculated in viscous hydrodynamics with Monte Carlo Glauber initial conditions and Cooper-Frye freeze-out on a hypersurface of constant temperature. Right: Ratio of $v_n$ coefficients for different values of effective shear viscosity over entropy ratio. [From
\cite{Schenke:2011bn}]}
\label{fig:vn_hydro}
\end{figure}

The anisotropic flow coefficients as defined in Subsection \ref{sec:coll_observables} contain interesting information about the properties of the quark-gluon plasma as well as the initial conditions. The angular modulations of collective flow in the plane transverse to the beam axis are analysed in a Fourier decomposition. Non-vanishing values of $v_n$ have been measured to high precision at RHIC and LHC as a function of transverse momentum, particle species and centrality. When hydrodynamic calculations are tuned to describe the yield and transverse momentum spectra for a system, the anisotropic flow can be very well predicted \cite{Kolb:2003dz, Huovinen:2001cy}. This central finding that lies at the foundation of our ``standard model'' of heavy ion collisions is based on a $\le 10$\% effect (for the average values)  on the background radial expansion discussed in the previous Section \ref{sec:trans_expansion}. The left panel of Fig.~\ref{fig:vn_hydro} shows one such example of a (3+1)-dimensional viscous hydrodynamic calculation for Au+Au collisions at $\sqrt{s_{\rm NN}} = 200$ GeV that fits the charged particle anisotropic flow as a function of transverse momentum.  

The agreement of hydrodynamic calculations with anisotropic flow measurements forms the basis for the paradigm that the quark-gluon plasma exhibits an extremely low ratio of shear viscosity to entropy density and is therefore one of the most ``perfect'' fluids in nature. Interestingly, this feature is shared with ultra-cold matter that can be probed in the laboratory, where atoms are confined in a trap in an almond-shaped configuration. Once released the particle stream along the pressure lines and the coordinate space anisotropy is transformed into a momentum space anisotropy \cite{OHara:2002pqs}. The complex response function of the nonlinear hydrodynamic evolution can be dissected into eccentricities of different order and combinations of eccentricities that lead to the final-state flow coefficients (see e.g. \cite{Teaney:2012ke}).  

\begin{figure}[ht]
\centering
\includegraphics[width=0.8\linewidth]{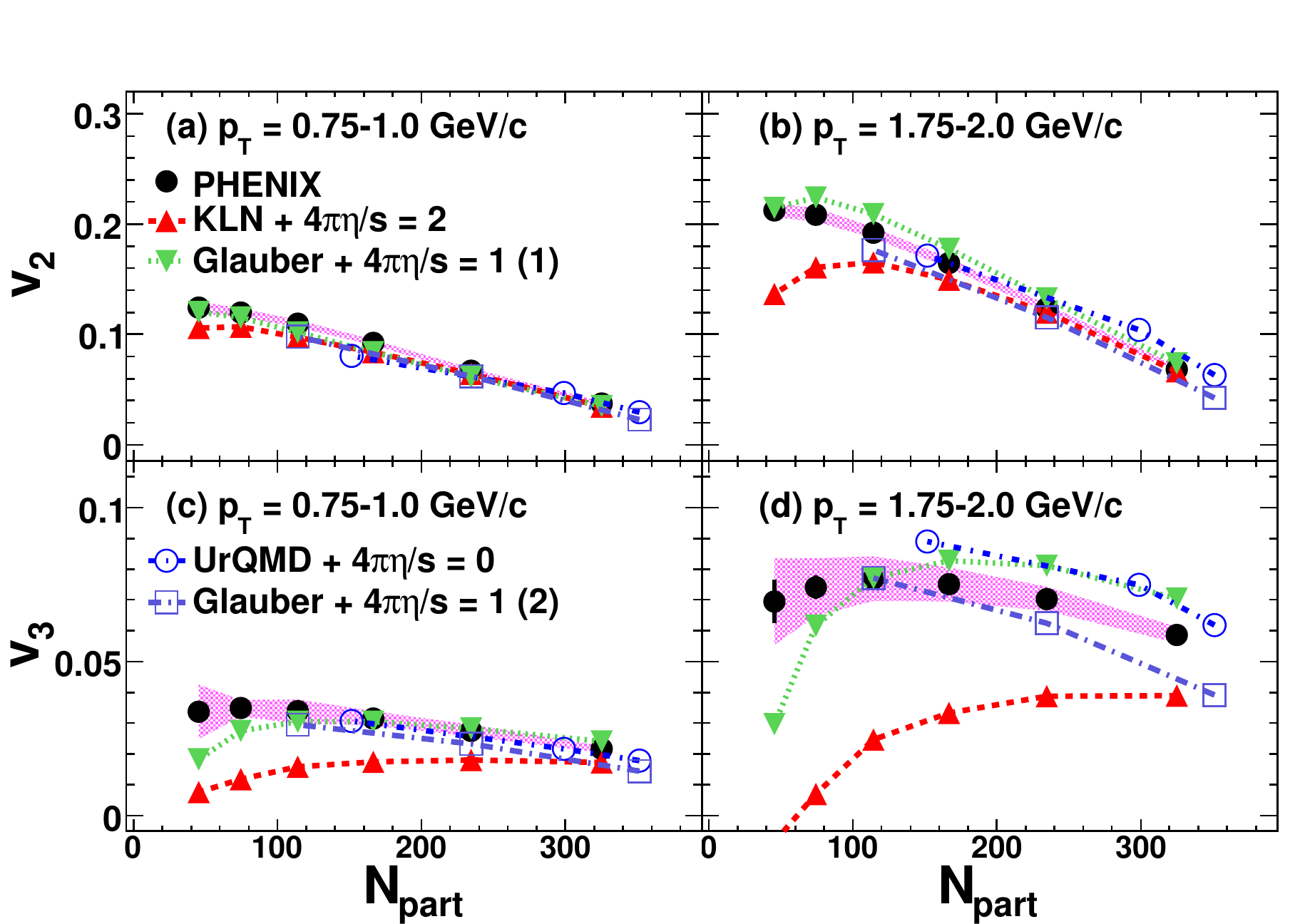}
\caption{Comparison of $v_2\{\Psi_2\}$ (panels (a) and (b)) and $v_3\{\Psi_3\}$ (panels (c) and (d)) as a function of centrality compared to different fluid dynamic calculations based on various initial conditions (``MC-KLN + $4\pi\eta/s = 2$" 
and ``Glauber + $4\pi\eta/s = 1$ (1)" \cite{Alver:2010dn}; 
``Glauber + $4\pi\eta/s = 1$ (2)" \cite{Schenke:2010rr}; 
and ``UrQMD" \cite{Petersen:2010cw}). [From
\cite{PHENIX:2011yyh}]}
\label{fig:phenix_v3_v2}
\end{figure}

Higher order flow coefficients probe smaller scale structures and are more sensitive to the properties of the plasma. Figure \ref{fig:vn_hydro}(right) demonstrates the effect of increasing the shear viscosity-over-entropy density ratio from 0 to 0.08 and 0.16. Higher viscosities lead to smaller flow coefficients since the initial state structures are diluted more quickly. Without initial state fluctuations the odd flow coefficients would be zero by symmetry. But since 2010 it has been recognized that triangular flow (and higher coefficients) are actually non-zero resulting from small scale structures in the initial state \cite{Alver:2010gr}. Figure \ref{fig:phenix_v3_v2} shows the first measurement of triangular flow as a function of the number of participants in Au+Au collisions at $\sqrt{s_{\rm NN}}=200$ GeV. While all calculations agree for elliptic flow in central collisions, the $v_3$ measurement is highly sensitive to the initial state structures (see \cite{Luzum:2013yya} for a review). On one hand, this poses the challenge that collective flow is not only sensitive to the medium properties but also to the details of the initial state. On the other hand, this also offers the opportunity to learn something about the initial state created by two nuclei colliding close to the speed of light.  

By systematic improvements of theory and experiment it has been confirmed that all observed collective flow data are consistent with a quark-gluon plasma that exhibits a very low specific shear viscosity that is close to the lower bound predicted by AdS/CFT for a strongly coupled gauge plasma of $\eta/s = (4\pi)^{-1} = 0.08$. The averaged flow coefficients as a function of centrality are the most sensitive observables for this purpose. The full event-by-event distributions of the flow coefficients are mainly sensitive to the initial state and its fluctuations, while the viscosity mainly influences the average. Nowadays, this complex many parameter-many observable inverse problem is attacked with Bayesian analysis techniques as discussed in Section \ref{sec:Bayesian}. 

The flow data for identified particles are sensitive probes of the hadronization mechanism and the dynamics of hadronic rescattering in the late stage that can be described by transitioning to a transport treatment for the fireball evolution (see \cite{Petersen:2014yqa} for a review). Additional information on the details of the initial conditions and the hydrodynamic response can be inferred from a vast amount of data on correlations of flow coefficients of different order, the correlation of the flow cofficients with transverse momentum and similar more advanced observables. 

Even at the highest currently available collision energies the hot matter produced in the collision is not entirely boost invariant. One way by which this shows up is the gradual decorrelation of the event plane angles $\Psi_n$ (\ref{eq:vn}) measured in different pseudorapidity windows $\Delta\eta$. The correlation coefficient $r_n(\eta)$ is defined as
\begin{equation}
    r_n(\eta) = \frac{\langle 
    {\bf q}_n(-\eta){\bf q}^*_n(\eta_0) + 
    {\bf q}_n(\eta){\bf q}^*_n(-\eta_0) \rangle}
    {\langle {\bf q}_n(\eta){\bf q}^*_n(\eta_0) + 
    {\bf q}_n(-\eta){\bf q}^*_n(-\eta_0) \rangle} ,
\end{equation}
where $\eta_0$ denotes a reference pseudo-rapidity window usually chosen at a far-forward or far-backward pseudorapidity.

\begin{figure}[ht]
\centering
\includegraphics[width=0.8\linewidth]{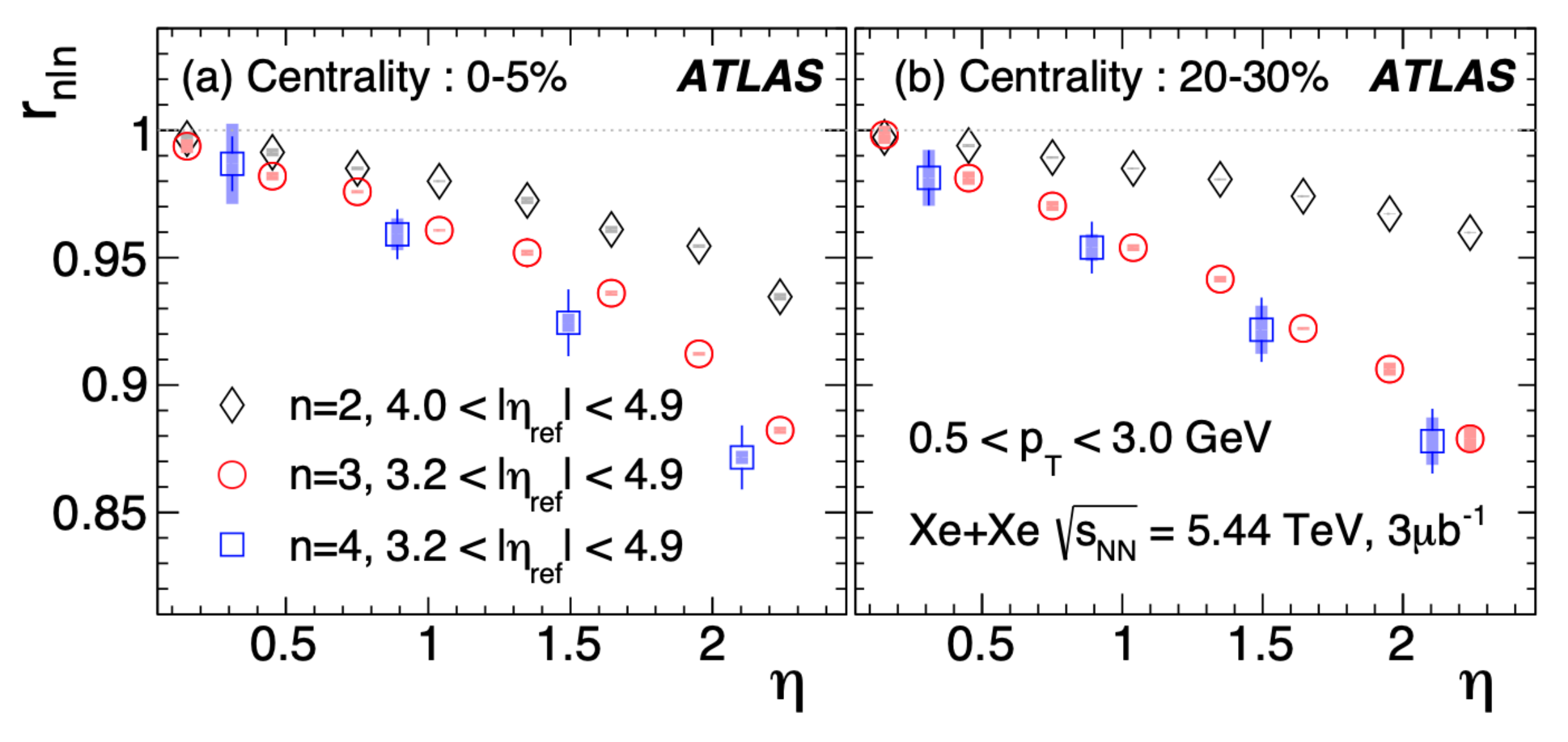}
\caption{Longitudinal decorrelation of the event plane $\Psi_n$ for $n=2,3,4$ in Xe+Xe collisions at the highest LHC energy measured by ATLAS. [From \cite{ATLAS:2020sgl}]}
\label{fig:Psi_R_LHC}
\end{figure}

\begin{figure}[ht]
\centering
\includegraphics[width=0.8\linewidth]{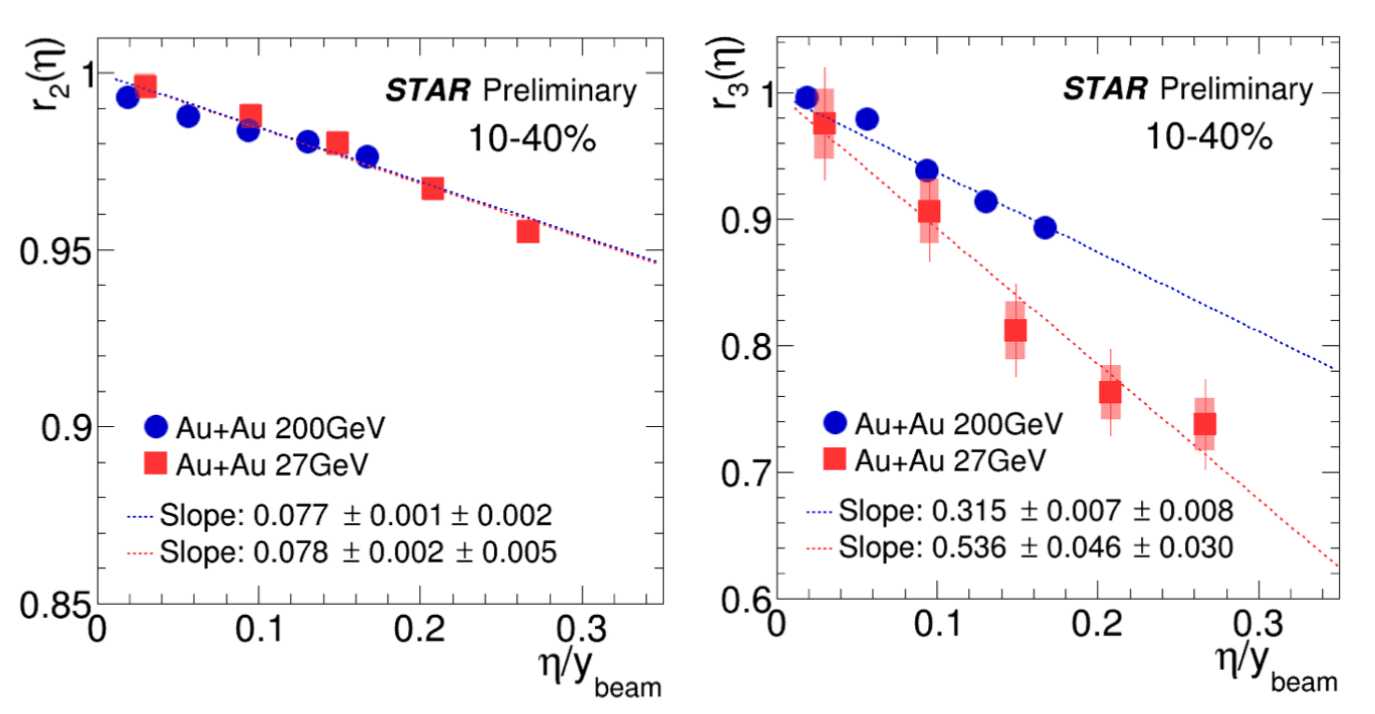}
\caption{Longitudinal decorrelation of the event plane $\Psi_n$ for $n=2,3$ in Au+Au collisions at RHIC at two different collision energies measured by STAR. [From \cite{Nie:2020trj}]}
\label{fig:Psi_R_RHIC}
\end{figure}

Figure \ref{fig:Psi_R_LHC} shows the longitudinal decorrelation of the event planes $\Psi_n$ for $n=2,3,4$ in Xe+Xe collisions at the highest LHC energy. The $n=2$ event plane is seen to decorrelate less rapidly than the $n=3,4$ planes. This phenomenon can be attributed to the predominantly geometric origin of elliptic flow, whereas the higher anisotropic flow coefficients are more sensitive to initial density fluctuations, for odd $n$ exclusively so.  Results for different energies at RHIC, shown in Fig.~\ref{fig:Psi_R_RHIC}, show that the decorrelation occurs much more rapidly at lower energies, in agreement with the expectation of larger deviations from boost invariance.


The current theoretical understanding is that the  decorrelation can be traced back to rapidity dependent fluctuations in the gluon densities of the colliding nuclei and to the gluon-gluon interactions that seed the initial state of the fireball. In addition, hydrodynamic fluctuations during the expansion of the quark-gluon plasma probably also contribute to the decorrelation. Theorists have successfully modeled these fluctuations and found rough agreement with the measured decorrelation (see e.~g.~\cite{Sakai:2021pev}). 

\begin{figure}[ht]
\centering
\includegraphics[width=0.6\linewidth]{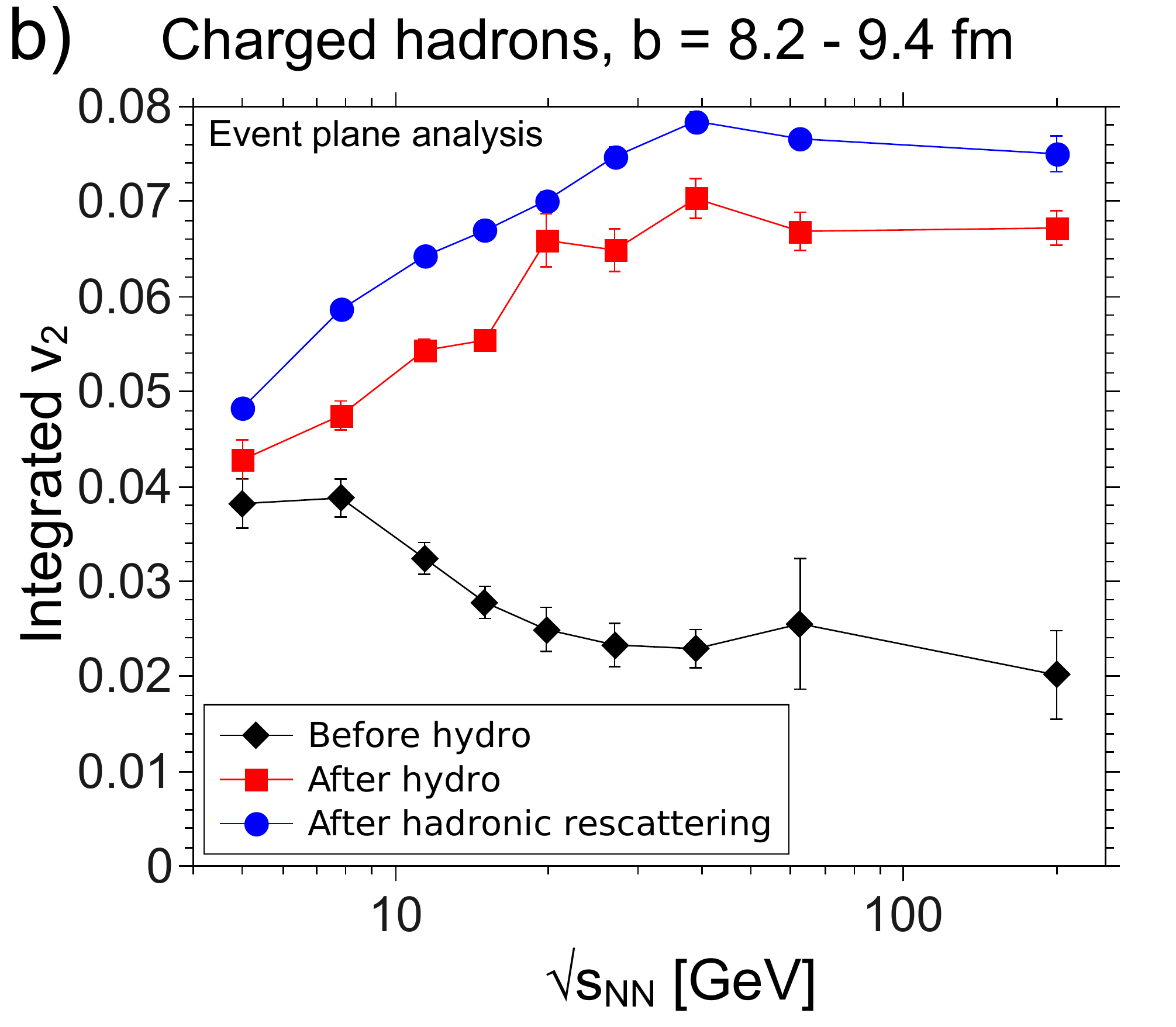}
\caption{Development of elliptic flow in the different stages of the reaction within the UrQMD hybrid approach [From
\cite{Auvinen:2013sba}]}
\label{fig:v2_beam_energy}
\end{figure}

In general, collective flow is among the main observables that are sensitive to the phase transition between the hadron gas and the quark-gluon plasma. This idea can be easily understood: Since the equation of state is encoded in the pressure as a function of the energy density (and possibly net baryon density), the anisotropic flow will react differently depending on the pressure profiles that are probed by heavy ion collisions as a function of time. In other words, the flow coefficients that are primarily built up in a quark-gluon plasma with its very low specific viscosity should eventually vanish when the beam energy is reduced and the plasma is no longer formed. Figure \ref{fig:v2_beam_energy} shows the integrated elliptic flow of charged particles as a function of the beam energy in Au+Au collisions. Interestingly, at low collision energies the elliptic flow is generated entirely by hadronic transport while at high energies the fraction of the flow built up during the hydrodynamic stage is more than 60\%. 
At lower collision energies, the so-called directed flow $v_1$ is an important observable as well. $v_1=\langle p_x/p_\perp \rangle$ quantifies the "bounce-off" of particles in the reaction plane. This rapidity-odd flow coefficient can be summarized by fitting the slope $dv_1/dy$ around midrapidity. The beam energy dependence of $dv_1/dy$ for protons is expected to show non-monotonic behavior in case of a first-order phase transition \cite{Brachmann:1999xt}. In fact, understanding the collective flow and, in particular, the directed flow poses a challenge, since it is also very sensitive to the treatment of the interactions with the spectators and the interface between hydrodynamics and transport in hybrid calculations \cite{STAR:2014clz}.

\subsection{Small droplets of quark-gluon plasma}
\label{sec:small_systems}

\begin{figure}[ht]
\centering
\includegraphics[width=0.95\linewidth]{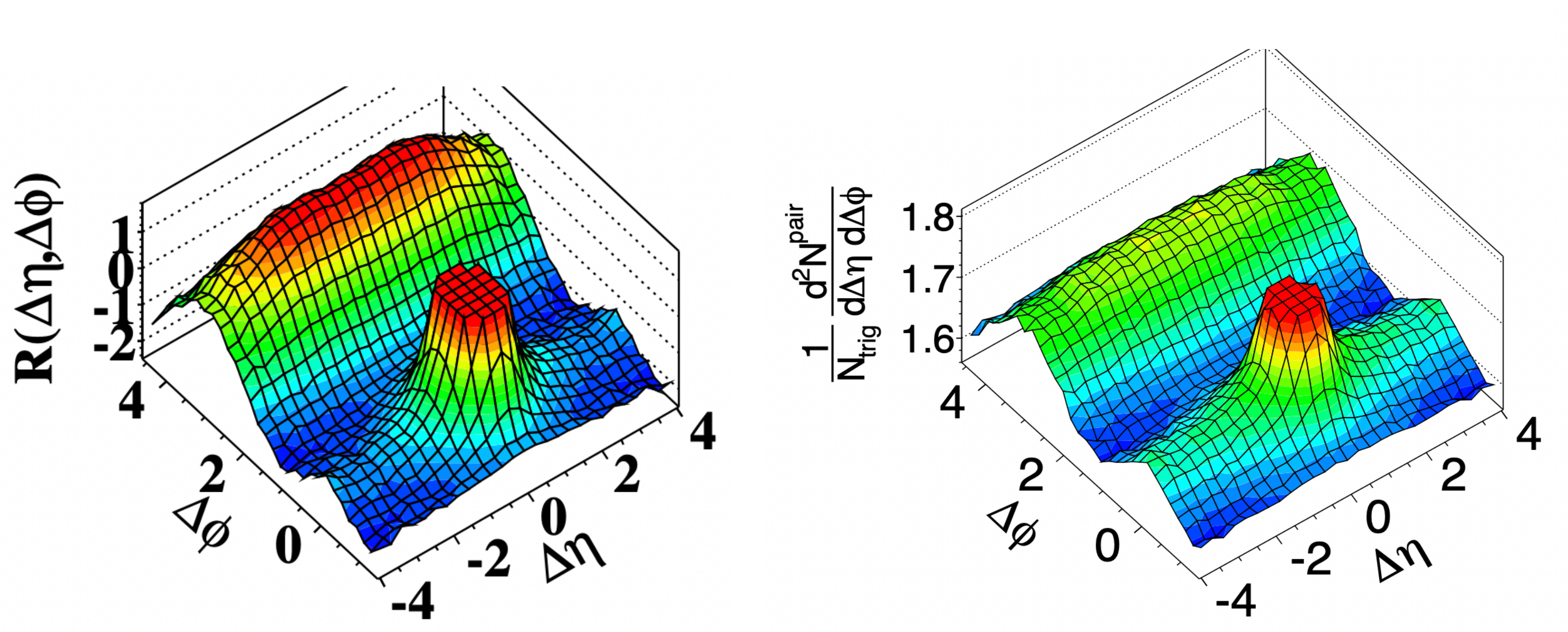}
\caption{Two-particle correlations of charged particles with transverse momenta in the range $1~{\rm GeV}/c < p_T < 3~{\rm GeV}/c$ versus azimuthal angle difference $\Delta\phi$ and pseudorapidity difference $\Delta\eta$. Only high-multiplicity events with more than 110 charged tracks were selected for this analysis. Left panel: p+p collisions; right panel: p+Pb collisions. [From \cite{CMS:2010ifv,CMS:2012qk}]}
\label{fig:CMSridge}
\end{figure}

Long-range correlations among emitted particles in rapidity, akin to the phenomenon of anisotropic collective flow, are also found in high-multiplicity p+p and p+Pb collisions at the LHC \cite{CMS:2010ifv,CMS:2012qk,ATLAS:2012cix,ALICE:2012eyl}. Figure \ref{fig:CMSridge} shows two-body correlations of charged particles with transverse momenta in the range $1~{\rm GeV}/c < p_T < 3~{\rm GeV}/c$ versus azimuthal angle difference $\Delta\phi$ and pseudorapidity difference $\Delta\eta$ in high-multiplicity ($N_{\rm ch} > 110$) events. The left panel is for p+p collisions, the right panel is for p+Pb collisions measured by CMS. The feature of interest in the near-side ``ridge'' visible at $\Delta\phi \approx 0$ and extending over the full pseudorapidity acceptance $|\Delta\eta| \leq 4$. The ridge is clearly more pronounced in p+Pb collisions than in p+p collisions, which may indicate a larger degree of collectivity. 

\begin{figure}[ht]
\centering
\includegraphics[width=0.95\linewidth]{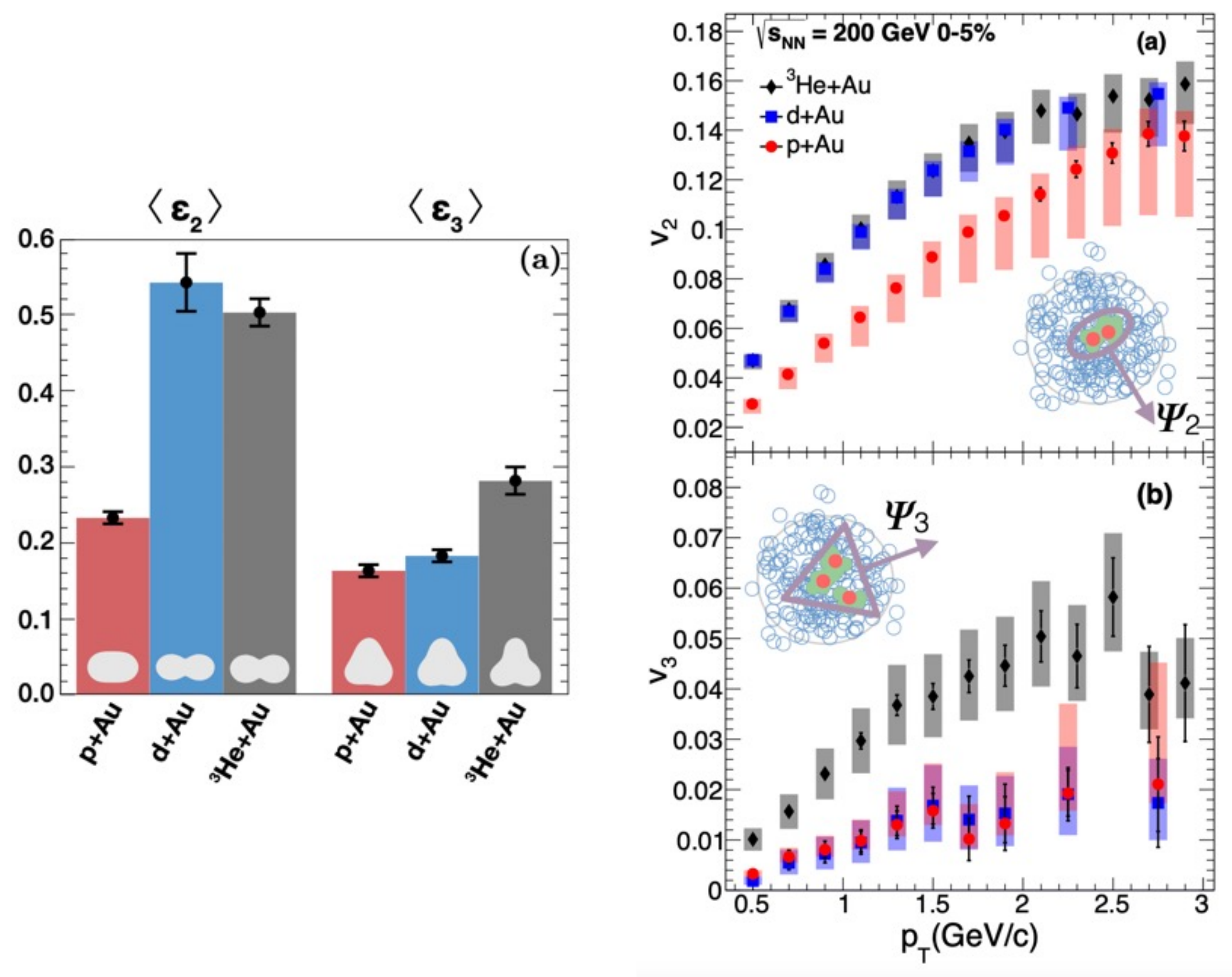}
\caption{Left panel: elliptic and triangular eccentricities of the ground states of proton (p), deuteron (d), and $^3$He. Right panel: Elliptic and triangular flow coefficients $v_2(p_T)$ and $v_3(p_T)$ for the p+Au, d+Au, and $^3$He+Au collision systems at the same energy $\sqrt{s_{\rm NN}} = 200$ GeV. [From \cite{PHENIX:2018lia}]}
\label{fig:PHENIX_pdHeAu}
\end{figure}

\begin{figure}[ht]
\centering
\includegraphics[width=0.6\linewidth]{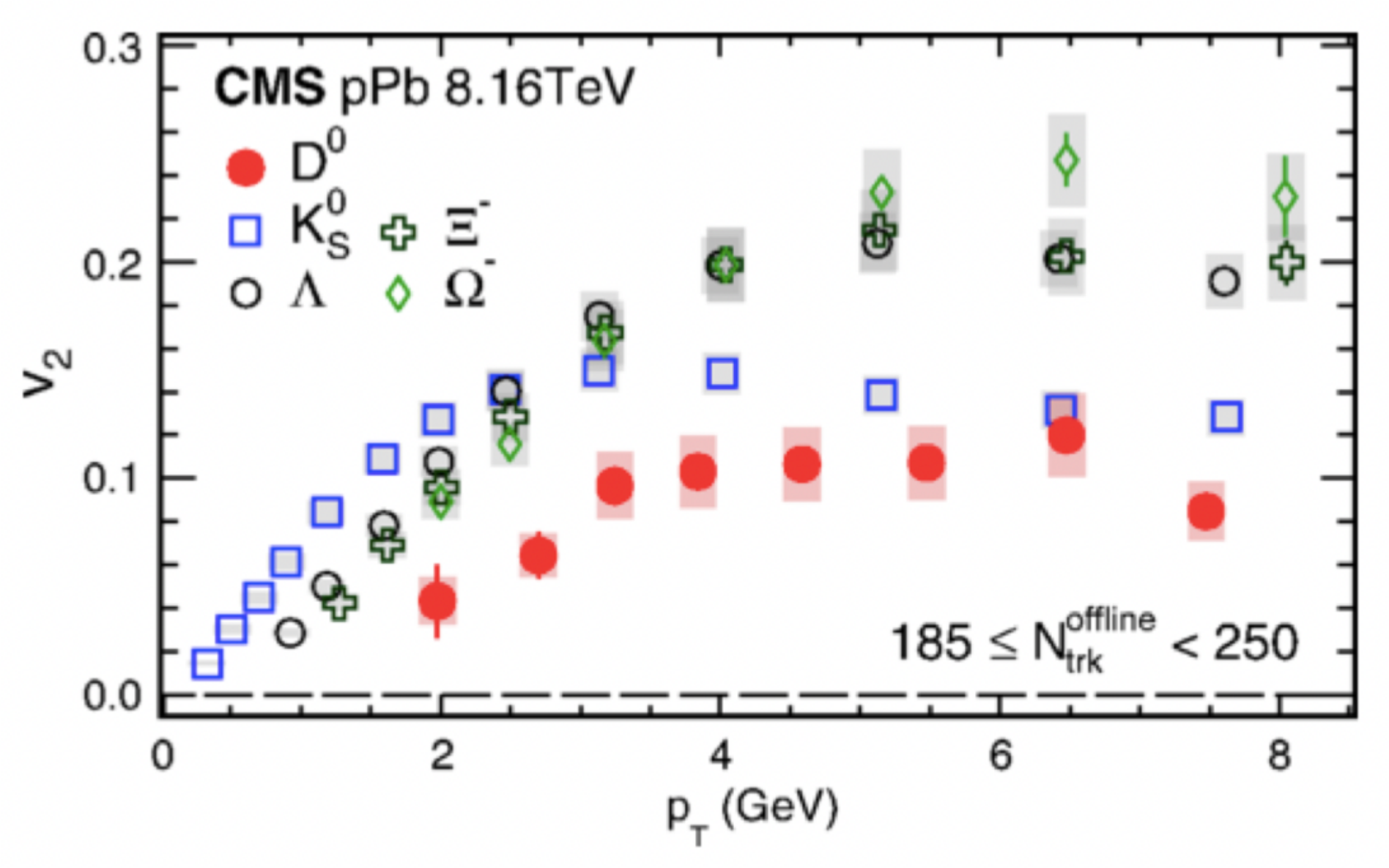}
\caption{Elliptic flow coefficients $v_2(p_T)$ for identified hadrons in high-multiplicity p+Pb collisions at $\sqrt{s_{\rm NN}} = 8.16$ TeV. The mass splitting between $K^0_s$ and $D^0$ mesons, as well as the usual baryon-meson splitting are clearly visible [From \cite{Bold:2021juc}]}
\label{fig:CMS_v2_pPb}
\end{figure}

A crucial test whether the ridge phenomenon observed in these small collision systems is caused by geometry driven collective flow was carried out by the PHENIX collaboration, which compared p+Au, d+Au, and $^3$He+Au collisions at RHIC. Because the geometric structure of the light ions (p, d, $^3$He) is distinctively different (see left panel of Fig.~\ref{fig:PHENIX_pdHeAu}), one expects significant, geometry driven differences in the elliptic and triangular anisotropic flow coefficients $v_2$ and $v_3$. This is, indeed, what was observed by the experiment, as seen in the right panel of Fig.~\ref{fig:PHENIX_pdHeAu}. In $v_2$ the p+Au system stands out as less elliptically deformed; in $v_3$ the $^3$He+Au system stands out as having a significantly larger triangular deformation.
Signatures of collective flow are also seen in the particle-specific elliptic flow $v_2(p_T)$ in high-multiplicity p+Pb events at $\sqrt{s_{\rm NN}} = 8.16$ TeV (see Fig.~\ref{fig:CMS_v2_pPb}). The $v_2$ of $K^0_s$ and $D^0$ show a pronounced mass splitting at low $p_T$, but become approximately equal for $p_T > 5$ GeV/c. The $v_2(p_T)$ for identified baryons and mesons exhibit the usual valence-quark number splitting.

\begin{figure}[ht]
\centering
\includegraphics[width=0.7\linewidth]{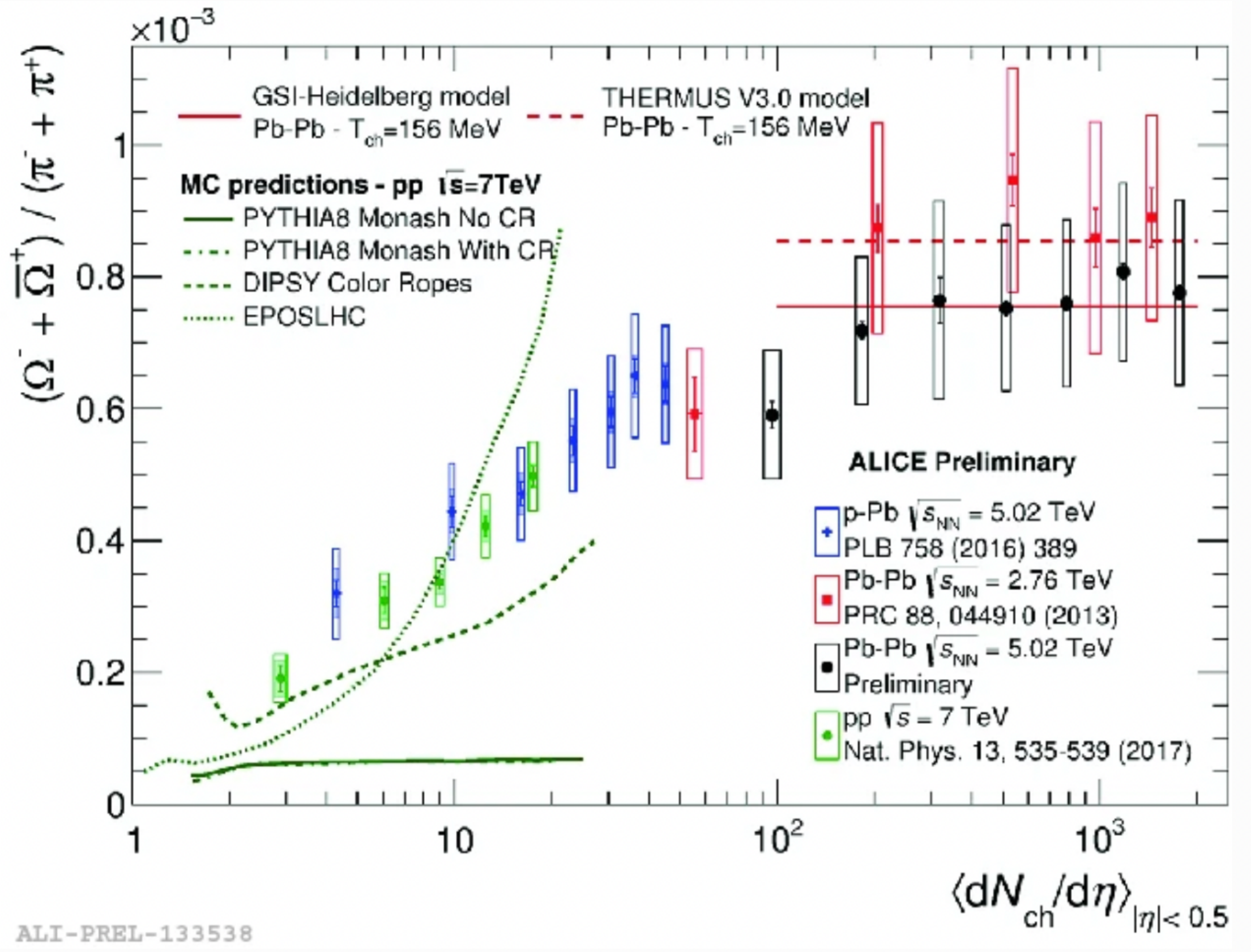}
\caption{Yield ratio of $\Omega$ and $\overline{\Omega}$ baryons relative to charged pions ($\pi^\pm$) in p+p, p+Pb, and Pb+Pb collisions at various energies versus charged particle multiplicity $dN_{\rm ch}/d\eta$. [From \cite{Sefcik:2020haw}]}
\label{fig:ALICE_Omega}
\end{figure}

Additional evidence for medium-like behavior in small systems comes from the gradual approach to full thermal (grand canonical) equilibrium of multi-strange baryon production \cite{ALICE:2016fzo}. Figure \ref{fig:ALICE_Omega} shows the ratio of $\Omega$ and $\overline{\Omega}$ baryons to charged pions in p+p, p+Pb, and Pb+Pb collisions at various LHC energies as function of $dN_{\rm ch}/d\eta$. The steady increase with multiplicity is interpreted as evidence that the volume and lifetime of the hot quark-gluon plasma are growing with multiplicity. It is difficult to imagine that this steady growth is not accompanied by collective flow in response to the thermal pressure inside the fireball. 

One often presented counter-indication against formation of a collectively behaving QGP in p+A collisions is that jet quenching has not been observed. Recent data from ALICE for $15~{\rm GeV}/c < p_T^{\rm ch} < 50~{\rm GeV}/c$ jets in high-multiplicity p+Pb collisions limit the energy loss for a $R=0.4$ jet cone to less than $0.4$ GeV/c, more than an order of magnitude less than the energy loss measured for $R=0.4$ jets in central Pb+Pb collisions \cite{ALICE:2017svf}. This can likely be understood as a consequence of the fact that hard scattered partons need time to evolve down to the virtuality scale $Q^2$ at which the surrounding medium interacts with them \cite{Tywoniuk:2014hta}. For a parton with energy $p_T$, the kinematic virtuality within a transverse distance $L$ from the scattering vertex is $Q^2 > p_T/L$. On the other hand, the virtuality scale associated with rescattering in the medium is 
\begin{equation}
Q^2_{\rm med} \leq \int_{\tau_0}^L d\tau \hat{q}(\tau) \sim (\hat{q}/T^3) T_0^3\tau_0 \ln(L/\tau_0)) .
\end{equation}
For $L \leq 3$ fm in p+Pb one estimates $Q^2 > 2 Q^2_{\rm med}$, which means that the parton has left the medium before it reaches the virtuality scale at which the medium can begin to influence its QCD evolution. 

The large flow gradients in the initial phase of the heavy ion reaction, especially for smaller collision system, severely strain the applicability of viscous hydrodynamics, which relies on a gradient expansion of the collective flow field. A measure of the applicability of hydrodynamics is the Knudsen number Kn, which describes the ratio of the mean free path $\lambda$ to the characteristic system size $L$:
\begin{equation}
    \textrm{Kn} = \lambda/L .
\end{equation}
Here $L$ could either be the size of the system or the typical length scale on which the fluid properties vary. A small Knudsen number, $\textrm{Kn} \ll 1$ implies that viscous hydrodynamics should provide a good description of the system. At early times during the collision and for small collision systems, such as p+A, this condition is violated. 

Numerical simulations have shown, however, that viscous hydrodynamics provides a remarkably robust description even under conditions where the Knudsen number is not small. This behavior has been traced to the existence of hydrodynamic attractors, flow patterns that asymptotically merge into hydrodynamic flow and serve as ``attractors'' of flow patterns for many different initial conditions and even different microscopic transport dynamics (see~\cite{Soloviev:2021lhs} for a review). The existence of these attractors ensures remarkable insensitivity of the expansion dynamics against many uncertainties about the initial state and against the not well understood dynamics of the fluid before equilibration. 

\begin{figure}[ht]
\flushright
\includegraphics[width=0.95\linewidth]{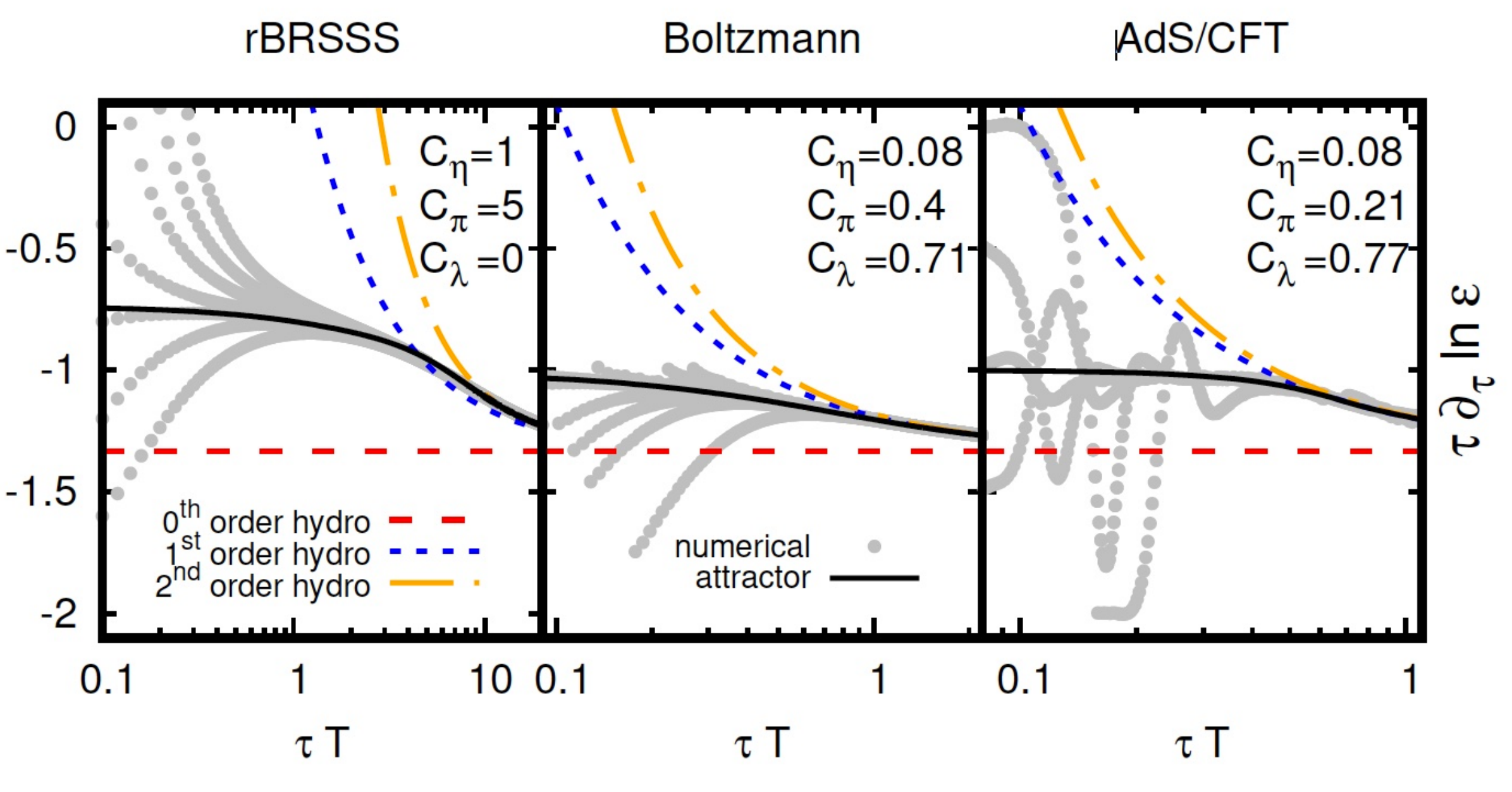}
\caption{Hydrodynamic attractors for three different microscopic dynamics in the Bjorken model, shown together with exact solutions for different initial conditions and first- and second-order hydrodynamics [From \cite{Romatschke:2017vte}]}
\label{fig:Hydro_attractor}
\end{figure}

Figure~\ref{fig:Hydro_attractor} shows the hydrodynamic attractors for three different exactly solvable microscopic transport dynamics in the boost invariant (Bjorken) scenario. Flows for widely different initial conditions are seen to rapidly converge on the attractor within $\tau < 1/T$, well before the hydrodynamic gradient expansion becomes reliable. 

\begin{figure}[ht]
\centering
\includegraphics[width=0.6\linewidth]{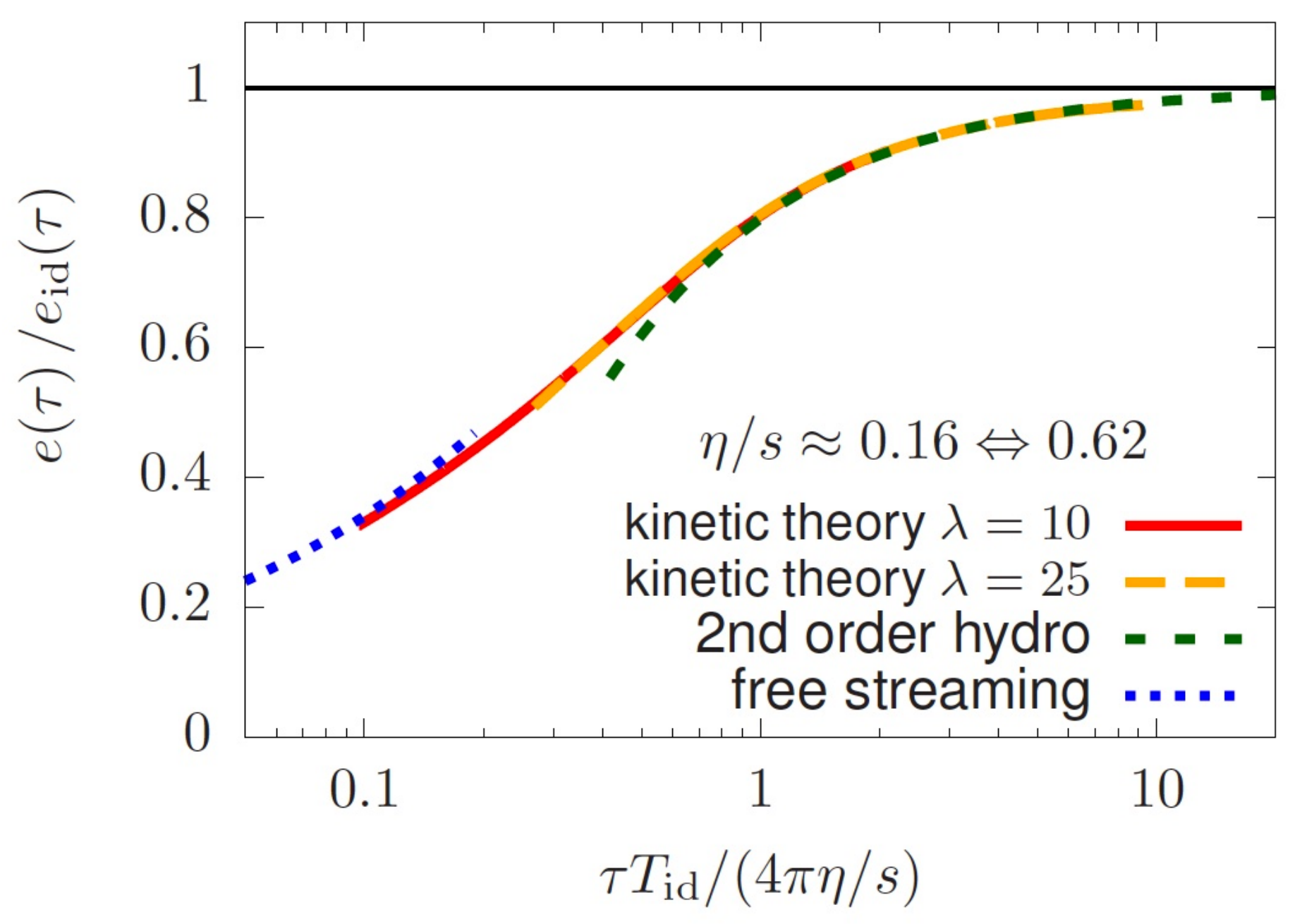}
\caption{Attractor for kinetic transport in the Bjorken model, shown together with early stage (free streaming) and late stage hydrodynamic flow. [From \cite{Kurkela:2018wud}]}
\label{fig:Global_attractor}
\end{figure}

Attractors not only apply to hydrodynamical flow patterns but also emerge in the framework of kinetic theory. In this setting, attractors have been found to smoothly interpolate between the very early stage of the collision where free streaming provides for a good approximation and the late stage when viscous hydrodynamics applies. This behavior is illustrated in Fig.~\ref{fig:Global_attractor}, where the evolution of the energy density is shown for two different values of the coupling constant, together with the early (free streaming) and late (hydrodynamic) evolution. Altogether, these discoveries have strengthened our confidence in the robustness of the existing framework for modeling relativistic heavy ion collisions over the entire duration of the collision.

\subsection{Quantitative constraints of equation of state and transport coefficients} 
\label{sec:Bayesian}

Heavy ion reactions follow a rather complex dynamical evolution that involves switching between non-equilibrium descriptions for the initial and final state and viscous hydrodynamic evolution during the intermediate stage. All the pieces of the theoretical framework come with their own parameters controlling, for example, the amount of initial state fluctuations, the transition criteria between descriptions, as well as the relevant physics properties, such as the equation of state and the transport coefficients. As discussed in the Section \ref{sec:aniso_flow} the observables are often sensitive to more than one parameter in a non-trivial interplay. Note that there is a difference between the model parameters and the parameters of the system under consideration. The beam energy, centrality and other choices are the controllable settings that have to be matched between theory and experiment for meaningful comparisons. 

One option to cope with such a multi-parameter, multi-observable problem is to apply the methodology of Bayesian multi-parameter analysis. The main advantage of such an approach is that one obtains a multi-dimensional sensitivity analysis as a by-product and statistically meaningful parameter extractions with quantified uncertainty. Important inputs are the statistical and systematic uncertainties of the measurements and of the theoretical models, which are sometimes difficult to assess. The pioneering studies that have applied Bayesian analysis for the first time in the field of heavy ion physics were carried out by the MADAI\footnote{\url{https://madai.phy.duke.edu}} collaboration \cite{Petersen:2010zt}. By now, many groups have adopted the procedures and there are further collaborative efforts between statisticians and heavy ion groups (e.g., JETSCAPE\footnote{\url{https://jetscape.org}}, MUSES\footnote{\url{https://icasu.illinois.edu/news/MUSES}} and BAND\footnote{\url{https://bandframework.github.io}} collaborations). The application of such advanced statistical techniques only makes sense, when the underlying theoretical description is established, otherwise no meaningful results can be obtained. If major physics components are missing (``unknown unknowns''), a Bayesian analysis will not produce valid results. 

The following steps are usually taken: 
\begin{enumerate}
\item Select the theoretical models and define the parameters to be varied.
\item Define the prior ranges for the parameters as large as practically feasible.
\item Run the full evolution model at enough points in the multi-parameter space, typically determined by Latin hypercube sampling.
\item Select the observables to be employed in the study.
\item Employ Principal Component Analysis (PCA) to determine orthogonal components of the observables.
\item Construct an emulator for each observable trained on the full model runs.
\item Use Markov Chain Monte Carlo (MCMC) techniques to determine the posterior distribution of the parameters.
\item Perform a closure test for verification that the posterior distribution covers results of full model runs that have not been used in the training of the emulator.
\end{enumerate}

While explaining all the details of such a multi-parameter Bayesian analysis goes beyond the purpose of this review, the most important concepts will be covered here. Bayes' theorem is formulated as follows 
\begin{equation}
P(A|B) = \frac{P(B|A) P(A)}{P(B)}   . 
\end{equation}
The posterior distribution $P(A|B)$ containing the information of how likely parameter set A is under the set of observations B is given by the inverse probability of observables B given parameters A ($P(B|A)$) with a prior distribution $P(A)$ and normalized by the integral $P(B) = \int dA\, P(B|A) P(A)$. In this manner, Bayes' relation permits to invert the conditional probabilities to one that is easier to determine. In other words, we can formulate the question which parameters fit the observations best based on the outcome of simulations for a certain set of parameters. Since the integral in the denominator is usually hard to calculate, one generally ignores the denominator and restricts the analysis to relative probabilities instead of absolute ones. 

\begin{figure}[ht]
\flushright
\includegraphics[width=0.8\linewidth]{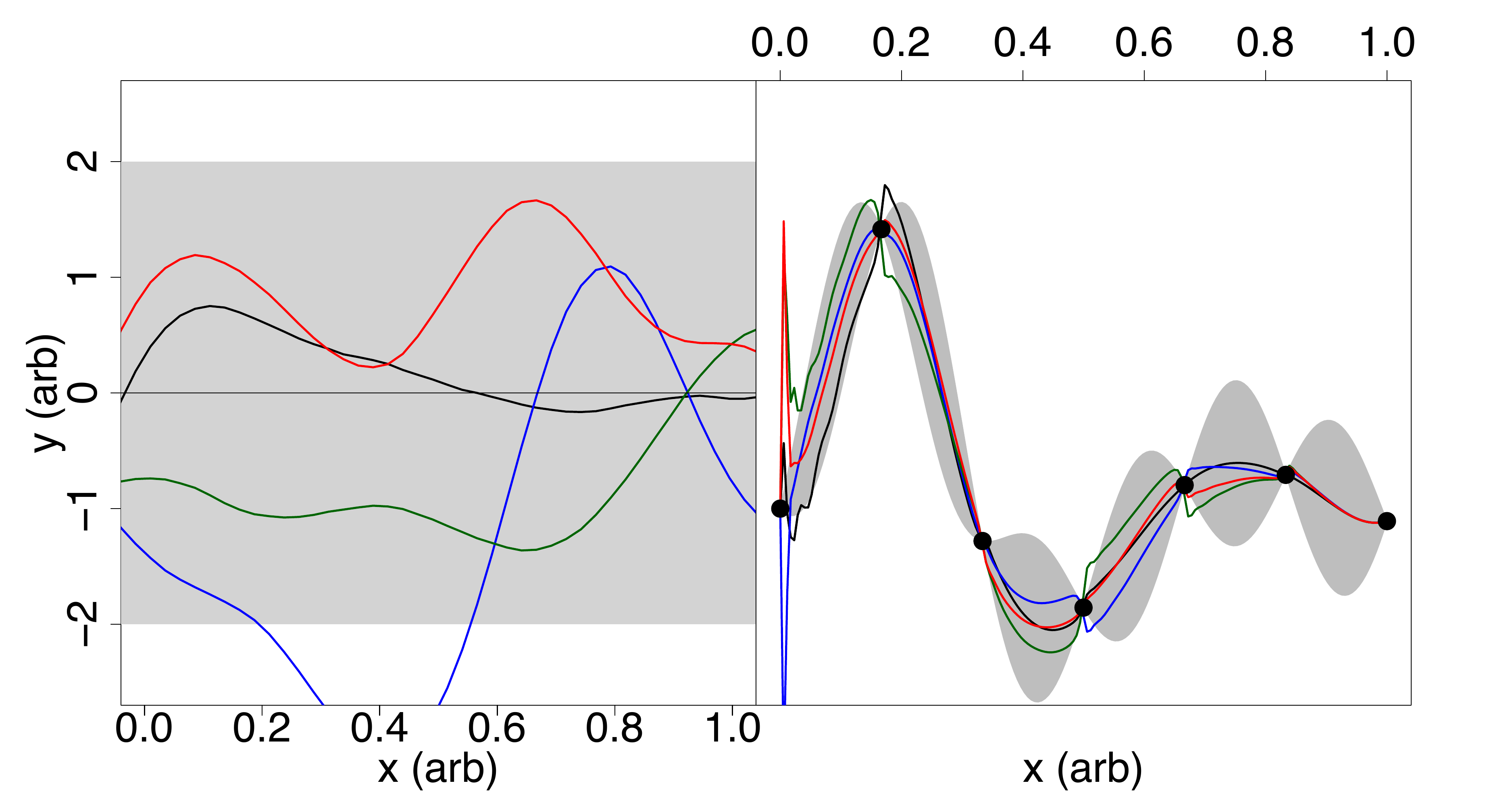}
\caption{Left: Unconditioned draws from a Gaussian Process with a mean of zero and constant unit variance. Right: Draws from the same process after conditioning on 7 training points (black circles). The gray band in both panels is a pointwise 95\% confidence interval. [From \cite{Novak:2013bqa}]}
\label{fig:Gaussian_process}
\end{figure}

Since evaluating the theoretical model for a multitude of parameter combinations becomes quickly too expensive even for current high-performance computing facilities, surrogate models are important to enable a wide parameter scan. Gaussian process emulators are constructed in the following way. Any arbitrary functional form describing the dependence of the observables on the parameters can be approximated with a Gaussian process $GP$ that encodes the mean and the covariance
\begin{equation}
f(x) \sim GP ({\rm mean}(x), {\rm cov}(x,x^\prime))    
\end{equation}
In general, one can envision such a Gaussian process emulator as a way to fit the training data points with associated quantified uncertainties for the values between training points. Therefore, a crucial input is the chosen kernel function
\begin{equation}
k(\vec{x_p},\vec{x_q}) = C^2\, \exp \left( - \frac{1}{2} \sum_{i=1}^s \frac{|x_{p,i}-x_{q,i}|^2}{l_i^2} \right)
\end{equation}
for parameter values $\vec{x}_p$ and PCA values of observables $\vec{x}_q$. The parameters $C$ and $l_i$ have to be chosen with care not to overfit the training data, but also to not leave too much freedom in the construction of the emulator. Figure \ref{fig:Gaussian_process} shows the behavior of unconditioned draws (left) as well as the result after proper training (right). The 95\% confidence interval indicated by the gray bands is fully constrained at the training points and increases further away from them as expected. Uncertainties at the training points can be incorporated by an additional white noise kernel into the Gaussian process. 

\begin{figure}[ht]
\flushright
\includegraphics[width=0.8\linewidth]{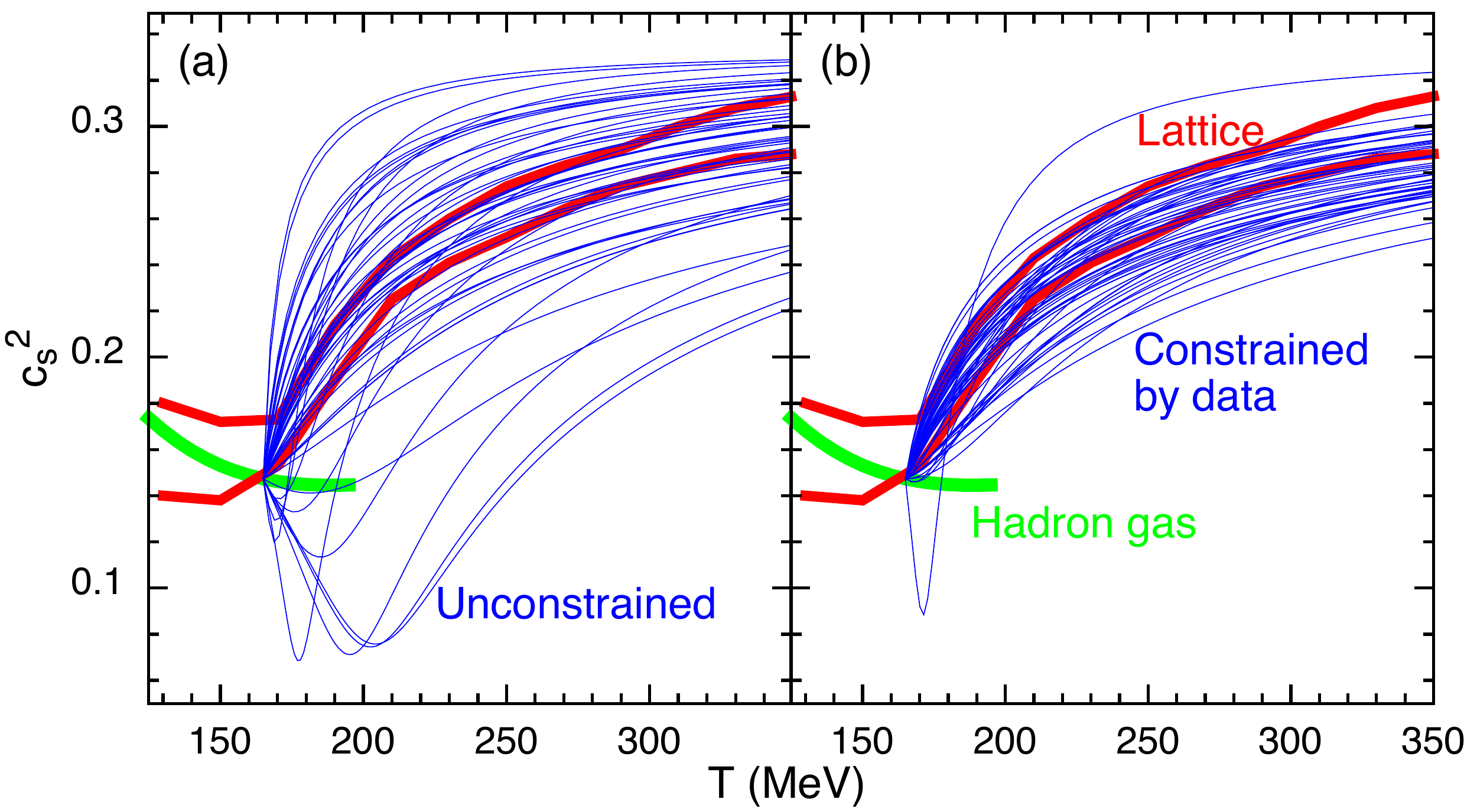}
\caption{Fifty equations of state were generated by randomly choosing equation of state parameters from the prior distribution and weighted by the posterior likelihood. The two upper thick lines in each figure represent the range of lattice equations of state shown in \cite{HotQCD:2014kol}, and the lower thick line shows the equation of state of a non-interacting hadron gas. [From \cite{Pratt:2015zsa}]}
\label{fig:Bayesian_eos}
\end{figure}

In Fig.~\ref{fig:Bayesian_eos} the first application of Bayesian methods to heavy ion collisions is shown. A (2+1)-dimensional hybrid approach was applied to Au+Au collisions at the highest RHIC energy and to Pb+Pb collisions at $\sqrt{s_{\rm NN}}= 2.76$ TeV. The results were compared to data for bulk observables including particle spectra, anisotropic flow and HBT correlations. When drawing parameters weighted with their posterior distribution, it is apparent that the experimental data prefer an equation of state that is consistent with lattice QCD calculations. This is a very nice confirmation that our dynamical description actually prefers an equation of state similar to lattice QCD results based on observables at two different collider energies. This is an important finding that confirms the predicted properties of hot and dense QCD matter from experimental data.

\begin{figure}[ht]
\flushright
\includegraphics[width=0.8\linewidth]{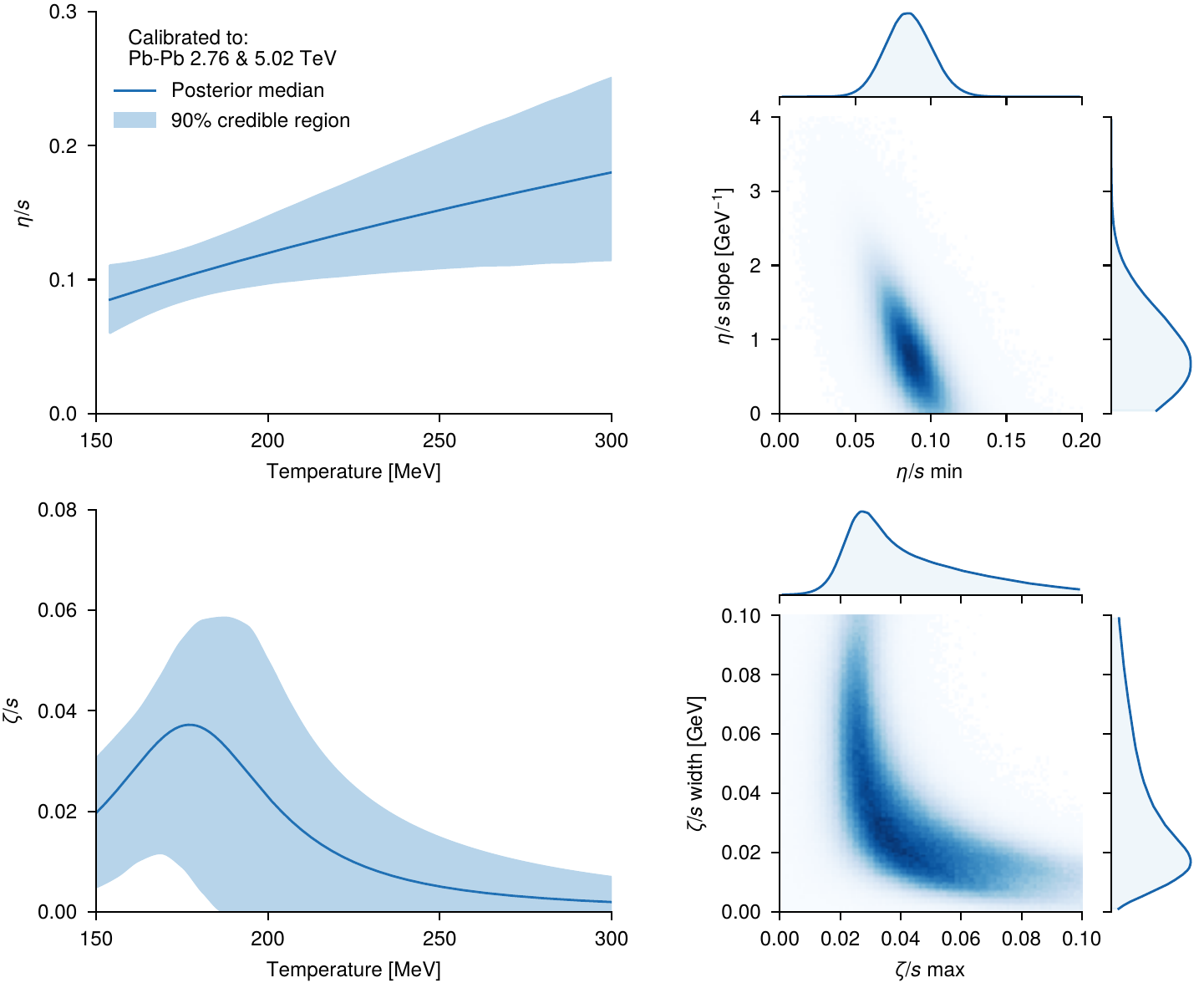}
\caption{Posterior distributions for the parameters determining the temperature dependence of shear and bulk viscosity as well as the 90\% confidence intervals shown for the transport properties of the quark-gluon plasma extracted with a Bayesian analysis of bulk properties at LHC energies within a hybrid approach. [From \cite{Bernhard:2019bmu}]}
\label{fig:Bayesian_transport_coefficients}
\end{figure}

Figure \ref{fig:Bayesian_transport_coefficients} shows the posterior distributions for shear and bulk viscosity obtained from a hybrid calculation compared to experimental data. Within this work, 14 parameters were varied within the Trento+VISH2+1+UrQMD hybrid approach and the bulk properties of hadrons measured in Pb+Pb collisions at 2 LHC energies were taken into account. The Bayesian analysis quantifies constraints on the temperature dependence of transport coefficients and properties of the initial state at the same time. Other analyses have focused on small systems \cite{Moreland:2018gsh} or employed different implementations for the hydrodynamic and hadronic transport evolution \cite{JETSCAPE:2020shq}. 

Extending the list of observables to transverse momentum fluctuations and anisotropic flow correlations has been the subject of recent studies (see e.g. \cite{Nijs:2020ors,Parkkila:2021yha}). Obtaining comprehensive quantitative conclusions on the QGP properties from all these efforts will be a major task for the future. While the different analyses agree on major features, details like the preferred size of hot spots in the initial state or the maximum value of the bulk viscosity differ substantially between different analyses. A very interesting further application of Bayesian analysis is demonstrated in \cite{Nijs:2021clz}, where the sensitivity of the parameters to potential future measurements in oxygen-oxygen collisions at LHC has been assessed based on ``best fit'' parameters from prior studies. 

While most Bayesian analysis have concentrated on the soft sector at RHIC and LHC energies, where the well established ``standard model'' for the dynamical evolution of heavy ion reactions is applicable, there are a few studies targeting hard probes and lower beam energies. In \cite{JETSCAPE:2021ehl} the jet quenching transport coefficient $\hat{q}$ governing the transverse momentum transport of a hard parton traveling through a medium has been quantified within the JETSCAPE framework \cite{Putschke:2019yrg}. Employing the MATTER and LBT energy loss modules for high and low virtuality partons, respectively, the temperature and energy dependence of $\hat{q}$ can be constrained by a comparison to data from RHIC and LHC. The results are compatible with prior constraints from the JET collaboration \cite{JET:2013cls}.  

In \cite{Xu:2017obm} a Bayesian analysis for the charm diffusion coefficient was reported, and in \cite{Auvinen:2017fjw} results for the beam energy dependence of the shear viscosity-over-entropy density were presented. Both results are consistent with prior works and expectations for the qualitative behaviour of the transport coefficients. As the range of Bayesian model-data comparison efforts grows, one must be careful to avoid applying Bayesian analyses in regimes where the theoretical model is under insufficient control and account properly for the variability in modeling choices (see e.g. \cite{JETSCAPE:2020shq}). In the future, it will be rewarding to see more and more observables from the hard and soft sector confronted with a unified theoretical description. A straightforward extension of previous work will be to do a Bayesian analysis for hard probes on a well-calibrated (by Bayesian methods) soft background.  

\subsection{Event-by-Event Fluctuations}
\label{sec:fluctuations}

Modern experiments with heavy ion collisions typically record many millions of events under identical conditions. Due to quantum fluctuations there will always be event-by-event fluctuations, even if the species, beam energy, and impact parameter selection is restricted. In heavy ion reactions there are many sources of fluctuations, some of them trivial (like statistical fluctuations), some of them far from trivial (like the dynamical fluctuations associated with a critical endpoint). Here, we will concentrate on fluctuations of conserved quantities at high beam energies as well as fluctuations associated with the quark-gluon plasma phase transition at lower beam energies. The event-by-event fluctuations in the initial state resulting in higher order flow coefficients have been addressed in Section \ref{sec:aniso_flow}.

As shown in Section \ref{sec:chem_fo} the system formed in heavy ion collisions can be regarded to first approximation as being in thermal equilibrium. In that case, the fluctuations of conserved charges follow the expectations from the grand canonical ensemble as known from statistical mechanics (see \cite{Jeon:2003gk} for a review). There is one caveat: When the charge under investigation is only produced in small quantities or the volume under consideration is close to the entire system, then exact conservation laws have to be taken into account in a canonical, instead of grand canonical, approach. 

\begin{figure}[htb]
\includegraphics[width=0.48\linewidth]{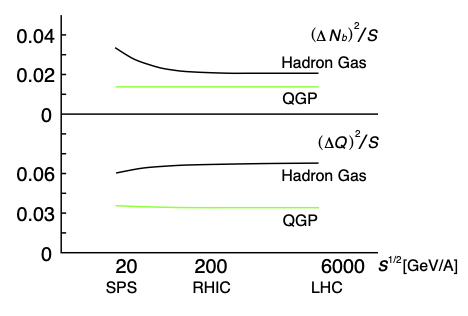}
\hspace{0.07\linewidth}
\includegraphics[width=0.48\linewidth]{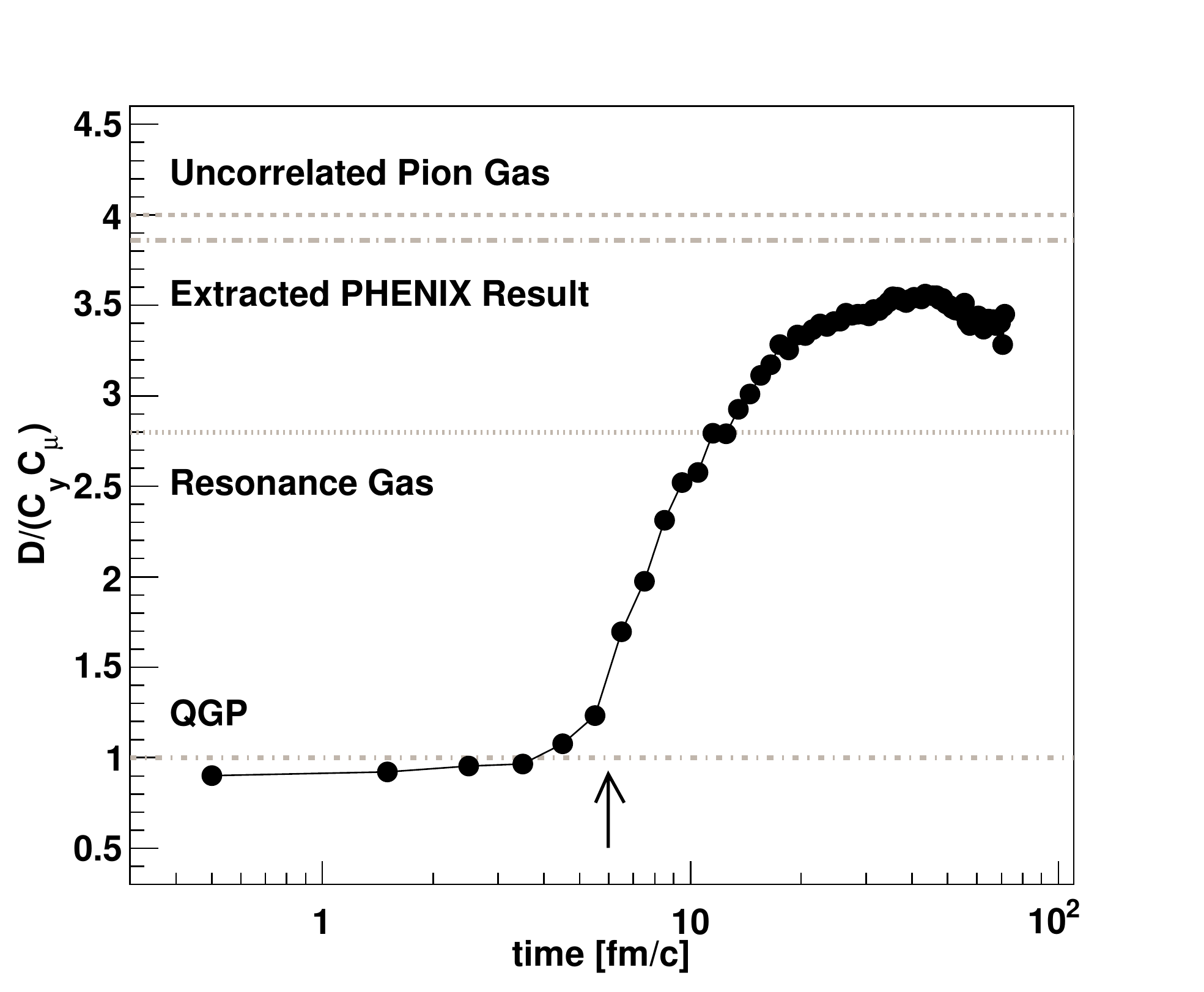}
\caption{Left: Schematic drawing of the beam energy dependence of the net
  baryon number and charge fluctuations per unit entropy
  for a hadronic gas and a quark-gluon plasma. [From \cite{Asakawa:2000wh}]. Right: Corrected charge fluctuations $\tilde{D}$ as a function of time within a hadronization model (arrow depicts time of hadronization) for Au+Au reaction at $\sqrt{s_{NN}}=200$~GeV (full symbols). Also shown are the values for an uncorrelated pion gas, a resonance gas and a quark-gluon plasma. [From \cite {Haussler:2007un}]}
\label{fig:charge_fluc}
\end{figure}

The idea to measure fluctuation observables originates from their association with properties of the system created in heavy ion collisions. For example, the fluctuations of mean transverse momentum are expected to reflect temperature fluctuations. If the system is hotter the particles are emitted with larger transverse momenta, while they obtain less transverse momentum in a colder fireball. For thermal fluctuations of conserved charges ($B,S,Q$) the expected size is sensitive to the degrees of freedom that are active in the system. Fig. \ref{fig:charge_fluc} depicts the expected differences in fluctuations of net baryon number and net charge as a function of beam energy for a quark-gluon plasma and a hadron gas. Since the quarks carry fractions of baryon number and electric charge the corresponding fluctuations are smaller. Experimental data are found to be mainly consistent with the fluctuations expected from a hadron resonance gas. One possible explanation is that the hadronization process washes the partonic fluctuations out, and finally only hadronic fluctuations are observed. This has been demonstrated in a dynamic coalescence approach, where the hadronization process was modeled microscopically in an expanding system (see the right panel of Fig. \ref{fig:charge_fluc}). Nowadays the interest in the mean number of pairwise produced conserved charges has shifted to correlation observables, such as balance functions for charged particles \cite{Pratt:2012dz,Pratt:2015jsa}. 

Another very intriguing application of fluctuation measurements is their direct comparison to lattice QCD calculations (see \cite{Ratti:2018ksb,Ratti:2021ubw} for a review and lecture notes on this topic). Fluctuations of conserved charges cannot be changed by local processes in the hot fireball, only by (slower) diffusion and thus reflect the conditions near the phase boundary. The thermal $\delta Q_i$ are related to susceptibilities $\chi_i$ by 
\begin{equation}
\label{eq:mom_susc}
    \langle{(\delta Q_i)^2}\rangle{} = T^2 \frac{\partial^2}{\partial\mu_i^2} \ln Z(T,\mu_i) = V T^3 \chi_2^{Q_i}
\end{equation}
where $Q_i$ is the conserved charge of interest, $T$ the temperature, $\mu_i$ the corresponding chemical potential and $Z$ the partition function. Experimentally, the moments of the conserved charge distribution can be measured as
\begin{eqnarray}
\mathrm{mean:}~~M_{Q_i}&&~~\mathrm{ variance:}~~\sigma^2_{Q_i}
\nonumber \\
\mathrm{skewness:}~~S_{Q_i}
~~&&~~
\mathrm{kurtosis:}~~\kappa_{Q_i}\,.
\end{eqnarray}
By associating each beam energy $\sqrt{s}$ with pairs of temperature and baryon chemical potential $(T,\mu_B)$, these moments correspond to certain susceptibilities as stated in Eq.~(\ref{eq:mom_susc}). To remove the volume dependence, one usually considers ratios
\begin{eqnarray}
~S\sigma=\chi_3/\chi_{2}
\quad&;&\quad
\kappa\sigma^2=\chi_4/\chi_{2}\nonumber\\
M/\sigma^2=\chi_1/\chi_2
\quad&;&\quad
S\sigma^3/M=\chi_3/\chi_1\,.
\label{moments}
\end{eqnarray}
Studies of the influence of volume fluctuations on such comparisons between experiment and lattice QCD calculations can be found in \cite{Skokov:2012ds,Alba:2015iva}.

\begin{figure}[htb]
\includegraphics[width=0.48\linewidth]{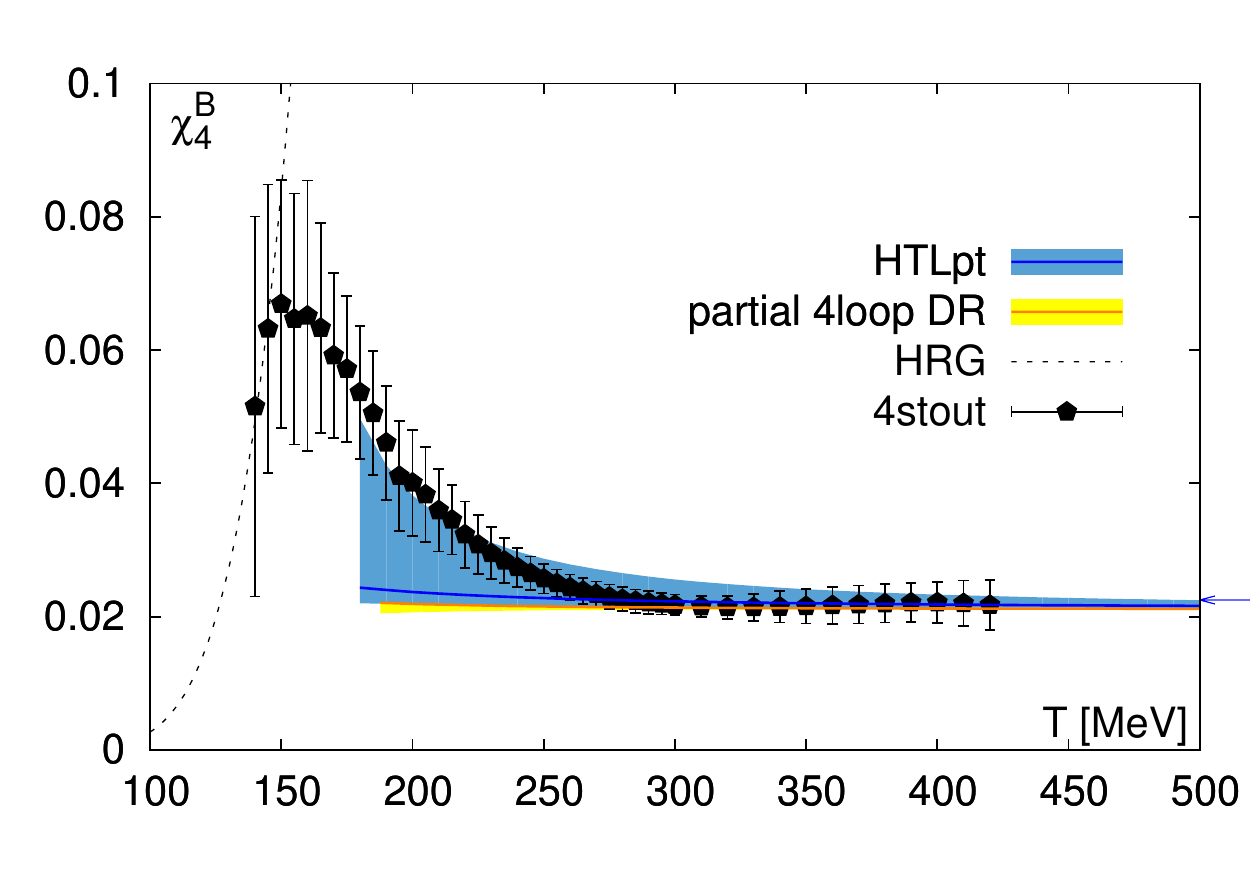}
\hspace{0.07\linewidth}
\includegraphics[width=0.48\linewidth]{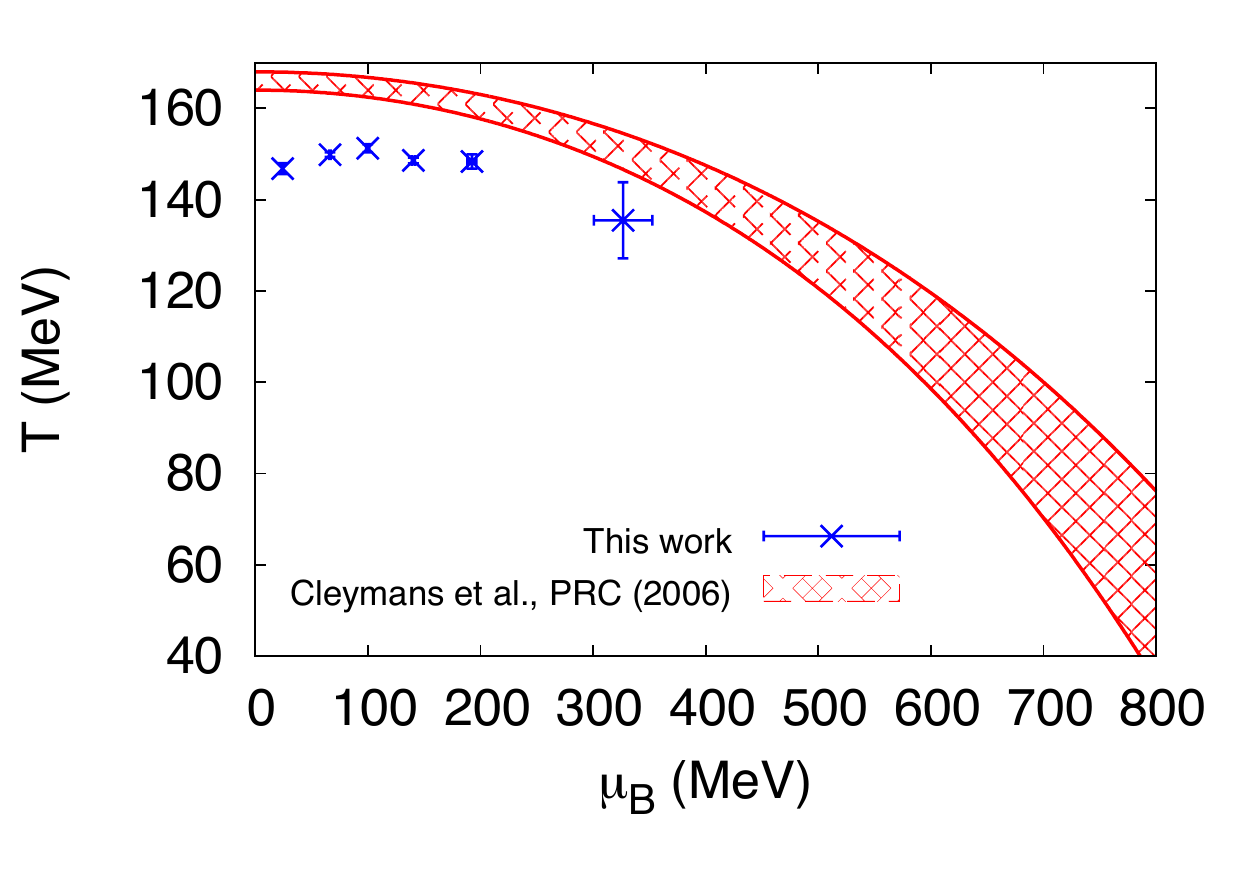}
\caption{Left: Fourth-order cumulants from lattice QCD calculations versus hard thermal loops \cite{Haque:2014rua} and the result from  dimensional reduction (DR) \cite{Mogliacci:2013mca}. The arrow on the right edge marks the Stefan-Boltzmann limit.  [From \cite{Bellwied:2015lba}]. Right: Freeze-out parameters in the ($T-\mu_B$) plane. The figure compares the chemical freeze-out curve \cite{Cleymans:2005xv} (red band) with the values obtained from the analysis of $\sigma^2/M$ for net electric charge and net protons (blue symbols) comparing a hadron resonance gas model to STAR data. [From \cite {Alba:2014eba}]}
\label{fig:susc_lattice}
\end{figure}

The comparison of fluctuation measurements to susceptibilities of conserved charges calculated from lattice QCD is a complementary method to determine the chemical freeze-out conditions in terms of temperature and chemical potential. The results for net baryon number and net charge fluctuations are in very good agreement with the findings according to the thermal model (see \ref{sec:chem_fo}). This is very much consistent since the lattice results in that regime correspond to a hadron-resonance gas. In \cite{Bellwied:2018tkc} it was suggested that comparisons between lattice calculations and strangeness fluctuations indicate a somewhat higher decoupling temperature than the one for light hadrons. This is in line with the expectations from microscopic models, where strange particles have a smaller cross section with other hadrons than the non-strange particles, most prominently protons and pions. 

While it is exciting to directly compare {\it ab initio} lattice calculations to experimental data, one has to be aware of the limitations: Depending on the kinematic cuts of the measurement, the comparison to a grand canonical ensemble calculation may be appropriate or not: Typically only net proton fluctuations are measured and the mapping to net baryon number fluctuations carries uncertainties; also final state interactions and non-equilibrium effects may affect fluctuations.  

The second important application of fluctuation observables is related to their expected sensitivity to the QCD critical endpoint. Finding signatures of the critical endpoint of the first-order phase transition between quark-gluon plasma and hadron gas is one of the main motivations for the heavy ion program at finite densities. The theory of phase transition dynamics predicts that the correlation length increases when the system passes through a critical region. In particular, higher moments of the distributions of conserved quantum numbers are related to a higher power to the increased correlation length. In the idealised equilibrium scenario, the correlation length  as well as the higher moments diverge. In heavy ion collisions this divergence is prevented by finite-size and finite-lifetime effects \cite{Berdnikov:1999ph}. 

\begin{figure}[htb]
\includegraphics[width=0.45\linewidth]{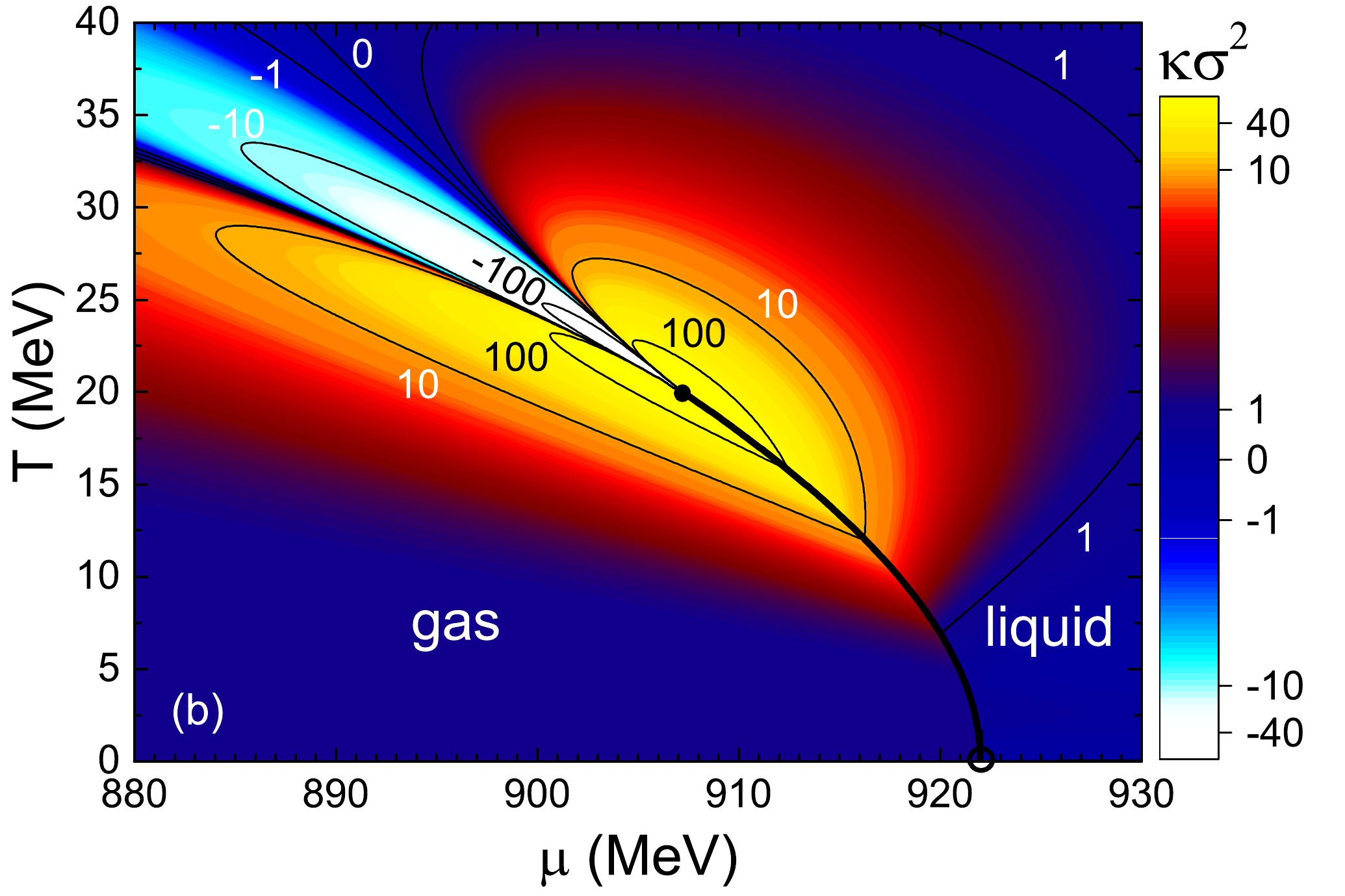}
\hspace{0.07\linewidth}
\includegraphics[width=0.48\linewidth]{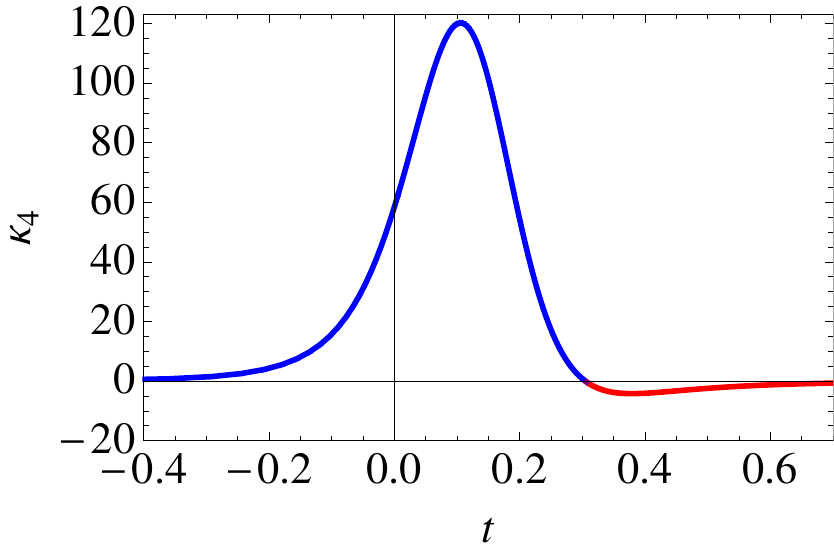}
\caption{Left: The scaled kurtosis $\kappa\sigma^2$ calculated for symmetric nuclear matter in $(T,\mu_B)$ coordinates within Van der Waals equation of state for fermions. [From \cite{Vovchenko:2015pya}]. 
Right: Dependence of the baryon number kurtosis $\kappa_4$ on the reduce temperature parameter $t$ the freeze-out curve (in arbitrary units). $t=0$ corresponds to the location of the critical endpoint; $t<0$ is the region where the phase transition is of first order. [From \cite{Stephanov:2011pb}]}
\label{fig:kurtosis}
\end{figure}

Figure \ref{fig:kurtosis} (left) shows the kurtosis $\kappa\sigma^2$ in the phase diagram of nuclear matter. There are interesting structures and sign changes expected in the region around the critical endpoint of the QCD phase transition, which are analogous to those known in standard liquid-gas phase transitions.  If one follows a typical freeze-out line in the phase diagram, Fig. \ref{fig:kurtosis} (right) depicts the expected beam energy dependence of the kurtosis involving a peak and then a dip structure when going from high to low beam energies.

\begin{figure}[htb]
\centering
\includegraphics[width=0.6\linewidth]{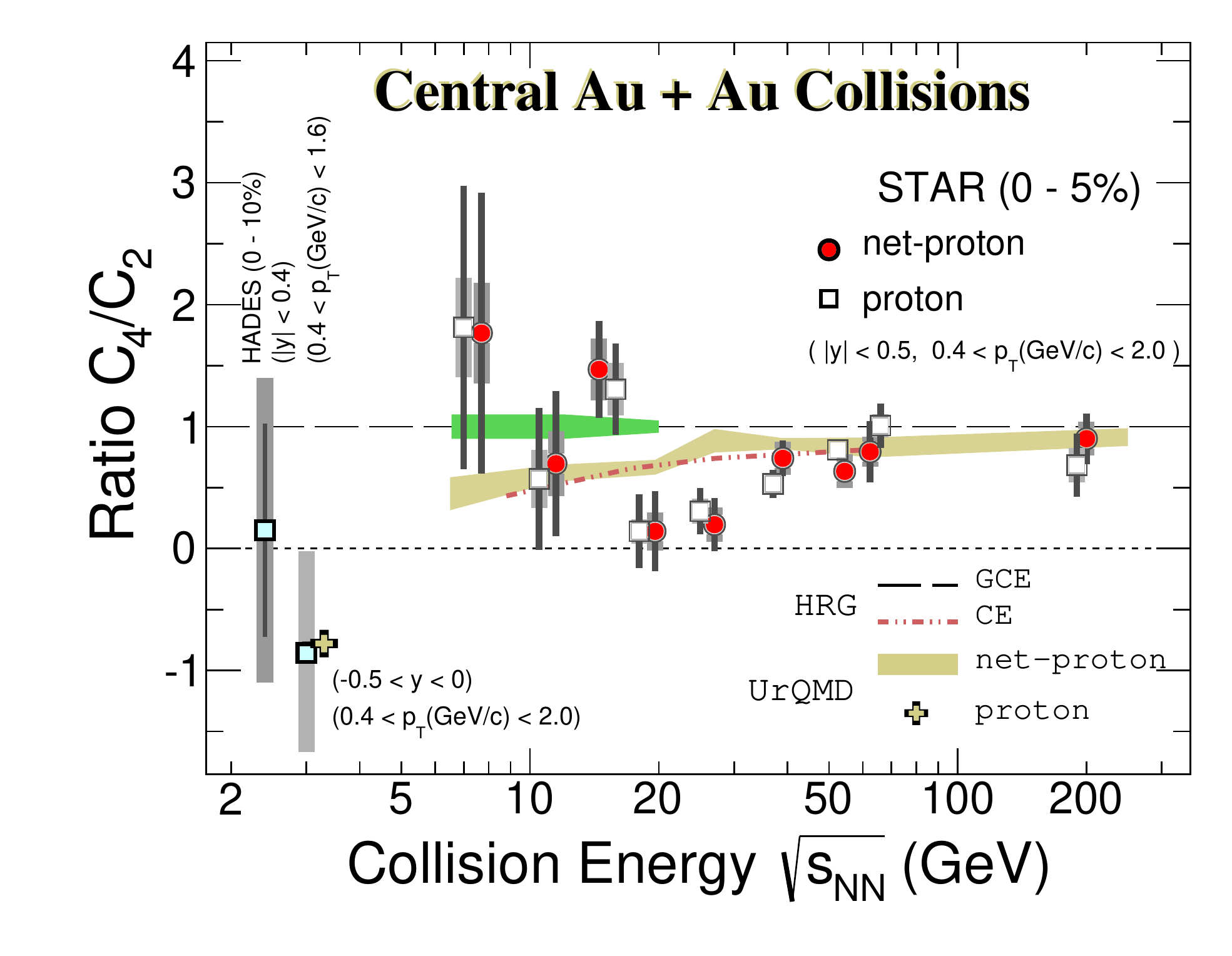}
\caption{Collision energy dependence of the ratios of cumulants, $C_4/C_2$, for proton (squares) and net-proton (red circles) from central Au+Au collisions \cite{STAR:2020tga,STAR:2021iop,HADES:2020wpc} (protons are shifted horizontally).[From
\cite{STAR:2021fge}]}
\label{fig:star_kurtosis}
\end{figure}

Experimentally, measurements of the excitation function of the kurtosis of the net proton distribution have been carried out within the (first) RHIC beam energy scan program by the STAR collaboration and by the HADES collaboration at GSI (see Fig.~\ref{fig:star_kurtosis} for a compilation of results). These are extremely challenging measurements since the impact of efficiencies and kinematic cuts for a multi-particle correlation measurement has to be controlled. As can be seen in Fig.~\ref{fig:star_kurtosis} the data indicate a non-trivial behaviour as function of collision energy that is not compatible with the prediction of a transport (UrQMD) calculation which does not contain information about a critical point. The global net baryon number conservation influences the results, but not enough to bring them into agreement with experimental data. NA61 has looked at lower-order fluctuation observables as a function of beam energy and system size but could, so far, find no sign of critical behaviour. In the future, higher precision data from the Beam Energy Scan II at RHIC, CBM at FAIR, MPD at NICA and other heavy ion physics programs at lower beam energies will complement the existing measurements. 

\begin{figure}[htb]
\includegraphics[width=0.4\linewidth]{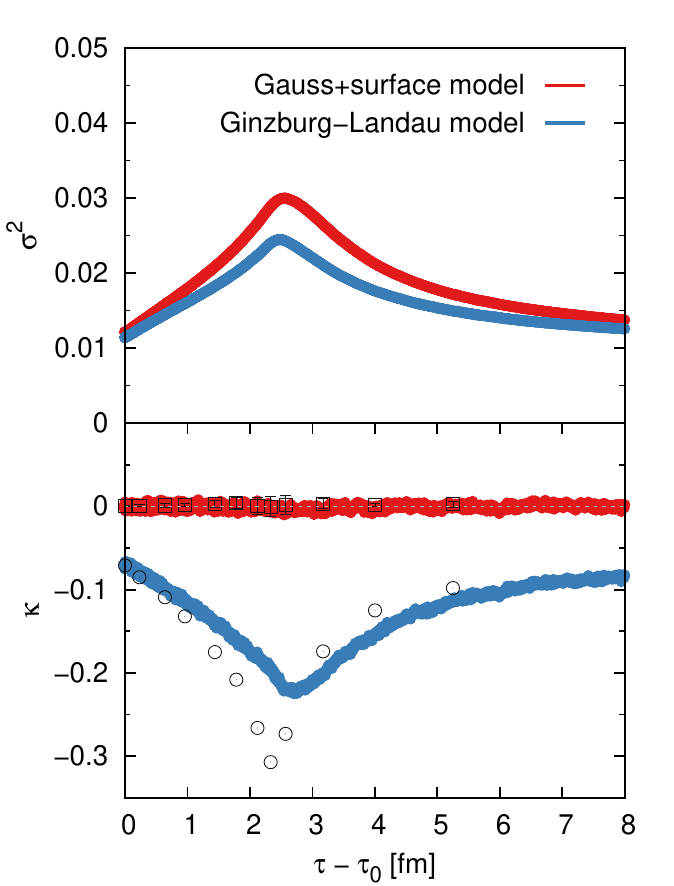}
\includegraphics[width=0.5\linewidth]{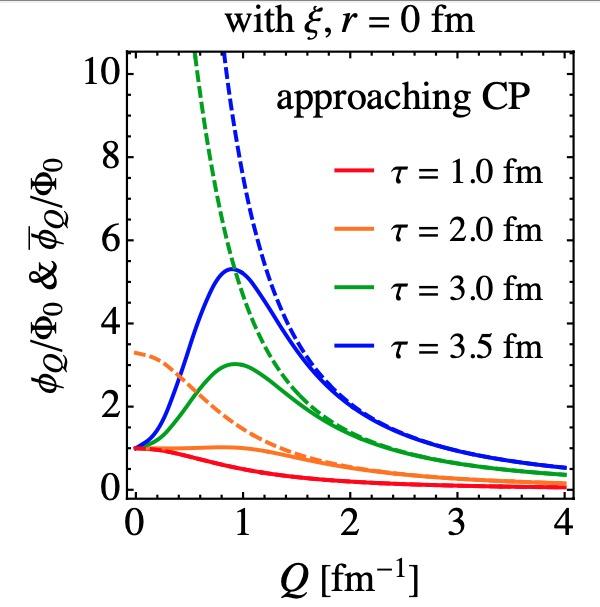}
\caption{Left: The scaled Gaussian cumulants, kurtosis of the net-baryon number as a function of scaled temperature $T/T_{c}$ from simulating stochastic diffusive equation with an Ising-like equation of state in $(1+1)d$ \cite{Nahrgang:2018afz}. Right: The magnitude of the critical fluctuations $\phi(Q)$  at a representative radial distance for different values of proper time $\tau$, where the dashed curve corresponds to the equilibrium and the full curve to the non-equilibrium expectation \cite{Du:2020bxp}. [From \cite{An:2021wof}]}
\label{fig:dyn_flucs}
\end{figure}

On the theory side, there have been many developments targeting a dynamic non-equilibrium evolution through a critical endpoint (see \cite{Bluhm:2020mpc} for a recent summary). All of these efforts are based on extending the fluid dynamics description to include thermal fluctuations. One approach is to add thermal noise to the relativistic hydrodynamic evolution, which is numerically challenging. One effort to systematically work towards a controlled description of the evolution of non-Gaussian fluctuations is shown in Fig. \ref{fig:dyn_flucs} (left). The lines depict the non-equilibrium calculation compared to the equilibrium expectation indicated by open symbols.

Figure \ref{fig:dyn_flucs} (right) shows one example of a complementary effort, where the two-particle correlations are propagated on top of a hydrodynamic background (Hydro+ formalism). Note that both calculations agree in their finding that the non-equilibrium evolution has significant effects on the magnitude and behaviour of the kurtosis, even though they are carried out in simplified settings. A full (3+1)-dimensional non-equilibrium evolution including the critical dynamics still remains a future challenge for a quantitative understanding of the experimental measurements. 

Event-by-event fluctuations in heavy ion physics can also be used to select events of interest. The ``event shape engineering'' technique groups, for example, events according to certain features like the magnitude of radial flow or certain anisotropic flow coefficients to provide further handles beyond centrality and beam energy. Machine learning techniques may make it possible to access interesting information by feeding information from single events into an artificial intelligence (AI) system. Of course this has to be done with great care, since the machine is not smarter than the best theoretical model on the market that was used to train the neural network. Systematic uncertainties inherent in such approaches are difficult to assess.

\subsection{Hadronization and Quark Collectivity}
\label{sec:ncq_scaling}

\begin{figure}[htb]
\centering
\includegraphics[width=0.75\linewidth]{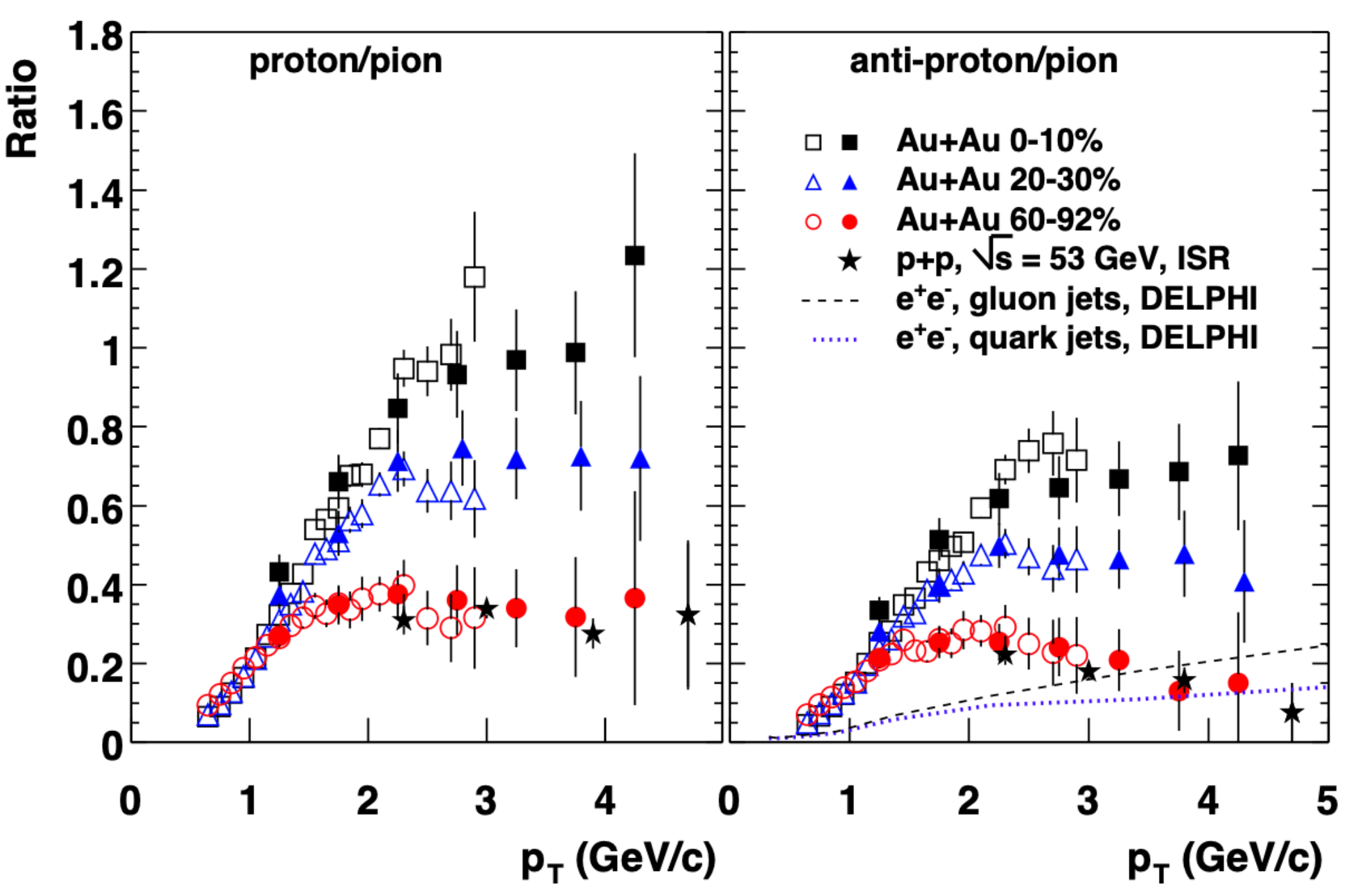}
\caption{Proton-to-pion ratio (left) and antiproton-to-pion ratio (right) for Au+Au collisions at $\sqrt{s_{\rm NN}} = 200$ GeV collisions for several centrality windows as function of transverse momentum $p_T$. [From \cite{PHENIX:2003tvk}]}
\label{fig:ptopi}
\end{figure}

\begin{figure}[htb]
\centering
\includegraphics[width=0.75\linewidth]{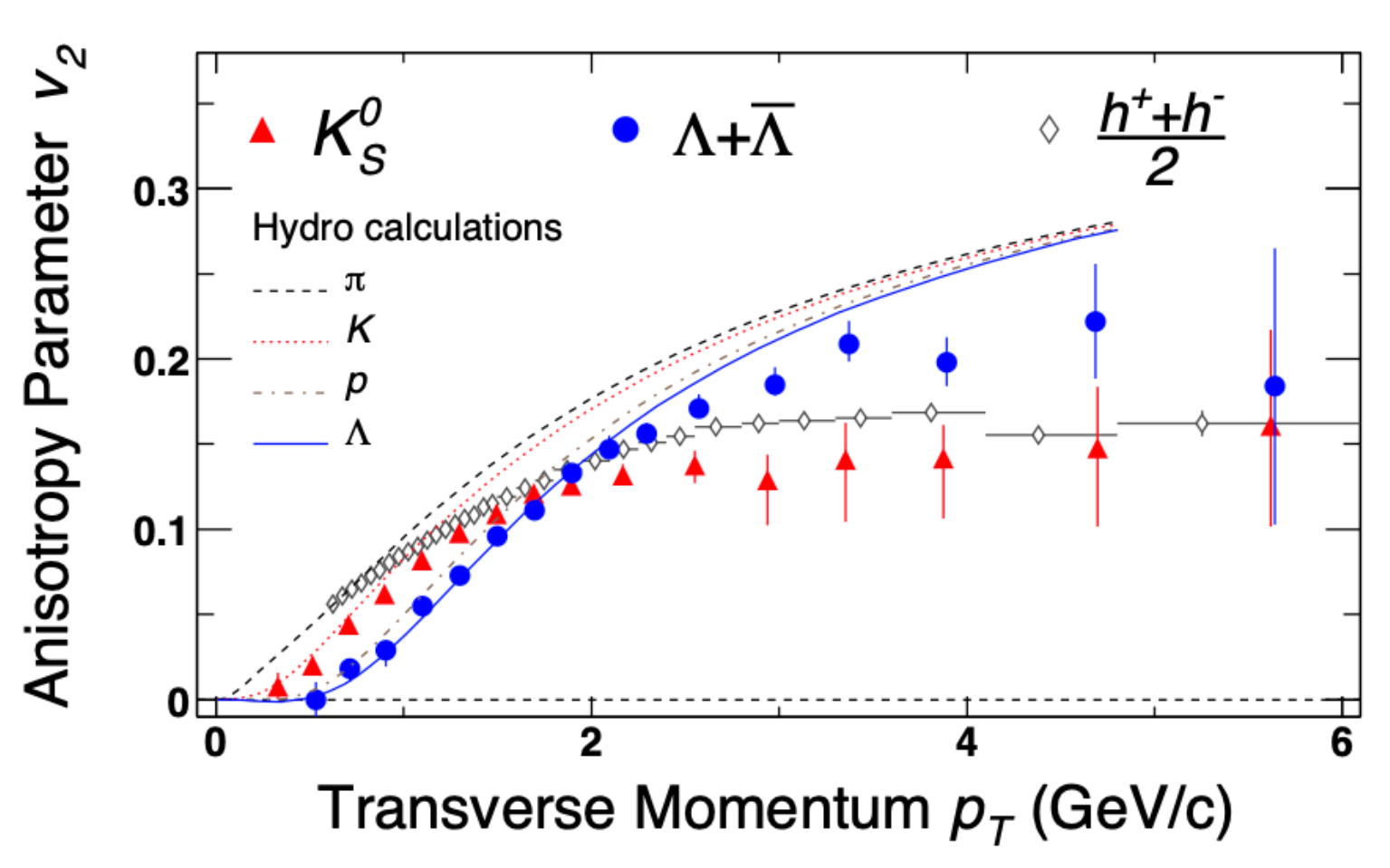}
\caption{Elliptic flow coefficient $v_2(p_T)$ for neutral K-mesons and $\Lambda$-hyperons in $\sqrt{s_{\rm NN}} = 200$ GeV Au+Au collisions. [From \cite{STAR:2003wqp}]}
\label{fig:v2_LambdaK}
\end{figure}

Hadron production from heavy ion collision in the transverse momentum region below a few GeV/c exhibits two striking features: (1) The baryon-to-meson ratio, for both protons and antiprotons, in central Au+Au grows steadily with $p_T$ for $p_T \leq 3$ GeV/c reaching a value three times as large as in peripheral collisions as shown in Fig.~\ref{fig:ptopi} \cite{PHENIX:2003tvk}. (2) The elliptic flow coefficient $v_2(p_T)$ for mesons and baryons shows a distinctly different behavior, with the baryon $v_2$ saturating at larger values than the meson $v_2$, as shown in Fig.~\ref{fig:v2_LambdaK} for $\Lambda$-hyperons and K-mesons \cite{STAR:2003wqp}.

Both phenomena can be explained by the mechanism of hadron formation by quark recombination (or coalescence) from the quark-gluon plasma, in which each valence quark inherits the collective flow properties of the QGP fluid. Since baryons contain three valence quarks whereas mesons contain only two, baryons experience a stronger push from collective flow towards higher $p_T$ than mesons \cite{Fries:2003vb,Greco:2003xt,Fries:2003kq}. For the elliptic flow $v_2(p_T)$ this mechanism implies a scaling law with valence quark number $n_q$ \cite{Molnar:2003ff}:
\begin{equation}
    v_2(p_T) = n_q v_2^{\rm (q)}(p_T/n_q) ,
\label{eq:nq-scaling}
\end{equation}
where $v_2^{\rm (q)}(p_T)$ denotes the elliptic flow coefficient for (anti-)quarks in the QGP. 

\begin{figure}[htb]
\centering
\includegraphics[width=0.85\linewidth]{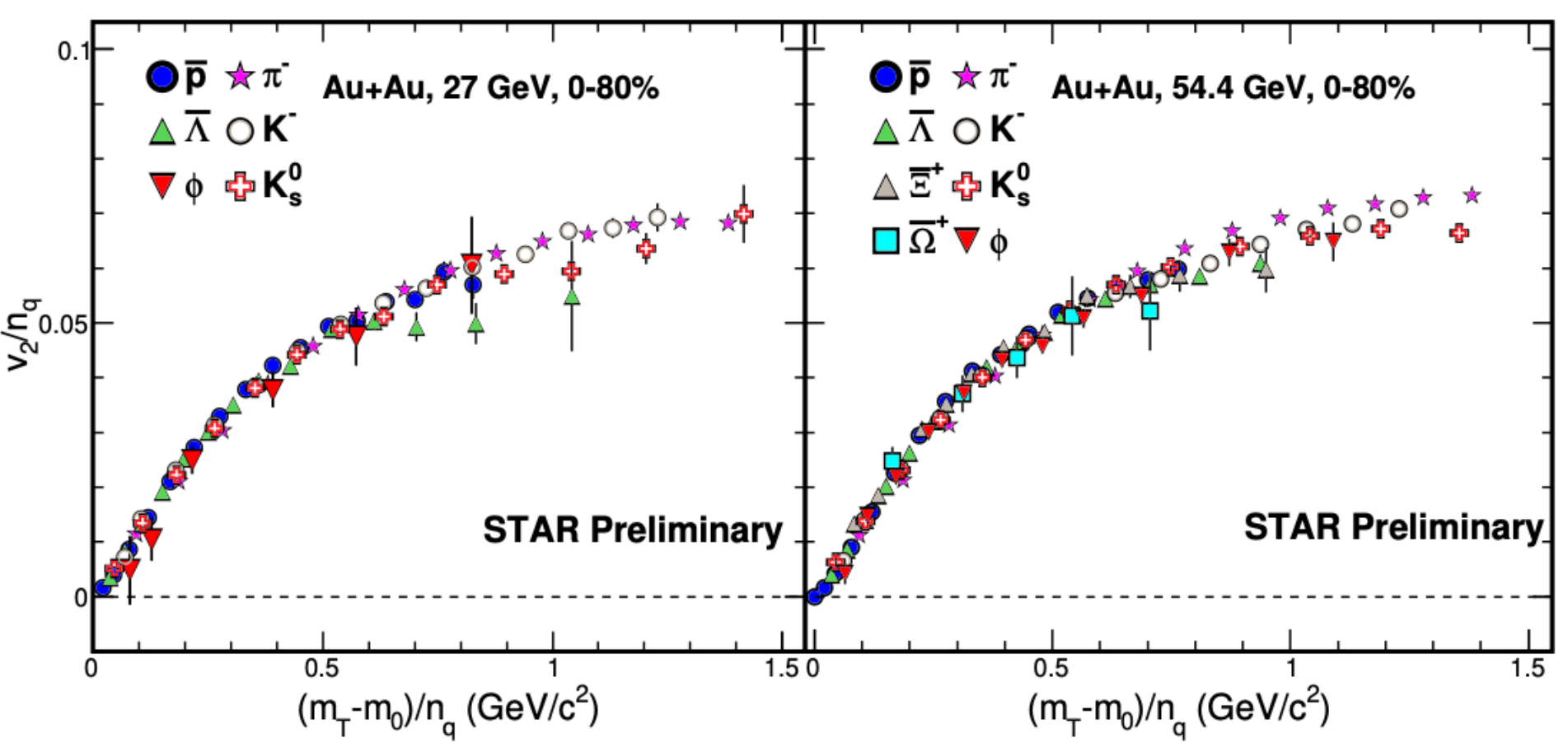}
\caption{Scaled elliptic flow coefficient $v_2/n_q$ for eight different hadron species in $\sqrt{s_{\rm NN}} = 27, 54.4$ GeV Au+Au collisions as a function of the scaling variable $(m_T-m_0)/n_q$. [From \cite{Dixit:2021qey}]}
\label{fig:v2_scaling}
\end{figure}

Strictly speaking, the scaling with $p_T/n_q$ can only be justified in the kinematic domain where hadron masses can either be neglected or described additively by constituent quark masses. In order to apply the scaling law heuristically over a wider momentum range, especially down to small momenta $p_T$, it is customary to compare the elliptic flow of different hadrons as a function of the scaled transverse mass $(m_T-m_0)/n_q$, where $m_0$ is the rest mass of the hadron and $m_T(p_T) = \sqrt{p_T^2+m_0^2}$. This version of the valence quark scaling law has been found to be remarkably well obeyed by a large number of hadron species over a wide collision energy range. Two examples from the recent RHIC beam energy scan are shown in Fig.~\ref{fig:v2_scaling}. A review of theoretical and experimental aspects of the quark recombination mechanism can be found in \cite{Fries:2008hs}.

\begin{figure}[htb]
\centering
\includegraphics[width=0.85\linewidth]{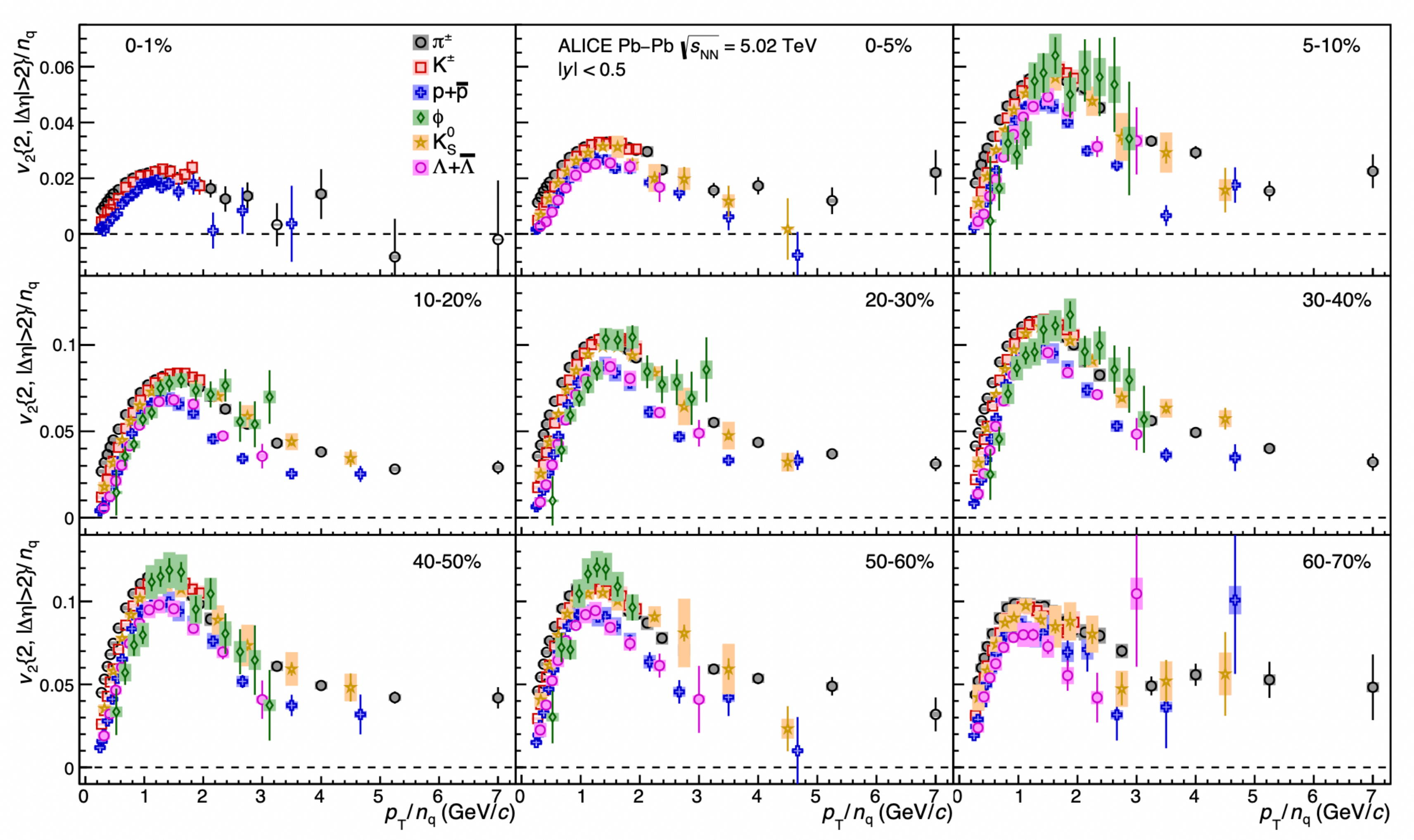}
\includegraphics[width=0.85\linewidth]{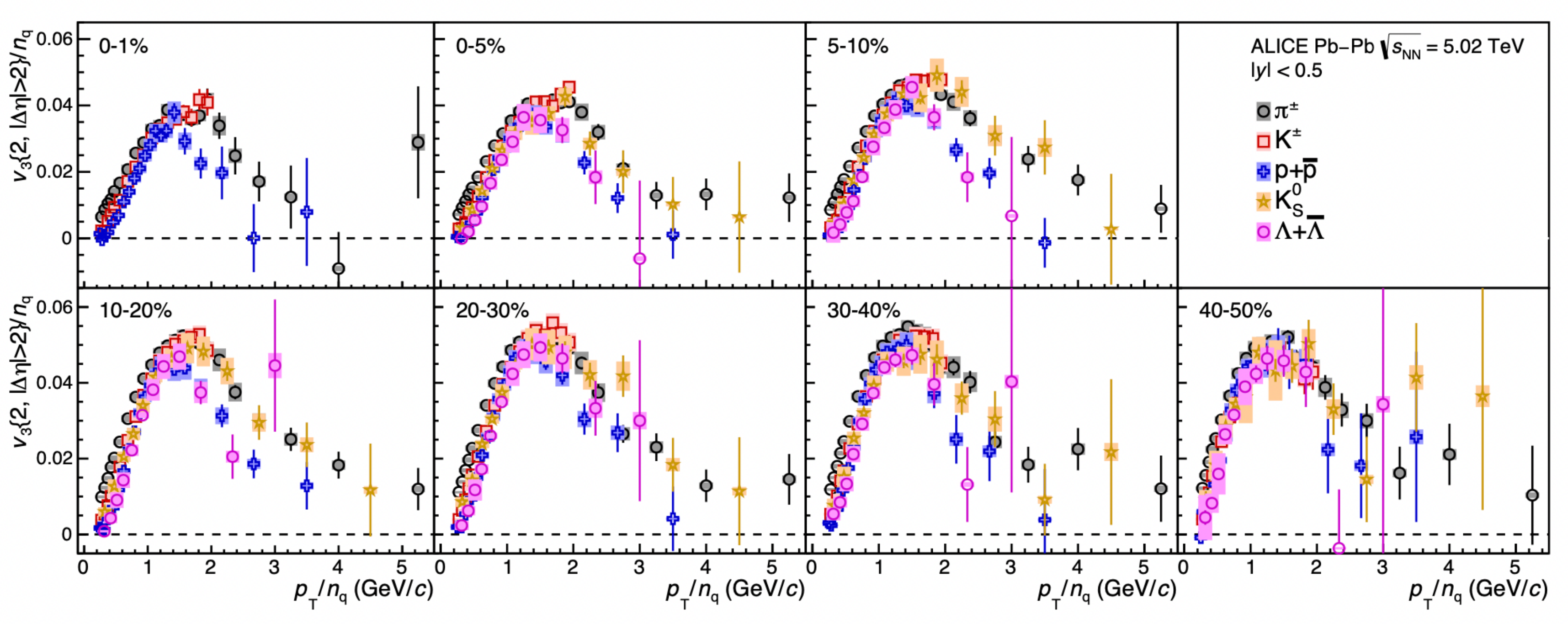}
\includegraphics[width=0.85\linewidth]{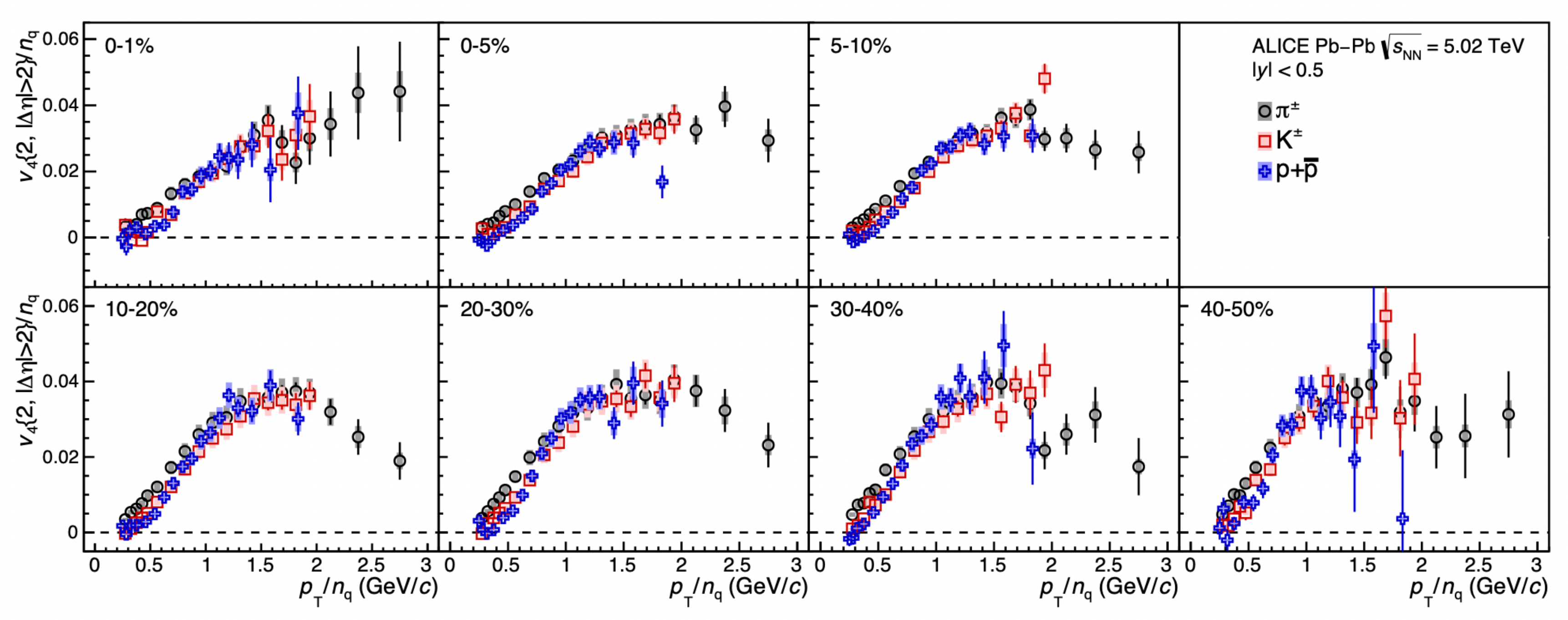}
\caption{Valence quark-number scaled anisotropic flow coefficients $v_n/n_q$ for six identified hadron species in $\sqrt{s_{\rm NN}} = 5.02$ TeV Pb+APb collisions as a function of the scaling variable $p_T/n_q$. Top panels: $n=2$; central panels: $n=3$; Bottom panels: $n=4$. Note that the momentum scale in the bottom set of panels is different. [From \cite{ALICE:2018yph}]}
\label{fig:ALICE_nq-scaling}
\end{figure}

The valence quark scaling (\ref{eq:nq-scaling}) has also been observed in identified particle emission patterns at the LHC, where the scaling is observed to hold even for the higher flow anisotropy coefficients $v_3$ and $v_4$ \cite{ALICE:2018yph}. Figure \ref{fig:ALICE_nq-scaling} shows valence quark-number scaled flow coefficients $v_n(p_T/n_q)/n_q$ for $n=2,3,4$ for several identified hadrons in Pb+Pb collisions at $\sqrt{s_{\rm NN}} = 5.02$ TeV. 

\begin{figure}[htb]
\centering
\includegraphics[width=0.23\linewidth]{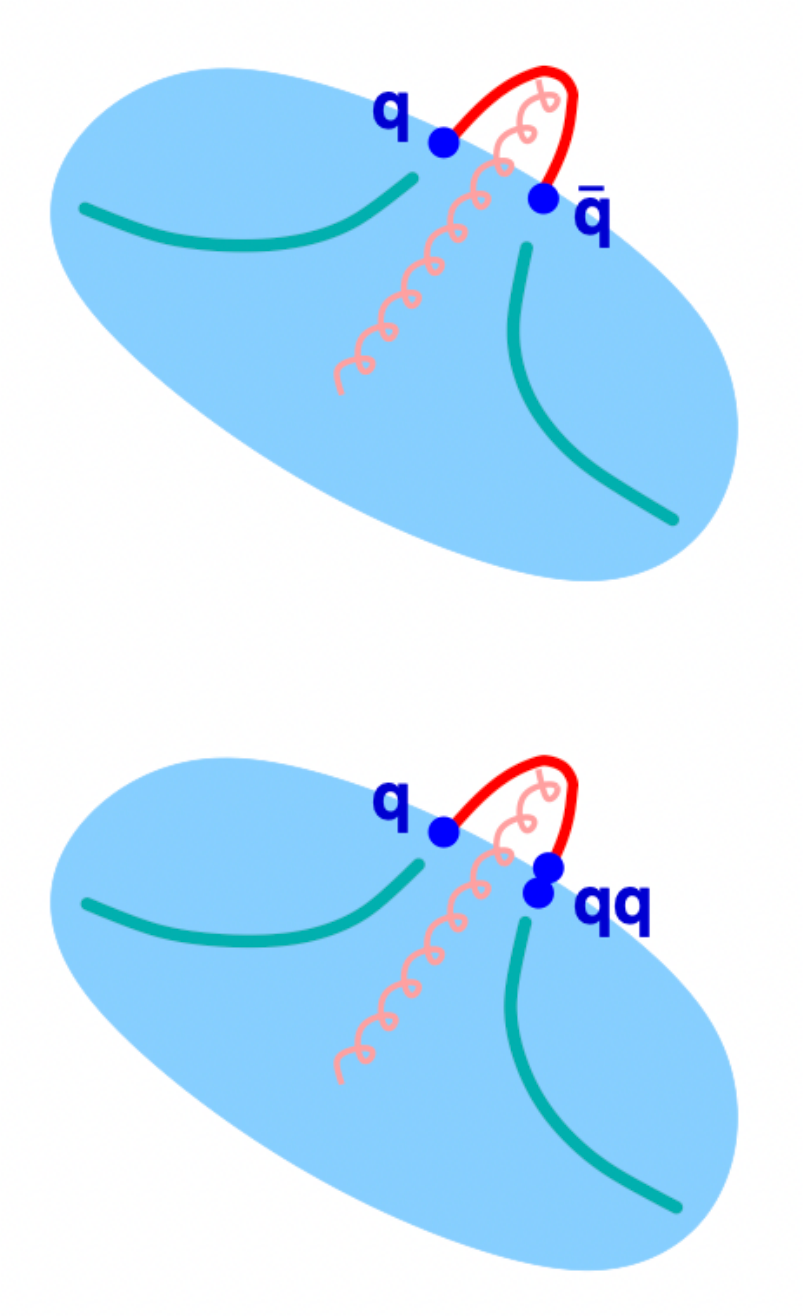}
\hspace{0.03\linewidth}
\includegraphics[width=0.6\linewidth]{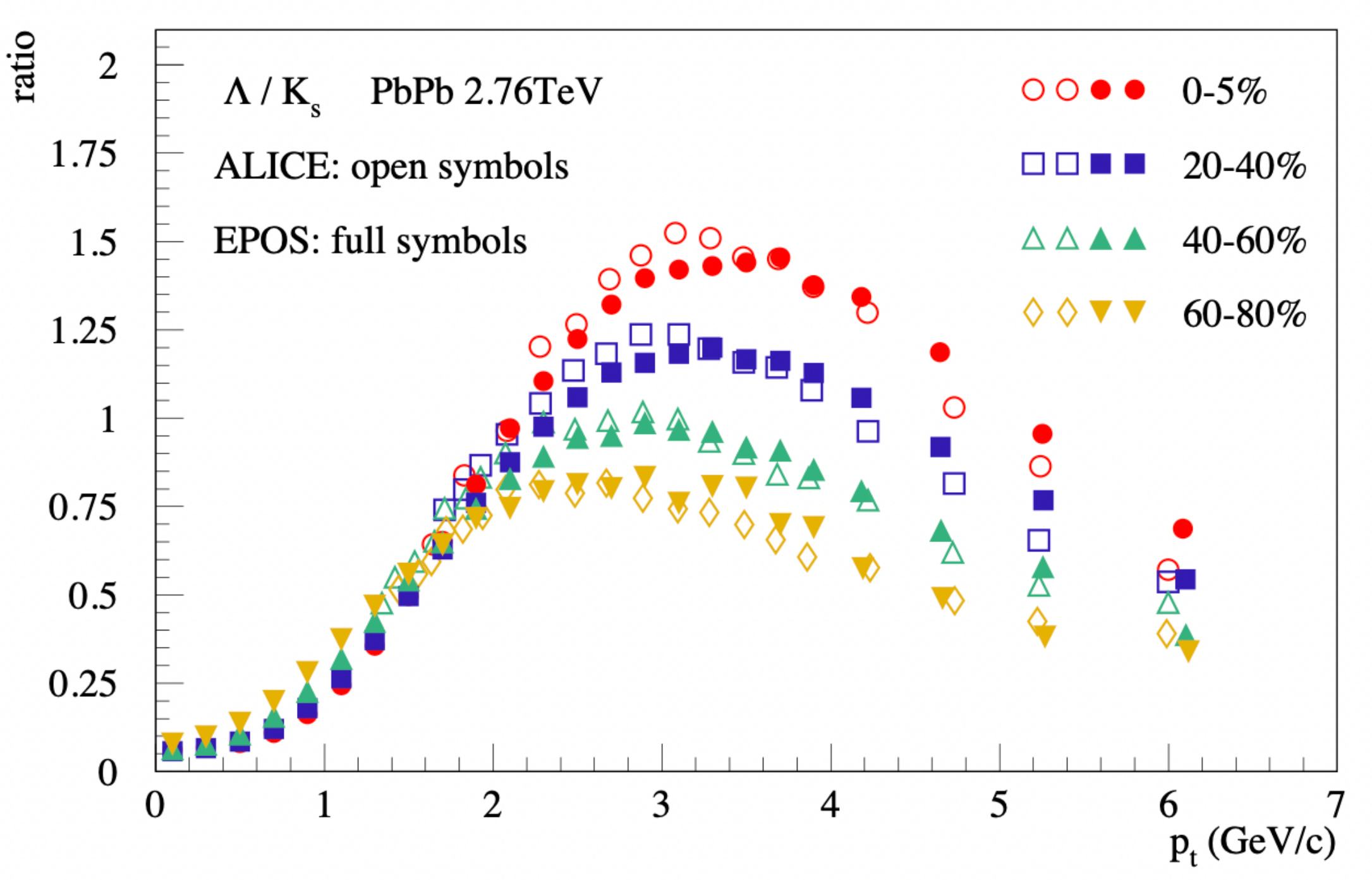}
\caption{Left panel: Recombination of thermal quarks from the QGP fluid with energetic string fragments. Right panel:  $p_T$-dependence of the hyperon-to-kaon ratio $N_\Lambda/N_K$ for different centrality windows in Pb+Pb collisions at $\sqrt{s_{\rm NN}} = 2.76$ TeV. ALICE data (open symbol) are shown together with results from the EPOS model (solid symbols). [From \cite{Werner:2012sv}]}
\label{fig:EPOS_K_Lambda}
\end{figure}

The scaled flow coefficients in Fig.~\ref{fig:ALICE_nq-scaling} exhibit broad peaks around $p_T/n_q \approx 1.5$ GeV/c. In the fragmentation-recombination scenario, this peak corresponds to a gradual transition to the fragmentation dominated regime. In the transition region it is plausible that quarks from a parton shower recombine with thermal, collectively flowing partons \cite{Hwa:2004ng}. An implementation of this idea in a dynamical model (EPOS) describes hadron formation at intermediate values of $p_T$ as string (color flux-tube) fragmentation in the presence of a thermal parton fluid \cite{Werner:2012sv,Werner:2012xh}. Figure \ref{fig:EPOS_K_Lambda} compares ALICE data \cite{Belikov:2011xk} for the $p_T$-dependence of the hyperon-to-kaon ratio $N_\Lambda/N_K$ for different centrality windows in Pb+Pb collisions at $\sqrt{s_{\rm NN}} = 2.76$ TeV with results from the EPOS model that accounts for recombination of the leading parton with thermal partons as well as fragmentation by quark-pair production from the vacuum \cite{Werner:2012xh}. 

\begin{figure}[htb]
\centering
\includegraphics[width=0.7\linewidth]{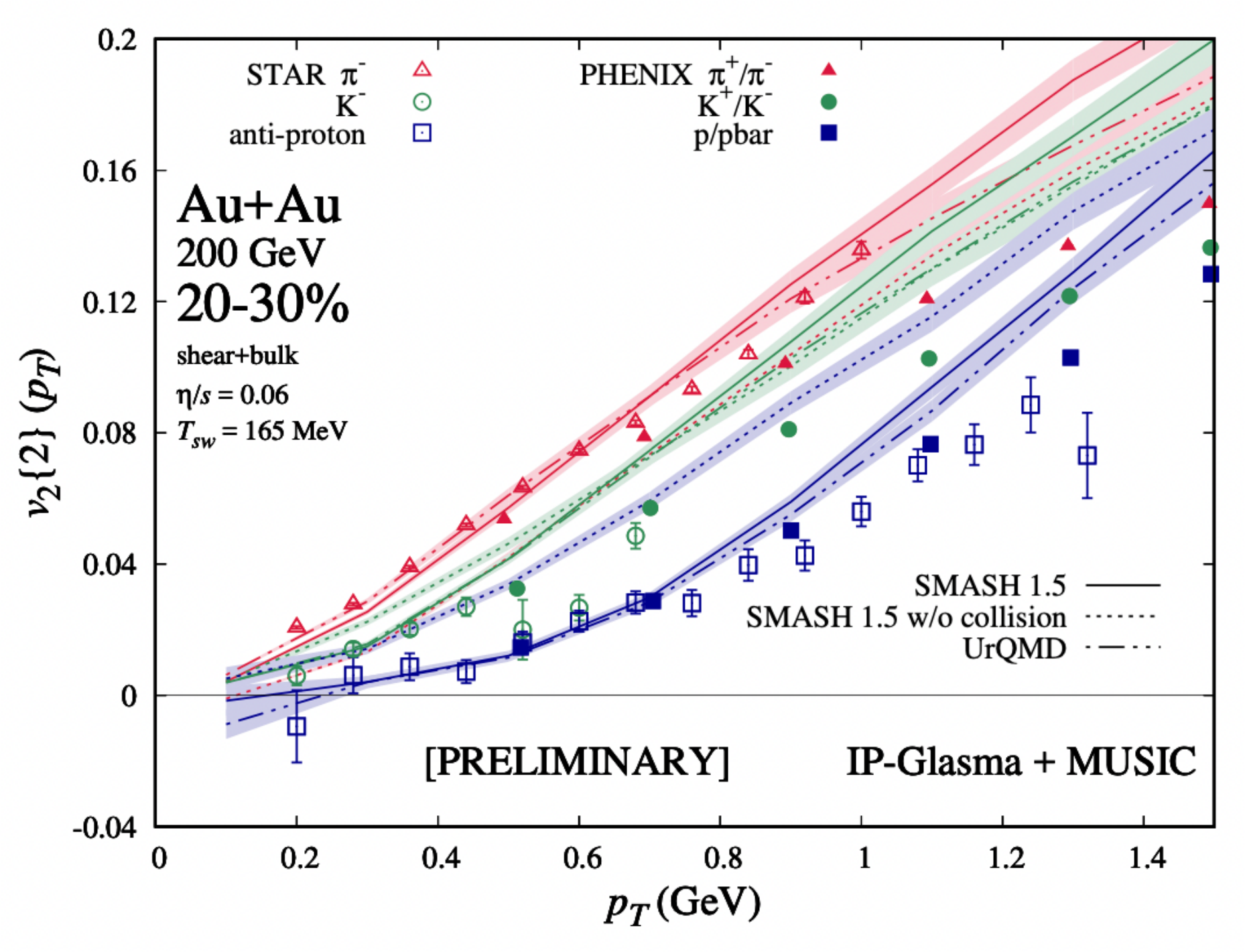}
\caption{Elliptic flow coefficient $v_2(p_T)$ for pions, kaons, and protons with and without hadronic rescattering in comparison with PHENIX data for Au+Au collisions at $\sqrt{s_{\rm NN}} = 200$ GeV. The solid and dash-dotted lines show calculations including hadronic rescattering; the dotted lines are calculated without rescattering. [From \cite{Ryu:2019atv}]}
\label{fig:v2-mass-splitting}
\end{figure}

At low momenta ($p_T < 1.5$ GeV/c) hadronic rescattering affects baryon and meson flow differently and amplifies the mass splitting observed in the unscaled elliptic flow $v_2(p_T)$ \cite{Lu:2006qn}. Pions move much faster than baryons and push them out to larger $p_T$ (``pion wind'', see also the end of Section \ref{sec:trans_expansion}), while heavy baryons have the opposite effect on pions and other mesons. Note that this mechanism does not work for $\phi$-mesons, as these do not have large cross sections with pions or nucleons \cite{Hirano:2007ei}. The increase of the mass splitting in $v_2(p_T)$ is clearly visible in Fig.~\ref{fig:v2-mass-splitting}, which shows $v_2(p_T)$ for pions, kaons, and protons calculated with and without hadronic rescattering in comparison with PHENIX data from Au+Au collisions at $\sqrt{s_{\rm NN}} = 200$ GeV \cite{Ryu:2019atv}.

\subsection{Jet Quenching and Parton Energy Loss}
\label{sec:jet_quenching}

The emission of hadrons at high transverse momentum ($p_T > 6$ GeV/c) in relativistic heavy ion collisions is suppressed because high-momentum quarks and gluons lose energy when they propagate through the quark-gluon plasma
\cite{Gyulassy:1990ye,Wang:1992qdg}. For light quarks or gluons the dominant energy mechanism is gluon radiation in association with scattering off a virtual gluon in the QGP. In the BDMPS-Z formalism of multiple scattering the scattering power of the QGP is encoded in the parameter $\hat{q}$, which describes the average squared momentum exchange with the medium per unit path length, $\hat{q} = d\langle q_T^2\rangle /dx$ \cite{Baier:1996sk,Zakharov:1996fv,Baier:2000mf}. The BDMPS-Z approach is well suited to describe energy loss in a thick QGP. The GLV formalism, which is based on an opacity expansion, is better suited for a thin QGP. The higher-twist formalism \cite{Guo:2000nz} aims at the description of the full virtuality evolution of a jet created by an energetic quark or gluon and is expected to apply to both, short and long path lengths (see also \cite{Caucal:2019uvr}).

\begin{figure}[htb]
\centering
\includegraphics[width=0.7\linewidth]{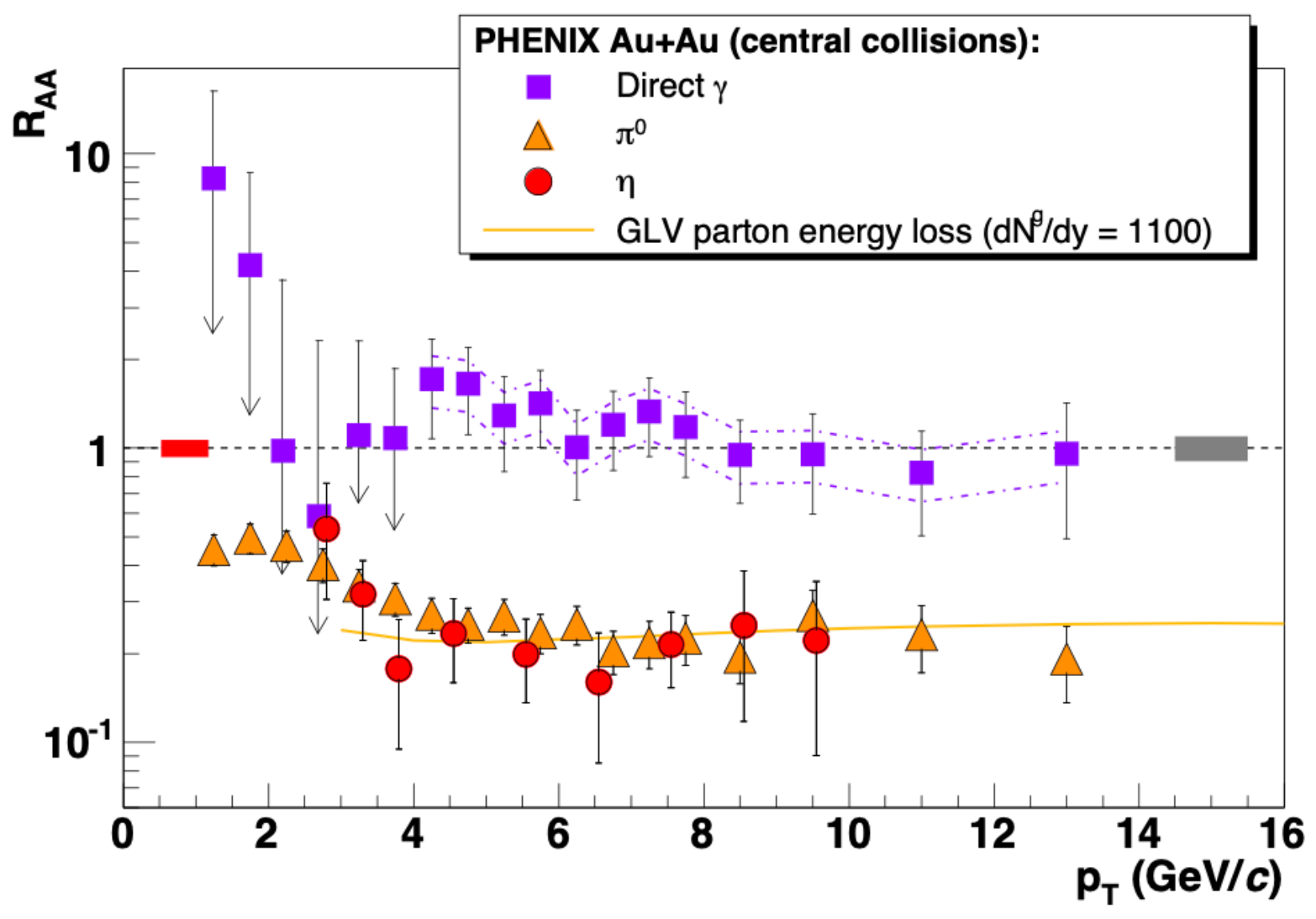}
\caption{Nuclear suppression factors $R_{\rm AA}(p_T)$ for direct photons, neutral pions and $\eta$-mesons in Au+Au collisions at $\sqrt{s_{\rm NN}} = 200$ GeV. Both meson species exhibit the same level of suppression, while photons are  not suppressed. [From \cite{PHENIX:2006ujp}]}
\label{fig:RAA_gpieta}
\end{figure}

Experimental evidence that high-$p_T$ hadron suppression is a final-state effect comes most directly from the comparison of the nuclear modification factors $R_{\rm AA}$ for hadrons with that for photons or $Z^0$ bosons. Figure \ref{fig:RAA_gpieta} shows that $R_{\rm AA}\approx 1$ for photons, whereas $R_{\rm AA}\approx 0.25$ for pions and $\eta$-mesons (see Eq.~(\ref{eq:RAA}) for the definition of $R_{AA}$).

\begin{figure}[htb]
\centering
\includegraphics[width=0.45\linewidth]{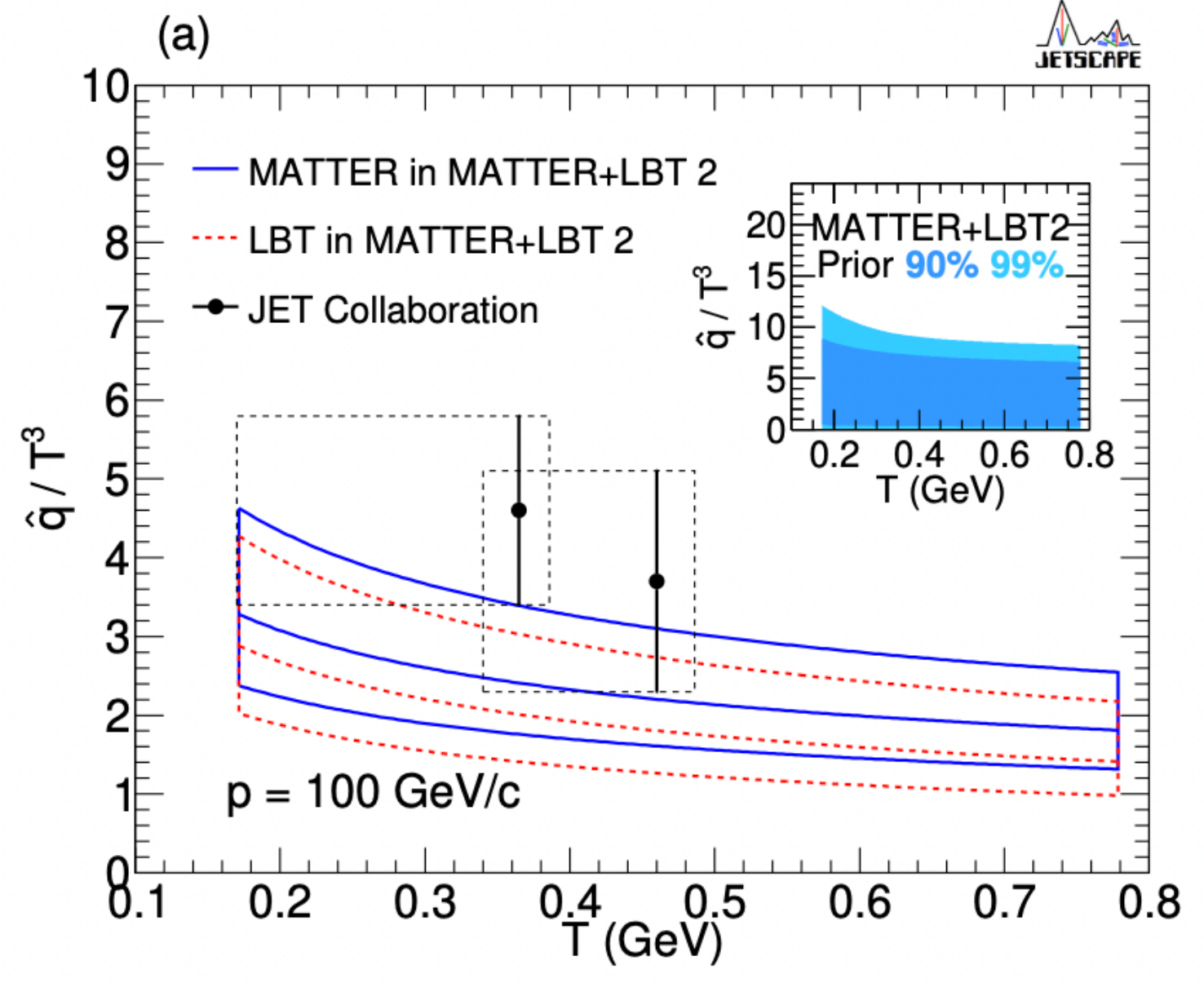}
\hspace{0.03\linewidth}
\includegraphics[width=0.45\linewidth]{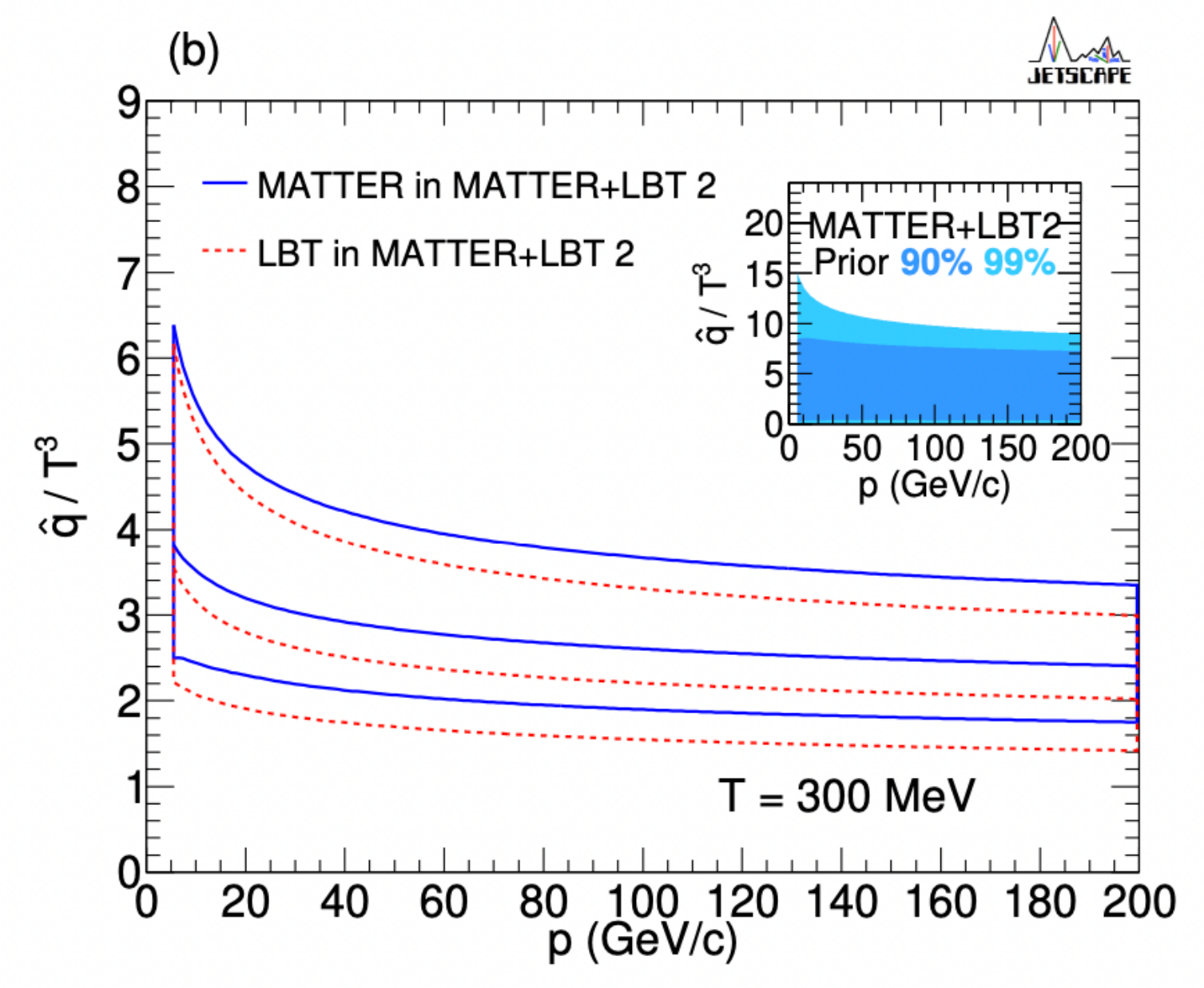}
\caption{Reduced jet quenching parameter $\hat{q}/T^3$ for quark-initiated jets in a quark-gluon plasma determined by a Bayesian analysis. $\hat{q}/T^3$ is shown as function of temperature $T$ (left panel) and quark-momentum $p$ (right panel). [From \cite{JETSCAPE:2021ehl}]}
\label{fig:qhat_Bayesian}
\end{figure}

Schematically, the hadron spectrum can be expressed as a convolution of the parton distribution functions $f_i^{(A)}(x)$ in the colliding nuclei with the hard QCD scattering cross section and a fragmentation function $D_{i\to h}^\mathrm{(med)}(z)$ that is modified by the medium
\begin{equation}
\frac{dN_h}{dp_T}(p_T) 
= f_i^{(A)}(x_1) \otimes f_j^{(A)}(x_2) \otimes \frac{d\sigma_{ij}}{dp_T}(p_T/z)
\otimes D_{i,j\to h}^\mathrm{(med)}(z) .
\end{equation}
The modified fragmentation function is related to the vacuum fragmentation function by
\begin{equation}
D_{i\to h}^\mathrm{(med)}(z) 
= \left\langle D_{i\to h}^\mathrm{(0)}\left(\frac{z}{1-z\Delta p_T/p_T}\right) \right\rangle ,
\end{equation}
where $\Delta p_T$ is the momentum loss of the hard parton in the medium and $\langle\cdots\rangle$ indicates an average over the position and orientation of the hard scattering event. 
\begin{figure}[htb]
\centering
\includegraphics[width=0.65\linewidth]{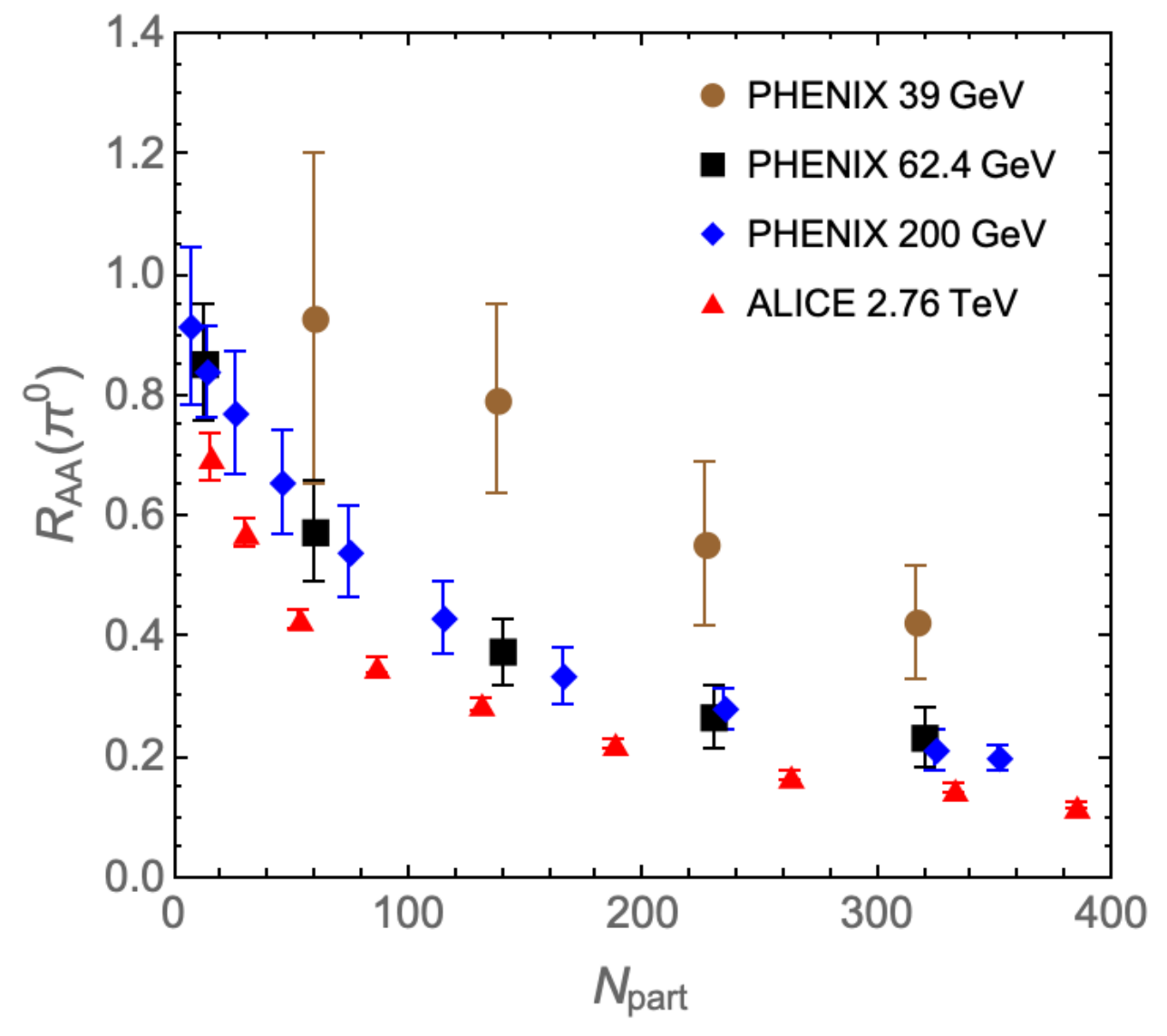}
\caption{Nuclear suppression factors $R_{\rm AA}$ for neutral pions in Au+Au collisions as a function of participant number $N_{\rm part}$ at $\sqrt{s_{\rm NN}} = 39, 62.4, 200$ GeV and Pb+Pb collisions at $\sqrt{s_{\rm NN}} = 2.76$ TeV.  [From \cite{PHENIX:2012oed,ALICE:2014hpa}]}
\label{fig:RAA_Npart}
\end{figure}
The magnitude of the suppression thus depends not only on the amount of energy loss but also on the steepness of the unmodified hadron spectrum. As the spectrum becomes flatter at high collision energy, the same energy loss causes less suppression. In spite of this effect, experimental data from RHIC and LHC confirms that the suppression increases with collision energy, as shown in Fig.~\ref{fig:RAA_Npart} for neutral pions with $p_T > 6$ GeV/c, implying a strong increase in the energy loss. This observation agrees with expectations, as the energy loss parameter $\hat{q}$ grows rapidly with temperature: $\hat{q}/T^3 \approx 2-5$ (see Fig.~\ref{fig:qhat_Bayesian}). 

\begin{figure}[htb]
\centering
\includegraphics[width=0.35\linewidth]{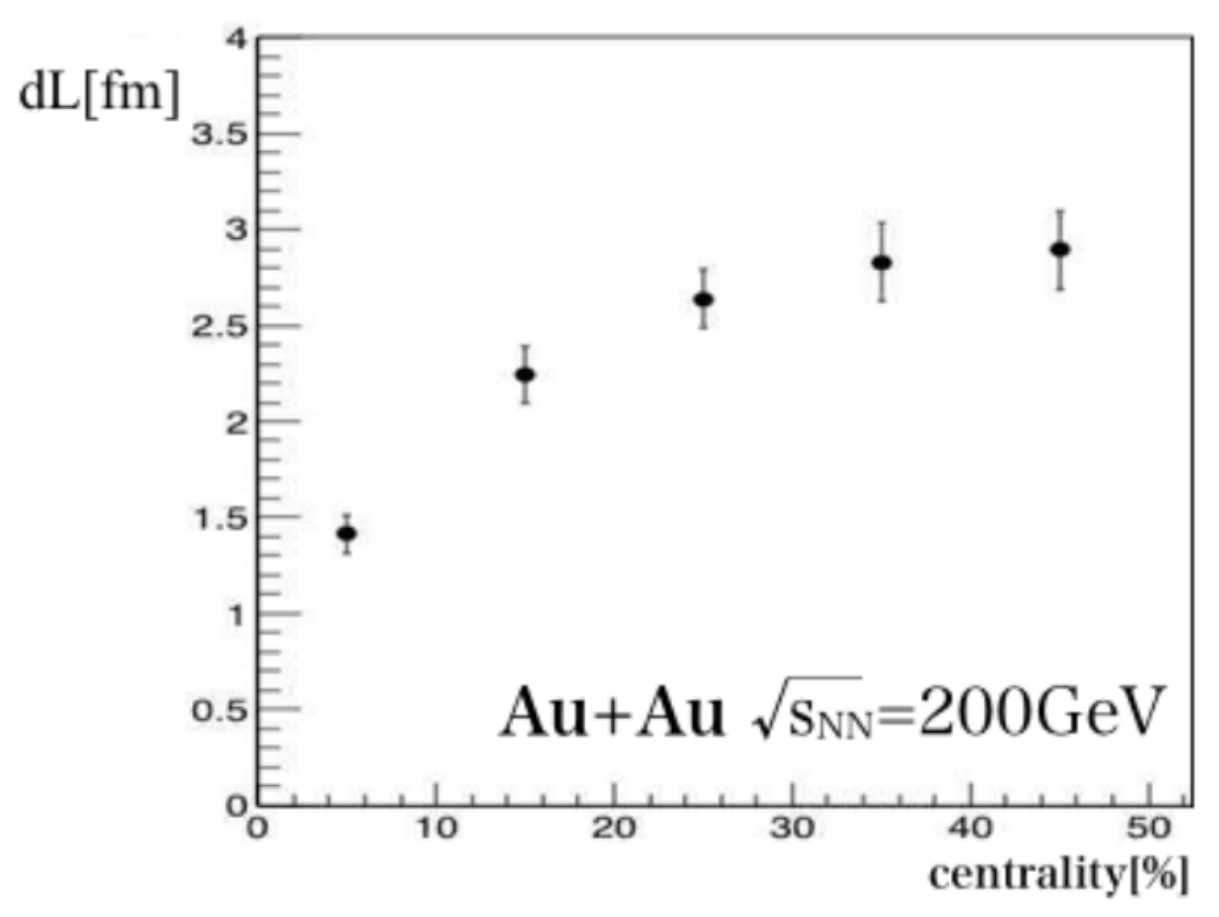}
\hspace{0.05\linewidth}
\includegraphics[width=0.55\linewidth]{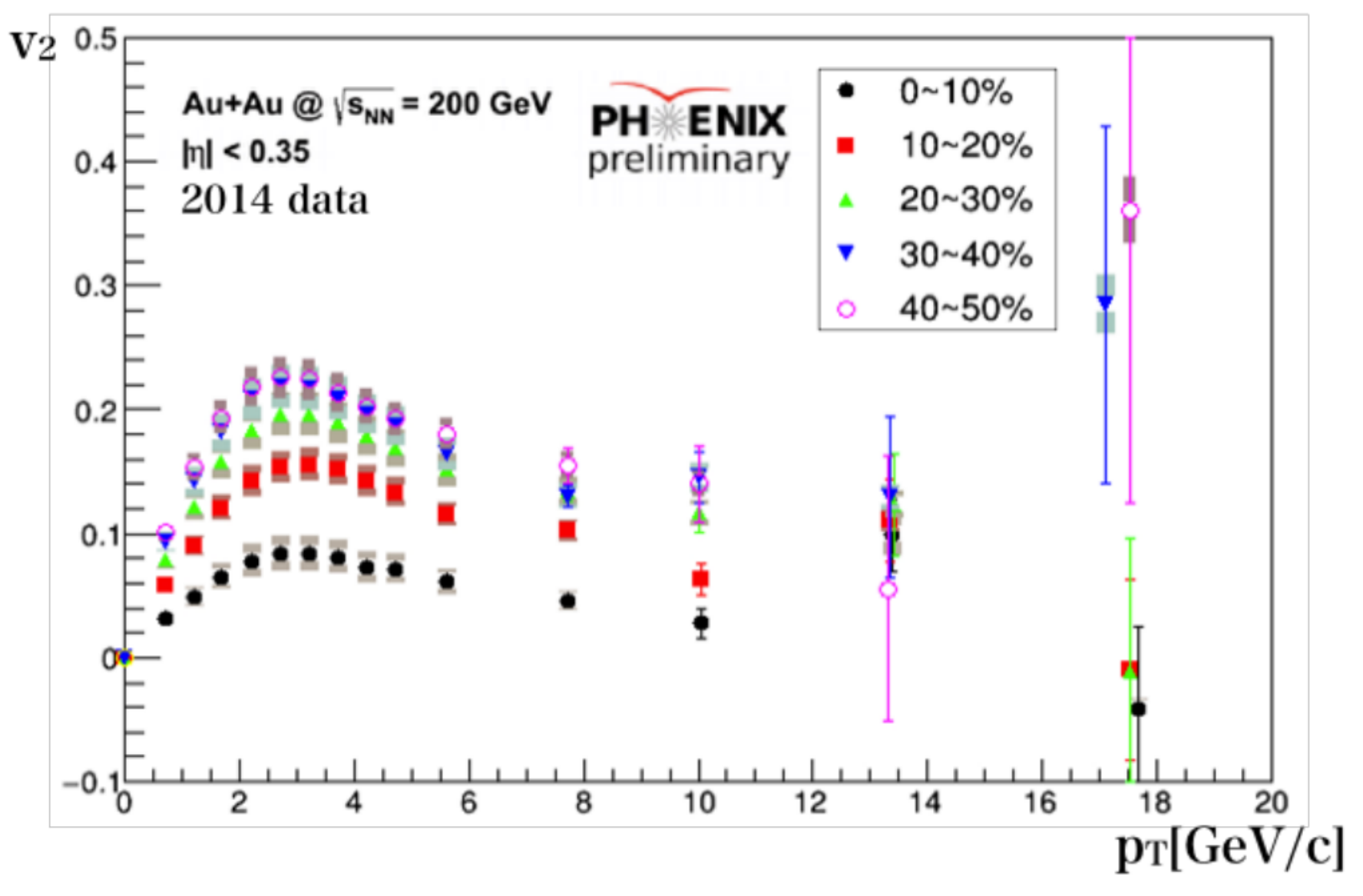}
\caption{Left panel: Average path-length difference $dL$ for a parton emitted perpendicular to the reaction plane compared with a parton emitted along the reaction plane. $dL$ does not vanish even in central collisions because of density fluctuations in the initial state, which generate an azimuthal anisotropy. Right panel: Azimuthal anisotropy coefficient $v_2(p_T)$ for charged hadrons in Au+Au collisions at $\sqrt{s_{\rm NN}} = 200$ GeV for different centralities.   [From \cite{Nishitani:2019tcy}]}
\label{fig:v2_highpT_PHENIX}
\end{figure}

\begin{figure}[htb]
\centering
\includegraphics[width=0.85\linewidth]{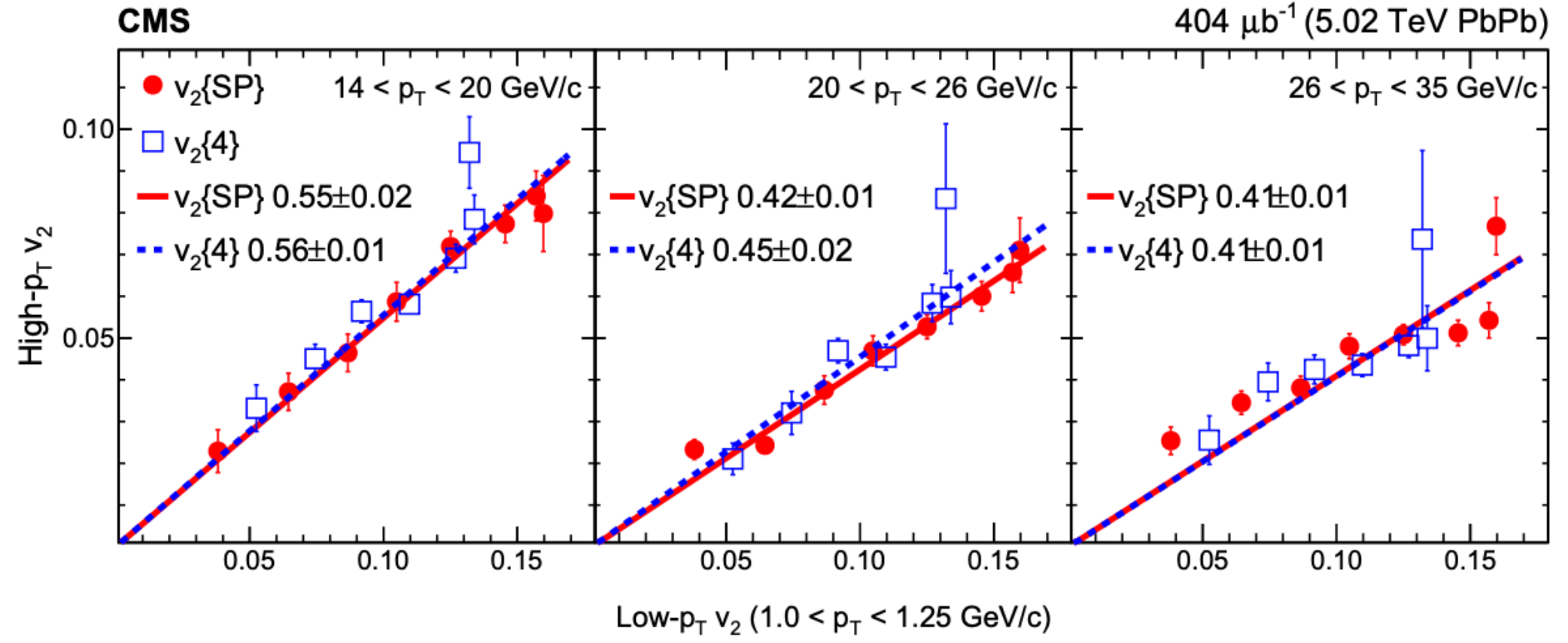}
\caption{Correlation between the azimuthal anisotropy coefficient $v_2(p_T)$ of charged hadrons in the high-$p_T$ region with the same coefficient measured in the low-$p_T$ region where it is considered a measure of the collective (``elliptic'') flow. The data are for Pb+Pb collisions at $\sqrt{s_{\rm NN}} = 5.02$ TeV. The linear correlation indicates that both phenomena have the same underlying geometric origin.  [From \cite{CMS:2017xgk}]}
\label{fig:v2_highpT_CMS}
\end{figure}

The increase with $N_{\rm part}$ reflects the strong path-length dependence of the radiative energy loss which, for a medium of constant density, is approximately given by \cite{Baier:2000mf}:
\begin{equation}
\Delta E = -\frac{1}{2} C_2 \alpha_s \hat{q} L^2 ,
\label{eq:DeltaE}
\end{equation}
where $C_2$ denotes the SU(3) Casimir operator for the energetic parton. An independent assessment of the path-length dependence of parton energy loss can be obtained by measuring the azimuthal anisotropy of the nuclear suppression with respect to the collision plane. This anisotropy can be expressed by the Fourier coefficient $v_2(p_T)$, defined as
\begin{equation}
\frac{dN(p_T)}{d\phi} \propto 1 + 2v_2(p_T)\cos(2\phi)
\end{equation}
as the ``elliptic flow'' coefficient. The average difference in path length $dL$ for partons emitted perpendicular to the collision plane compared with those emitted along the plane is quite large as shown in the left panel of Fig.~\ref{fig:v2_highpT_PHENIX}, implying that a significant dependence on the emission angle relative to the collision plane is to be expected. This expectation is confirmed by data from Au+Au collisions at RHIC, see the right panel of Fig.~\ref{fig:v2_highpT_PHENIX} which shows a sizable value of $v_2$ for hadrons up to 10 GeV/c momentum. Data from Pb+Pb collisions at LHC for much higher $p_T$, shown in Fig.~\ref{fig:v2_highpT_CMS}, reveal a strong correlation with the elliptic flow coefficient $v_2$ measured in the low-$p_T$ region and indicate that the anisotropy at high $p_T$ has the same geometric origin.

\begin{figure}[htb]
\centering
\includegraphics[width=0.65\linewidth]{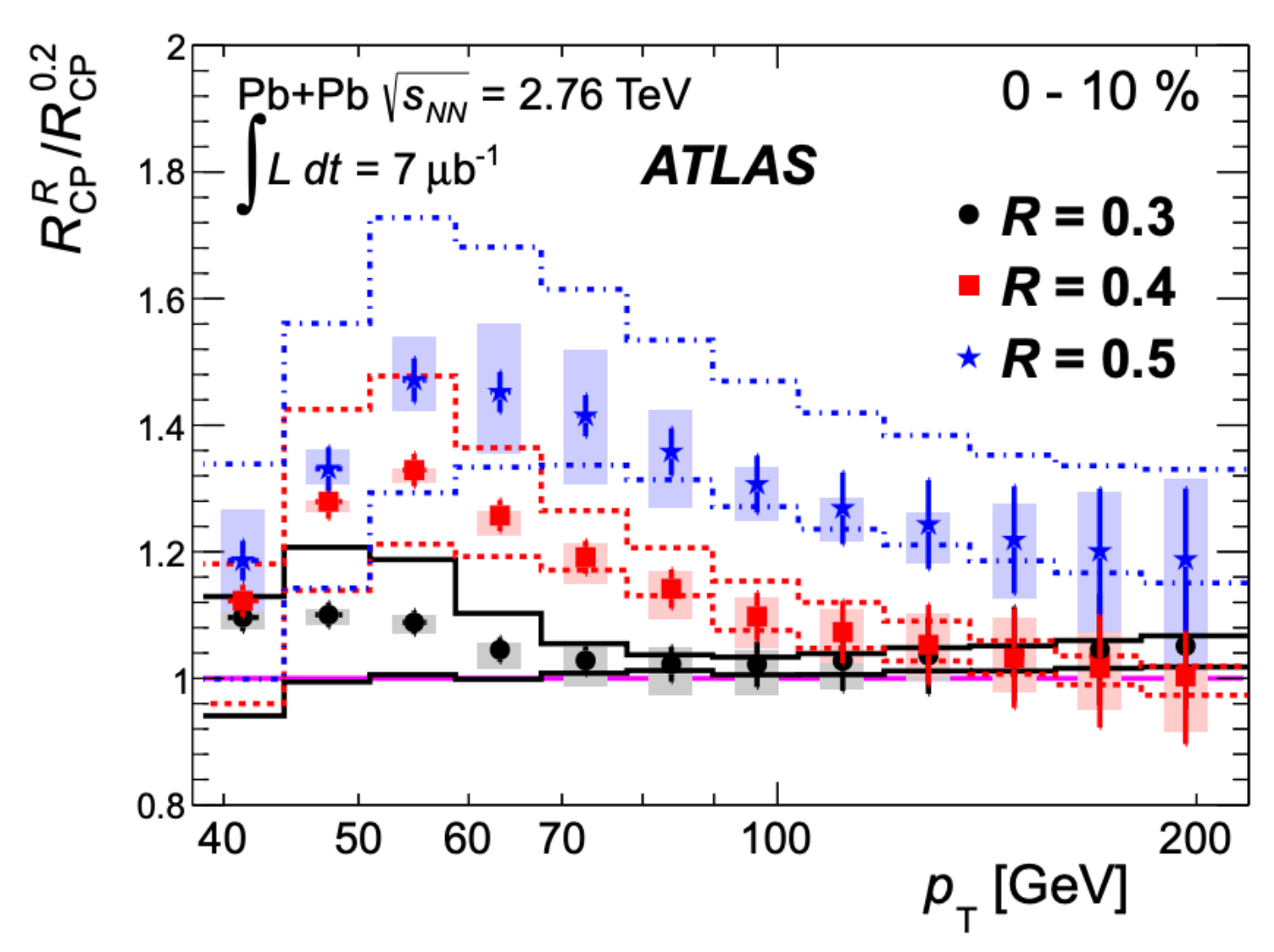}
\caption{Relative suppression factor $R_{\rm CP}^R(p_T)/R_{\rm CP}^0.2(p_T)$ for jets in Pb+Pb collisions at $\sqrt{s_{\rm NN}} = 2.76$ TeV. A larger value of $R_{\rm CP}$ means less suppression. The figure shows that more in-medium radiated energy is recovered for larger opening angles $R$.  [From \cite{ATLAS:2012tjt}]}
\label{fig:RCPvsR_ATLAS}
\end{figure}

Unless they are absorbed by the QGP, the gluons that are radiated by a fast parton in the medium, remain part of the full jet. What fraction of the radiated energy is recovered in measurements of the full jet energy depends on the opening angle $R$ (the ``jet radius'') that is used to define the jet. Typical values used for such studies are $R = 0.2-0.6$. Smaller jet radii imply a larger energy loss. This trend is readily apparent in Fig.~\ref{fig:RCPvsR_ATLAS}, which shows the relative magnitude of $R_{\rm CP}(p_T)$ for jets with total transverse momentum $p_T$ and radius $0.2 \leq R \leq 0.5$. 

Full jet quenching exhibits many of the same phenomena as the suppression of single hadrons at high $p_T$, except that all results depend quantitatively on the jet radius $R$. This means that full jet quenching measurements are not only sensitive to longitudinal energy loss, but also to the angular redistribution of the energy within the jet \cite{Casalderrey-Solana:2010bet}. Measurements of full jet suppression thus enable more differential measurements, e.~g., the study of how quenching modifies the jet shape in terms of the longitudinal momentum fraction of a hadron within the jet, $z_h=p_T^h/p_T^{\rm jet}$, and the relative angle $r<R$ of the momentum of a hadron with respect to the jet axis (see \cite{Connors:2017ptx} for a review on jet measurements). 

\begin{figure}[htb]
\centering
\includegraphics[width=0.4\linewidth]{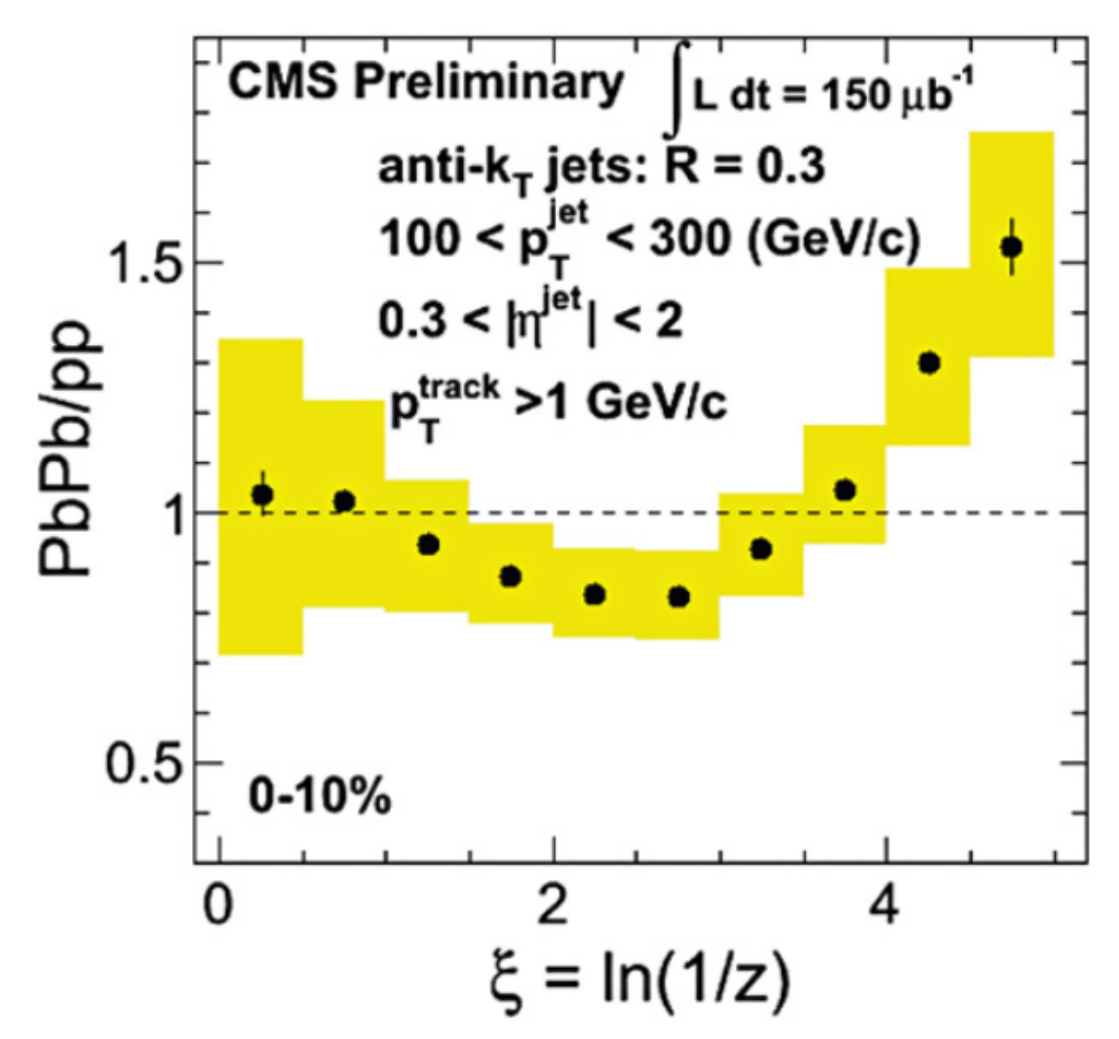}
\hspace{0.05\linewidth}
\includegraphics[width=0.4\linewidth]{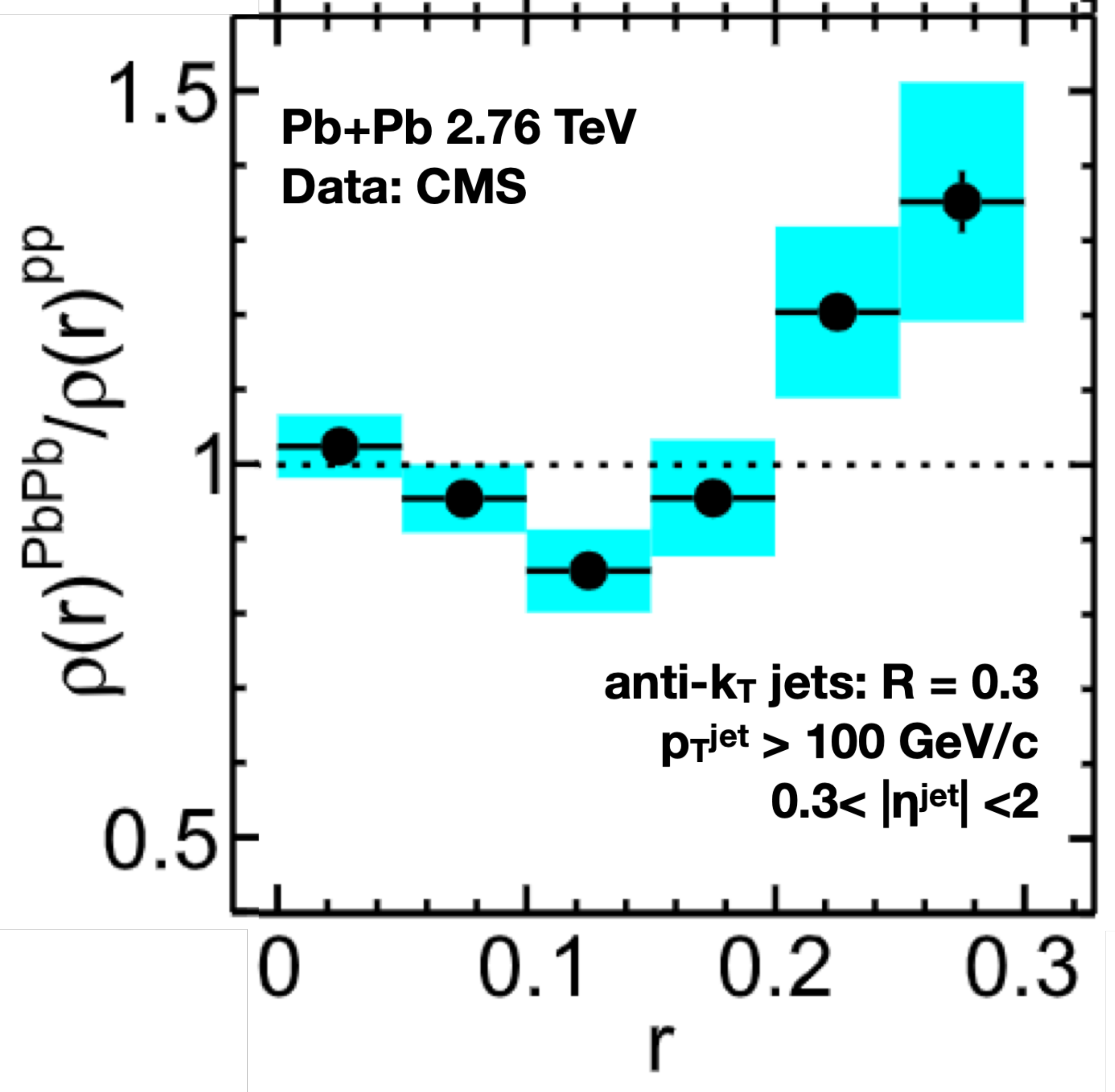}
\caption{Left panel: Modification of the longitudinal jet shape $\rho(\xi)$ in Pb+Pb collisions at $\sqrt{s_{\rm NN}} = 2.76$ TeV relative to $pp$ collisions. The hadron density in the jet is suppressed at moderate values of $z=e^{-\xi}$ and enhanced at small values, reflecting the increase in the soft components of the jet caused by additional gluon radiation in the QGP. 
Right panel: Modification of the transverse jet shape $\rho(r)$ in Pb+Pb collisions at $\sqrt{s_{\rm NN}} = 2.76$ TeV relative to $pp$ collisions for jets with $R=0.3$. The hadron density in the jet is shifted to larger angles $r>0.2$, reflecting the redistribution of energy within the jet cone by gluon radiation in the QGP.
[From \cite{Mao:2014jja}]}
\label{fig:jet_shape}
\end{figure}

Often the jet shape $\rho(\xi,r)$ is expressed in terms of the variables $\xi=\ln(1/z)$ and $r$. Examples of the modification of the jet shape in Pb+Pb collisions compared with p+p collisions are shown in Fig.~\ref{fig:jet_shape}. The data show ``softening'' of the shape of the jet in terms of a redistribution of the energy in the jet to smaller $z$ and larger angles $r$. This is precisely the pattern expected from in-medium gluon radiation, which involves lower parton virtuality than vacuum radiation and does not exhibit the same angular ordering that suppresses low-$z$, large-angle radiation.

\begin{figure}[htb]
\centering
\includegraphics[width=0.65\linewidth]{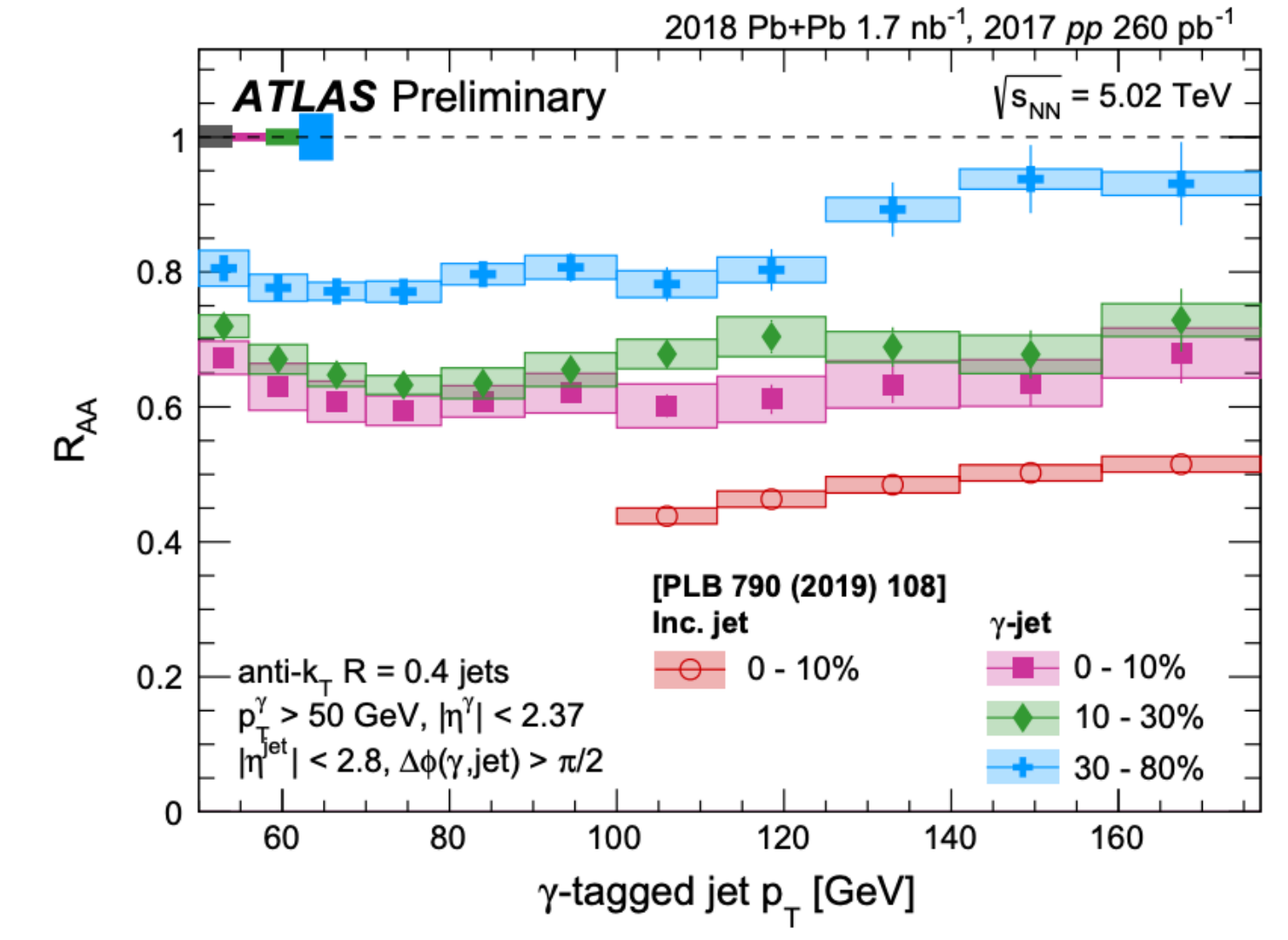}
\caption{Comparison of the suppression factor $R_{\rm AA}(p_T)$ of $R=0.4$ jets in Pb+Pb collisions at $\sqrt{s_{\rm NN}} = 5.02$ TeV for inclusive jets and photon-tagged jets. The observation that $R_{\rm AA}$ is larger for photon-tagged jets can be attributed to the fact that these contain a much smaller fraction (20--30\%) of gluon-initiated jets than inclusive jets (50-65\%).  [From \cite{ATLAS:2022cim}]}
\label{fig:RAA_taggedjets}
\end{figure}

Photon-tagged jets enhance the fraction of jets initiated by hard scattered quarks over those initiated by gluons, from 35--50\% to 70-80\% at LHC energies \cite{ATLAS:2022cim}. A comparison between photon-tagged jets and inclusive jets thus allows to probe the color charge dependence of parton energy loss expressed by the dependence of the energy loss (\ref{eq:DeltaE}) on the color-SU(3) Casimir operator ($C_2 = 4/3$ for quarks and $C_2 = 3$ for gluons). Data from ATLAS shown in Fig.~\ref{fig:RAA_taggedjets} confirm the expectation that jets initiated by quarks are less suppressed than those initiated by gluons, manifested in a larger suppression of inclusive jets than of photon-tagged jets.

Jets generally occur in pairs (di-jets) where one jet balances the transverse momentum of the other. This means that the relative di-jet distribution is strongly peaked at $180^\circ$ in azimuthal angle. While highly correlated in azimuthal angle $\phi$, di-jets are not strongly correlated in pseudorapidity $\eta$ but separated by a variable gap $\Delta\eta \sim \ln(x_2/x_1)$, where $x_i$ are the momentum fractions of the colliding partons that produce the di-jet. In order to localize both partners of the dijet, one employs two trigger particles, one for each jet, or two calorimeter-based triggers. The particle distribution in each jet is then measured relative to the ($\eta$,$\phi$) coordinates of the respective trigger and denoted as \cite{STAR:2012civ}
\begin{equation}
    \frac{1}{N_{\rm trig}} \frac{d^2N}{d(\Delta\eta) d(\Delta\phi)} .
\end{equation}
In order to isolate di-jets in heavy ion collisions one typically also imposes a lower $p_T$-cutoff on the included particles and subtracts the randomized background from mixed minimum-bias events \cite{STAR:2012civ}.

\begin{figure}[htb]
\centering
\includegraphics[width=0.48\linewidth]{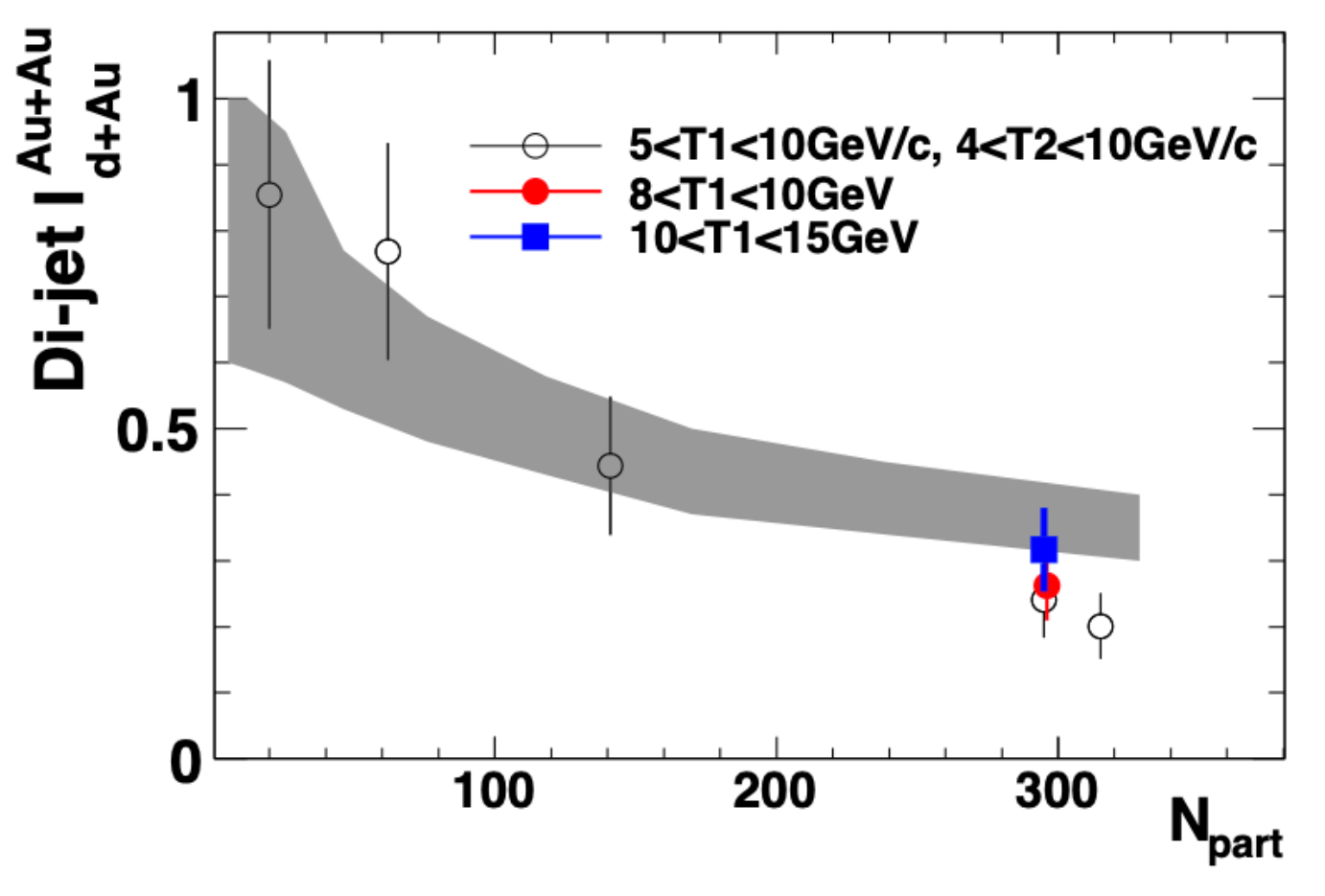}
\hspace{0.03\linewidth}
\includegraphics[width=0.45\linewidth]{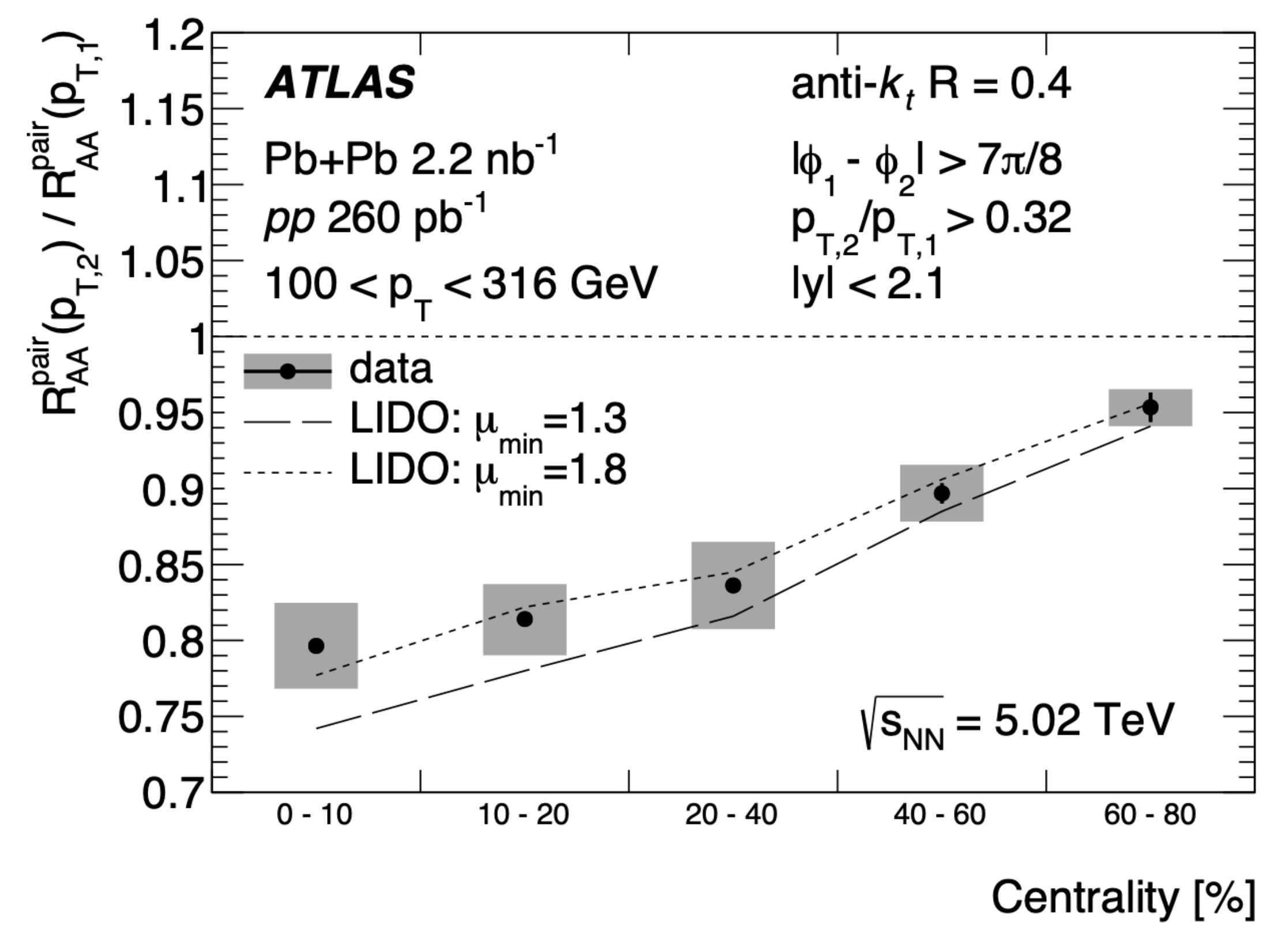}
\caption{Left panel: Di-hadron suppression factor $I_{\rm AA}$ versus centrality (participant number $N_{\rm part}$) for Au+Au collisions at $\sqrt{s_{\rm NN}} = 200$ GeV measured by STAR. The different symbols indicate different trigger selections. The grey band shows the expectation for di-jet surface emission. See \cite{STAR:2012civ} for details.
Right panel: Di-jet suppression factor $I_{\rm AA}^{\rm dijet}$ measured by ATLAS in $\sqrt{s_{\rm NN}} = 5.02$ TeV Pb+Pb collisions. For details see \cite{ATLAS:2022zbu}.}
\label{fig:dijets}
\end{figure}

There are two main observables that have been studied for di-jets in $A+A$ collisions. One is the additional nuclear suppression of high-$p_T$ hadron pairs (di-hadrons) relative to the suppression of single inclusive hadrons. Such a suppression is to be expected, because both di-jet precursor partons propagate through the quark-gluon plasma and lose energy. The additional suppression for inclusive hadron pairs is expressed in terms of the quantity
\begin{equation}
    I_{\rm AA} = \frac{R_{\rm AA}^{\rm di-jet~triggers}}{R_{\rm AA}^{\rm single~triggers}} .
\end{equation}
The left panel of Fig.~\ref{fig:dijets} shows the $I_{\rm AA}$ in Au+Au collisions at $\sqrt{s_{\rm NN}} = 200$ GeV, relative to the baseline from d+Au collisions at the same energy. As one can see, the additional suppression in central collisions is comparable to the single suppression factor $R_{\rm AA}$ shown in Fig.~\ref{fig:RAA_Npart}. This is to be expected as the average path lengths of both scattered partons in the medium are comparable, and thus both partons suffer similar energy loss.
The right panel of Fig.~\ref{fig:dijets} shows the analogous di-jet suppression factor $I_{\rm AA}^{\rm dijet}$ for the complete jets measured in Pb+Pb collisions at $\sqrt{s_{\rm NN}} = 5.02$ TeV. This quantity is to be compared with the single jet $R_{AA}^{\rm jet}$ shown in Fig.~\ref{fig:RAA_taggedjets}. The additional suppression grows with centrality, but is generally less severe than the suppression observed for single jets.

\begin{figure}[htb]
\centering
\includegraphics[width=0.45\linewidth]{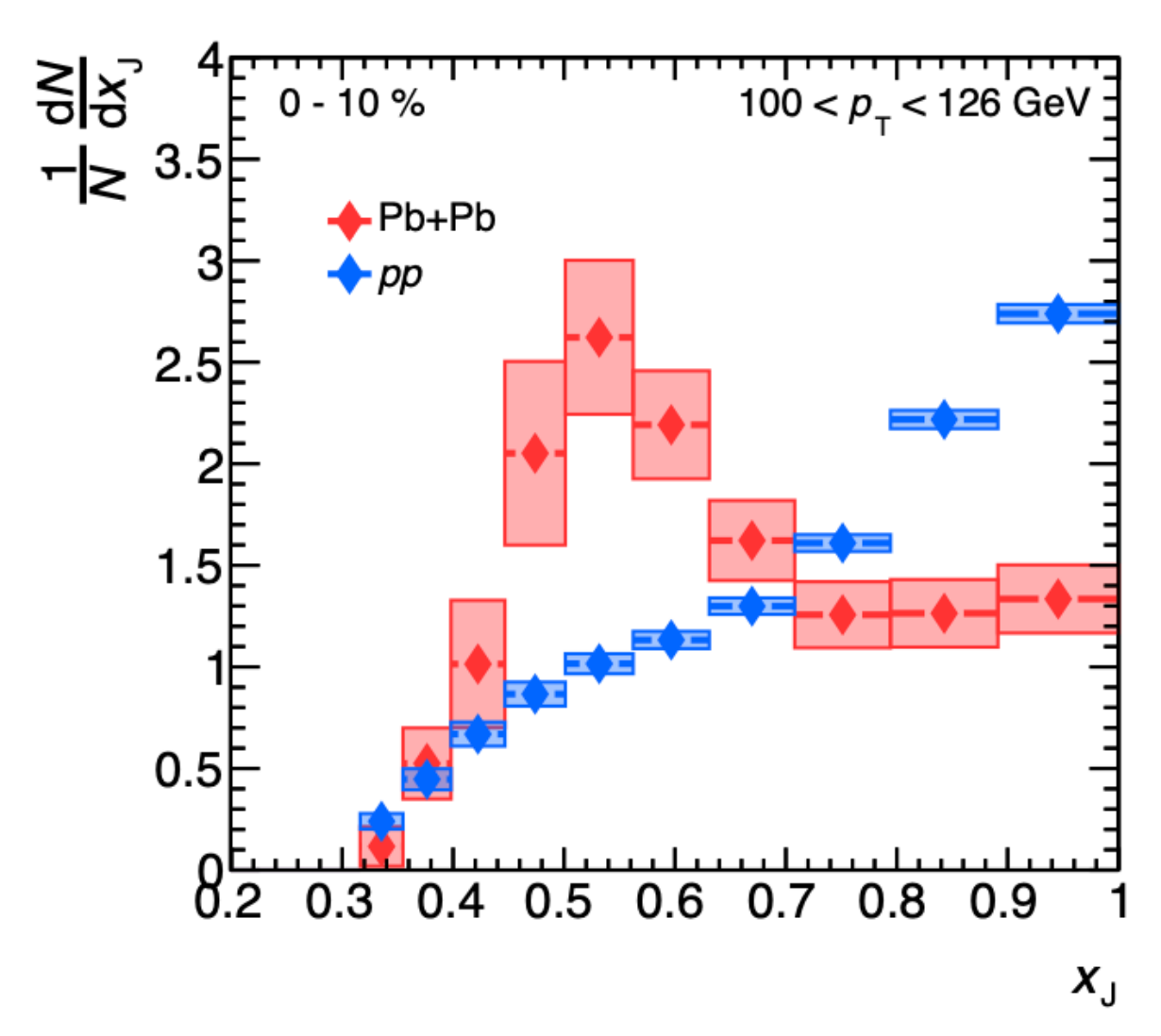}
\hspace{0.05\linewidth}
\includegraphics[width=0.45\linewidth]{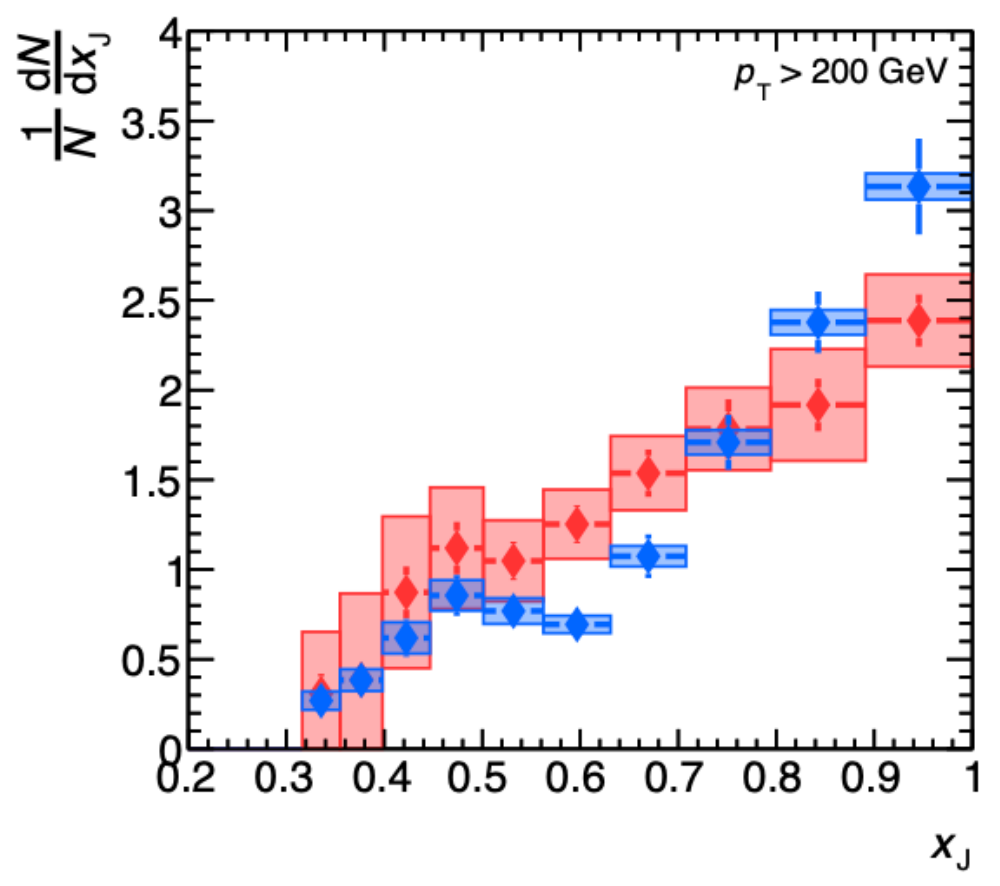}
\caption{Di-jet balance ratio $x_{\rm J}$ in Pb+Pb collisions at $\sqrt{s_{\rm NN}} = 2.76$ TeV measured by ATLAS. Left panel: $p_{T,1} > 100$ GeV; right panel: $p_{T,1} > 200$ GeV. (From \cite{Perepelitsa:2016zbe})}
\label{fig:dijets_xJ}
\end{figure}

The other observable is the di-jet asymmetry $A_{\rm J}$ or, equivalently, the di-jet imbalance ratio $x_{\rm J}$, defined as
\begin{equation}
    a_{\rm J} = \frac{p_{T,1}-p_{T,2}}{p_{T,1}+p_{T,2}}, \qquad
    x_{\rm J} = \frac{p_{T,2}}{p_{T,1}},
\end{equation}
where $p_{T,1}$ and $p_{T,2}$ denote the transverse momenta of the leading and sub-leading jet, respectively. The di-jet balance ratio $x_{\rm J}$ in central Pb+Pb collisions at $\sqrt{s_{\rm NN}} = 2.76$ TeV is shown in Fig.~\ref{fig:dijets_xJ} in comparison with the same ratio in p+p collisions. The trigger conditions were either $p_{T,1} > 100$ GeV (left panel) or $p_{T,1} > 200$ GeV (right panel) and $p_{T,2} > 25$ GeV for jets within $|\eta| < 2.1$. For the lower trigger energy (left panel) the di-jet balance ratio distribution in Pb+Pb collisions is peaked around $x_{\rm J} \approx 0.5$ and differs strongly from the distribution observed in p+p collisions. 

This behavior can be interpreted as follows: When the jet is produced well outside the center of the fireball, one of the jets traverses a substantially shorter distance through the medium than the other. This causes a larger energy loss, which is reflected in a ratio $x_{\rm J}$ significantly smaller than unity. Interestingly, the difference between Pb+Pb and p+p collisions shrinks with increasing trigger threshold until the distributions are statistically indistinguishable for $p_{T,1} > 200$ GeV. The same trend is found when one goes from central to peripheral collisions \cite{Perepelitsa:2016zbe,ATLAS:2022zbu}.

\subsection{Heavy Quark Probes}
\label{sec:heavy_quarks}

Hadrons containing heavy quarks ($c$ or $b$ quarks) are of interest both theoretically and experimentally. One distinguishes hadrons with open heavy flavor, such as $D$- and $B$-mesons or $\Lambda_c$ baryons, and those with hidden heavy flavor, such as charmonium ($c\overline{c}$) and bottomonium ($b\overline{b}$). Open heavy flavor hadrons in their ground states decay weakly and live long enough so that their decay can be identified by micro-vertex detectors. Hadrons with hidden heavy flavor can decay either electromagnetically or by gluon-mediated strong interaction. For certain states (especially those with quantum number $J^{PC} = 1^{--}$, e.~g.\ $J/\psi$ and $\Upsilon$) the strong decay channels are so strongly suppressed that the decay into lepton pairs dominates, making them readily detectable. An extensive survey of heavy quark physics in relativistic heavy ion collisions (as of 2015) can be found in \cite{Andronic:2015wma}.

On the theoretical side, heavy quarks are interesting because they are almost exclusively produced during the initial stage of the reaction by hard QCD processes, mainly $g+g \rightarrow Q+\overline{Q}$. They may or may not subsequently thermalize, which is an interesting question by itself, but their number remains essentially conserved from then until hadronic freeze-out. Inside the QGP, when light quarks are deconfined, hadrons containing both heavy and light quarks cannot exist. During hadronization, such hadrons are created by recombination of deconfined heavy quarks with light quarks or antiquarks. However, hadrons containing solely heavy quarks may survive under conditions not too far above the deconfinement threshold because their binding radii are small and their binding energies are large compared with the temperature. 

The question, above which temperature $T_\mathrm{m}^{(H)}$ a specific heavy heavy quark bound state $H$ ``melts'', has been studied in great detail using lattice QCD. Initial investigations focused on static color screening studies \cite{Karsch:1990wi} but more recently the focus has shifted to the investigation of dynamic properties encoded in the spectral functions \cite{Jakovac:2006sf,Petreczky:2021zmz}, which include non-static effects, such as ionization by thermal gluons. More generally, the heavy quark mass provides for a large momentum scale $M_Q \gg \Lambda_\mathrm{QCD}$ that enables various effective field theory approaches to QCD, known as Heavy Quark Effective Theory (HQET) or nonrelativistic QCD (NRQCD). In combination with HTL perturbation theory techniques, these approaches form a rigorous theoretical framework for the calculation of transport properties of heavy quarks in the QGP, including the formation and destruction of quarkonia \cite{Rothkopf:2019ipj}. 

Experimentally, the inclusive study of the transport properties of heavy quarks in the QGP relies on the measurement of single leptons ($e, \mu$) emitted in their semi-leptonic weak decays \cite{PHENIX:2005nhb}. The discrimination between leptons from $b$-decays and those of $c$-decays requires, in addition, the identification of the decay vertex where one uses, on a statistical basis, the property that $b$-quarks have a longer average lifetime than $c$-quarks. The comparison of the inclusive lepton spectrum measured in A+A collision with the binary collision-scaled spectrum measured in p+p collisions yields information about the transport of heavy quarks in the QGP. This information is usually presented as nuclear modification factor $R_{\rm AA}$ plotted as a function of $p_T$. In order to relate to nuclear modification of the heavy quark spectrum, the lepton spectrum requires unfolding with the decay spectrum of the parent hadrons in their rest frame, which is an ill-defined procedure. One therefore usually compares the data with calculations that include the weak decays of open heavy flavor hadrons. 

\begin{figure}[htb]
\centering
\includegraphics[width=0.45\linewidth]{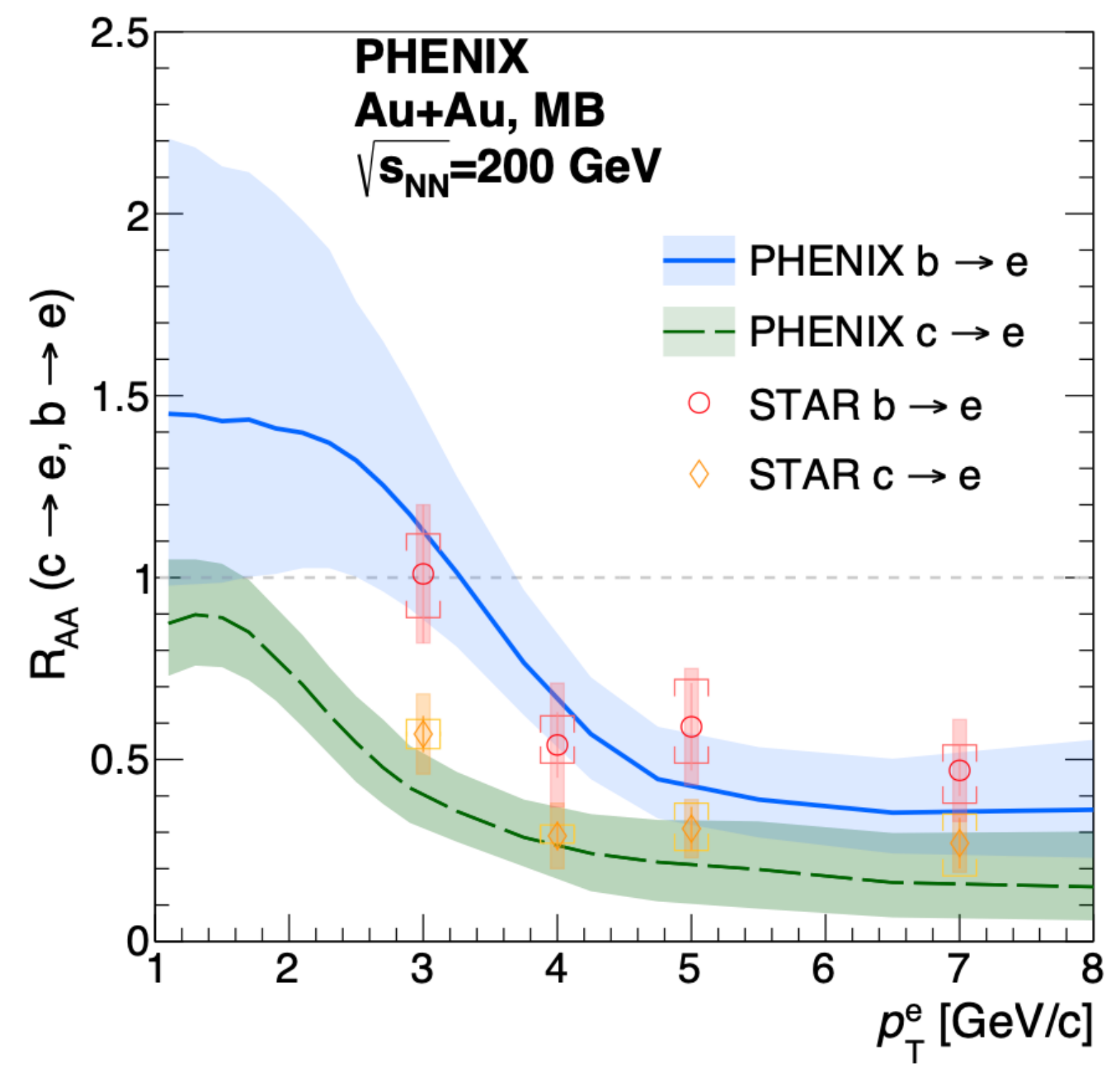}
\hspace{0.05\linewidth}
\includegraphics[width=0.45\linewidth]{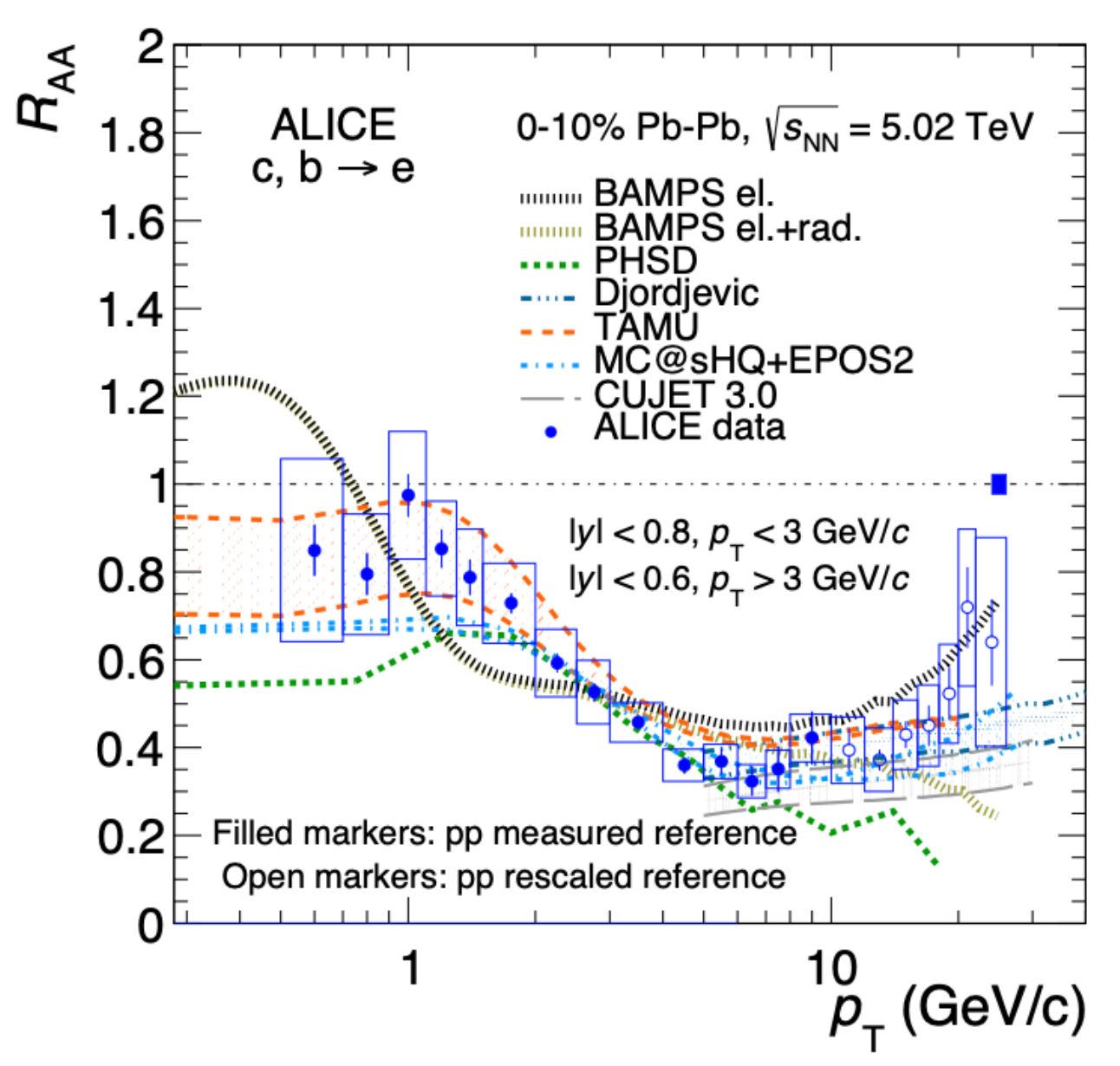}
\caption{Nuclear suppression factor $R_{\rm AA}(p_T)$ for single electrons from semi-leptonic heavy quark decays. Left panel: PHENIX and STAR results for vertex separated electrons from $b$- and $c$-quark decays measured in central Au+Au collisions at $\sqrt{s_\mathrm{NN}} = 200$ GeV.  [From \cite{PHENIX:2022wim}] Right panel: Results from ALICE for electrons from $b,C$ decays in $\sqrt{s_\mathrm{NN}} = 5.02$ TeV central Pb+Pb collisions. [From \cite{ALICE:2019nuy}]}
\label{fig:bc_RAA}
\end{figure}

Figure \ref{fig:bc_RAA} shows that lepton spectra from heavy flavor decays exhibit similar nuclear modification features as those of light hadrons. The left panel, which shows flavor separated $R_{\rm AA}$ for leptons from $c$ versus $b$ decays in Au+Au collisions at RHIC \cite{STAR:2021uzu,PHENIX:2022wim}, provides evidence that $c$-quarks experience strong rescattering  in the QGP, resulting in a suppression which is comparable to that of light quarks. The suppression effect for leptons from $b$-quark decays is significantly smaller. This is expected, as the energy loss of a $b$-quark in collisions with thermal partons is reduced by a factor $m_c/m_b$ when compared with that of $c$-quarks. In addition, radiative energy loss by heavy quarks exhibits a dead-cone effect \cite{Dokshitzer:1991fd,ALICE:2021aqk} that increases with quark mass. Although medium-induced gluon radiation is predicted to partially fill the radiation dead cone, a mass dependent reduction of the radiative energy loss is predicted \cite{Armesto:2003jh}.

The right panel of Fig.~\ref{fig:bc_RAA} shows the lepton $R_{\rm AA}$ for unseparated $b$- and $c$-decays in central Pb+Pb collisions at the LHC \cite{ALICE:2019nuy}. Again, the nuclear modification exhibits similar features as that measured for light charged hadrons with a minimum around $p_T \sim 10$ GeV/c. The observed trend is generally well reproduced by theoretical calculations that include collisional and radiative energy loss.

\begin{figure}[htb]
\centering
\includegraphics[width=0.45\linewidth]{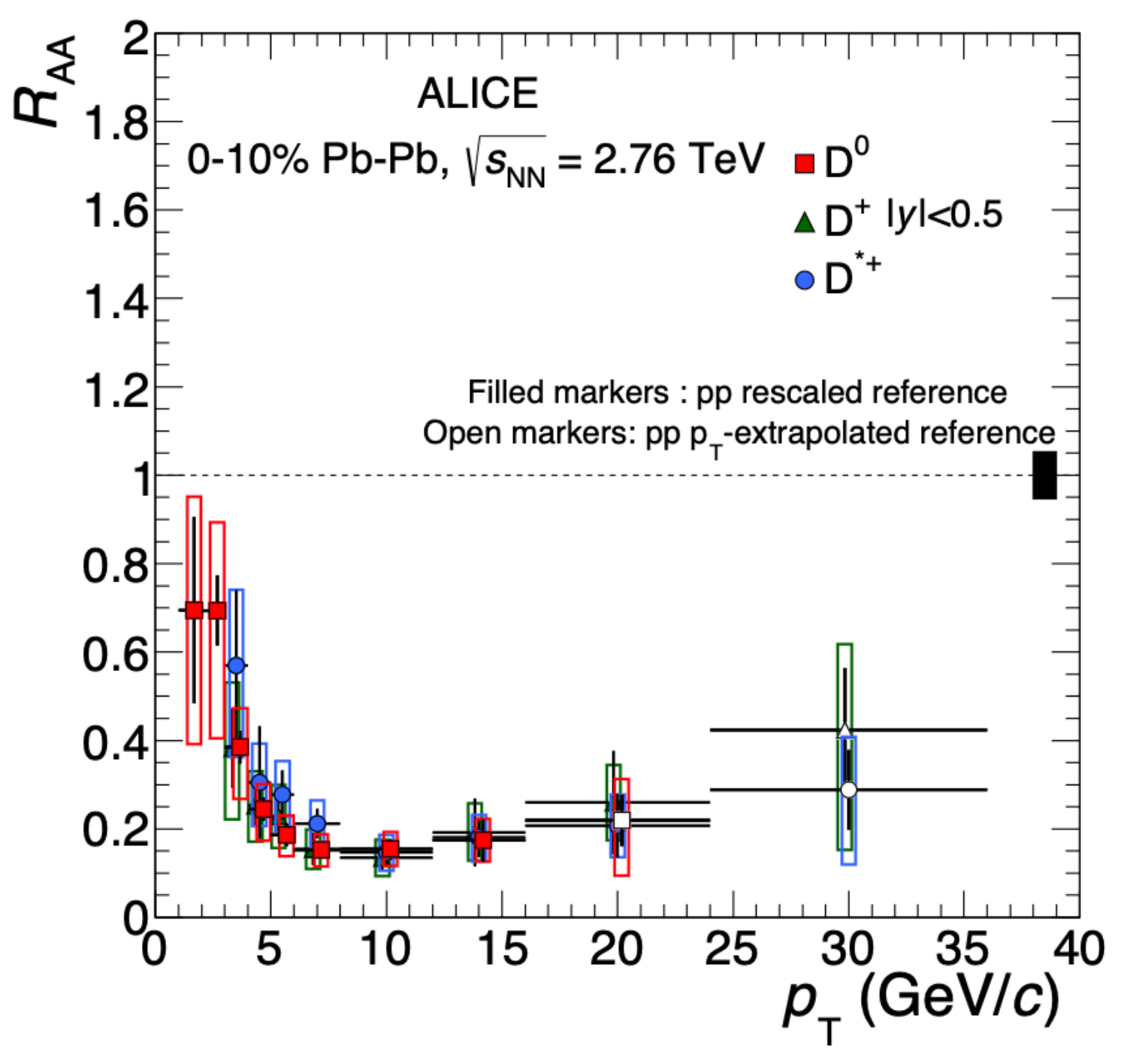}
\hspace{0.05\linewidth}
\includegraphics[width=0.45\linewidth]{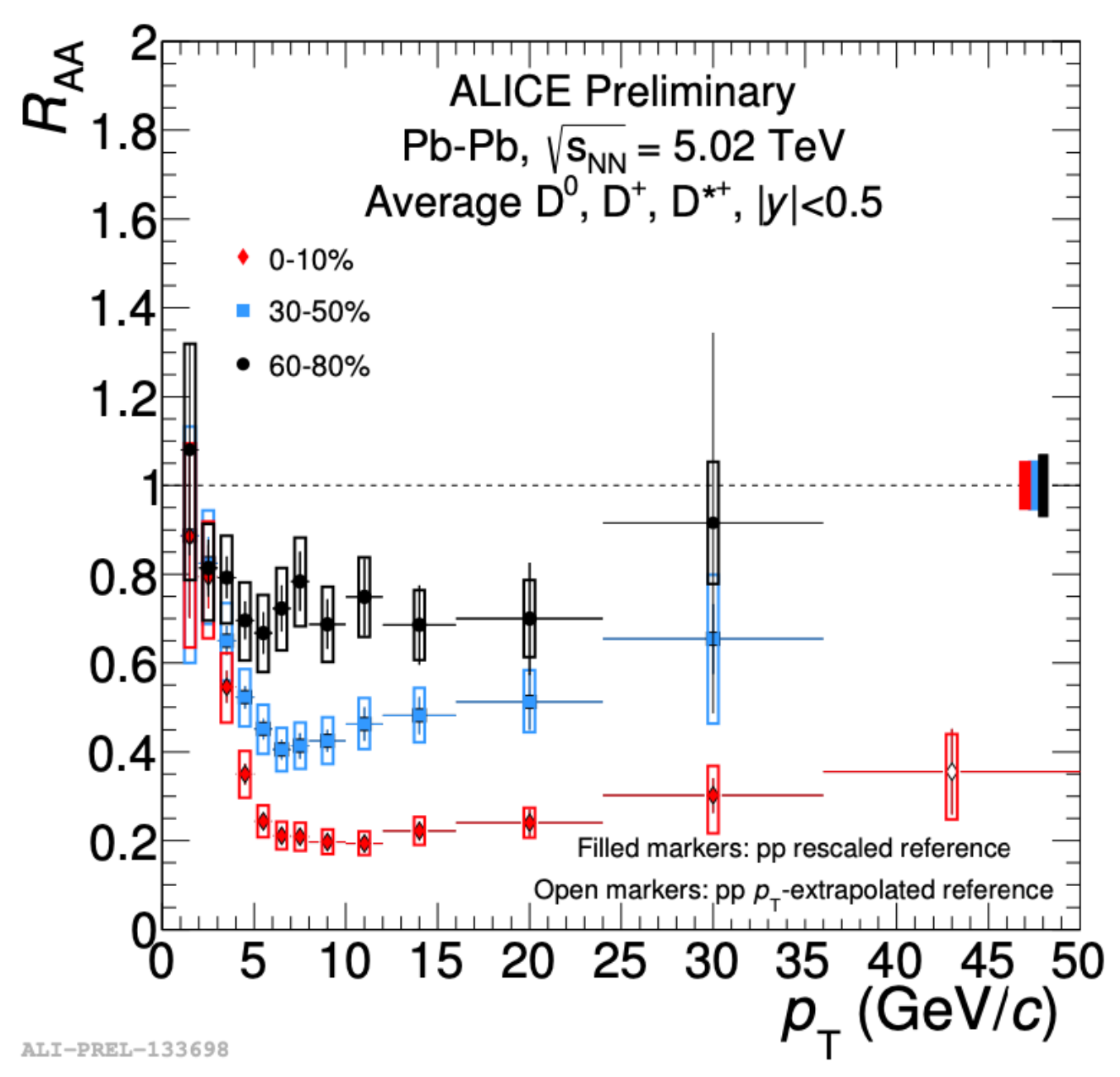}
\caption{Nuclear suppression factor $R_{\rm AA}(p_T)$ for identified $D$-mesons in Pb+Pb collisions at the LHC measured by ALICE. Left panel: $R_{\rm AA}(p_T)$ for $D^0$, $D^+$, and $D^{*+}$ mesons in central collisions at $\sqrt{s_\mathrm{NN}} = 2.76$ TeV. [From \cite{ALICE:2015vxz}] Right panel: $R_{\rm AA}(p_T)$ for all identified $D$-mesons in three centrality windows at $\sqrt{s_\mathrm{NN}} = 5.02$ TeV central Pb+Pb collisions. [From \cite{Grosa:2018zix}]}
\label{fig:Dmeson_RAA}
\end{figure}

For open charm hadrons that exhibit a characteristic hadronic weak decay mode, such as $D^0$, $D^+$, and $\Lambda_c$, it has been possible to measure identified particle $p_T$ spectra that do not require unfolding. The best data are available for $D$-mesons for Pb+Pb collisions at LHC \cite{ALICE:2015vxz}.  Figure \ref{fig:Dmeson_RAA} shows $R_{\rm AA}(p_T)$ for identified $D$-mesons measured by ALICE. The left panel confirms that all species of $D$-mesons are equally suppressed, in agreement with the hypothesis that the suppression mainly reflects the energy loss of $c$-quarks in the QGP. The right panel shows that the suppression is strongly centrality dependent \cite{Grosa:2018zix} as would be expected from the path-length dependence of the energy loss. 

\begin{figure}[htb]
\centering
\includegraphics[width=0.9\linewidth]{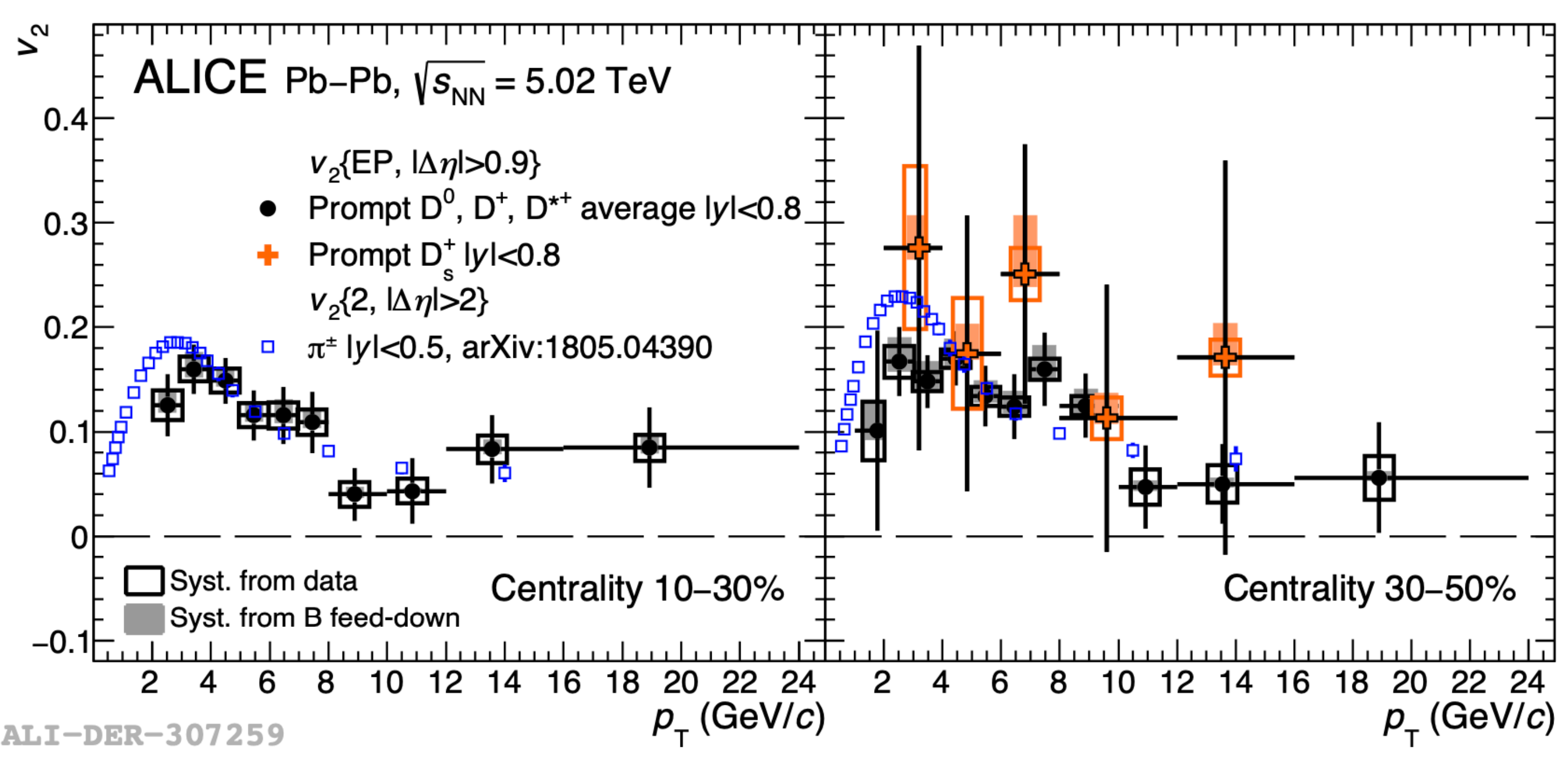}
\caption{Elliptic flow $v_2(p_T)$ of identified $D$-mesons in Pb+Pb collisions at $\sqrt{s_\mathrm{NN}} = 5.02$ TeV measured by ALICE. The two panels show results for two different centrality windows, in comparison with the $v_2$ for charged pions. The right panel also shows the $v_2$ for $D_s$ mesons. [From \cite{Grosa:2018zix}]}
\label{fig:Dmeson_v2}
\end{figure}

The strong suppression of $D$-mesons, similar to the pattern observed for light hadrons, raises the question whether $c$-quarks thermalize in the QGP and participate in the collective flow and if so, to what degree. A partial answer to this question is afforded by the measurement of the elliptic flow coefficient $v_2$ for $D$-mesons. Results for $v_2(p_T)$ of identified $D$-mesons measured by ALICE are shown in Fig.~\ref{fig:Dmeson_v2} in comparison with $v_2(p_T)$ for charged pions in two centrality windows. Except at the lowest measured $p_T$ the $D$-mesons exhibit almost the same amount of elliptic flow as pions, which indicates that they participate in the overall collective flow of the QGP. The reduced $v_2$ at low $p_T$ is expected because $D$-mesons have almost twice the mass of a proton and thus should show an even stronger kinematic reduction of $v_2$ at low $p_T$ than protons.

Bound states of heavy quarks (charmonium, Upsilon, and $B_c$), generically called quarkonia, are sensitive to the color screening length in the QGP \cite{Matsui:1986dk}. When the color screening length $r_D$, also called the color Debye length, is shorter than the radius of the heavy quark bound state, the bound state dissolves (``melts'') in the QGP and becomes a broad resonance. For each bound state there is a characteristic threshold temperature \cite{Karsch:1990wi}, also called the Mott temperature $T_{\rm M}$ \cite{Hufner:1996pq}. In addition to color screening, the other important contribution to this process is ionization by thermal gluons in the QGP. 

The two determinants of the Mott temperature are the radius of the quarkonium bound state and its binding energy. In the non-relativistic limit, these are given by $R_{Q\bar{Q}} = N^2(\alpha_s m_{q\bar{Q}}c/\hbar)^{-1}$ and $B_{Q\bar{Q}} = N^2 \alpha_s^2 m_{Q\bar{Q}} c^2$, where $m_{Q\bar{Q}}$ is the reduced mass and $N \ge 1$ is the principal quantum number of the Coulombic bound state. For charmonium the 1s and 2s states are bound in the vacuum ($J/\psi$ and $\psi'$); for bottomonium the 3s state is also bound ($\Upsilon$, $\Upsilon'$, and $\Upsilon''$.) One thus expects that when the temperature is raised above the critical temperature $T_c$, first the $\psi'$ and $\Upsilon''$ states melt (close to $T_c$), then the $J/\psi$ and $\Upsilon'$ states (around $1.5 T_x$) and eventually the $\Upsilon$ ground state (slightly above $2T_c$). This predicted phenomenon is known as ``sequential melting''.

\begin{figure}[htb]
\centering
\includegraphics[width=0.45\linewidth]{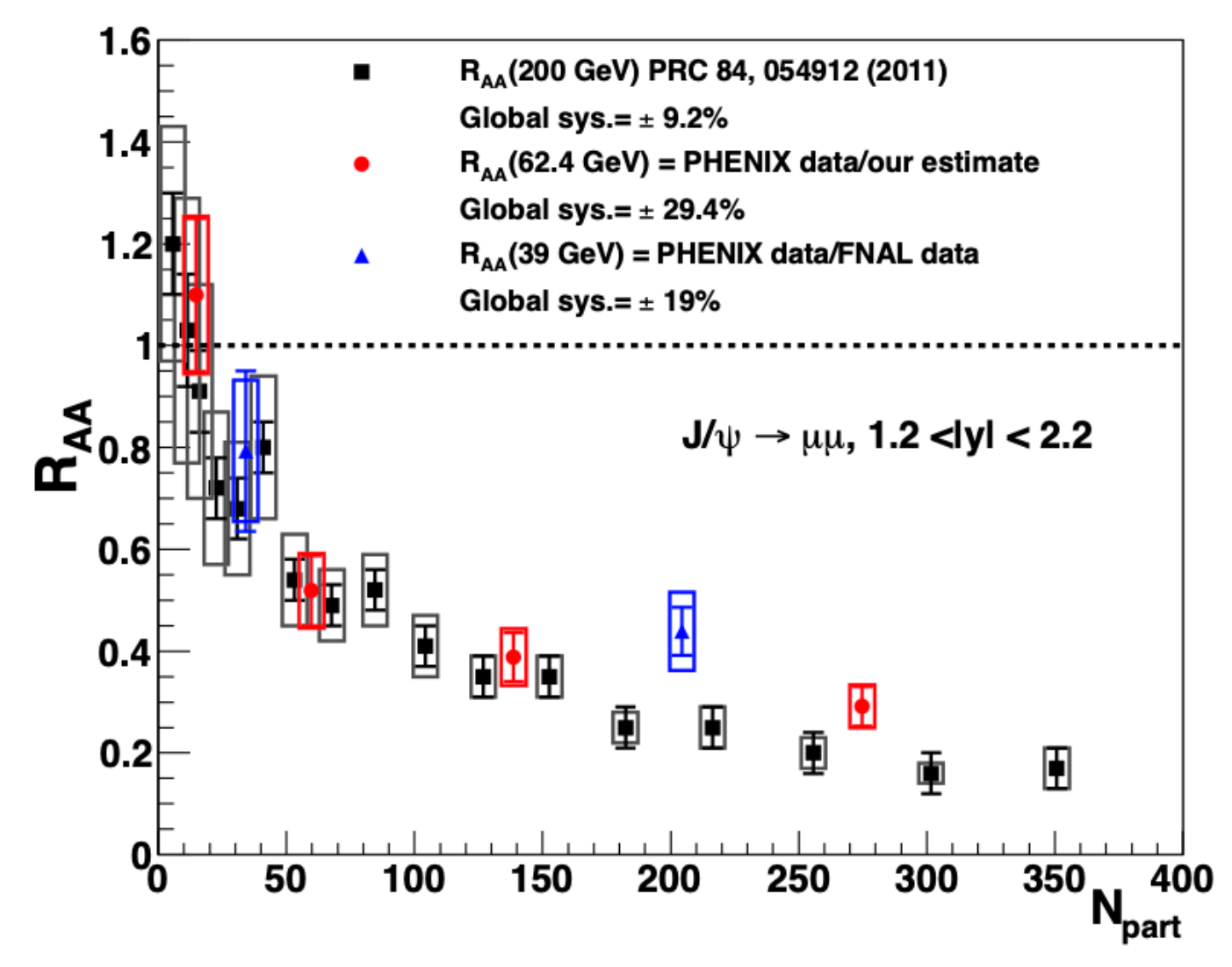}
\hspace{0.05\linewidth}
\includegraphics[width=0.45\linewidth]{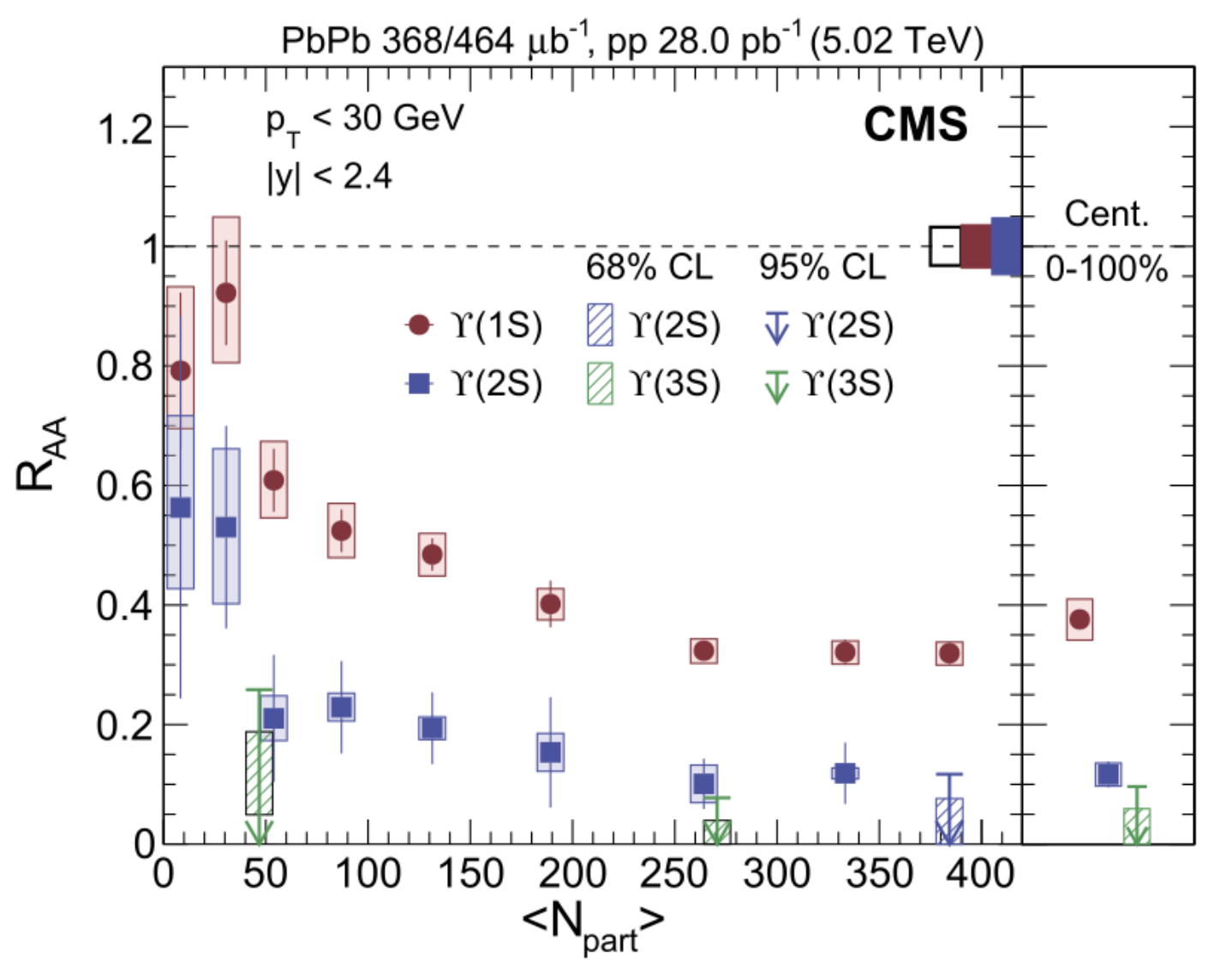}
\caption{Nuclear suppression factor $R_{\rm AA}$ for quarkonium production as function of centrality, measured by the particpant number $N_{\rm part}$. Left panel: Au+Au collisions at $\sqrt{s_{\rm NN}} = 200, 62.4, 39$ GeV (from \cite{PHENIX:2012xtg}). Right panel: Pb+Pb collisions at $\sqrt{s_{\rm NN}} = 5.02$ TeV (from \cite{CMS:2018zza}).}
\label{fig:QQsuppression}
\end{figure}

Over the past three decades the suppression of quarkonium production in heavy ion collisions, compared with scaled proton-proton collisions, has been measured in great detail. The suppression is commonly expressed in terms of the ratio $R_{\rm AA}$, similar to the suppression of jet production. Results for charmonium suppression in Au+Au collisions at RHIC and bottomonium suppression in Pb+Pb collisions at LHC are shown in Fig.~\ref{fig:QQsuppression}.

\begin{figure}[htb]
\centering
\includegraphics[width=0.45\linewidth]{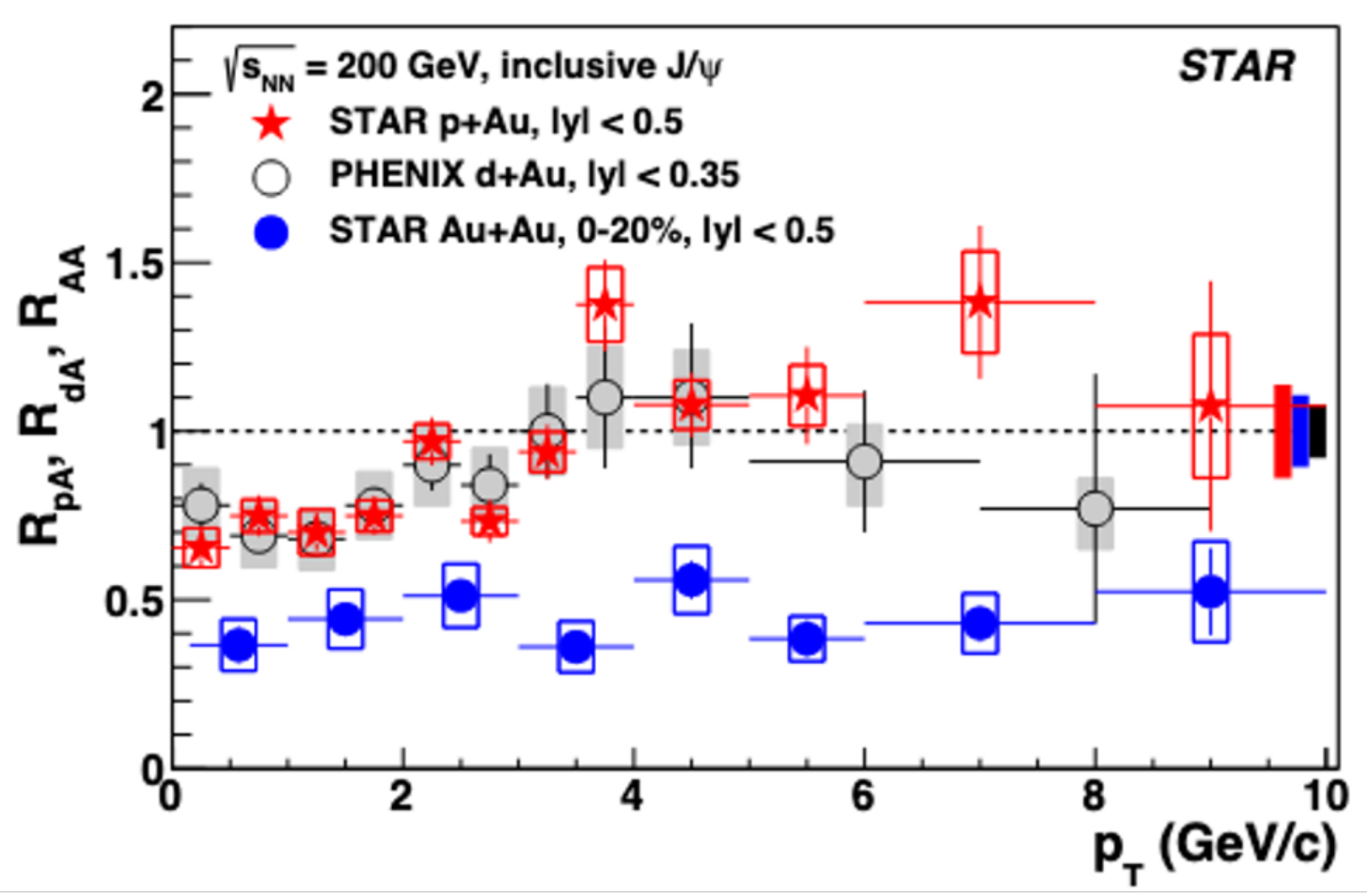}
\hspace{0.05\linewidth}
\includegraphics[width=0.45\linewidth]{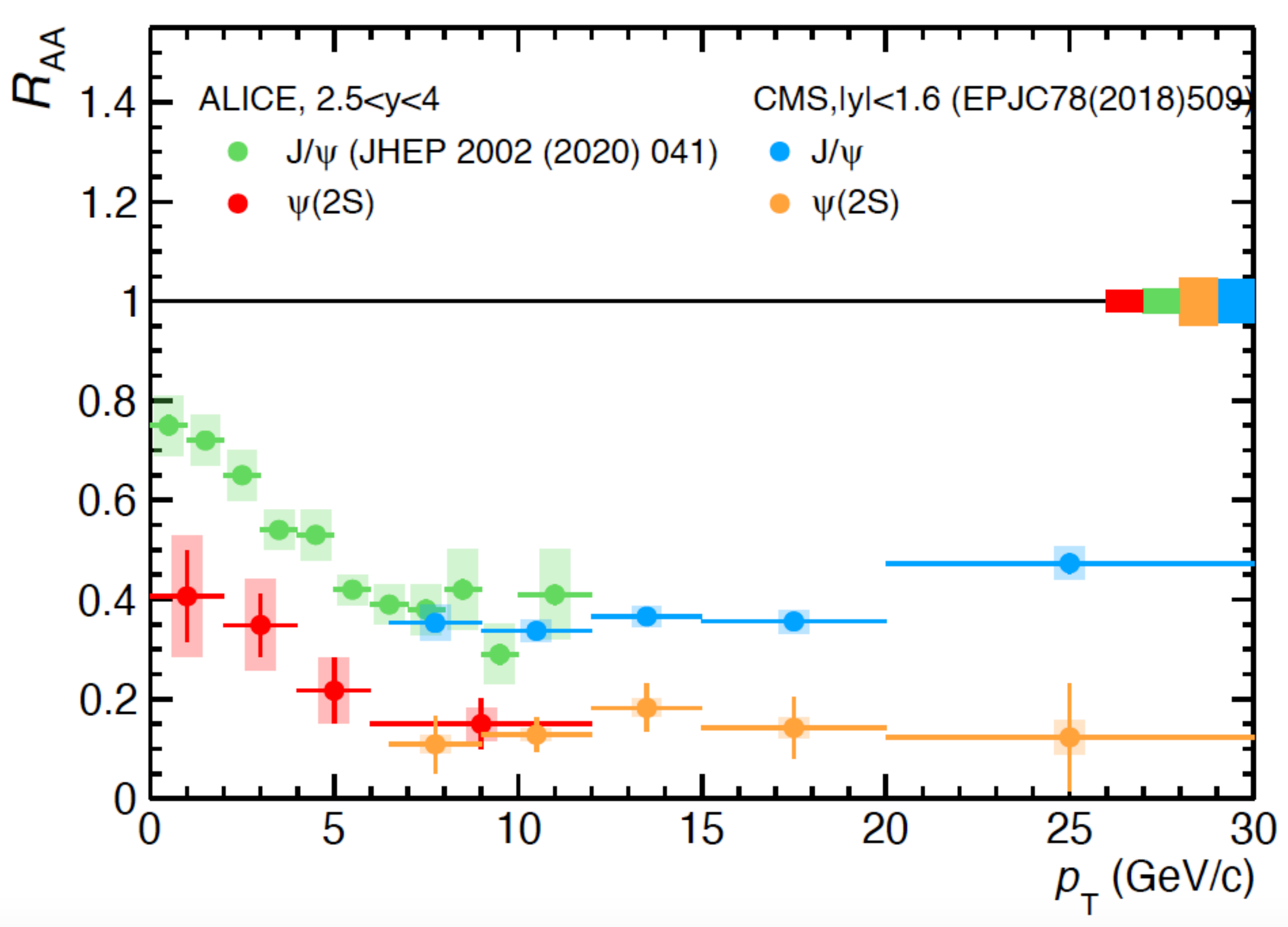}
\caption{Left panel: Nuclear suppression factor $R_{\rm (p/d)A}(p_T)$ for $J/\psi$ production in $p+Au$ and $d+Au$ collisions at $\sqrt{s_{\rm NN}} = 200$ GeV (from \cite{STAR:2021zvb}). The approximately 30\% suppression at small $p_T$ is mainly due to gluon shadowing in the Au nucleus. Right panel: Nuclear suppression factor $R_{\rm AA}(p_T)$ for $J/\psi$ and $\psi'$ production in Pb+Pb $\sqrt{s_{\rm NN}} = 5.02$ TeV (Fig. 59 (right panel) from ALICE Whitepaper (to be published)) \he{reference has to be added}. The enhancement at small $p_T$ is attributed to recombination of $c$ and $\bar{c}$ quarks during hadronization (regeneration).}
\label{fig:Jpsi_pT}
\end{figure}

The interpretation of the $R_{\rm AA}$ data is complicated by several effects. The two most important ones are:
\begin{itemize}
    \item 
    The primary production process for heavy quark pairs, $gg \rightarrow Q\bar{Q}$, is suppressed in nuclear collisions because the nuclear gluon distribution at small Bjorken-$x$ is screened (``shadowed''). This effect, which can be studied in $p(d)+A$ collisions, is mainly observed at low transverse momentum $p_T$, as is visible in Fig.~\ref{fig:Jpsi_pT} (left panel). 
    \item
    In Pb+Pb collisions at LHC energies, $c\bar{c}$ pairs are copiously produced and are thought to be thermalized in the QGP. These pairs can coalesce into $J/\psi$ and $\psi'$ mesons when the QGP hadronizes. This regeneration mechanism leads to an enhancement of charmonium production at low $p_T$, which can be seen in the right panel of Fig.~\ref{fig:Jpsi_pT}. The same mechanism may already be visible in Au+Au collisions at RHIC at central rapidity ($|y|<0.35$), where $J/\psi$ production is found to be less suppressed than at forward rapidity ($|y|>1.2$). The stronger suppression of $\psi'$ compared with $J/\psi$ seen in this figure is also evidence for the sequential melting concept.
\end{itemize}

\subsection{Electromagnetic Probes}
\label{sec:em_probes}

Electromagnetic probes are, at the same time, theoretically interesting and experimentally challenging because they do not interact strongly and therefore are emitted from all stages of the reaction. Many hadrons and other particles are also produced during the whole evolution of the heavy ion reaction, but the main advantage of electromagnetic probes is that they reach the detector undisturbed. A dilepton pair or a photon escapes even the hot and dense reaction zone without further interaction. The purely electromagnetic interaction mechanism leads to very small production rates and therefore extremely challenging measurements. In addition, there typically exists a huge background created by weak decays of hadrons that create photons and dilepton pairs in the final state (e.g. $\pi^0\rightarrow\gamma\gamma$). Any interpretation of experimental results relies therefore on theoretical input on the origin of the contributions from the different sources and stages of the reaction.

\begin{figure}[htb]
\centering
\includegraphics[width=0.5\linewidth]{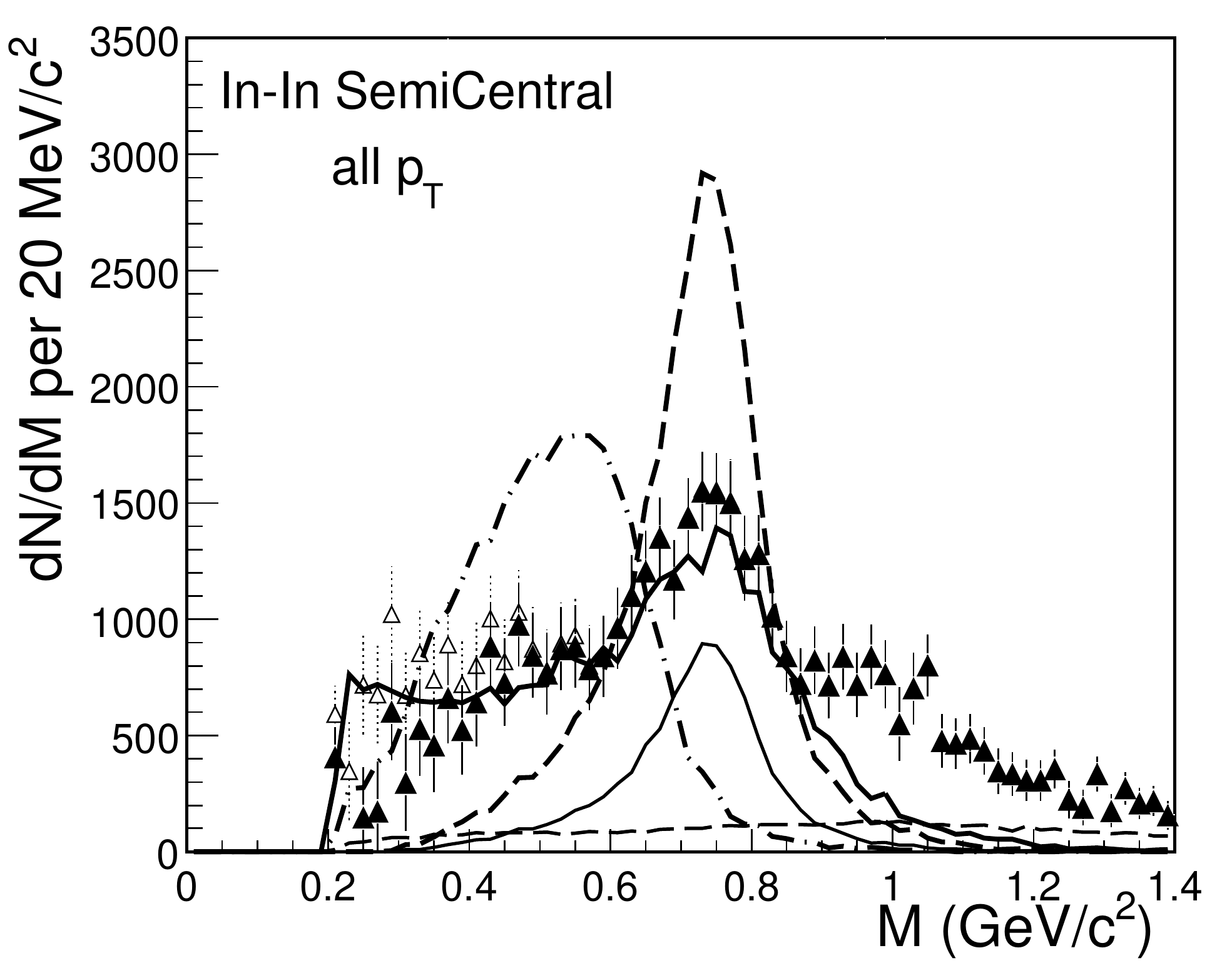}
\includegraphics[width=0.4\linewidth]{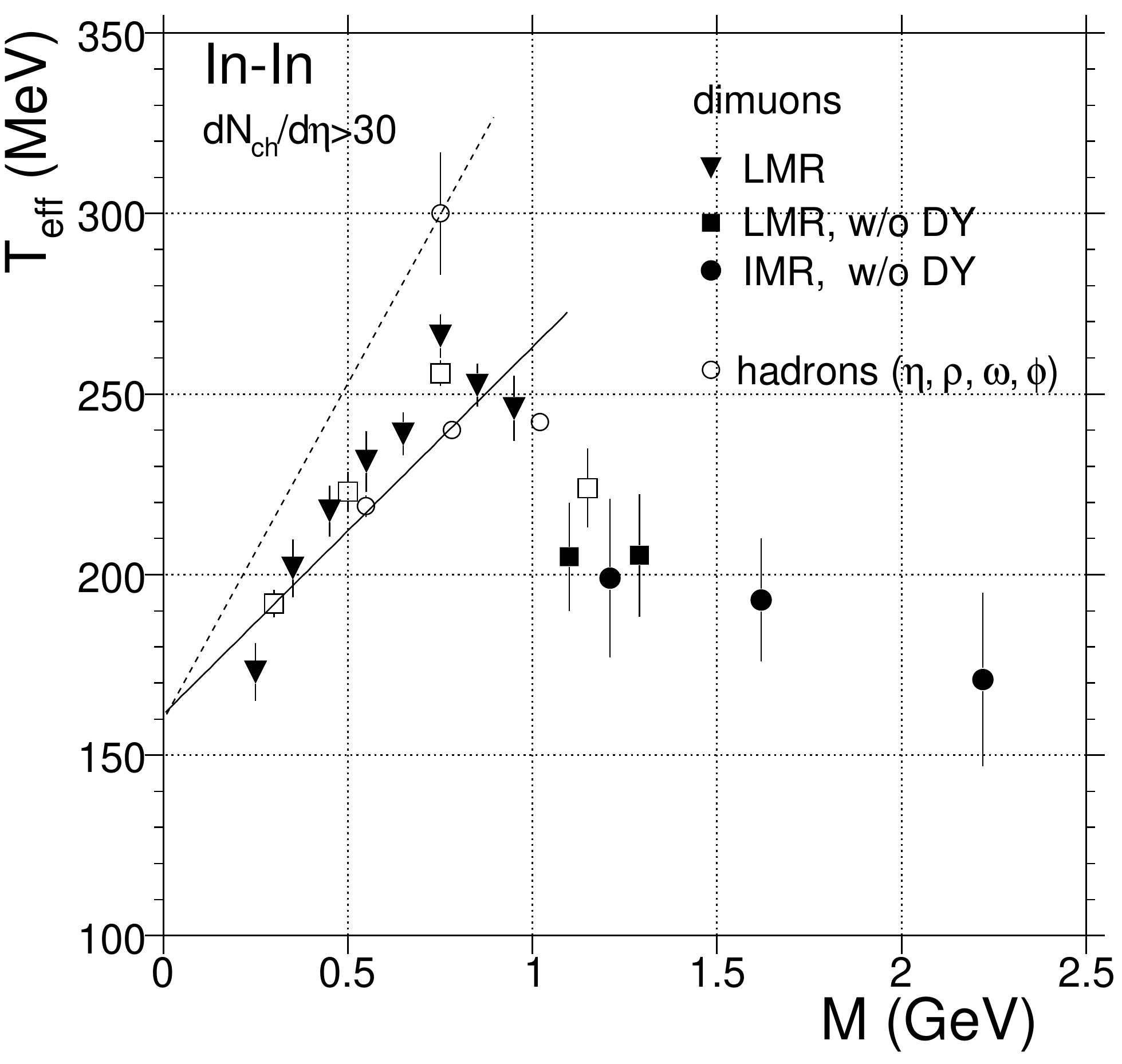}
\caption{Left: Comparison of the excess mass spectrum for In-In at
$dN_{ch}/d\eta$=140 to the cocktail $\rho$ (thin solid), unmodified $\rho$
(dashed), in-medium broadening $\rho$~\cite{Chanfray:1995jgo,Rapp:1999ej}
(thick solid), in-medium moving $\rho$ related to~\cite{Brown:2001nh}
(dashed-dotted). [From \cite{NA60:2006ymb}] Right: Inverse slope parameter $T_\mathrm{eff}$ vs. dimuon mass $M$ for $dN_{ch}/d\eta>30$ with open charm subtraction. [From \cite{NA60:2007lzy}]}
\label{fig:na60_dileptons}
\end{figure}

There are two main physics questions that can be addressed with dilepton and photon measurements. By investigating the invariant mass spectrum of dileptons emitted from vector mesons one can get insight about the properties of the spectral functions of resonances in the medium (see \cite{Salabura:2020tou} for a recent review). The $\rho$-meson is of special interest, since the idea is to study chiral symmetry restoration by observing the spectral functions of the $\rho$ and its chiral partner $a_1$ become degenerate. More recently, the focus has shifted to signatures of chiral mixing, since the measurement of the $a_1$ spectral function is out of reach. The second main topic is the idea of extracting a temperature of the quark-gluon plasma from the thermal radiation. For this purpose, photons might seem more straightforward, but one has to account for a blue shift in the photon spectrum due to radial flow. The slope of dilepton spectra in the invariant mass region from $m_{\rm inv}=1-3$ GeV provides a more direct measure of the thermal radiation from the plasma. 

Fig. \ref{fig:na60_dileptons} depicts the most precise dilepton measurement in a heavy ion environment to date. The invariant mass spectrum of dimuons was recorded for In-In collisions at $E_{\rm lab}=158A$ GeV at the CERN-SPS. After contribution from the ``cocktail'' of known hadronic sources has been subtracted the remaining excess yield (see the left panel of Fig.~\ref{fig:na60_dileptons}) provides a measure of the spectral function of the $\rho$-meson inside the hot and dense medium. From these results, it can be concluded that the $\rho$-meson is strongly modified in the medium and mainly broadened, while a mass shift of the pole position has not been observed. The differential dimoun measurements make it possible to fit transverse momentum spectra in different invariant mass bins and extract the effective temperature as the inverse slope shown in the right panel of Fig.~\ref{fig:na60_dileptons}. In the low mass region (LMR) the extracted values agree with the ones from the hadronic spectra, consistent with a hadronic origin, while above 1 GeV in invariant mass the thermal slope saturates and suggests emission from an equilibrated quark-gluon plasma. In this mass region, there is a correlated background from heavy quark decays that needs to be carefully subtracted.

\begin{figure}[htb]
\centering
\includegraphics[width=0.48\linewidth]{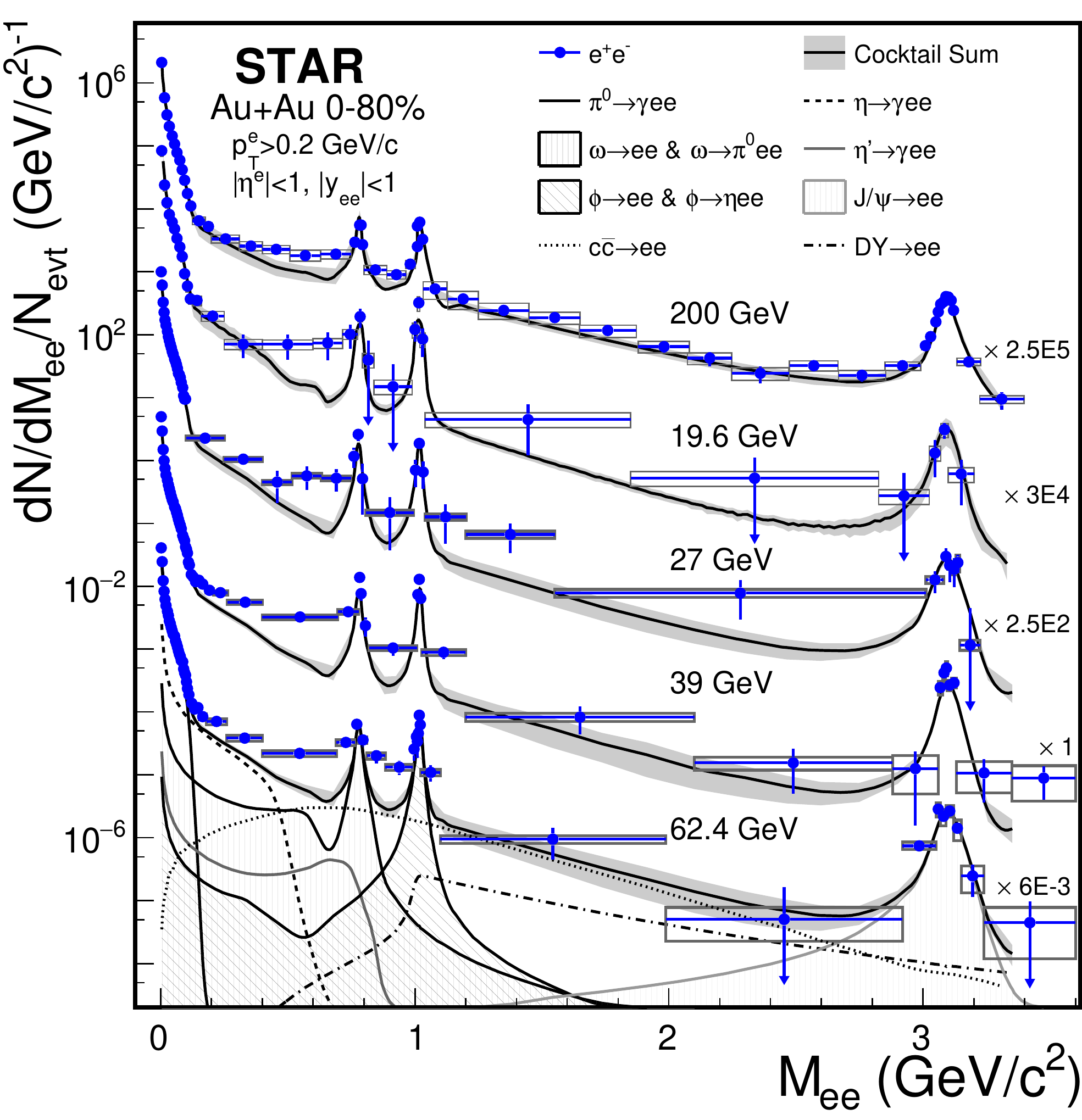}
\hspace{0.02\linewidth}
\includegraphics[width=0.48\linewidth]{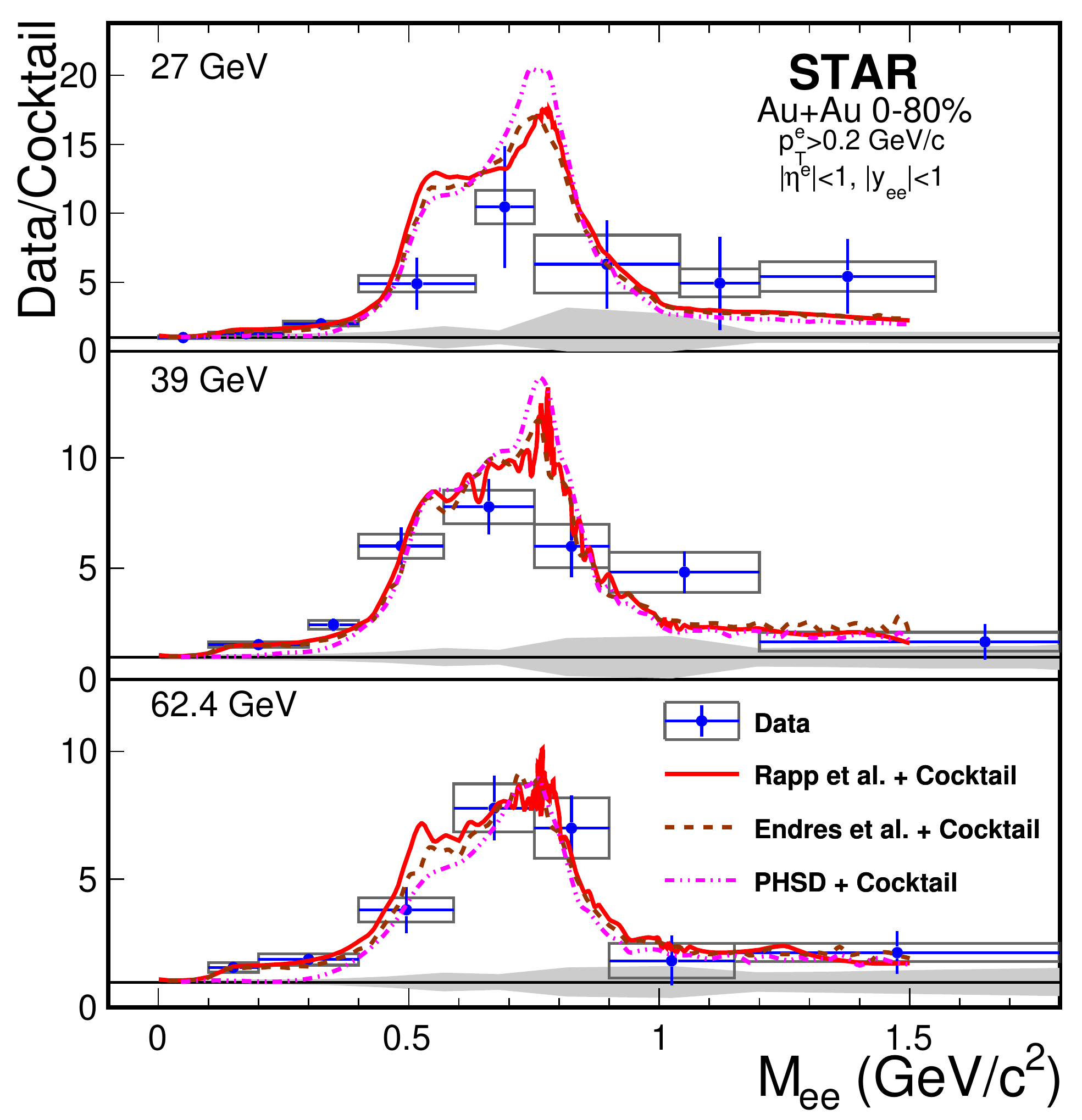}
\caption{Left: Background subtracted dielectron invariant mass spectra within the STAR acceptance from $\sqrt{s_{\rm NN}}$ = 19.6, 27, 39, 62.4, and 200~GeV 0$-$80\% most-central Au$+$Au collisions (scaled for visibility). Right: The ratio of the invariant mass spectra to the cocktail with the $\omega$ and $\phi$ 
yields removed compared to model calculations.  [From \cite{STAR:2018xaj}]}
\label{fig:star_dileptons}
\end{figure}

Moving to higher beam energies the STAR collaboration has measured di-electron invariant mass spectra as shown in Fig. \ref{fig:star_dileptons}. The excess yield above the hadronic cocktail emission indicates that the spectral function of the $\rho$-meson is also broadened at these higher beam energies. Theoretical calculations include thermal dilepton rates from effective field theory folded with a fireball model \cite{vanHees:2006ng, Rapp:2013nxa}, coarse-grained UrQMD transport calculations with the same thermal dilepton rates \cite{Endres:2016tkg,Endres:2014zua} and fully microscopic non-equilibrium calculations by within the PHSD approach \cite{Cassing:2009vt,Linnyk:2015rco}. In the future, ALICE is expected to be also able to measure precise, background subtracted dilepton spectra in Pb+Pb collisions at LHC energies. 

\begin{figure}[htb]
\centering
\includegraphics[width=0.48\linewidth]{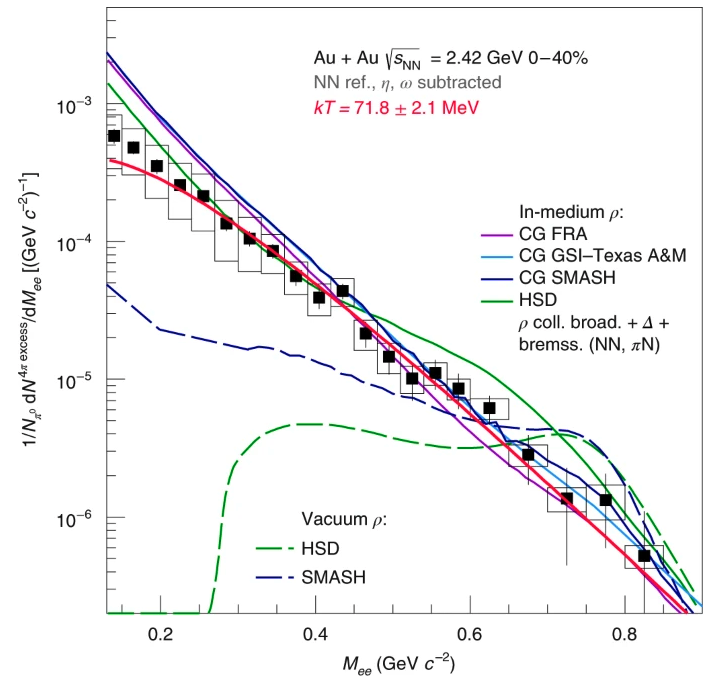}
\hspace{0.02\linewidth}
\includegraphics[width=0.48\linewidth]{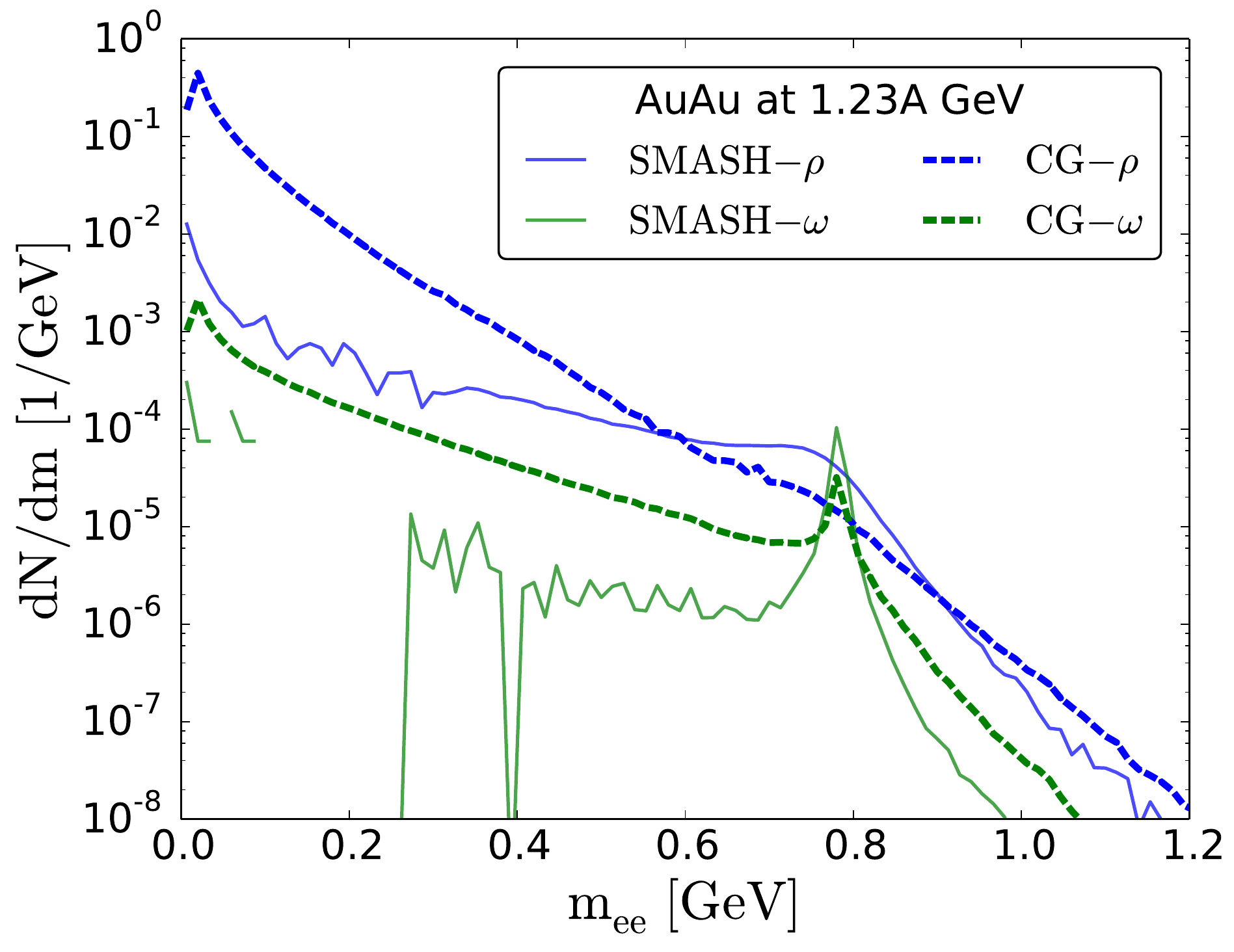}
\caption{Left: Excess yield of dileptons extracted by subtracting $\eta, \omega$ contributions as well as the $NN$ reference normalized to the number of neutral pions, the red curve shows a thermal $dN/dM_{ee} \propto (M_{ee})^{3/2} \exp(-M_{ee}/T)$ fit. [From \cite{HADES:2019auv}] Right: Comparison of invariant mass spectra of dielectrons produced by $\rho$ and $\omega$ in Au+Au collisions at $E_{\rm kin} = 1.23~A\,\textrm{GeV}$ within the coarse-graining approach versus the default SMASH dilepton production. [From \cite{Staudenmaier:2017vtq}]}
\label{fig:hades_dileptons}
\end{figure}

At GSI, the HADES experiment is dedicated to investigating dilepton emission from elementary and heavy ion reactions. In this baryon-dominated environment the $\rho$-meson is mainly modified due to its interactions with the baryonic resonances. Fig.~\ref{fig:hades_dileptons} shows the extracted thermal emission from Au+Au collisions at $E_{\rm lab}= 1.23A$ GeV. There are clear indications of medium modifications in comparison with the vacuum environment indicated by the agreement with coarse-grained (CG) transport calculations including medium-modified spectral functions for vector mesons. The calculation (see right panel of Fig.~\ref{fig:hades_dileptons}) shows this difference explicitly for the radiation from the $\rho$- and $\omega$-mesons, the main vector mesons contributing in this mass range. In the future, HADES and the CBM experiment will measure the excitation function of thermal dilepton emission with the goal to identify signatures of a first-order phase transition. 

\begin{figure}[htb]
\centering
\includegraphics[width=1.0\linewidth]{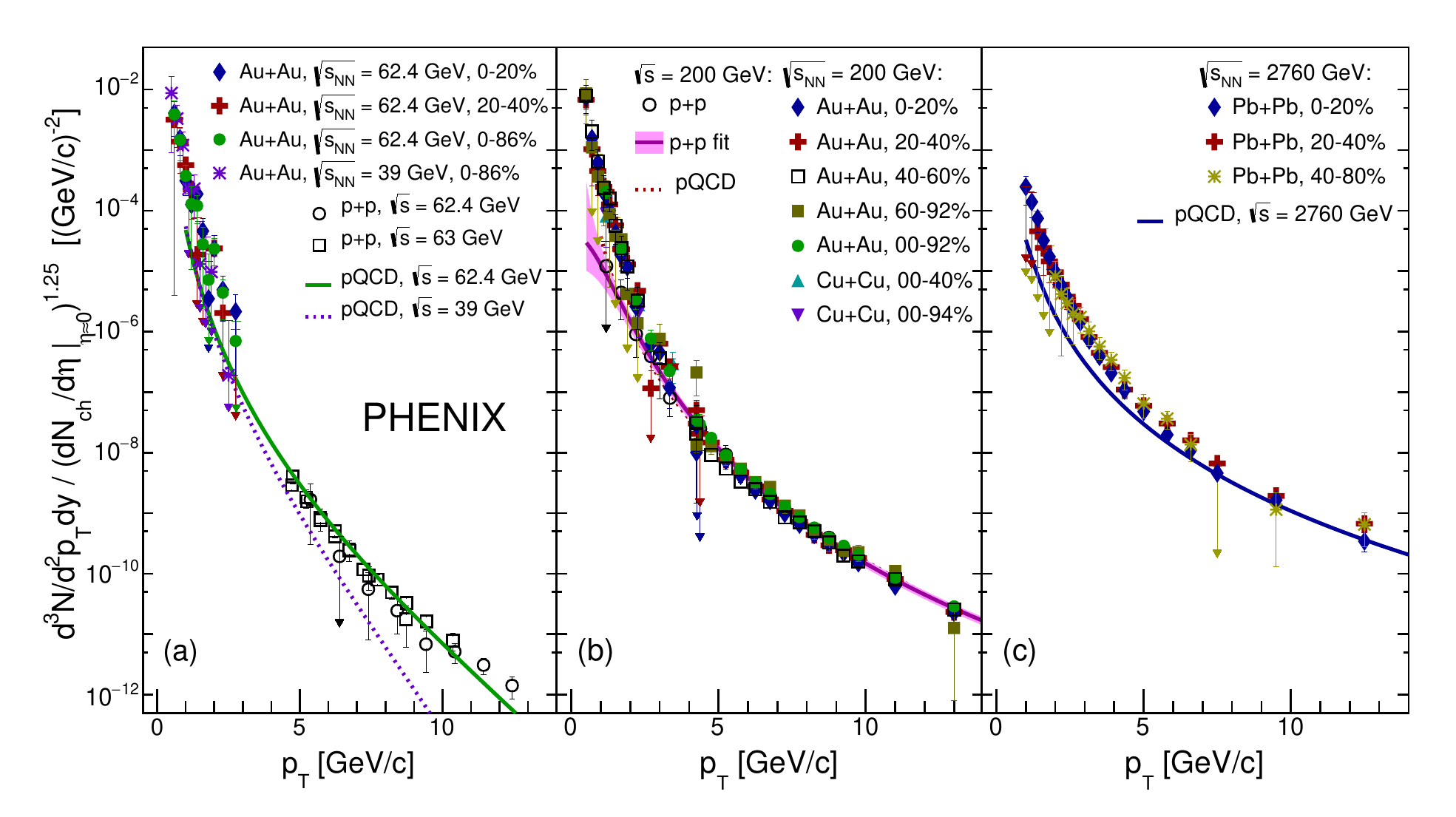}
\caption{Direct-photon $p_{T}$-spectra normalized by $(dN_{\rm ch}/dy)^{1.25}$ for (a) the minimum bias Au+Au 39 and 62.4~GeV data sets, (b) various centrality selected 200~GeV Au+Au~\cite{PHENIX:2012jbv,PHENIX:2008uif,PHENIX:2014nkk} and Cu+Cu~\protect\cite{PHENIX:2018che} data sets, and (c) various centrality selected Pb+Pb 2760~GeV data sets~\cite{ALICE:2015xmh}. Also shown are (a) $pp$ data from the ISR~\cite{CMOR:1989qzc,AxialFieldSpectrometer:1989nag} and (b) $pp$ 200~GeV data~\cite{PHENIX:2012krx}. [From \cite{PHENIX:2022qfp}]}
\label{fig:photon_spectra}
\end{figure}

\begin{figure}[htb]
\centering
\includegraphics[width=0.6\linewidth]{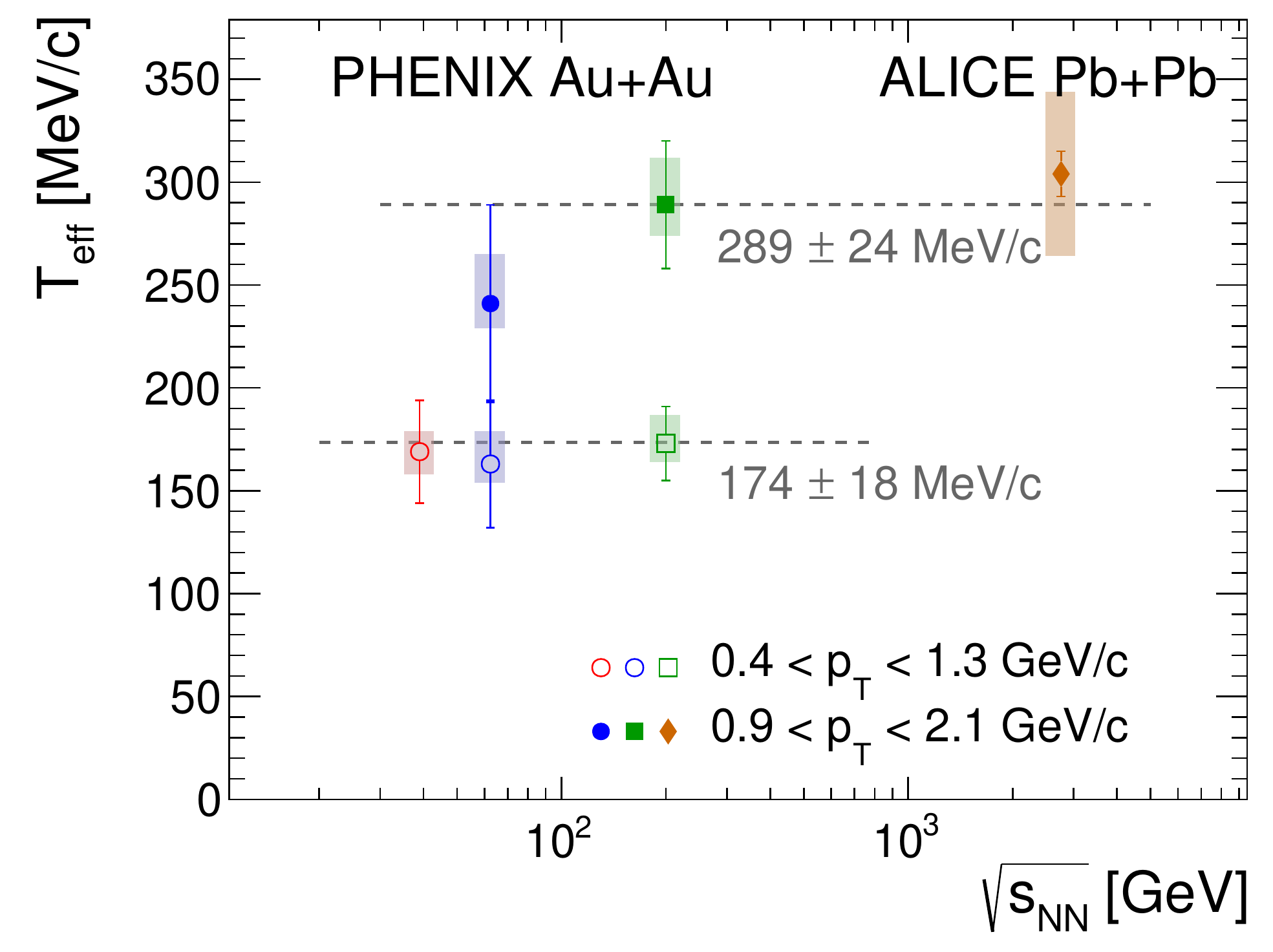}
\caption{Inverse slopes, $T_{eff}$, obtained from fitting the combined 
data from central collisions is compared to the fit results of the individual data sets at 62.4, 200, and 2760 GeV.  Also included is the value for $\sqrt{s_{\rm NN}}=39$ GeV obtained from fitting the minimum bias data set in the lower-$p_\perp$ range. [From \cite{PHENIX:2022qfp}]}
\label{fig:photon_effectivetemp}
\end{figure}

Photon production from heavy ion collisions is dominated by sources from hadronic decays, most prominently the $\pi^0$ that decays to photons with an almost 100\% branching ratio. Therefore, it is very challenging to experimentally extract the primary photons, usually called ``direct photons''. The first measurement of a direct photon spectrum was accomplished by the WA98 collaboration at SPS \cite{WA98:2000vxl}. More recently, PHENIX and ALICE have published transverse momentum spectra and flow measurements for direct photons. A compilation of the spectra is shown in Fig. \ref{fig:photon_spectra}. At higher transverse momenta the spectra match the expectations from perturbative QCD calculations \cite{Paquet:2015lta}, while at lower momenta an exponential behavior is observed.  
By fitting the slope of the transverse momentum spectra of direct photons, one can infer an effective temperature of the quark-gluon plasma as depicted in Fig. \ref{fig:photon_effectivetemp}. 

\begin{figure}[htb]
\centering
\includegraphics[width=0.52\linewidth]{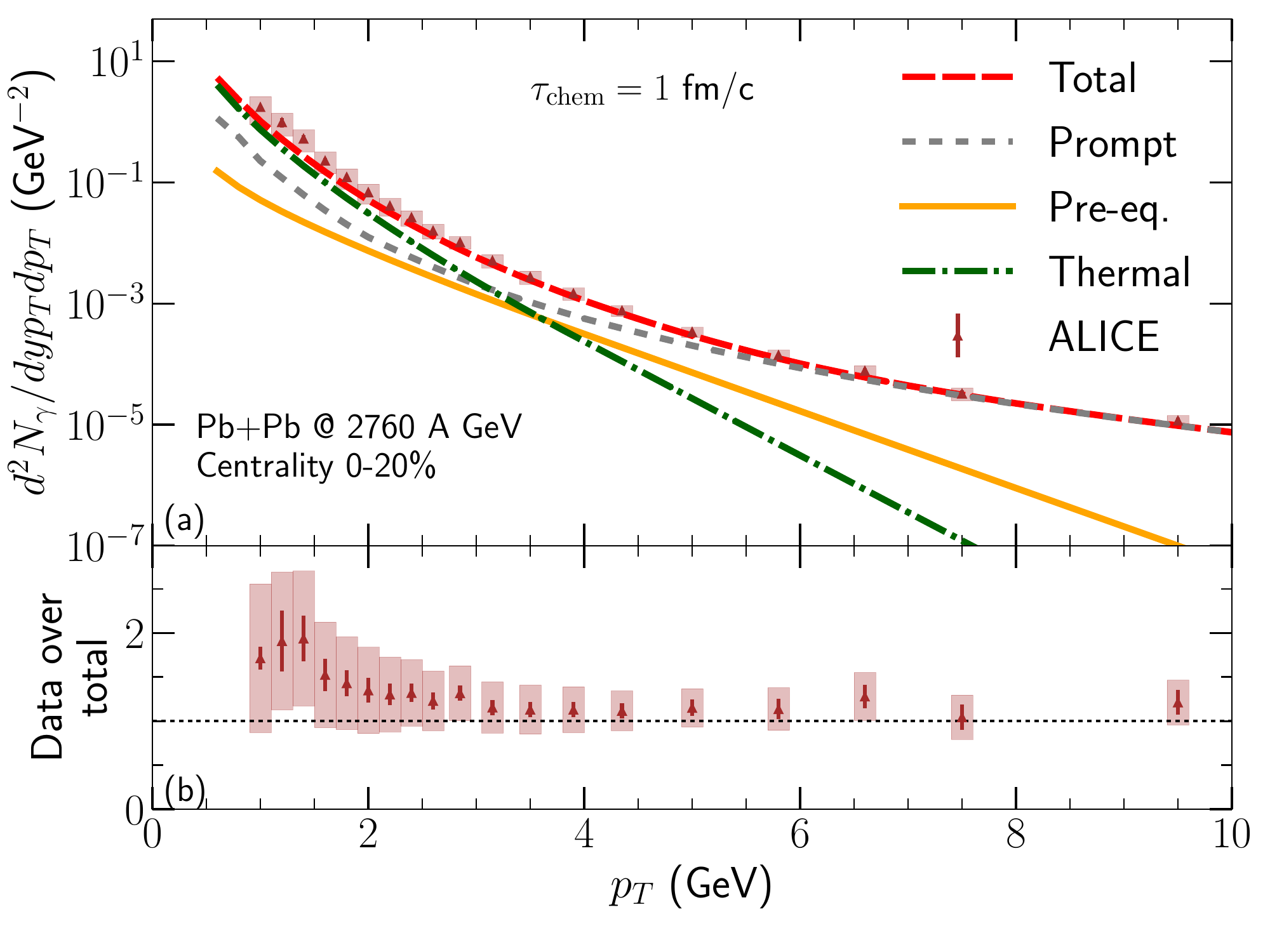}
\hspace{0.03\linewidth}
\includegraphics[width=0.4\linewidth]{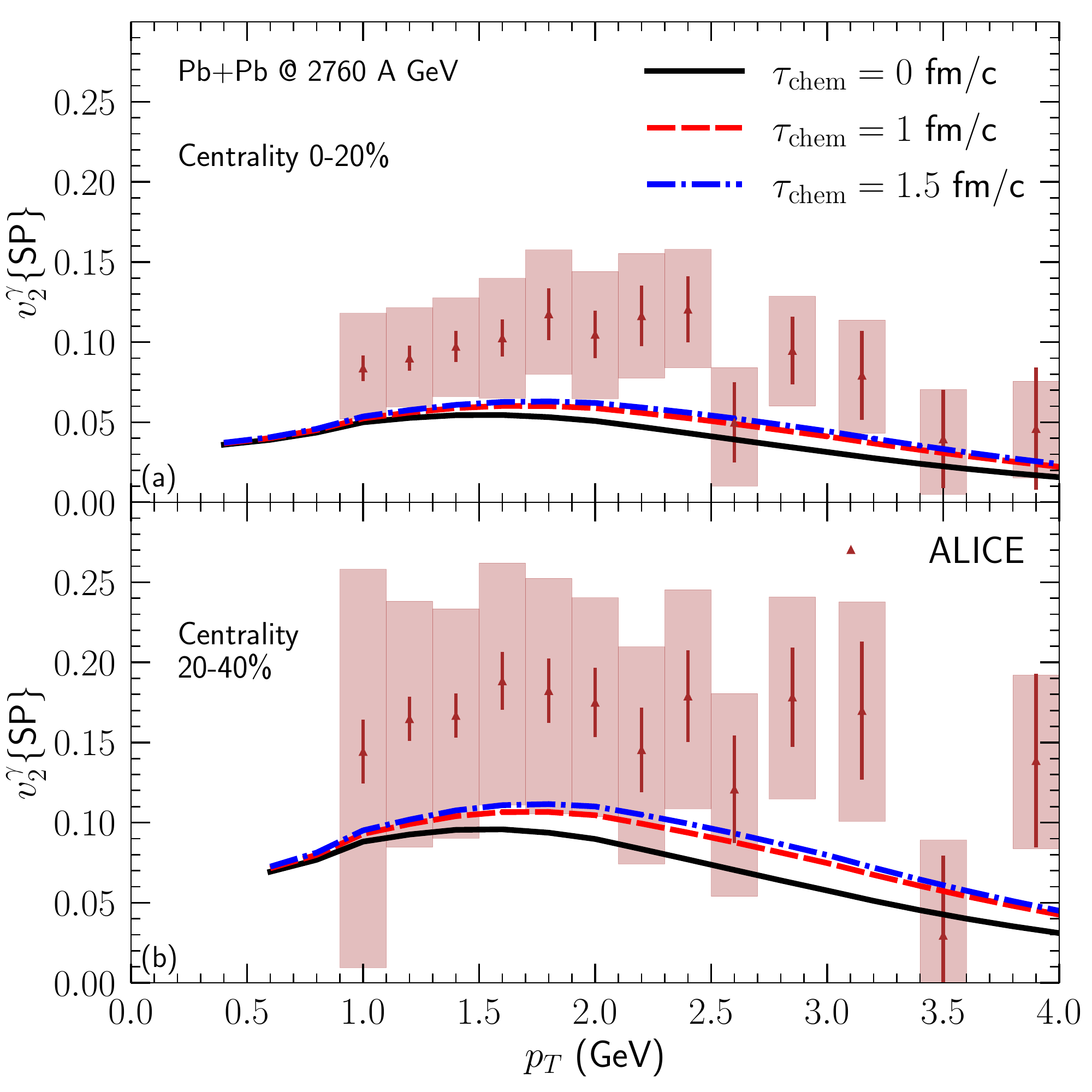}
\caption{Left: (a) Direct photon yield in Pb+Pb collisions at $\sqrt{s_{\rm NN}}= 2.76$ TeV, in the $0 - 20\%$ centrality class with depicting the different components compared to data from the ALICE Collaboration \cite{Adam:2015lda}. Right: Direct photon elliptic flow $v_2^\gamma (p_T)$ in $\sqrt{s_{\rm NN}}= 2.76$ TeV Pb+Pb collisions for different chemical equilibration times compared to experimental data from the ALICE collaboration \cite{ALICE:2018dti}.  [From \cite{Gale:2021emg}]}
\label{fig:photons_lhc}
\end{figure}

In Fig.~\ref{fig:photons_lhc} state-of-the-art calculations considering direct photon emission from all stages of the reaction are compared with spectra and elliptic flow data from the ALICE collaboration in Pb+Pb collisions at $\sqrt{s_{\rm NN}}= 2.76$ TeV. Again the prompt photons from the early hard collisions dominate the high $p_T$ region. At lower transverse photon momenta the thermal emission from the hydrodynamic medium dominates. The photons emitted during the pre-equilibrium stage are depicted by the full line. The magnitude of photon elliptic flow it depends on the time at which full chemical equilibrium between quarks and gluons is achieved (indicated by the different values of $\tau_{\rm chem}$). 

In general, it is still hard to explain the photon production yield and elliptic flow at the same time consistently in one theoretical calculation. The intuitive reason is as follows: Processes that increase the yield need to occur early in the evolution when the plasma is hotter, while larger elliptic flow is reached in the later stages of the evolution. Therefore, elliptic flow is increased, if later sources are enhanced, for example in \cite{Schafer:2021slz} the contribution of the non-equilibrium hadronic stage has been shown to be significant for low transverse momenta. In the future, it is going to be crucial to investigate emission of electromagnetic probes from well-calibrated models for the bulk evolution \cite{Gale:2021emg}.

\subsection{Production of light nuclei and exotic hadrons}
\label{sec:light_nuclei}

There are two different mechanism that lead to the production of light nuclei in heavy ion reactions. In the fragmentation regions, at very forward and backward rapidity, the spectator remnants can fragment and reach the detector as a multitude of smaller and larger nuclei. This can happen because the spectator remnants emerge from the collision in a highly excited state, and collisions among nucleons within these spectator remnants often result in their disintegration. We will not discuss this phenomenon further as the main interest here is on the production of light (anti-)nuclei at midrapidity within the hot and dense collision region. The production of nuclei is interesting because of its dependence on the nucleon-nucleon interaction, but also as a probe of possible differences in the properties of matter and antimatter. 

Fig.~\ref{fig:lhc_fits} shows the production yields of many particle species including light nuclei in Pb+Pb collisions at $\sqrt{s_{\rm NN}}= 2.76$ TeV \cite{Andronic:2017pug}. Due to the vanishing chemical potential at this high energy the yields of deuterons and anti-deuterons, helium-3, hypertriton and helium-4 and their anti-nuclei are pairwise identical. Generally, the predictions within the statistical hadronization model agree very well with the measurements for the same temperature as for all other hadron species. This poses immediately a question of current debate: How can particles that have small binding energies of a few MeV freeze out chemically from a fireball of hot and dense strongly-interacting matter at a temperature that is many times higher ($T_{\rm ch} \sim 150$ MeV)?

\begin{figure}[htb]
\centering
\includegraphics[width=0.43\linewidth]{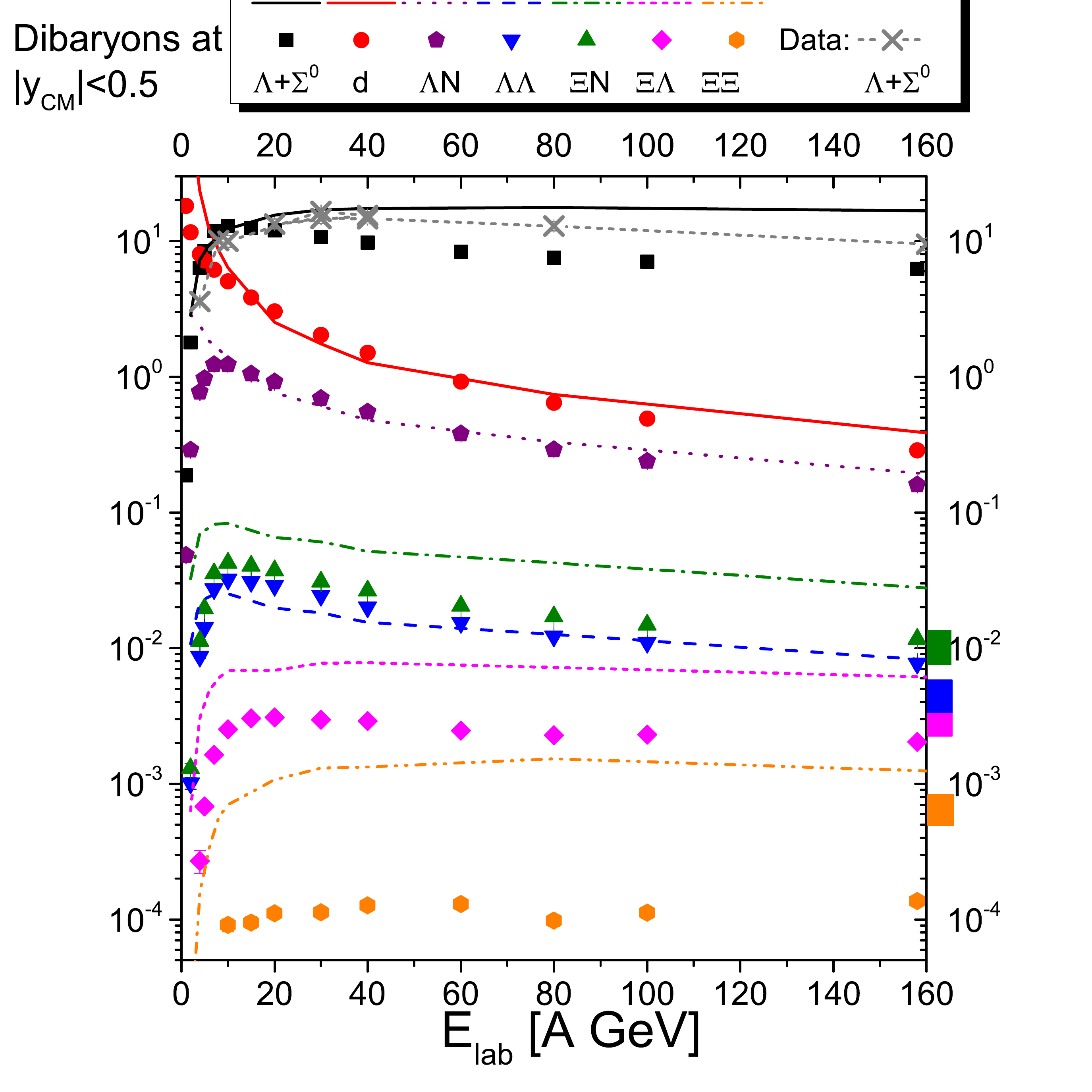}
\includegraphics[width=0.45\linewidth]{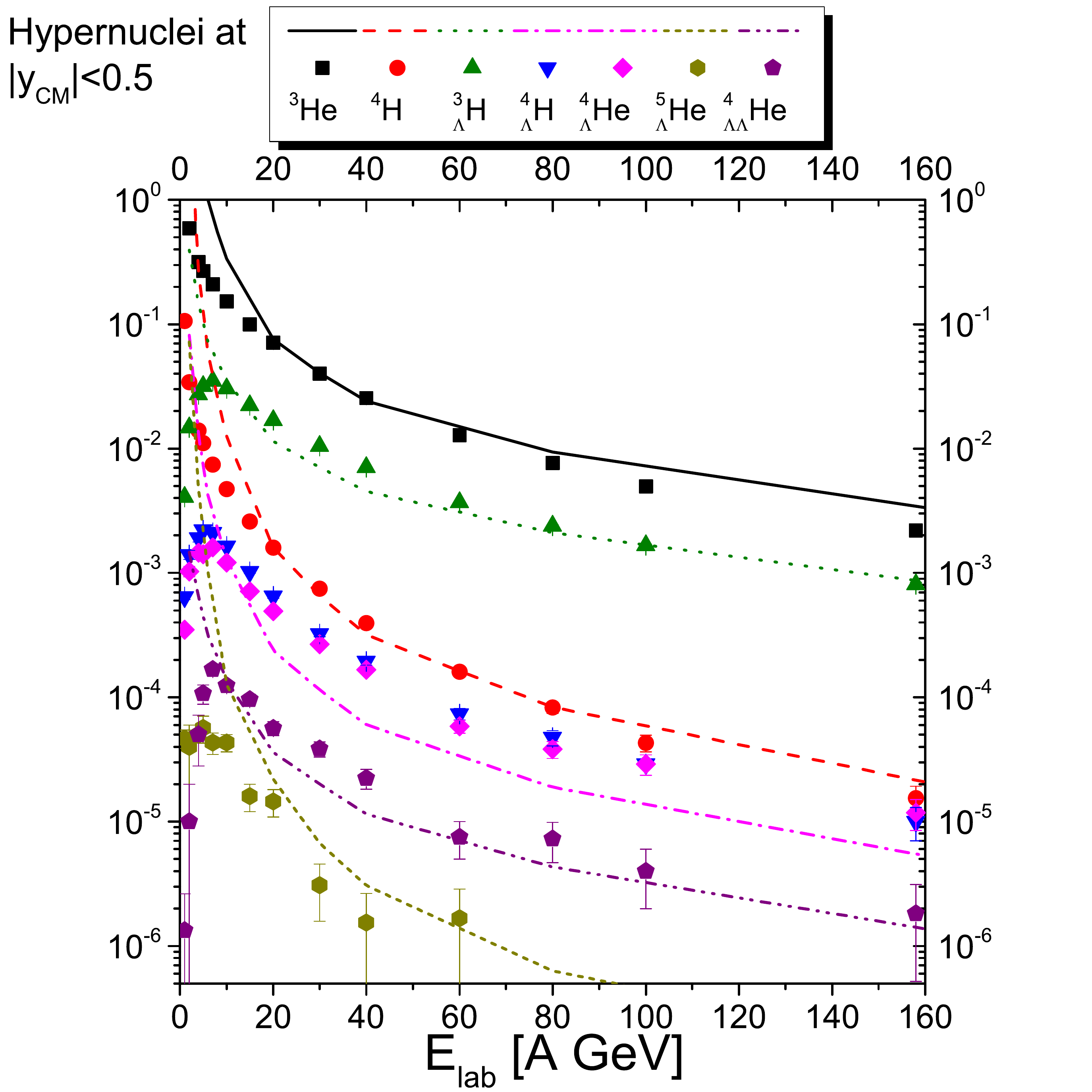}
\caption{Excitation function of dibaryons (left) and hypernuclei (right) calculated within a thermal production model from the UrQMD hybrid approach (full lines) compared to a coalescence approach (symbols).  [From \cite{Steinheimer:2012tb}]}
\label{fig:light_nuclei_excfct}
\end{figure}

Besides the thermal production of light nuclei, another production mechanism is proposed. The idea is that only individual hadrons are produced from the fireball, but later on the nucleons or antinucleons combine to form light nuclei. This coalescence picture \cite{Csernai:1986qf, Scheibl:1998tk} involves calculating the overlap in phase-space of all nucleons and drawing conclusions about the abundance of light nuclei from there. In such a picture it is expected that the transverse momentum spectra of light nuclei and their anisotropic flow coefficients $v_n$ scale according to the number of nucleons contained in a nucleus, in analogy to the recombination approach and the partonic quark number scaling discussed above (see Section \ref{sec:ncq_scaling}). Fig.~\ref{fig:light_nuclei_excfct} depicts the expected yields at midrapidity in a thermal and a coalescence production approach. Due to the change from a baryon dominated to a meson dominated system the highest yields are expected in the beam energy range of $E_{\rm lab}=10-20A$ GeV. 

Due to recent increased interest in the topic, several models have been developed that aim to explain the mechanism behind this behavior. One model attempts to describe the early chemical freeze-out via the Saha equation in analogy to cosmology \cite{Vovchenko:2019aoz} or rate equations \cite{Neidig:2021bal}. Another approach involves microscopic calculations of the non-equilibrium dynamics of light nuclei in the hadronic stage of the reaction \cite{Danielewicz:1991dh}. At low beam energies in a baryon-rich environment, the main reactions are the nucleon catalysis reactions, while at high beam energies the pion catalysis is more important. Calculations within a hadronic transport approach suggest that the chemical equilibrium is maintained due to the high reaction cross sections during the rescattering phase in nuclear collisions \cite{Oliinychenko:2018ugs,Staudenmaier:2021lrg}. The kinetic and chemical decoupling almost coincide in such an approach. 

\begin{figure}[htb]
\centering
\includegraphics[width=0.7\linewidth]{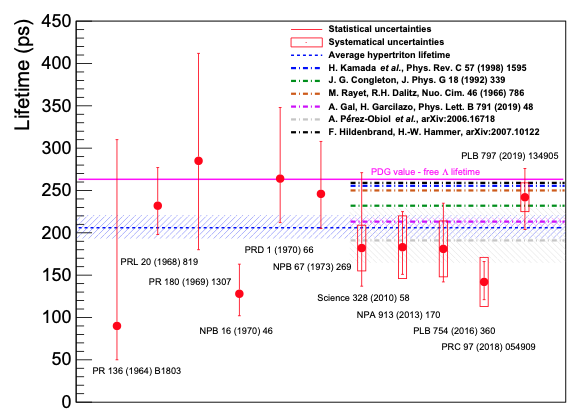}
\caption{Comparison of measurements of the hypertriton lifetime in heavy ion reactions to theoretical calculations and the free value from the PDG. [From \cite{Donigus:2020fon}]}
\label{fig:hypertriton}
\end{figure}

Hypernuclei and their properties are of particular interest since one might be able to infer knowledge about the $\Lambda-N$ interaction. For a while, the hypertriton ($^3_\Lambda{\rm H}$ ``lifetime puzzle'' attracted much interest, since the first measurements of the lifetime of the hypertriton seemed to deviate from its free value. Recent more recent measurements from ALICE (see Fig.~\ref{fig:hypertriton}), STAR, and HADES point instead to good agreement with the world data. 

A rather recent idea that has already been applied with considerable success, is to measure the femtoscopic correlations of exotic hadrons in elementary and heavy ion reactions. These final-state correlation measurements can be connected to the hadronic interactions of those particles. In this manner it is possible to extract quantitative information about the interaction of $\Xi, \Omega$ and other exotic hadrons containing one or more strange quarks with other hadrons (see the recent reviews \cite{Tolos:2020aln,Fabbietti:2020bfg}).

\subsection{Vorticity and polarization} 
\label{sec:vorticity}

The initial state of a non-central heavy ion collision is characterized by a very large angular momentum in the center-of-mass frame. For example, a collision between two $^{208}$Pb nuclei at $\sqrt{s_{\rm NN}} = 5.02$ TeV with impact parameter $b = 7.5$ fm commands an angular momentum of nearly $20,000\hbar$. Only a small fraction of this angular momentum ends up in the central rapidity region where most observed particles are produced. However, even this small fraction endows the QGP in non-central collisions with a sizable vorticity. 

Assuming thermal equilibrium, the probability for a certain spin orientation of $\Lambda$-hyperon within the final-state hadronic gas is \cite{Becattini:2020ngo}:
\begin{equation}
P_\Lambda(\hat{S}) \propto 
\exp\left( \frac{(\vec{\omega}+\mu_\Lambda\vec{B}) \cdot\hat{S}}{T} \right) ,
\label{eq:Lambda_pol}
\end{equation}
where $\vec{\omega}$ is the vorticity vector of the matter, $\mu_\Lambda = (-0.6138 \pm 0.0047)\mu_N$ is the $\Lambda$ magnetic moment in nuclear magnetons $\mu_N$, and $\vec{B}$ is the magnetic field present at emission. Since the magnetic moments of the $\Lambda$ and $\overline\Lambda$ differ by their sign, a magnetic field would cause them to be oppositely polarized, whereas vorticity of the medium results in identical polarizations. $\Lambda$-hyperons are ideal probes of polarization, because their parity violating weak decay $\Lambda \rightarrow p+\pi^-$ is self-analyzing. The angular distribution of the decay proton momentum $\vec{p}=p\hat{n}$ in the $\Lambda$ rest frame is given by \cite{Becattini:2020ngo}
\begin{equation}
\frac{dw}{d\Omega} = \frac{1+\alpha_\Lambda \vec{P}_\Lambda \cdot \hat{n}}{4\pi} 
\end{equation}
with $\alpha_\Lambda = 0.732 \pm 0.014$ \cite{ParticleDataGroup:2020ssz}.

\begin{figure}[htb]
\centering
\includegraphics[width=0.65\linewidth]{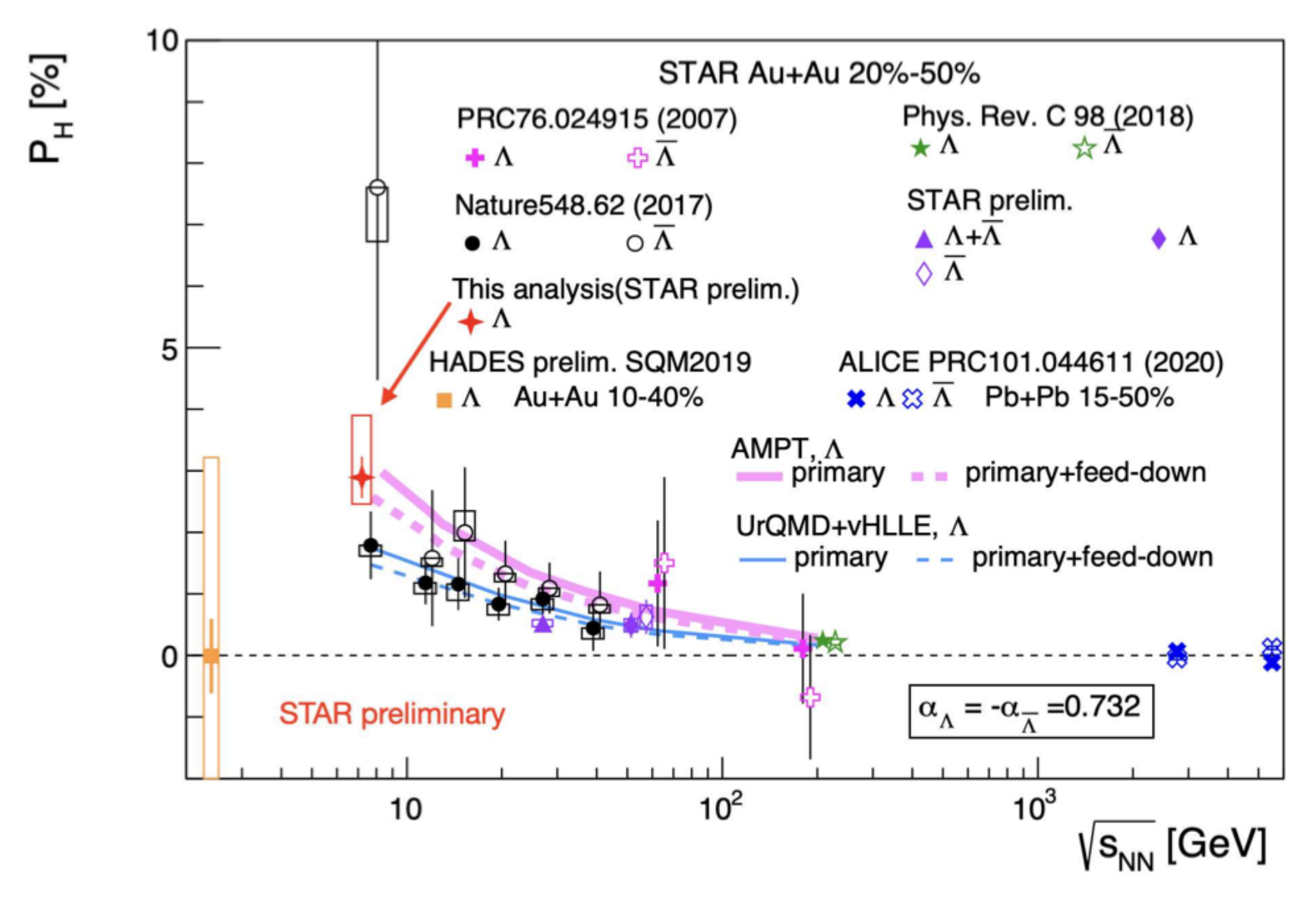}
\caption{Global $\Lambda$ and $\overline\Lambda$ polarization with respect to the collision plane as a function of collision energy.  [From \cite{Okubo:2021dbt}]}
\label{fig:Lambda_pol}
\end{figure}

The global polarization of $\Lambda$ and $\overline\Lambda$ have been measured in Au+Au collisions at RHIC over a wide energy range \cite{STAR:2017ckg,STAR:2021beb} and in Pb+Pb collisions at LHC \cite{ALICE:2019onw}. The polarization is found to be along the direction of the angular momentum in the collision, perpendicular to the reaction plane, and to grow with decreasing collision energy, presumably because a larger fraction of the angular momentum carried by the incident nucleons ends up at midrapidity. A recent compilation of data is shown in Fig.~\ref{fig:Lambda_pol}.
The measured average polarizations can be converted into an estimate of the vorticity. From (\ref{eq:Lambda_pol}) one finds
\begin{equation}
\langle\omega\rangle = (\langle P_\Lambda \rangle + \langle P_{\overline{\Lambda}} \rangle )\, T ,
\end{equation}
which gives $\langle\omega\rangle \approx 10^{22}~{\rm s}^{-1}$ for $\langle P_\Lambda \rangle \approx \langle P_{\overline{\Lambda}} \rangle \approx 0.02$. STAR has also measured the global polarizations of $\Xi$ and $\Omega$ hyperons which are of similar magnitude, confirming the interpretation of the phenomenon as an effect of QGP vorticity \cite{STAR:2020xbm}.

The average polarizations $\langle P_\Lambda \rangle$ and $\langle P_{\overline{\Lambda}} \rangle$ agree with each other at all collision energies within the experimental errors. The most precise values have been measured in Au+Au at $\sqrt{s_{\rm NN}} = 200$ GeV, where $\langle P_\Lambda \rangle - \langle P_{\overline{\Lambda}} \rangle = 0.037 \pm 0.07$ \cite{STAR:2018gyt}. This measurement allows to set an upper limit on the magnetic field at the emission time: $|B| < 10^{12}$ T \cite{Muller:2018ibh}. This is less than $10^{-3}$ of the maximal magnetic field generated during the collision of the two nuclei.

In addition to the global polarization of  $\Lambda$ and $\overline{\Lambda}$, the experiments have also observed local polarization of hyperons along the beam direction. The orientation of the polarization vector depends on the direction of the hyperon transverse momentum and shows a quadrupole pattern with respect to the beam axis \cite{STAR:2019erd}. This effect is now understood as a result of the shear caused by the anisotropy of the transverse flow (see \cite{Becattini:2022zvf} for a review).

Vector meson alignment is another phenomenon related to spin that has been experimentally observed. Both STAR \cite{Singha:2020qns} and ALICE \cite{Mohanty:2020bqq} have found evidence for a global alignment of the spins of $K^{*0}$ and $\phi$ mesons with respect to the collision plane. Alignment is defined as the deviation of the $m=m'=0$ component of the spin--1 density matrix $\rho_{m'm} = \langle m'|\rho|m\rangle$ from its equilibrium value $\rho_{00}=1/3$ when all spin orientations are equally likely. Different from the concept of polarization, which distinguishes between spin orientation up and down with respect to the collision plane, alignment makes no such distinction. The mechanisms that can cause a nonzero aligment are thus less constrained by symmetry than those that can cause polarization. Indeed, the experiments find much larger values $|\rho_{00}-1/3| \approx 0.1-0.2$ (compared with $~10^{-2}$ for global polarization), but there is currently no generally accepted explanation for the origin of this effect. 

\begin{figure}
\centering
\includegraphics[width=0.5\linewidth]{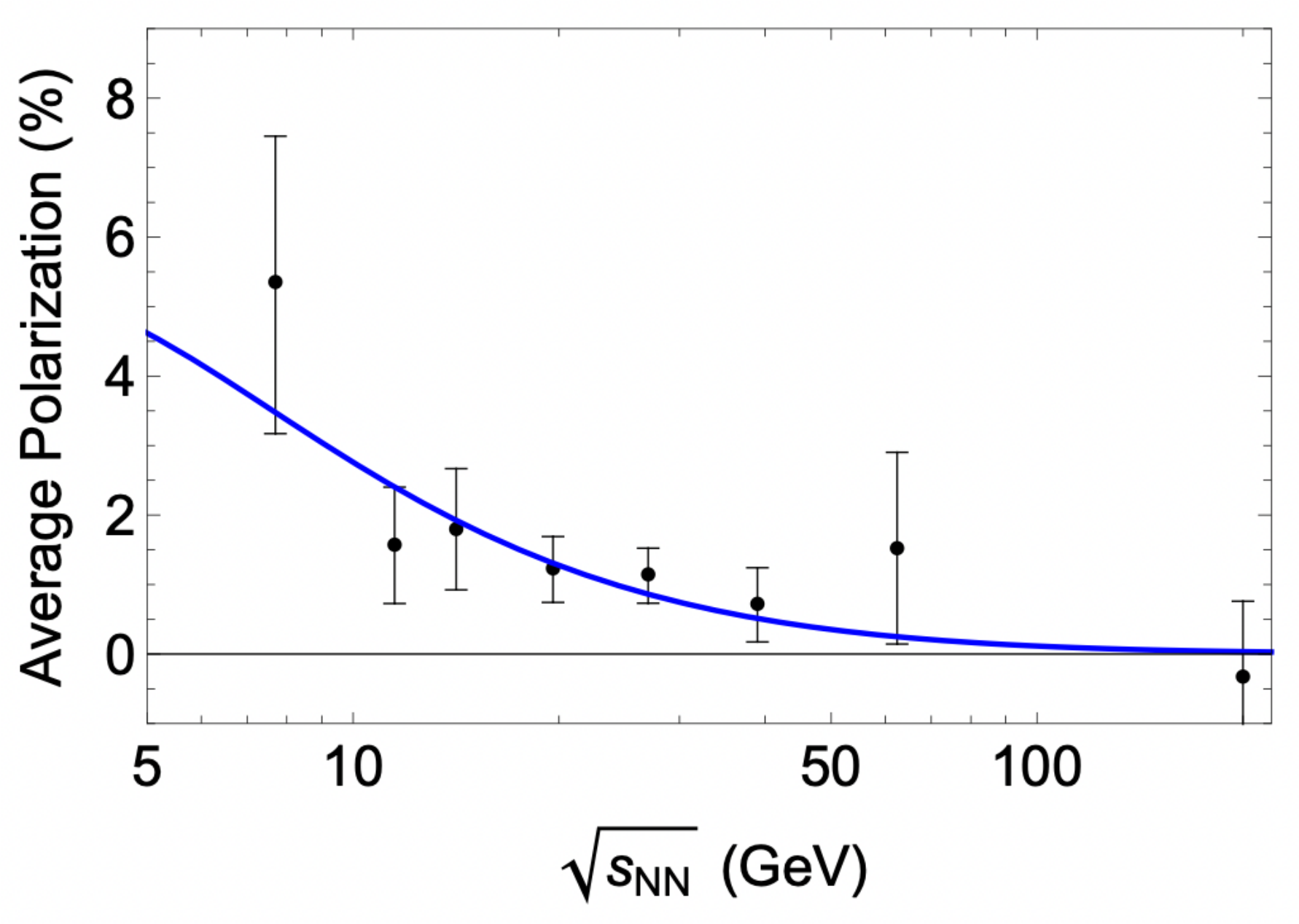}
\caption{Estimate of the collision energy dependence of the average hyperon polarization to the STAR measurement [1]. The solid blue curve represents the average value of the polarization of $\Lambda$ and $\bar\Lambda$.  [From \cite{Gallegos:2021bzp}]}
\label{fig:Pol_vs_E}
\end{figure}

Particle spin can be added to the hydrodynamical model of quark-gluon plasma expansion by introducing spin degrees of freedom to the fluid \cite{Florkowski:2018fap}. The equations for such a spinning, viscous fluid can be derived from kinetic theory in the usual way by applying a reduction to a limited number of moments of the momentum space distribution \cite{Weickgenannt:2019dks,Weickgenannt:2022zxs} or by symmetry-based analysis of the allowed perturbations of the energy-momentum tensor and the spin current around equilibrium \cite{Gallegos:2021bzp}. In such an approach the observed collision energy dependence of the global $\Lambda$ polarization can be explained with reasonable assumptions about the initial conditions (see Fig.~\ref{fig:Pol_vs_E}).

\subsection{Chiral magnetic effect} 
\label{sec:chiral_magnetic}

QCD gauge fields are characterized by a topological quantum number, called winding number, that changes due to the action of quantum mechanical fluctuations called instantons. While field configurations with a definite winding number break CP invariance, the QCD vacuum realized in nature contains a superposition of such field configurations that conserves the global CP symmetry of QCD for yet unknown reasons. (An axion field, if it exists, would explain this mystery in a natural way through the so-called Peccei-Quinn mechanism \cite{Peccei:1977hh}). In the presence of electromagnetic fields, however, these winding number fluctuations can generate local CP violations via the chiral anomaly \cite{Kharzeev:2007jp} of the axial current
\begin{equation}
    \partial_\mu j_5^\mu = -\frac{N_c}{2\pi^2}\sum_f Q_f^2 \vec{E}\cdot\vec{B} ,
\label{eq:chir-anom}
\end{equation}
where $Q_f$ are the electric charges of the light quark flavors. Anomalous hydrodynamics \cite{Newman:2005hd,Son:2009tf} adds (\ref{eq:chir-anom}) to the conservation laws for energy-monentum, baryon number, and electric charge as a fourth macroscopic equation. As usual, the conservation laws must be supplemented with constitutive equations, in this case for the vector and axial vector currents:
\begin{eqnarray}
    j^\mu = n u^\mu + \sigma E^\mu + \sigma_B B^\mu
    \nonumber \\
    j_5^\mu = n_5 u^\mu + \xi_E E^\mu + \xi_B B^\mu ,
\label{eq:j-j5}
\end{eqnarray}
where $\sigma, \sigma_B, \xi_E, \xi_B$ are transport coefficients ($\sigma$ is the usual electric conductivity), $n, n_5$ denote the usual and axial quark densities, and $E^\mu = F^{\mu\nu}u_\nu$, $B^\mu = \varepsilon^{\mu\nu\alpha\beta}u_{\nu}F_{\alpha\beta}$ are Lorentz covariant expressions for the electric and magnetic field. It is convenient to express these transport coefficients through vector and axial vector chemical potentials, $\mu$ and $\mu_5$ \cite{Son:2009tf}. 

In the absence of explicit parity violation, the global axial chemical potential $\mu_5=0$, which implies that the magnetic conductivity $\sigma_B$ vanishes. However, the presence of instantons implies that $\mu_5$ fluctuates locally, which means that the electric current receives locally fluctuating contributions from the magnetic field. Owing to the motion of the colliding nuclei heavy ion collisions generate very strong, short-lived magnetic fields of the order $eB \sim m_\pi^2$ that point perpendicular to the collision plane. One thus expects electric current fluctuations perpendicular to the collision plane, which result in event-by-event separation of the net electric charge of particles emitted into the upper and lower hemispheres. This is called the chiral magnetic effect (CME) \cite{Kharzeev:2007jp}. 

\begin{figure}[htb]
\centering
\includegraphics[width=0.55\linewidth]{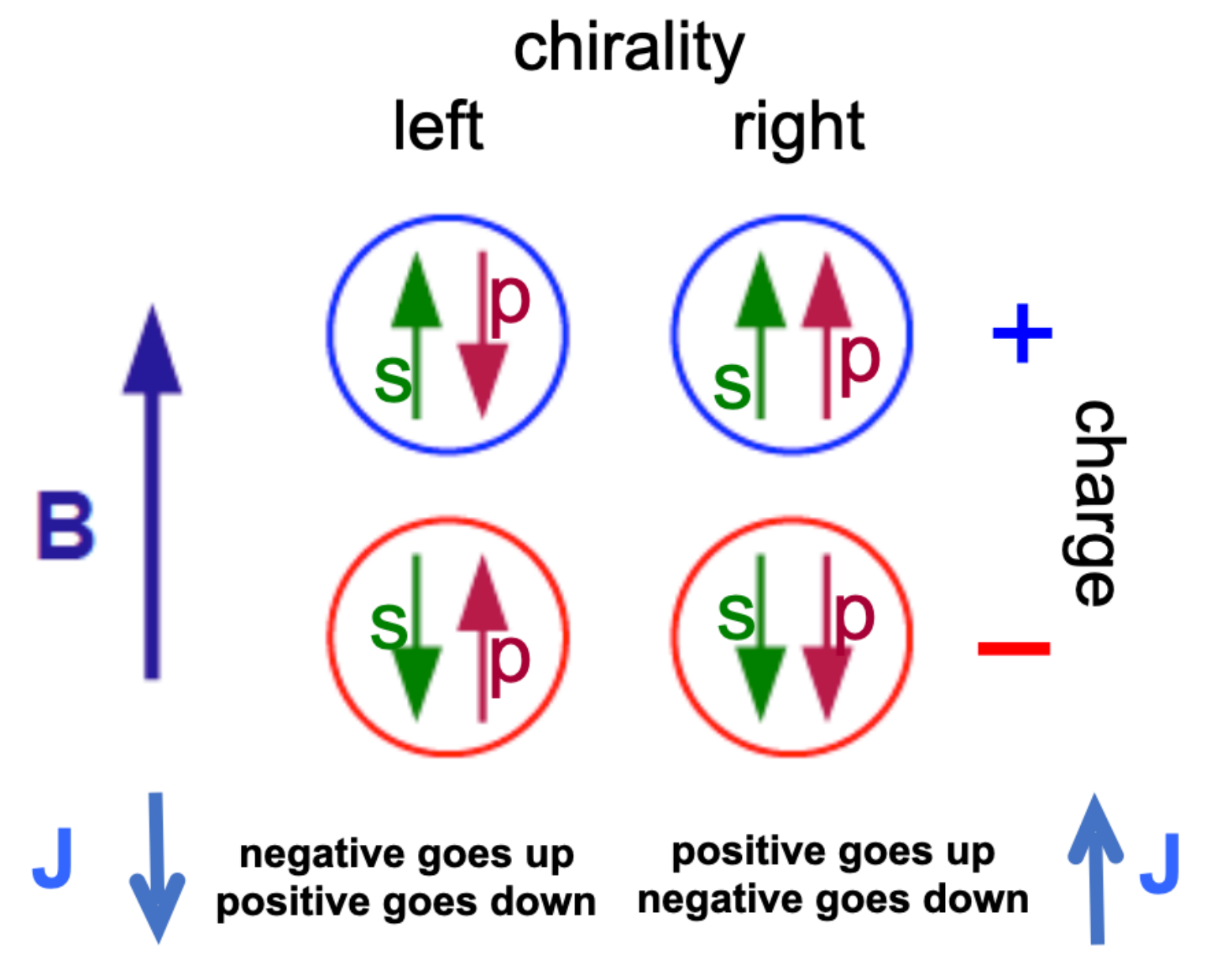}
\caption{Schematic illustration of the mechanism behind the chiral magnetic effect. The magnetic field aligns the quark spins ($s$) along the field lines according to the quark charge. Chirality associates associates a momentum direction $p$ with the quark spin; positive chirality (right-handed) quark spins are alinged with the momentum, negative (left-handed) quarks spins are anti-aligned. This creates a net electric current if the QGP contains a different number of left- and right-handed quarks.}
\label{fig:CME-schematic}
\end{figure}

The charge separation can be understood as a direct kinematic consequence of the alignment of quark spins along (or against) the magnetic field and the alignment of spin and momentum encoded in the chirality of the quark. If the magnetic field aligns a quark spin in the direction of the field, then a right-handed quark (positive chirality) will move in the direction of the field, and a left-handed quark (negative chirality) will move against the field. A positive axial density implies a preponderance of quarks with positive chirality and vice versa. Since the magnetic moment of a quark depends on the sign of its electric charge, this leads to a current in the direction of the magnetic field, if the axial density $n_5>0$, and against the field if $n_5<0$, as is illustrated in Fig.~\ref{fig:CME-schematic}.

Since the coefficients $\sigma_B$ and $\xi_B$ in (\ref{eq:j-j5}) are proportional to the densities $n_5$ and $n$, respectively, the anomalous hydrodynamic equations sustain a low-energy propagating mode corresponding to propagating coupled density-axial density fluctuations similar to the sound mode \cite{Kharzeev:2010gd}. This excitation, which propagates with a speed proportional to the magnetic field $B$, is known under the name ``chiral magnetic wave'' (CMW). Like the current fluctuations induced by the CME, the CMW could be seeded by the topological charge density fluctuations in the glasma during the earliest stage of the collision. 

Another phenomenon worth mentioning is the ``chiral vortical effect'' (CVE). A complete analysis shows that the constitutive equations (\ref{eq:j-j5}) also receive contributions proportional to the vorticity vector $\omega^\mu$ of the QGP. Their effect is similar to the chiral magnetic effect with axial charge fluctuations driving a fluctuating electric current along the direction of the vorticity vector. It can be similarly understood as the CME as a result of the alignment of quark spins along the QGP vorticity vector in thermal equilibrium, which we discussed in the context of global hyperon polarization. Thus, the illustration in Fig.~\ref{fig:CME-schematic} applies, except that the vorticity $\omega$ replaces the magnetic field $B$. In an off-central heavy ion collision both vectors, $\vec{\omega}$ and $\vec{B}$, point in the same direction perpendicular to the reaction plane. An overview of these phenomena can be found in \cite{Kharzeev:2015znc}.

A number of observables are specifically sensitive to such fluctuating electric charge separation phenomena. A possible search strategy in the context of known background effects is discussed in \cite{Voloshin:2018qsm}. Quantitative predictions based on solutions of the anomalous hydrodynamical equations with reasonable assumptions for the initial axial density fluctuations can be found in \cite{Shi:2017cpu}. The magnitude of the expected event-by-event fluctuations also depends strongly on the longevity of the magnetic field. For the parameters used in \cite{Shi:2017cpu} the experimental signals for the chiral magnetic effect are in the range of $10^{-4}$, but other assumptions may lead to much smaller predicted values \cite{Muller:2010jd}.

As theoretical predictions of the magnitude of observables for the chiral magnetic effect are beset with large uncertainties, experimental searches for it are of paramount importance. Owing to the parity conserving nature of QCD, the anomalous electric current must fluctuate event by event, thus all signals involve the measurement of event-by-event fluctuations. This means that other ``normal'' sources that are sensitive to the orientation of the reaction plane can contribute, in particular, those involving charged resonance decays modulated by the elliptic flow of the final-state hadron distribution (see \cite{Voloshin:2018qsm} for an in-depth discussion and references). Experimental studies at RHIC \cite{STAR:2021pwb} and LHC \cite{ALICE:2020siw} have concluded that at most a small fraction (less than 10\% for RHIC) of the observed signals can be attributed to the CME.

The comparison of measurements of observables sensitive to the chiral magnetic effect in p+Pb and Pb+Pb collisions is another way to assess the size of background effects. Any magnetic field-driven effect, such as the CME, should be greatly suppressed in p+Pb collisions in comparison with Pb+Pb collisions (by a factor $(1/Z_{\rm Pb}^2=1.5\times 10^{-4}$). Data from the CMS experiment constrain possible contributions of the CME to the Pb+Pb data to less than 7\% \cite{CMS:2017lrw}.

A more sensitive search for CME signals requires a suppression or cancellation of such background effects. This motivated a comparative study of two collision systems involving nuclear isobars, $^{96}$Zr and $^{96}$Ru, which was carried out at RHIC. The magnetic field produced in Ru+Ru collisions is larger than that produced in Zr+Zr collisions under otherwise identical conditions, because a $^{96}$Zr nucleus contains 40 protons, while a $^{96}$Ru nucleus contains 44 protons. As the CME observables are proportional to the square of the magnetic field, one expects roughly a 15\% difference between the two systems for the CME contribution to any observable. Great care was taken to ensure that the experimental conditions for collisions in the two isobar systems were identical, the data were subjected to a sophisticated blind analysis protocol \cite{STAR:2019bjg}, the first of its kind in the field of relativistic heavy ion physics, and the analysis was performed independently by several groups.

\begin{figure}[htb]
\centering
\includegraphics[width=0.95\linewidth]{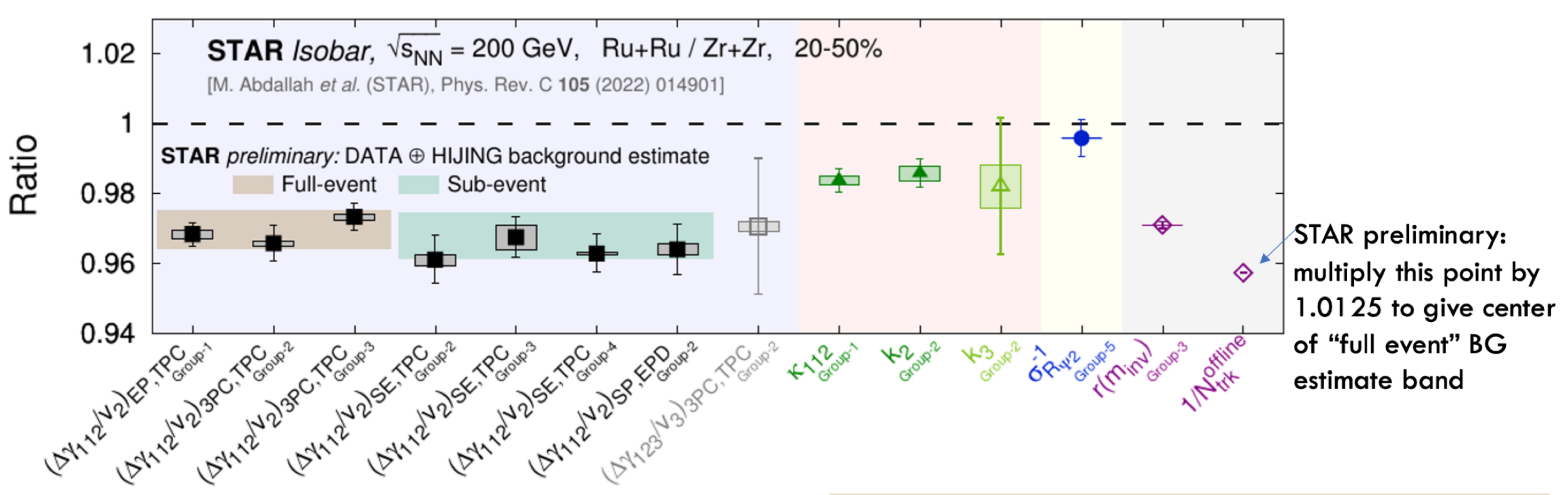}
\caption{Results of the blind analysis of observables $S$ sensitive to the CME in the $^{96}$Zr$-$$^{96}$Ru isobar system. Shown is the ratio $S$(Ru)/$S$(Zr). The horizontal shaded bars indicate the baseline without CME contribution, corrected for the nuclear shape difference and minor efficiency effects. A possible CME contribution would cause this ratio to be higher than the baseline. For details see \cite{STAR:2021mii}.}
\label{fig:isobar}
\end{figure}

The results published by the STAR collaboration \cite{STAR:2021mii} showed no evidence for the presence of a CME contribution to any of the predefined observables with an experimental precision of $\pm 4\times 10^{-3}$. Figure \ref{fig:isobar} shows the measured ratios $S$(Ru)/$S$(Zr) for each of the signature observables $S$ considered in the analysis. All signals are in some way related to the difference $\Delta\gamma$ for same-sign and opposite-sign charged pairs of particles of the quantity $\gamma$ defined in (\ref{eq:gamma}), normalized to the elliptic flow anisotropy $v_2$ that drives the background effects. A contribution from the CME would cause this ratio to be larger than unity. Clearly, all measured ratios lie well below one, which means that they do not provide evidence for a CME contribution. 

The fact that all ratios related to $\Delta\gamma/v_2$ cluster around a value of 0.97 suggests that they have a common source that can be traced to a difference between the two isobars, which is not related to the nuclear charge. Indeed, such differences are known to exist: $^{96}$Zr has a thicker neutron skin than $^{96}$Ru (the $^{96}$Zr nucleus has four additional neutrons), and the two nuclei have different quadrupole deformations. The experiment clearly revealed these differences in the centrality dependence of the final-state multiplicity distributions. This means that the shape differences of the two isobar nuclei are the source of the largest systematic uncertainty as background effects associated with elliptic flow do not cancel exactly. Whether this uncertainty, which lies in the percent range, can be further reduced, is an open question.

\section{Future Opportunities}
\label{sec:opps}

The physics of heavy ion collisions is a vibrant field of research, and there are many different avenues for future progress in understanding strongly interacting matter under extreme conditions. In general, the field is driven by experimental measurements and their theoretical interpretation, since only very few predictions from first principle calculations have been possible. It is important to identify questions that require more precise or more comprehensive data on well-studied observables or completely new types of measurements and analysis. 

There are four areas which are going to be explored in the near-to-midterm future: 

\begin{itemize}
    \item Precision measurements with ultra-relativistic nucleus-nucleus collisions at RHIC with sPHENIX \cite{PHENIX:2015siv} and STAR and at LHC with ALICE, ATLAS, and CMS and LHCb; 
    \item A whole new level of heavy flavor and electromagnetic probes measurements with ALICE3 \cite{Adamova:2019vkf}; 
    \item Ultraperipheral collisions and the transition to the electron-ion collider \cite{Accardi:2012qut}; 
    \item Measurements at low beam energies with final results from the RHIC beam energy scan, HADES at GSI and future FAIR. 
\end{itemize}

Let us point out the main physics case and opportunities for each of the four broad directions mentioned above. 

The upcoming LHC runs 3 and 4, where much higher luminosities are going to be achieved, will generate excellent data sets recorded for elementary particle collisions as well as heavy ion collisions in the TeV energy range. In addition to Pb+Pb collisions, there is the proposal to add O+O collisions to enhance the understanding of the transition between very small systems and large collision systems. Collisions of different species of ions might be employed to explore nuclear structure in heavy ion collisions. 

All four experimental collaborations at the LHC (ALICE, ATLAS, CMS, and LHCb) are planning to participate in the heavy ion runs. Very precise measurements of bulk observables and hard probes are expected. In a complementary campaign the sPHENIX and STAR will be taking data at the lower beam energy of RHIC, but with comparable precision for hard probes. To understand the temperature dependence of transport coefficients like $\hat{q}$ the lever arm in beam energy between $\sqrt{s_{\rm NN}}=200$ GeV and a few TeV is going to be important. 

For Run 5 expected in the 2030's the ALICE collaboration has proposed to construct a completely new detector named ALICE3. This device will allow for an entirely new level of precision for measurements of heavy flavor probes in the soft and hard sector as well as electromagnetic probes down to very low transverse momenta. In addition, new capabilities to identify multi-charm hadronic states will become available. ALICE3 will provide detailed constraints for the more refined theoretical modeling that is going to ready by the time the new detector will come online. 

In addition to the study of the midrapidity region, where the quark-gluon plasma production can be studied, it is going to be of interest to explore the whole longitudinal phase space. For certain investigations a fixed target setup at collider facilities provides advantages and is being discussed at LHC to study the quark-gluon plasma in the fragmentation region. Forward rapidities may also be useful to investigate the dependence on net baryon content of the system. 

Extremely peripheral (``ultraperipheral'') events also receive increasing attention. The idea is that in such collisions the nuclei pass without coming into direct contact but interact via electromagnetic interactions that are enhanced by the nuclear charge $Z$ and the Lorentz factor $\gamma$. Exploring the interplay between electromagnetic and strong interactions under such conditions paves the path to the planned electron-ion collider (EIC). In 2020, the decision was been made to construct such a facilty at Brookhaven National Laboratory, which will allows for collisions of electrons with light and heavy ions at $\sqrt{s_{\rm eN}}$ up to 140 GeV. The electron-ion collider will generate unprecedented insights into the quark-gluon structure of the proton and complex nuclei. This will provide very detailed information for the initial state of a heavy ion collision. 

Moving forward with the exploration of the QCD phase diagram at nonzero net baryon chemical potential, several experimental efforts are worth mentioning. The beam energy scan program at RHIC has just finished Phase 2, and precision measurements are expected to be published by the STAR collaboration in the near future for energies down to $\sqrt{\rm {NN}} = 3$ GeV. The HADES experiment \cite{HADES:2009aat} at GSI is running within the FAIR Phase-0 program and will measure the excitation function of observables in the fixed target beam energy regime around $E_{\rm lab}/A \approx 1$ GeV per nucleon ($\sqrt{s_{\rm NN}} \approx 2.4$ GeV). In the later part of the decade, SIS-100 will be completed and the CBM (Compressed Baryonic Matter) experiment \cite{Friman:2011zz} will measure rare probes with unprecedented precision in the beam energy region from $E_{\rm lab}/A = 3.4-12$ GeV ($\sqrt{s_{\rm NN}} \approx 3.3-5.3$ GeV). There are also other projects proposed around the world, like a heavy ion extension of J-PARC in Japan \cite{Hachiya:2020bjg}, HIAF in China \cite{Xiaohong:2018weu}, and NICA in Dubna \cite{Kekelidze:2017ual}. 

These lower-energy facilities aim to determine the structure of the QCD phase diagram, in particular, to find out whether there is a first-order phase transition between the hadron gas and the quark-gluon plasma at nonzero baryon chemical potential with a critical endpoint or whether the transition is a cross-over everywhere. Knowing the nuclear matter equation of state at high net baryon density is also important for the understanding of neutron star mergers. Since the first detection of gravitational waves from such a merger event in 2017, there has been increasing interaction between astrophysicists and the heavy ion physics community, which is likely to further intensity in the years to come.  

All of the expected measurements will need to be accompanied by theoretical progress in the understanding of QCD matter and high energy nuclear reactions. Such progress relies on fundamental theory developments based on lattice QCD techniques, functional methods, and effective field theories as well as on sophisticated dynamical modeling that connects properties of strongly interacting matter to experimental measurements. More standardized ways to compare data to calculations will be helpful for quantitative conclusions, e.g. based on HepMC and RIVET adapted for heavy ions. Modern analysis tools based on machine learning and deep learning methods as well as potential applications of quantum computing will complement the more traditional efforts.

\section{Summary} 
\label{sec:sum}

Relativistic heavy ion collisions produce matter with the highest energy density known in nature, thereby recreating conditions similar to those in the early universe or in neutron star mergers. We now know that this matter, the quark-gluon plasma, is also the most ``perfect'' fluid and endowed with a high vorticity. This conclusion has been reached by a concerted theoretical and experimental effort over the last three decades. Many detailed measurements and sophisticated calculations enabled by technological advances have led to a ``standard model'' for relativistic heavy ion collisions that is based on non-equilibrium initial conditions, viscous hydrodynamics and hadronic transport. While the quantitative insight into the properties of the quark-gluon plasma has lately seen remarkable progress by the application of Bayesian multi-parameter model-to-data comparisons, a more complete understanding of the structures in the QCD phase diagram -- a potential first order phase transition and critical endpoint -- requires further theoretical developments and a new level of experimental precision. 

Given the multitude of available beam energies, collision systems, and experimental probes it is important not to lose one's overview. Everyone working in this field must from time to time ask themselves how their current project connects to the major physics questions of the field and what can be learned by looking from a broader perspective. This is especially true for scientists at the beginning of their career, who have not yet developed the breadth of knowledge and insight of more experienced scientists. In a field as complex as this it is easy to be misled to conclusions based on a limited set of data and observables. To avoid going down the wrong path, the sophisticated dynamical models now available need to be consistently applied to as many observables as possible with the same parameter settings. In doing so, it is crucial to apply the right methods in the right places and to be aware of the limits of applicability of each of them. Getting a prediction from a complex numerical code does not guarantee that it is physically meaningful!

Relativistic heavy ion physics is especially attractive to young researchers owing to its mode of international collaboration, on the experimental as well as increasingly on the theoretical side. The multitude of different methods that are being applied to further our knowledge makes it an ideal training ground. 

There are close connections to other fields within nuclear physics, for example hadron physics, nuclear structure physics, and nuclear astrophysics. More recently, the connection to the astrophysics community has intensified since it was realised that heavy ion collisions at low beam energy allows us to produce and study conditions in the laboratory that resemble those occurring in neutron star mergers . The hadronic interactions that are relevant in heavy ion physics, are also of interest to the astroparticle physics community for the description of cosmic air showers. The non-equilibrium phase transitions and chiral phenomena encountered in heavy ion collisions have connections to phenomena of interest to the condensed matter physics community. Last but not least, the small system debate has revitalized the connection to the high energy physics community. There is also some exchange of technology for model-to-data comparisons. 

We hope that this review will help beginning and more experienced scientists alike to get a more complete appreciation for the wealth of phenomena and approaches that are currently available to study and understand relativistic heavy ion collisions. Impressive progress is being made, and the future opportunities that await those who venture into this field of research are great.

\section{Acknowledgements}
We thank Niklas G{\"o}tz, Andrew Gordeev, Renan Hirayama, Reed Hodges, and Derek Soeder for valuable feedback on a draft version of this manuscript.
BM acknowledges support by a grant (DO-FG02-05ER41367) from the Office of Science of the U.~S.~Department of Energy, as well as support by Yale University during a sabbatical stay in Spring 2022. HE acknowledges support by by the State of Hesse within the Research Cluster ELEMENTS (Project ID 500/10.006) and the Deutsche Forschungsgemeinschaft (DFG, German Research Foundation) – Project number 315477589 – TRR 211.

\vspace{1cm}


\bibliographystyle{h-physrev5}
\bibliography{HEBMReview}

\begin{thebibliography}{100}

\bibitem{Schwarz:2003du}
D.~J. Schwarz,
\newblock Annalen Phys. {\bf 12}, 220 (2003), astro-ph/0303574.

\bibitem{Aoki:2006br}
Y.~Aoki, Z.~Fodor, S.~D. Katz, and K.~K. Szabo,
\newblock Phys. Lett. {\bf B643}, 46 (2006), hep-lat/0609068.

\bibitem{Aoki:2006we}
Y.~Aoki, G.~Endrodi, Z.~Fodor, S.~D. Katz, and K.~K. Szabo,
\newblock Nature {\bf 443}, 675 (2006), hep-lat/0611014.

\bibitem{Fukushima:2003fw}
K.~Fukushima,
\newblock Phys. Lett. {\bf B591}, 277 (2004), hep-ph/0310121.

\bibitem{Rossner:2007ik}
S.~Roessner, T.~Hell, C.~Ratti, and W.~Weise,
\newblock Nucl. Phys. {\bf A814}, 118 (2008), arXiv:0712.3152.

\bibitem{Arsene:2004fa}
BRAHMS, I.~Arsene {\em et~al.},
\newblock Nucl. Phys. {\bf A757}, 1 (2005), nucl-ex/0410020.

\bibitem{Adcox:2004mh}
PHENIX, K.~Adcox {\em et~al.},
\newblock Nucl. Phys. {\bf A757}, 184 (2005), nucl-ex/0410003.

\bibitem{Back:2004je}
B.~B. Back {\em et~al.},
\newblock Nucl. Phys. {\bf A757}, 28 (2005), nucl-ex/0410022.

\bibitem{Adams:2005dq}
STAR, J.~Adams {\em et~al.},
\newblock Nucl. Phys. {\bf A757}, 102 (2005), nucl-ex/0501009.

\bibitem{Muller:2006ee}
B.~M{\"u}ller and J.~L. Nagle,
\newblock (2006), nucl-th/0602029.

\bibitem{Bannur:1998nq}
V.~M. Bannur,
\newblock Eur. Phys. J. {\bf C11}, 169 (1999), arXiv:hep-ph/9811397.

\bibitem{Gyulassy:2004zy}
M.~Gyulassy and L.~McLerran,
\newblock Nucl. Phys. {\bf A750}, 30 (2005), nucl-th/0405013.

\bibitem{Schukraft:2011cz}
J.~Schukraft and A.~Collaboration,
\newblock J.Phys.G {\bf G38}, 124003 (2011), arXiv:1106.5620.

\bibitem{Steinberg:2011qq}
P.~Steinberg and A.~Collaboration,
\newblock J. Phys. G {\bf 38}, 124004 (2011), arXiv:1110.3352.

\bibitem{Wyslouch:2011zz}
CMS, B.~Wyslouch,
\newblock J. Phys. G {\bf 38}, 124005 (2011), arXiv:1107.2895.

\bibitem{Muller:2012zq}
B.~M{\"u}ller, J.~Schukraft, and B.~Wyslouch,
\newblock Ann. Rev. Nucl. Part. Sci. {\bf 62}, 361 (2012), arXiv:1202.3233.

\bibitem{Braun-Munzinger:2015hba}
P.~Braun-Munzinger, V.~Koch, T.~Sch\"afer, and J.~Stachel,
\newblock Phys. Rept. {\bf 621}, 76 (2016), arXiv:1510.00442.

\bibitem{Busza:2018rrf}
W.~Busza, K.~Rajagopal, and W.~van~der Schee,
\newblock Ann. Rev. Nucl. Part. Sci. {\bf 68}, 339 (2018), arXiv:1802.04801.

\bibitem{Rafelski:2003zz}
J.~Kapusta, B.~M{\"u}ller, and J.~Rafelski.

\bibitem{Yagi:2005yb}
K.~Yagi, T.~Hatsuda, and Y.~Miake,
\newblock Camb. Monogr. Part. Phys. Nucl. Phys. Cosmol. {\bf 23}, 1 (2005).

\bibitem{Letessier:2002gp}
J.~Letessier and J.~Rafelski,
\newblock Camb. Monogr. Part. Phys. Nucl. Phys. Cosmol. {\bf 18}, 1 (2002).

\bibitem{Czajka:2017bod}
A.~Czajka and S.~Jeon,
\newblock Phys. Rev. C {\bf 95}, 064906 (2017), arXiv:1701.07580.

\bibitem{Sveshnikov:1995vi}
N.~A. Sveshnikov and F.~V. Tkachov,
\newblock Phys. Lett. B {\bf 382}, 403 (1996), arXiv:hep-ph/9512370.

\bibitem{Komiske:2022enw}
P.~T. Komiske, I.~Moult, J.~Thaler, and H.~X. Zhu,
\newblock (2022), arXiv:2201.07800.

\bibitem{Chang:2013rca}
H.-M. Chang, M.~Procura, J.~Thaler, and W.~J. Waalewijn,
\newblock Phys. Rev. Lett. {\bf 111}, 102002 (2013), arXiv:1303.6637.

\bibitem{Chang:2013iba}
H.-M. Chang, M.~Procura, J.~Thaler, and W.~J. Waalewijn,
\newblock Phys. Rev. D {\bf 88}, 034030 (2013), arXiv:1306.6630.

\bibitem{Li:2021zcf}
Y.~Li, I.~Moult, S.~S. van Velzen, W.~J. Waalewijn, and H.~X. Zhu,
\newblock Phys. Rev. Lett. {\bf 128}, 182001 (2022), arXiv:2108.01674.

\bibitem{Kaplan:1992bt}
D.~B. Kaplan,
\newblock Phys. Lett. {\bf B288}, 342 (1992), arXiv:hep-lat/9206013.

\bibitem{Narayanan:1993sk}
R.~Narayanan and H.~Neuberger,
\newblock Nucl. Phys. {\bf B412}, 574 (1994), arXiv:hep-lat/9307006.

\bibitem{Kuti:1980gh}
J.~Kuti, J.~Polonyi, and K.~Szlachanyi,
\newblock Phys. Lett. {\bf B98}, 199 (1981).

\bibitem{McLerran:1981pb}
L.~D. McLerran and B.~Svetitsky,
\newblock Phys. Rev. {\bf D24}, 450 (1981).

\bibitem{Borsanyi:2010cj}
S.~Borsanyi {\em et~al.},
\newblock JHEP {\bf 11}, 077 (2010), arXiv:1007.2580.

\bibitem{Borsanyi:2013bia}
S.~Borsanyi {\em et~al.},
\newblock Phys. Lett. B {\bf 730}, 99 (2014), arXiv:1309.5258.

\bibitem{HotQCD:2014kol}
HotQCD, A.~Bazavov {\em et~al.},
\newblock Phys. Rev. D {\bf 90}, 094503 (2014), arXiv:1407.6387.

\bibitem{Allton:2005gk}
C.~R. Allton {\em et~al.},
\newblock Phys. Rev. {\bf D71}, 054508 (2005), hep-lat/0501030.

\bibitem{deForcrand:2002ci}
P.~de~Forcrand and O.~Philipsen,
\newblock Nucl. Phys. {\bf B642}, 290 (2002), hep-lat/0205016.

\bibitem{Fodor:2001pe}
Z.~Fodor and S.~D. Katz,
\newblock JHEP {\bf 03}, 014 (2002), hep-lat/0106002.

\bibitem{deForcrand:2007rq}
P.~de~Forcrand, S.~Kim, and O.~Philipsen,
\newblock PoS {\bf LAT2007}, 178 (2007), arXiv:0711.0262 [hep-lat].

\bibitem{Borsanyi:2010bp}
Wuppertal-Budapest, S.~Borsanyi {\em et~al.},
\newblock JHEP {\bf 09}, 073 (2010), arXiv:1005.3508.

\bibitem{Borsanyi:2020fev}
S.~Borsanyi {\em et~al.},
\newblock Phys. Rev. Lett. {\bf 125}, 052001 (2020), arXiv:2002.02821.

\bibitem{Majumder:2010ik}
A.~Majumder and B.~M{\"u}ller,
\newblock Phys. Rev. Lett. {\bf 105}, 252002 (2010), arXiv:1008.1747.

\bibitem{Bazavov:2014xya}
A.~Bazavov {\em et~al.},
\newblock Phys. Rev. Lett. {\bf 113}, 072001 (2014), arXiv:1404.6511.

\bibitem{Andronic:2018qqt}
A.~Andronic {\em et~al.},
\newblock Phys. Lett. B {\bf 792}, 304 (2019), arXiv:1808.03102.

\bibitem{Andersen:2010wu}
J.~O. Andersen, L.~E. Leganger, M.~Strickland, and N.~Su,
\newblock Phys. Lett. {\bf B696}, 468 (2011), arXiv:1009.4644.

\bibitem{Andersen:2011sf}
J.~O. Andersen, L.~E. Leganger, M.~Strickland, and N.~Su,
\newblock JHEP {\bf 08}, 053 (2011), arXiv:1103.2528.

\bibitem{Peshier:1999ww}
A.~Peshier, B.~Kampfer, and G.~Soff,
\newblock Phys. Rev. C {\bf 61}, 045203 (2000), arXiv:hep-ph/9911474.

\bibitem{Ivanov:2004gq}
Y.~B. Ivanov, V.~V. Skokov, and V.~D. Toneev,
\newblock Phys. Rev. D {\bf 71}, 014005 (2005), arXiv:hep-ph/0410127.

\bibitem{Braaten:1989mz}
E.~Braaten and R.~D. Pisarski,
\newblock Nucl. Phys. {\bf B337}, 569 (1990).

\bibitem{Kapusta:2006pm}
J.~I. Kapusta and C.~Gale,
\newblock {\em {Finite-temperature field theory: Principles and
  applications}}Cambridge Monographs on Mathematical Physics (Cambridge
  University Press, 2011).

\bibitem{Lebellac:2000}
M.~Le~Bellac,
\newblock {\em Thermal Field Theory} (Cambridge University Press, 2000).

\bibitem{Arnold:2000dr}
P.~Arnold, G.~D. Moore, and L.~G. Yaffe,
\newblock JHEP {\bf 11}, 001 (2000), hep-ph/0010177.

\bibitem{Arnold:2003zc}
P.~Arnold, G.~D. Moore, and L.~G. Yaffe,
\newblock JHEP {\bf 05}, 051 (2003), hep-ph/0302165.

\bibitem{Ghiglieri:2018dib}
J.~Ghiglieri, G.~D. Moore, and D.~Teaney,
\newblock JHEP {\bf 03}, 179 (2018), arXiv:1802.09535.

\bibitem{Caron-Huot:2008zna}
S.~Caron-Huot,
\newblock Phys. Rev. D {\bf 79}, 065039 (2009), arXiv:0811.1603.

\bibitem{Ghiglieri:2015zma}
J.~Ghiglieri and D.~Teaney,
\newblock Int. J. Mod. Phys. E {\bf 24}, 1530013 (2015), arXiv:1502.03730.

\bibitem{Ghiglieri:2015ala}
J.~Ghiglieri, G.~D. Moore, and D.~Teaney,
\newblock JHEP {\bf 03}, 095 (2016), arXiv:1509.07773.

\bibitem{Thoma:1991ea}
M.~H. Thoma,
\newblock Phys. Lett. {\bf B273}, 128 (1991).

\bibitem{Caron-Huot:2007rwy}
S.~Caron-Huot and G.~D. Moore,
\newblock Phys. Rev. Lett. {\bf 100}, 052301 (2008), arXiv:0708.4232.

\bibitem{Arnold:2001ms}
P.~Arnold, G.~D. Moore, and L.~G. Yaffe,
\newblock JHEP {\bf 12}, 009 (2001), arXiv:hep-ph/0111107.

\bibitem{Ghiglieri:2013gia}
J.~Ghiglieri {\em et~al.},
\newblock JHEP {\bf 05}, 010 (2013), arXiv:1302.5970.

\bibitem{Braaten:1990wp}
E.~Braaten, R.~D. Pisarski, and T.-C. Yuan,
\newblock Phys. Rev. Lett. {\bf 64}, 2242 (1990).

\bibitem{Ghiglieri:2014kma}
J.~Ghiglieri and G.~D. Moore,
\newblock JHEP {\bf 12}, 029 (2014), arXiv:1410.4203.

\bibitem{Ghiglieri:2015nba}
J.~Ghiglieri,
\newblock Nucl. Part. Phys. Proc. {\bf 276-278}, 305 (2016), arXiv:1510.00525.

\bibitem{Maldacena:1997re}
J.~M. Maldacena,
\newblock Adv. Theor. Math. Phys. {\bf 2}, 231 (1998), arXiv:hep-th/9711200.

\bibitem{Aharony:1999ti}
O.~Aharony, S.~S. Gubser, J.~M. Maldacena, H.~Ooguri, and Y.~Oz,
\newblock Phys. Rept. {\bf 323}, 183 (2000), arXiv:hep-th/9905111.

\bibitem{Casalderrey-Solana:2011dxg}
J.~Casalderrey-Solana, H.~Liu, D.~Mateos, K.~Rajagopal, and U.~A. Wiedemann,
\newblock {\em {Gauge/String Duality, Hot QCD and Heavy Ion Collisions}}
  (Cambridge University Press, 2014), arXiv:1101.0618.

\bibitem{Natsuume:2014sfa}
M.~Natsuume,
\newblock {\em {AdS/CFT Duality User Guide}} (Springer Tokyo, 2015),
  arXiv:1409.3575.

\bibitem{Gursoy:2007cb}
U.~Gursoy and E.~Kiritsis,
\newblock JHEP {\bf 02}, 032 (2008), arXiv:0707.1324.

\bibitem{Karch:2006pv}
A.~Karch, E.~Katz, D.~T. Son, and M.~A. Stephanov,
\newblock Phys. Rev. D {\bf 74}, 015005 (2006), arXiv:hep-ph/0602229.

\bibitem{deTeramond:2008ht}
G.~F. de~Teramond and S.~J. Brodsky,
\newblock Phys. Rev. Lett. {\bf 102}, 081601 (2009), arXiv:0809.4899.

\bibitem{Son:2002sd}
D.~T. Son and A.~O. Starinets,
\newblock JHEP {\bf 09}, 042 (2002), arXiv:hep-th/0205051.

\bibitem{Balasubramanian:2010ce}
V.~Balasubramanian {\em et~al.},
\newblock Phys. Rev. Lett. {\bf 106}, 191601 (2011), arXiv:1012.4753.

\bibitem{Balasubramanian:2011ur}
V.~Balasubramanian {\em et~al.},
\newblock Phys. Rev. D {\bf 84}, 026010 (2011), arXiv:1103.2683.

\bibitem{Price:1986yy}
R.~H. Price and K.~S. Thorne,
\newblock Phys. Rev. D {\bf 33}, 915 (1986).

\bibitem{Policastro:2001yc}
G.~Policastro, D.~T. Son, and A.~O. Starinets,
\newblock Phys. Rev. Lett. {\bf 87}, 081601 (2001), arXiv:hep-th/0104066.

\bibitem{Kovtun:2004de}
P.~Kovtun, D.~T. Son, and A.~O. Starinets,
\newblock Phys. Rev. Lett. {\bf 94}, 111601 (2005), arXiv:hep-th/0405231.

\bibitem{Ecker:2016thn}
C.~Ecker, D.~Grumiller, P.~Stanzer, S.~A. Stricker, and W.~van~der Schee,
\newblock JHEP {\bf 11}, 054 (2016), arXiv:1609.03676.

\bibitem{Herzog:2006gh}
C.~P. Herzog, A.~Karch, P.~Kovtun, C.~Kozcaz, and L.~G. Yaffe,
\newblock JHEP {\bf 07}, 013 (2006), arXiv:hep-th/0605158.

\bibitem{Liu:2006ug}
H.~Liu, K.~Rajagopal, and U.~A. Wiedemann,
\newblock Phys. Rev. Lett. {\bf 97}, 182301 (2006), arXiv:hep-ph/0605178.

\bibitem{Chesler:2008uy}
P.~M. Chesler, K.~Jensen, A.~Karch, and L.~G. Yaffe,
\newblock Phys. Rev. D {\bf 79}, 125015 (2009), arXiv:0810.1985.

\bibitem{Janik:2005zt}
R.~A. Janik and R.~B. Peschanski,
\newblock Phys. Rev. D {\bf 73}, 045013 (2006), arXiv:hep-th/0512162.

\bibitem{Heller:2011ju}
M.~P. Heller, R.~A. Janik, and P.~Witaszczyk,
\newblock Phys. Rev. Lett. {\bf 108}, 201602 (2012), arXiv:1103.3452.

\bibitem{Shuryak:2011aa}
E.~Shuryak,
\newblock J. Phys. G {\bf 39}, 054001 (2012), arXiv:1112.2573.

\bibitem{Chesler:2009cy}
P.~M. Chesler and L.~G. Yaffe,
\newblock Phys. Rev. D {\bf 82}, 026006 (2010), arXiv:0906.4426.

\bibitem{Chesler:2010bi}
P.~M. Chesler and L.~G. Yaffe,
\newblock Phys. Rev. Lett. {\bf 106}, 021601 (2011), arXiv:1011.3562.

\bibitem{Chesler:2013lia}
P.~M. Chesler and L.~G. Yaffe,
\newblock JHEP {\bf 07}, 086 (2014), arXiv:1309.1439.

\bibitem{Casalderrey-Solana:2013aba}
J.~Casalderrey-Solana, M.~P. Heller, D.~Mateos, and W.~van~der Schee,
\newblock Phys. Rev. Lett. {\bf 111}, 181601 (2013), arXiv:1305.4919.

\bibitem{Muller:2020ziz}
B.~M\"uller, A.~Rabenstein, A.~Sch\"afer, S.~Waeber, and L.~G. Yaffe,
\newblock Phys. Rev. D {\bf 101}, 076008 (2020), arXiv:2001.07161.

\bibitem{vanderSchee:2013pia}
W.~van~der Schee, P.~Romatschke, and S.~Pratt,
\newblock Phys. Rev. Lett. {\bf 111}, 222302 (2013), arXiv:1307.2539.

\bibitem{Shukla:2001mb}
P.~Shukla,
\newblock (2001), arXiv:nucl-th/0112039.

\bibitem{Stock:2020blh}
R.~Stock,
\newblock Springer Proc. Phys. {\bf 250}, 3 (2020).

\bibitem{Kharzeev:2000ph}
D.~Kharzeev and M.~Nardi,
\newblock Phys. Lett. B {\bf 507}, 121 (2001), arXiv:nucl-th/0012025.

\bibitem{Blaizot:2014wba}
J.-P. Blaizot, W.~Broniowski, and J.-Y. Ollitrault,
\newblock Phys. Rev. C {\bf 90}, 034906 (2014), arXiv:1405.3274.

\bibitem{dEnterria:2020dwq}
D.~d'Enterria and C.~Loizides,
\newblock Ann. Rev. Nucl. Part. Sci. {\bf 71}, 315 (2021), arXiv:2011.14909.

\bibitem{Alver:2010gr}
B.~Alver and G.~Roland,
\newblock Phys. Rev. C {\bf 81}, 054905 (2010), arXiv:1003.0194,
\newblock [Erratum: Phys.Rev.C 82, 039903 (2010)].

\bibitem{Gelis:2010nm}
F.~Gelis, E.~Iancu, J.~Jalilian-Marian, and R.~Venugopalan,
\newblock Ann. Rev. Nucl. Part. Sci. {\bf 60}, 463 (2010), arXiv:1002.0333.

\bibitem{McLerran:1993ni}
L.~D. McLerran and R.~Venugopalan,
\newblock Phys. Rev. D {\bf 49}, 2233 (1994), arXiv:hep-ph/9309289.

\bibitem{McLerran:1993ka}
L.~D. McLerran and R.~Venugopalan,
\newblock Phys. Rev. D {\bf 49}, 3352 (1994), arXiv:hep-ph/9311205.

\bibitem{Lappi:2007ku}
T.~Lappi,
\newblock Eur. Phys. J. C {\bf 55}, 285 (2008), arXiv:0711.3039.

\bibitem{Mueller:2001uk}
A.~H. Mueller,
\newblock Phys. Lett. B {\bf 523}, 243 (2001), arXiv:hep-ph/0110169.

\bibitem{Rummukainen:2003ns}
K.~Rummukainen and H.~Weigert,
\newblock Nucl. Phys. A {\bf 739}, 183 (2004), arXiv:hep-ph/0309306.

\bibitem{Praszalowicz:2011zza}
M.~Praszalowicz,
\newblock Phys. Lett. B {\bf 704}, 566 (2011), arXiv:1101.6012.

\bibitem{Kowalski:2003hm}
H.~Kowalski and D.~Teaney,
\newblock Phys. Rev. D {\bf 68}, 114005 (2003), arXiv:hep-ph/0304189.

\bibitem{Gelis:2009wh}
F.~Gelis, T.~Lappi, and L.~McLerran,
\newblock Nucl. Phys. A {\bf 828}, 149 (2009), arXiv:0905.3234.

\bibitem{Schenke:2012wb}
B.~Schenke, P.~Tribedy, and R.~Venugopalan,
\newblock Phys. Rev. Lett. {\bf 108}, 252301 (2012), arXiv:1202.6646.

\bibitem{Moreland:2014oya}
J.~S. Moreland, J.~E. Bernhard, and S.~A. Bass,
\newblock Phys. Rev. C {\bf 92}, 011901 (2015), arXiv:1412.4708.

\bibitem{Baier:2000sb}
R.~Baier, A.~H. Mueller, D.~Schiff, and D.~T. Son,
\newblock Phys. Lett. B {\bf 502}, 51 (2001), arXiv:hep-ph/0009237.

\bibitem{Kurkela:2018vqr}
A.~Kurkela, A.~Mazeliauskas, J.-F. Paquet, S.~Schlichting, and D.~Teaney,
\newblock Phys. Rev. C {\bf 99}, 034910 (2019), arXiv:1805.00961.

\bibitem{Huovinen:2001cy}
P.~Huovinen, P.~F. Kolb, U.~W. Heinz, P.~V. Ruuskanen, and S.~A. Voloshin,
\newblock Phys. Lett. B {\bf 503}, 58 (2001), arXiv:hep-ph/0101136.

\bibitem{Kolb:2003dz}
P.~F. Kolb and U.~W. Heinz,
\newblock p. 634 (2003), arXiv:nucl-th/0305084.

\bibitem{Denicol:2014vaa}
G.~S. Denicol, S.~Jeon, and C.~Gale,
\newblock Phys. Rev. C {\bf 90}, 024912 (2014), arXiv:1403.0962.

\bibitem{Ollitrault:2007du}
J.-Y. Ollitrault,
\newblock Eur. J. Phys. {\bf 29}, 275 (2008), arXiv:0708.2433.

\bibitem{Romatschke:2009im}
P.~Romatschke,
\newblock Int. J. Mod. Phys. E {\bf 19}, 1 (2010), arXiv:0902.3663.

\bibitem{Gale:2013da}
C.~Gale, S.~Jeon, and B.~Schenke,
\newblock Int. J. Mod. Phys. A {\bf 28}, 1340011 (2013), arXiv:1301.5893.

\bibitem{Huovinen:2009yb}
P.~Huovinen and P.~Petreczky,
\newblock Nucl. Phys. A {\bf 837}, 26 (2010), arXiv:0912.2541.

\bibitem{Noronha-Hostler:2019ayj}
J.~Noronha-Hostler, P.~Parotto, C.~Ratti, and J.~M. Stafford,
\newblock Phys. Rev. C {\bf 100}, 064910 (2019), arXiv:1902.06723.

\bibitem{Monnai:2019hkn}
A.~Monnai, B.~Schenke, and C.~Shen,
\newblock Phys. Rev. C {\bf 100}, 024907 (2019), arXiv:1902.05095.

\bibitem{Grefa:2021qvt}
J.~Grefa {\em et~al.},
\newblock Phys. Rev. D {\bf 104}, 034002 (2021), arXiv:2102.12042.

\bibitem{Bjorken:1982qr}
J.~D. Bjorken,
\newblock Phys. Rev. D {\bf 27}, 140 (1983).

\bibitem{Gale:2020dum}
C.~Gale, J.-F. Paquet, B.~Schenke, and C.~Shen,
\newblock PoS {\bf HardProbes2020}, 039 (2021), arXiv:2009.07841.

\bibitem{Fotakis:2019nbq}
J.~A. Fotakis, M.~Greif, C.~Greiner, G.~S. Denicol, and H.~Niemi,
\newblock Phys. Rev. D {\bf 101}, 076007 (2020), arXiv:1912.09103.

\bibitem{Schenke:2010nt}
B.~Schenke, S.~Jeon, and C.~Gale,
\newblock Phys. Rev. C {\bf 82}, 014903 (2010), arXiv:1004.1408.

\bibitem{Karpenko:2013wva}
I.~Karpenko, P.~Huovinen, and M.~Bleicher,
\newblock Comput. Phys. Commun. {\bf 185}, 3016 (2014), arXiv:1312.4160.

\bibitem{Shen:2014vra}
C.~Shen {\em et~al.},
\newblock Comput. Phys. Commun. {\bf 199}, 61 (2016), arXiv:1409.8164.

\bibitem{Pang:2018zzo}
L.-G. Pang, H.~Petersen, and X.-N. Wang,
\newblock Phys. Rev. C {\bf 97}, 064918 (2018), arXiv:1802.04449.

\bibitem{Denicol:2012cn}
G.~S. Denicol, H.~Niemi, E.~Molnar, and D.~H. Rischke,
\newblock Phys. Rev. D {\bf 85}, 114047 (2012), arXiv:1202.4551,
\newblock [Erratum: Phys.Rev.D 91, 039902 (2015)].

\bibitem{McNelis:2018jho}
M.~McNelis, D.~Bazow, and U.~Heinz,
\newblock Phys. Rev. C {\bf 97}, 054912 (2018), arXiv:1803.01810.

\bibitem{Alqahtani:2017mhy}
M.~Alqahtani, M.~Nopoush, and M.~Strickland,
\newblock Prog. Part. Nucl. Phys. {\bf 101}, 204 (2018), arXiv:1712.03282.

\bibitem{Weickgenannt:2022zxs}
N.~Weickgenannt, D.~Wagner, E.~Speranza, and D.~Rischke,
\newblock (2022), arXiv:2203.04766.

\bibitem{Bhadury:2020cop}
S.~Bhadury, W.~Florkowski, A.~Jaiswal, A.~Kumar, and R.~Ryblewski,
\newblock Phys. Rev. D {\bf 103}, 014030 (2021), arXiv:2008.10976.

\bibitem{Ammon:2020rvg}
M.~Ammon {\em et~al.},
\newblock JHEP {\bf 04}, 078 (2021), arXiv:2012.09183.

\bibitem{Xu:2004mz}
Z.~Xu and C.~Greiner,
\newblock Phys. Rev. C {\bf 71}, 064901 (2005), arXiv:hep-ph/0406278.

\bibitem{Lin:2004en}
Z.-W. Lin, C.~M. Ko, B.-A. Li, B.~Zhang, and S.~Pal,
\newblock Phys. Rev. C {\bf 72}, 064901 (2005), arXiv:nucl-th/0411110.

\bibitem{Cassing:2009vt}
W.~Cassing and E.~L. Bratkovskaya,
\newblock Nucl. Phys. A {\bf 831}, 215 (2009), arXiv:0907.5331.

\bibitem{Petersen:2014yqa}
H.~Petersen,
\newblock J. Phys. G {\bf 41}, 124005 (2014), arXiv:1404.1763.

\bibitem{Huovinen:2012is}
P.~Huovinen and H.~Petersen,
\newblock Eur. Phys. J. A {\bf 48}, 171 (2012), arXiv:1206.3371.

\bibitem{Cooper:1974mv}
F.~Cooper and G.~Frye,
\newblock Phys. Rev. D {\bf 10}, 186 (1974).

\bibitem{JETSCAPE:2020shq}
JETSCAPE, D.~Everett {\em et~al.},
\newblock Phys. Rev. Lett. {\bf 126}, 242301 (2021), arXiv:2010.03928.

\bibitem{Bass:1998ca}
S.~A. Bass {\em et~al.},
\newblock Prog. Part. Nucl. Phys. {\bf 41}, 255 (1998), arXiv:nucl-th/9803035.

\bibitem{Bleicher:1999xi}
M.~Bleicher {\em et~al.},
\newblock J. Phys. G {\bf 25}, 1859 (1999), arXiv:hep-ph/9909407.

\bibitem{Weil:2016zrk}
J.~Weil {\em et~al.},
\newblock Phys. Rev. C {\bf 94}, 054905 (2016), arXiv:1606.06642.

\bibitem{Plumberg:2015eia}
C.~Plumberg and U.~Heinz,
\newblock Phys. Rev. C {\bf 91}, 054905 (2015), arXiv:1503.05605.

\bibitem{Oliinychenko:2014tqa}
D.~Oliinychenko, P.~Huovinen, and H.~Petersen,
\newblock Phys. Rev. C {\bf 91}, 024906 (2015), arXiv:1411.3912.

\bibitem{Oliinychenko:2019zfk}
D.~Oliinychenko and V.~Koch,
\newblock Phys. Rev. Lett. {\bf 123}, 182302 (2019), arXiv:1902.09775.

\bibitem{Majumder:2009cf}
A.~Majumder, B.~M\"uller, and S.~Mrowczynski,
\newblock Phys. Rev. D {\bf 80}, 125020 (2009), arXiv:0903.3683.

\bibitem{Guo:2000nz}
X.-f. Guo and X.-N. Wang,
\newblock Phys. Rev. Lett. {\bf 85}, 3591 (2000), arXiv:hep-ph/0005044.

\bibitem{Fries:2002kt}
R.~J. Fries, B.~M{\"u}ller, and D.~K. Srivastava,
\newblock Phys. Rev. Lett. {\bf 90}, 132301 (2003), arXiv:nucl-th/0208001.

\bibitem{Zapp:2013vla}
K.~C. Zapp,
\newblock Eur. Phys. J. C {\bf 74}, 2762 (2014), arXiv:1311.0048.

\bibitem{Putschke:2019yrg}
J.~H. Putschke {\em et~al.},
\newblock (2019), arXiv:1903.07706.

\bibitem{Dokshitzer:1991fd}
Y.~L. Dokshitzer, V.~A. Khoze, and S.~I. Troian,
\newblock J. Phys. G {\bf 17}, 1602 (1991).

\bibitem{Petreczky:2005nh}
P.~Petreczky and D.~Teaney,
\newblock Phys. Rev. D {\bf 73}, 014508 (2006), arXiv:hep-ph/0507318.

\bibitem{Casalderrey-Solana:2006fio}
J.~Casalderrey-Solana and D.~Teaney,
\newblock Phys. Rev. D {\bf 74}, 085012 (2006), arXiv:hep-ph/0605199.

\bibitem{Ke:2018tsh}
W.~Ke, Y.~Xu, and S.~A. Bass,
\newblock Phys. Rev. C {\bf 98}, 064901 (2018), arXiv:1806.08848.

\bibitem{Dai:2020rlu}
T.~Dai, J.-F. Paquet, D.~Teaney, and S.~A. Bass,
\newblock Phys. Rev. C {\bf 105}, 034905 (2022), arXiv:2012.03441.

\bibitem{JET:2013cls}
JET, K.~M. Burke {\em et~al.},
\newblock Phys. Rev. C {\bf 90}, 014909 (2014), arXiv:1312.5003.

\bibitem{Majumder:2010qh}
A.~Majumder and M.~Van~Leeuwen,
\newblock Prog. Part. Nucl. Phys. {\bf 66}, 41 (2011), arXiv:1002.2206.

\bibitem{Neufeld:2010tz}
R.~B. Neufeld and T.~Renk,
\newblock Phys. Rev. C {\bf 82}, 044903 (2010), arXiv:1001.5068.

\bibitem{Tachibana:2017syd}
Y.~Tachibana, N.-B. Chang, and G.-Y. Qin,
\newblock Phys. Rev. C {\bf 95}, 044909 (2017), arXiv:1701.07951.

\bibitem{KunnawalkamElayavalli:2017hxo}
R.~Kunnawalkam~Elayavalli and K.~C. Zapp,
\newblock JHEP {\bf 07}, 141 (2017), arXiv:1707.01539.

\bibitem{Floerchinger:2021xhb}
S.~Floerchinger, C.~Gebhardt, and K.~Reygers,
\newblock (2021), arXiv:2112.12497.

\bibitem{Gale:2003iz}
C.~Gale and K.~L. Haglin,
\newblock p. 364 (2003), arXiv:hep-ph/0306098.

\bibitem{Rapp:2016xzw}
R.~Rapp and H.~van Hees,
\newblock Eur. Phys. J. A {\bf 52}, 257 (2016), arXiv:1608.05279.

\bibitem{Ding:2016hua}
H.-T. Ding, O.~Kaczmarek, and F.~Meyer,
\newblock Phys. Rev. D {\bf 94}, 034504 (2016), arXiv:1604.06712.

\bibitem{Ghiglieri:2016tvj}
J.~Ghiglieri, O.~Kaczmarek, M.~Laine, and F.~Meyer,
\newblock Phys. Rev. D {\bf 94}, 016005 (2016), arXiv:1604.07544.

\bibitem{Kapusta:1993hq}
J.~I. Kapusta and E.~V. Shuryak,
\newblock Phys. Rev. D {\bf 49}, 4694 (1994), arXiv:hep-ph/9312245.

\bibitem{Holt:2012wr}
N.~P.~M. Holt, P.~M. Hohler, and R.~Rapp,
\newblock Phys. Rev. D {\bf 87}, 076010 (2013), arXiv:1210.7210.

\bibitem{ALICE:2017ban}
ALICE, J.~Adam {\em et~al.},
\newblock Phys. Rev. C {\bf 95}, 064606 (2017), arXiv:1702.00555.

\bibitem{Becattini:2008tx}
F.~Becattini, P.~Castorina, J.~Manninen, and H.~Satz,
\newblock Eur. Phys. J. C {\bf 56}, 493 (2008), arXiv:0805.0964.

\bibitem{Andronic:2017pug}
A.~Andronic, P.~Braun-Munzinger, K.~Redlich, and J.~Stachel,
\newblock Nature {\bf 561}, 321 (2018), arXiv:1710.09425.

\bibitem{Zhang:2018euf}
Z.~Zhang and C.~M. Ko,
\newblock Phys. Lett. B {\bf 780}, 191 (2018).

\bibitem{Muller:2017vnp}
B.~M\"uller and A.~Sch\"afer,
\newblock (2017), arXiv:1712.03567.

\bibitem{Cleymans:2005xv}
J.~Cleymans, H.~Oeschler, K.~Redlich, and S.~Wheaton,
\newblock Phys. Rev. C {\bf 73}, 034905 (2006), arXiv:hep-ph/0511094.

\bibitem{HADES:2015oef}
HADES, G.~Agakishiev {\em et~al.},
\newblock Eur. Phys. J. A {\bf 52}, 178 (2016), arXiv:1512.07070.

\bibitem{Bellwied:2018tkc}
R.~Bellwied {\em et~al.},
\newblock Phys. Rev. C {\bf 99}, 034912 (2019), arXiv:1805.00088.

\bibitem{Rafelski:1982pu}
J.~Rafelski and B.~M{\"u}ller,
\newblock Phys. Rev. Lett. {\bf 48}, 1066 (1982),
\newblock [Erratum: Phys.Rev.Lett. 56, 2334 (1986)].

\bibitem{NA49:2002pzu}
NA49, S.~V. Afanasiev {\em et~al.},
\newblock Phys. Rev. C {\bf 66}, 054902 (2002), arXiv:nucl-ex/0205002.

\bibitem{Steinheimer:2011mp}
J.~Steinheimer and M.~Bleicher,
\newblock Phys. Rev. C {\bf 84}, 024905 (2011), arXiv:1104.3981.

\bibitem{Petersen:2009mz}
H.~Petersen, J.~Steinheimer, M.~Bleicher, and H.~Stocker,
\newblock J. Phys. G {\bf 36}, 055104 (2009), arXiv:0902.4866.

\bibitem{Vovchenko:2019pjl}
V.~Vovchenko and H.~Stoecker,
\newblock Comput. Phys. Commun. {\bf 244}, 295 (2019), arXiv:1901.05249.

\bibitem{Gorenstein:2007mw}
M.~I. Gorenstein, M.~Hauer, and O.~N. Moroz,
\newblock Phys. Rev. C {\bf 77}, 024911 (2008), arXiv:0708.0137.

\bibitem{Steinheimer:2012rd}
J.~Steinheimer, J.~Aichelin, and M.~Bleicher,
\newblock Phys. Rev. Lett. {\bf 110}, 042501 (2013), arXiv:1203.5302.

\bibitem{Blume:2005ru}
C.~Blume,
\newblock J. Phys. G {\bf 31}, S57 (2005).

\bibitem{Anishetty:1980zp}
R.~Anishetty, P.~Koehler, and L.~D. McLerran,
\newblock Phys. Rev. D {\bf 22}, 2793 (1980).

\bibitem{ALICE:2013mez}
ALICE, B.~Abelev {\em et~al.},
\newblock Phys. Rev. C {\bf 88}, 044910 (2013), arXiv:1303.0737.

\bibitem{STAR:2017sal}
STAR, L.~Adamczyk {\em et~al.},
\newblock Phys. Rev. C {\bf 96}, 044904 (2017), arXiv:1701.07065.

\bibitem{Schnedermann:1993ws}
E.~Schnedermann, J.~Sollfrank, and U.~W. Heinz,
\newblock Phys. Rev. C {\bf 48}, 2462 (1993), arXiv:nucl-th/9307020.

\bibitem{Song:2011hk}
H.~Song, S.~A. Bass, U.~Heinz, T.~Hirano, and C.~Shen,
\newblock Phys. Rev. C {\bf 83}, 054910 (2011), arXiv:1101.4638,
\newblock [Erratum: Phys.Rev.C 86, 059903 (2012)].

\bibitem{Ryu:2015vwa}
S.~Ryu {\em et~al.},
\newblock Phys. Rev. Lett. {\bf 115}, 132301 (2015), arXiv:1502.01675.

\bibitem{Ryu:2017qzn}
S.~Ryu {\em et~al.},
\newblock Phys. Rev. C {\bf 97}, 034910 (2018), arXiv:1704.04216.

\bibitem{JETSCAPE:2022cob}
JETSCAPE, D.~Everett {\em et~al.},
\newblock (2022), arXiv:2203.08286.

\bibitem{STAR:2014shf}
STAR, L.~Adamczyk {\em et~al.},
\newblock Phys. Rev. C {\bf 92}, 014904 (2015), arXiv:1403.4972.

\bibitem{Lisa:2005dd}
M.~A. Lisa, S.~Pratt, R.~Soltz, and U.~Wiedemann,
\newblock Ann. Rev. Nucl. Part. Sci. {\bf 55}, 357 (2005),
  arXiv:nucl-ex/0505014.

\bibitem{ALICE:2015hvw}
ALICE, J.~Adam {\em et~al.},
\newblock Phys. Rev. C {\bf 92}, 054908 (2015), arXiv:1506.07884.

\bibitem{ALICE:2020mkb}
ALICE, S.~Acharya {\em et~al.},
\newblock Phys. Lett. B {\bf 813}, 136030 (2021), arXiv:2007.08315.

\bibitem{Pratt:2008qv}
S.~Pratt,
\newblock Phys. Rev. Lett. {\bf 102}, 232301 (2009), arXiv:0811.3363.

\bibitem{Li:2008qm}
Q.-f. Li, J.~Steinheimer, H.~Petersen, M.~Bleicher, and H.~Stocker,
\newblock Phys. Lett. B {\bf 674}, 111 (2009), arXiv:0812.0375.

\bibitem{Rischke:1996em}
D.~H. Rischke and M.~Gyulassy,
\newblock Nucl. Phys. A {\bf 608}, 479 (1996), arXiv:nucl-th/9606039.

\bibitem{Heinz:2002un}
U.~W. Heinz and P.~F. Kolb,
\newblock {Two RHIC puzzles: Early thermalization and the HBT problem},
\newblock in {\em {18th Winter Workshop on Nuclear Dynamics}}, 2002,
  arXiv:hep-ph/0204061.

\bibitem{OHara:2002pqs}
K.~M. O'Hara, S.~L. Hemmer, M.~E. Gehm, S.~R. Granade, and J.~E. Thomas,
\newblock Science {\bf 298}, 2179 (2002), arXiv:cond-mat/0212463.

\bibitem{Teaney:2012ke}
D.~Teaney and L.~Yan,
\newblock Phys. Rev. C {\bf 86}, 044908 (2012), arXiv:1206.1905.

\bibitem{Schenke:2011bn}
B.~Schenke, S.~Jeon, and C.~Gale,
\newblock Phys. Rev. C {\bf 85}, 024901 (2012), arXiv:1109.6289.

\bibitem{Luzum:2013yya}
M.~Luzum and H.~Petersen,
\newblock J. Phys. G {\bf 41}, 063102 (2014), arXiv:1312.5503.

\bibitem{Alver:2010dn}
B.~H. Alver, C.~Gombeaud, M.~Luzum, and J.-Y. Ollitrault,
\newblock Phys. Rev. C {\bf 82}, 034913 (2010), arXiv:1007.5469.

\bibitem{Schenke:2010rr}
B.~Schenke, S.~Jeon, and C.~Gale,
\newblock Phys. Rev. Lett. {\bf 106}, 042301 (2011), arXiv:1009.3244.

\bibitem{Petersen:2010cw}
H.~Petersen, G.-Y. Qin, S.~A. Bass, and B.~M{\"u}ller,
\newblock Phys. Rev. C {\bf 82}, 041901 (2010), arXiv:1008.0625.

\bibitem{PHENIX:2011yyh}
PHENIX, A.~Adare {\em et~al.},
\newblock Phys. Rev. Lett. {\bf 107}, 252301 (2011), arXiv:1105.3928.

\bibitem{ATLAS:2020sgl}
ATLAS, G.~Aad {\em et~al.},
\newblock Phys. Rev. Lett. {\bf 126}, 122301 (2021), arXiv:2001.04201.

\bibitem{Nie:2020trj}
STAR, M.~Nie,
\newblock Nucl. Phys. A {\bf 1005}, 121783 (2021), arXiv:2005.03252.

\bibitem{Sakai:2021pev}
A.~Sakai, K.~Murase, and T.~Hirano,
\newblock (2021), arXiv:2111.08963.

\bibitem{Auvinen:2013sba}
J.~Auvinen and H.~Petersen,
\newblock Phys. Rev. C {\bf 88}, 064908 (2013), arXiv:1310.1764.

\bibitem{Brachmann:1999xt}
J.~Brachmann {\em et~al.},
\newblock Phys. Rev. C {\bf 61}, 024909 (2000), arXiv:nucl-th/9908010.

\bibitem{STAR:2014clz}
STAR, L.~Adamczyk {\em et~al.},
\newblock Phys. Rev. Lett. {\bf 112}, 162301 (2014), arXiv:1401.3043.

\bibitem{CMS:2010ifv}
CMS, V.~Khachatryan {\em et~al.},
\newblock JHEP {\bf 09}, 091 (2010), arXiv:1009.4122.

\bibitem{CMS:2012qk}
CMS, S.~Chatrchyan {\em et~al.},
\newblock Phys. Lett. B {\bf 718}, 795 (2013), arXiv:1210.5482.

\bibitem{ATLAS:2012cix}
ATLAS, G.~Aad {\em et~al.},
\newblock Phys. Rev. Lett. {\bf 110}, 182302 (2013), arXiv:1212.5198.

\bibitem{ALICE:2012eyl}
ALICE, B.~Abelev {\em et~al.},
\newblock Phys. Lett. B {\bf 719}, 29 (2013), arXiv:1212.2001.

\bibitem{PHENIX:2018lia}
PHENIX, C.~Aidala {\em et~al.},
\newblock Nature Phys. {\bf 15}, 214 (2019), arXiv:1805.02973.

\bibitem{Bold:2021juc}
ALICE, ATLAS, CMS, T.~Bold,
\newblock PoS {\bf LHCP2021}, 058 (2021).

\bibitem{ALICE:2016fzo}
ALICE, J.~Adam {\em et~al.},
\newblock Nature Phys. {\bf 13}, 535 (2017), arXiv:1606.07424.

\bibitem{Sefcik:2020haw}
ALICE, M.~\v{S}ef\v{c}\'\i{}k,
\newblock Springer Proc. Phys. {\bf 250}, 173 (2020).

\bibitem{ALICE:2017svf}
ALICE, S.~Acharya {\em et~al.},
\newblock Phys. Lett. B {\bf 783}, 95 (2018), arXiv:1712.05603.

\bibitem{Tywoniuk:2014hta}
K.~Tywoniuk,
\newblock Nucl. Phys. A {\bf 926}, 85 (2014).

\bibitem{Soloviev:2021lhs}
A.~Soloviev,
\newblock (2021), arXiv:2109.15081.

\bibitem{Romatschke:2017vte}
P.~Romatschke,
\newblock Phys. Rev. Lett. {\bf 120}, 012301 (2018), arXiv:1704.08699.

\bibitem{Kurkela:2018wud}
A.~Kurkela, A.~Mazeliauskas, J.-F. Paquet, S.~Schlichting, and D.~Teaney,
\newblock Phys. Rev. Lett. {\bf 122}, 122302 (2019), arXiv:1805.01604.

\bibitem{Petersen:2010zt}
H.~Petersen, C.~Coleman-Smith, S.~A. Bass, and R.~Wolpert,
\newblock J. Phys. G {\bf 38}, 045102 (2011), arXiv:1012.4629.

\bibitem{Novak:2013bqa}
J.~Novak {\em et~al.},
\newblock Phys. Rev. C {\bf 89}, 034917 (2014), arXiv:1303.5769.

\bibitem{Pratt:2015zsa}
S.~Pratt, E.~Sangaline, P.~Sorensen, and H.~Wang,
\newblock Phys. Rev. Lett. {\bf 114}, 202301 (2015), arXiv:1501.04042.

\bibitem{Bernhard:2019bmu}
J.~E. Bernhard, J.~S. Moreland, and S.~A. Bass,
\newblock Nature Phys. {\bf 15}, 1113 (2019).

\bibitem{Moreland:2018gsh}
J.~S. Moreland, J.~E. Bernhard, and S.~A. Bass,
\newblock Phys. Rev. C {\bf 101}, 024911 (2020), arXiv:1808.02106.

\bibitem{Nijs:2020ors}
G.~Nijs, W.~van~der Schee, U.~G\"ursoy, and R.~Snellings,
\newblock Phys. Rev. Lett. {\bf 126}, 202301 (2021), arXiv:2010.15130.

\bibitem{Parkkila:2021yha}
J.~E. Parkkila {\em et~al.},
\newblock (2021), arXiv:2111.08145.

\bibitem{Nijs:2021clz}
G.~Nijs and W.~van~der Schee,
\newblock (2021), arXiv:2110.13153.

\bibitem{JETSCAPE:2021ehl}
JETSCAPE, S.~Cao {\em et~al.},
\newblock Phys. Rev. C {\bf 104}, 024905 (2021), arXiv:2102.11337.

\bibitem{Xu:2017obm}
Y.~Xu, J.~E. Bernhard, S.~A. Bass, M.~Nahrgang, and S.~Cao,
\newblock Phys. Rev. C {\bf 97}, 014907 (2018), arXiv:1710.00807.

\bibitem{Auvinen:2017fjw}
J.~Auvinen, J.~E. Bernhard, S.~A. Bass, and I.~Karpenko,
\newblock Phys. Rev. C {\bf 97}, 044905 (2018), arXiv:1706.03666.

\bibitem{Jeon:2003gk}
S.~Jeon and V.~Koch,
\newblock {Event by event fluctuations},
\newblock in {\em {Quark-gluon plasma 3}}, edited by R.~C. Hwa and X.~N. Wang,
  pp. 430--490, {World Scientific}, 2004, arXiv:hep-ph/0304012.

\bibitem{Asakawa:2000wh}
M.~Asakawa, U.~W. Heinz, and B.~M{\"u}ller,
\newblock Phys. Rev. Lett. {\bf 85}, 2072 (2000), arXiv:hep-ph/0003169.

\bibitem{Haussler:2007un}
S.~Haussler, S.~Scherer, and M.~Bleicher,
\newblock Phys. Lett. B {\bf 660}, 197 (2008), arXiv:hep-ph/0702188.

\bibitem{Pratt:2012dz}
S.~Pratt,
\newblock Phys. Rev. Lett. {\bf 108}, 212301 (2012), arXiv:1203.4578.

\bibitem{Pratt:2015jsa}
S.~Pratt, W.~P. McCormack, and C.~Ratti,
\newblock Phys. Rev. C {\bf 92}, 064905 (2015), arXiv:1508.07031.

\bibitem{Ratti:2018ksb}
C.~Ratti,
\newblock Rept. Prog. Phys. {\bf 81}, 084301 (2018), arXiv:1804.07810.

\bibitem{Ratti:2021ubw}
C.~Ratti and R.~Bellwied,
\newblock {\em {The Deconfinement Transition of QCD: Theory Meets Experiment}},
  Lecture Notes in Physics Vol. 981 (Springer-Verlag, 2021).

\bibitem{Skokov:2012ds}
V.~Skokov, B.~Friman, and K.~Redlich,
\newblock Phys. Rev. C {\bf 88}, 034911 (2013), arXiv:1205.4756.

\bibitem{Alba:2015iva}
P.~Alba {\em et~al.},
\newblock Phys. Rev. C {\bf 92}, 064910 (2015), arXiv:1504.03262.

\bibitem{Haque:2014rua}
N.~Haque {\em et~al.},
\newblock JHEP {\bf 05}, 027 (2014), arXiv:1402.6907.

\bibitem{Mogliacci:2013mca}
S.~Mogliacci, J.~O. Andersen, M.~Strickland, N.~Su, and A.~Vuorinen,
\newblock JHEP {\bf 12}, 055 (2013), arXiv:1307.8098.

\bibitem{Bellwied:2015lba}
R.~Bellwied {\em et~al.},
\newblock Phys. Rev. D {\bf 92}, 114505 (2015), arXiv:1507.04627.

\bibitem{Alba:2014eba}
P.~Alba {\em et~al.},
\newblock Phys. Lett. B {\bf 738}, 305 (2014), arXiv:1403.4903.

\bibitem{Berdnikov:1999ph}
B.~Berdnikov and K.~Rajagopal,
\newblock Phys. Rev. D {\bf 61}, 105017 (2000), arXiv:hep-ph/9912274.

\bibitem{Vovchenko:2015pya}
V.~Vovchenko, D.~V. Anchishkin, M.~I. Gorenstein, and R.~V. Poberezhnyuk,
\newblock Phys. Rev. C {\bf 92}, 054901 (2015), arXiv:1506.05763.

\bibitem{Stephanov:2011pb}
M.~A. Stephanov,
\newblock Phys. Rev. Lett. {\bf 107}, 052301 (2011), arXiv:1104.1627.

\bibitem{STAR:2020tga}
STAR, J.~Adam {\em et~al.},
\newblock Phys. Rev. Lett. {\bf 126}, 092301 (2021), arXiv:2001.02852.

\bibitem{STAR:2021iop}
STAR, M.~Abdallah {\em et~al.},
\newblock Phys. Rev. C {\bf 104}, 024902 (2021), arXiv:2101.12413.

\bibitem{HADES:2020wpc}
HADES, J.~Adamczewski-Musch {\em et~al.},
\newblock Phys. Rev. C {\bf 102}, 024914 (2020), arXiv:2002.08701.

\bibitem{STAR:2021fge}
STAR, M.~S. Abdallah {\em et~al.},
\newblock (2021), arXiv:2112.00240.

\bibitem{Bluhm:2020mpc}
M.~Bluhm {\em et~al.},
\newblock Nucl. Phys. A {\bf 1003}, 122016 (2020), arXiv:2001.08831.

\bibitem{Nahrgang:2018afz}
M.~Nahrgang, M.~Bluhm, T.~Schaefer, and S.~A. Bass,
\newblock Phys. Rev. D {\bf 99}, 116015 (2019), arXiv:1804.05728.

\bibitem{Du:2020bxp}
L.~Du, U.~Heinz, K.~Rajagopal, and Y.~Yin,
\newblock Phys. Rev. C {\bf 102}, 054911 (2020), arXiv:2004.02719.

\bibitem{An:2021wof}
X.~An {\em et~al.},
\newblock Nucl. Phys. A {\bf 1017}, 122343 (2022), arXiv:2108.13867.

\bibitem{PHENIX:2003tvk}
PHENIX, S.~S. Adler {\em et~al.},
\newblock Phys. Rev. Lett. {\bf 91}, 172301 (2003), arXiv:nucl-ex/0305036.

\bibitem{STAR:2003wqp}
STAR, J.~Adams {\em et~al.},
\newblock Phys. Rev. Lett. {\bf 92}, 052302 (2004), arXiv:nucl-ex/0306007.

\bibitem{Fries:2003vb}
R.~J. Fries, B.~M{\"u}ller, C.~Nonaka, and S.~A. Bass,
\newblock Phys. Rev. Lett. {\bf 90}, 202303 (2003), arXiv:nucl-th/0301087.

\bibitem{Greco:2003xt}
V.~Greco, C.~M. Ko, and P.~Levai,
\newblock Phys. Rev. Lett. {\bf 90}, 202302 (2003), arXiv:nucl-th/0301093.

\bibitem{Fries:2003kq}
R.~J. Fries, B.~Muller, C.~Nonaka, and S.~A. Bass,
\newblock Phys. Rev. C {\bf 68}, 044902 (2003), arXiv:nucl-th/0306027.

\bibitem{Molnar:2003ff}
D.~Molnar and S.~A. Voloshin,
\newblock Phys. Rev. Lett. {\bf 91}, 092301 (2003), arXiv:nucl-th/0302014.

\bibitem{Fries:2008hs}
R.~J. Fries, V.~Greco, and P.~Sorensen,
\newblock Ann. Rev. Nucl. Part. Sci. {\bf 58}, 177 (2008), arXiv:0807.4939.

\bibitem{Dixit:2021qey}
STAR, P.~Dixit,
\newblock {Invariant yield and azimuthal anisotropy measurements of strange and
  multi-strange hadrons in Au+Au collision at $\sqrt{s_{NN}}$ = 27 and 54.4 GeV
  at STAR},
\newblock in {\em {International Conference on Critical Point and Onset of
  Deconfinement}}, 2021, arXiv:2111.04674.

\bibitem{ALICE:2018yph}
ALICE, S.~Acharya {\em et~al.},
\newblock JHEP {\bf 09}, 006 (2018), arXiv:1805.04390.

\bibitem{Hwa:2004ng}
R.~C. Hwa and C.~B. Yang,
\newblock Phys. Rev. C {\bf 70}, 024905 (2004), arXiv:nucl-th/0401001.

\bibitem{Werner:2012sv}
K.~Werner,
\newblock Phys. Rev. Lett. {\bf 109}, 102301 (2012), arXiv:1204.1394.

\bibitem{Werner:2012xh}
K.~Werner, I.~Karpenko, M.~Bleicher, T.~Pierog, and S.~Porteboeuf-Houssais,
\newblock Phys. Rev. C {\bf 85}, 064907 (2012), arXiv:1203.5704.

\bibitem{Belikov:2011xk}
ALICE, I.~Belikov,
\newblock J. Phys. G {\bf 38}, 124078 (2011), arXiv:1109.4807.

\bibitem{Lu:2006qn}
Y.~Lu {\em et~al.},
\newblock J. Phys. G {\bf 32}, 1121 (2006), arXiv:nucl-th/0602009.

\bibitem{Hirano:2007ei}
T.~Hirano, U.~W. Heinz, D.~Kharzeev, R.~Lacey, and Y.~Nara,
\newblock Phys. Rev. C {\bf 77}, 044909 (2008), arXiv:0710.5795.

\bibitem{Ryu:2019atv}
S.~Ryu, J.~Staudenmaier, and H.~Elfner,
\newblock MDPI Proc. {\bf 10}, 44 (2019).

\bibitem{Gyulassy:1990ye}
M.~Gyulassy and M.~Plumer,
\newblock Phys. Lett. B {\bf 243}, 432 (1990).

\bibitem{Wang:1992qdg}
X.-N. Wang and M.~Gyulassy,
\newblock Phys. Rev. Lett. {\bf 68}, 1480 (1992).

\bibitem{Baier:1996sk}
R.~Baier, Y.~L. Dokshitzer, A.~H. Mueller, S.~Peigne, and D.~Schiff,
\newblock Nucl. Phys. B {\bf 484}, 265 (1997), arXiv:hep-ph/9608322.

\bibitem{Zakharov:1996fv}
B.~G. Zakharov,
\newblock JETP Lett. {\bf 63}, 952 (1996), arXiv:hep-ph/9607440.

\bibitem{Baier:2000mf}
R.~Baier, D.~Schiff, and B.~G. Zakharov,
\newblock Ann. Rev. Nucl. Part. Sci. {\bf 50}, 37 (2000), arXiv:hep-ph/0002198.

\bibitem{Caucal:2019uvr}
P.~Caucal, E.~Iancu, and G.~Soyez,
\newblock JHEP {\bf 10}, 273 (2019), arXiv:1907.04866.

\bibitem{PHENIX:2006ujp}
PHENIX, S.~S. Adler {\em et~al.},
\newblock Phys. Rev. Lett. {\bf 96}, 202301 (2006), arXiv:nucl-ex/0601037.

\bibitem{PHENIX:2012oed}
PHENIX, A.~Adare {\em et~al.},
\newblock Phys. Rev. Lett. {\bf 109}, 152301 (2012), arXiv:1204.1526,
\newblock [Erratum: Phys.Rev.Lett. 125, 049901 (2020)].

\bibitem{ALICE:2014hpa}
ALICE, B.~B. Abelev {\em et~al.},
\newblock Eur. Phys. J. C {\bf 74}, 3108 (2014), arXiv:1405.3794.

\bibitem{Nishitani:2019tcy}
PHENIX, R.~Nishitani,
\newblock MDPI Proc. {\bf 10}, 42 (2019).

\bibitem{CMS:2017xgk}
CMS, A.~M. Sirunyan {\em et~al.},
\newblock Phys. Lett. B {\bf 776}, 195 (2018), arXiv:1702.00630.

\bibitem{ATLAS:2012tjt}
ATLAS, G.~Aad {\em et~al.},
\newblock Phys. Lett. B {\bf 719}, 220 (2013), arXiv:1208.1967.

\bibitem{Casalderrey-Solana:2010bet}
J.~Casalderrey-Solana, J.~G. Milhano, and U.~A. Wiedemann,
\newblock J. Phys. G {\bf 38}, 035006 (2011), arXiv:1012.0745.

\bibitem{Connors:2017ptx}
M.~Connors, C.~Nattrass, R.~Reed, and S.~Salur,
\newblock Rev. Mod. Phys. {\bf 90}, 025005 (2018), arXiv:1705.01974.

\bibitem{Mao:2014jja}
CMS, Y.~Mao,
\newblock Nucl. Phys. A {\bf 932}, 88 (2014).

\bibitem{ATLAS:2022cim}
ATLAS, A.~Collaboration,
\newblock (2022).

\bibitem{STAR:2012civ}
STAR, L.~Adamczyk {\em et~al.},
\newblock Phys. Rev. C {\bf 87}, 044903 (2013), arXiv:1212.1653.

\bibitem{ATLAS:2022zbu}
ATLAS, A.~Collaboration,
\newblock (2022), arXiv:2205.00682.

\bibitem{Perepelitsa:2016zbe}
ATLAS, D.~V. Perepelitsa,
\newblock Nucl. Phys. A {\bf 956}, 653 (2016).

\bibitem{Andronic:2015wma}
A.~Andronic {\em et~al.},
\newblock Eur. Phys. J. C {\bf 76}, 107 (2016), arXiv:1506.03981.

\bibitem{Karsch:1990wi}
F.~Karsch and H.~Satz,
\newblock Z. Phys. C {\bf 51}, 209 (1991).

\bibitem{Jakovac:2006sf}
A.~Jakovac, P.~Petreczky, K.~Petrov, and A.~Velytsky,
\newblock Phys. Rev. D {\bf 75}, 014506 (2007), arXiv:hep-lat/0611017.

\bibitem{Petreczky:2021zmz}
P.~Petreczky, S.~Sharma, and J.~H. Weber,
\newblock Phys. Rev. D {\bf 104}, 054511 (2021), arXiv:2107.11368.

\bibitem{Rothkopf:2019ipj}
A.~Rothkopf,
\newblock Phys. Rept. {\bf 858}, 1 (2020), arXiv:1912.02253.

\bibitem{PHENIX:2005nhb}
PHENIX, S.~S. Adler {\em et~al.},
\newblock Phys. Rev. Lett. {\bf 96}, 032301 (2006), arXiv:nucl-ex/0510047.

\bibitem{PHENIX:2022wim}
PHENIX, U.~A. Acharya {\em et~al.},
\newblock (2022), arXiv:2203.17058.

\bibitem{ALICE:2019nuy}
ALICE, S.~Acharya {\em et~al.},
\newblock Phys. Lett. B {\bf 804}, 135377 (2020), arXiv:1910.09110.

\bibitem{STAR:2021uzu}
STAR, M.~S. Abdallah {\em et~al.},
\newblock (2021), arXiv:2111.14615.

\bibitem{ALICE:2021aqk}
ALICE, S.~Acharya {\em et~al.},
\newblock Nature {\bf 605}, 440 (2022), arXiv:2106.05713.

\bibitem{Armesto:2003jh}
N.~Armesto, C.~A. Salgado, and U.~A. Wiedemann,
\newblock Phys. Rev. D {\bf 69}, 114003 (2004), arXiv:hep-ph/0312106.

\bibitem{ALICE:2015vxz}
ALICE, J.~Adam {\em et~al.},
\newblock JHEP {\bf 03}, 081 (2016), arXiv:1509.06888.

\bibitem{Grosa:2018zix}
ALICE, F.~Grosa,
\newblock PoS {\bf HardProbes2018}, 138 (2018), arXiv:1812.06188.

\bibitem{Matsui:1986dk}
T.~Matsui and H.~Satz,
\newblock Phys. Lett. {\bf B178}, 416 (1986).

\bibitem{Hufner:1996pq}
J.~Hufner, S.~P. Klevansky, and P.~Rehberg,
\newblock Nucl. Phys. A {\bf 606}, 260 (1996).

\bibitem{PHENIX:2012xtg}
PHENIX, A.~Adare {\em et~al.},
\newblock Phys. Rev. C {\bf 86}, 064901 (2012), arXiv:1208.2251.

\bibitem{CMS:2018zza}
CMS, A.~M. Sirunyan {\em et~al.},
\newblock Phys. Lett. B {\bf 790}, 270 (2019), arXiv:1805.09215.

\bibitem{STAR:2021zvb}
STAR, M.~Abdallah {\em et~al.},
\newblock Phys. Lett. B {\bf 825}, 136865 (2022), arXiv:2110.09666.

\bibitem{Chanfray:1995jgo}
G.~Chanfray, R.~Rapp, and J.~Wambach,
\newblock Phys. Rev. Lett. {\bf 76}, 368 (1996), arXiv:hep-ph/9508353.

\bibitem{Rapp:1999ej}
R.~Rapp and J.~Wambach,
\newblock Adv. Nucl. Phys. {\bf 25}, 1 (2000), arXiv:hep-ph/9909229.

\bibitem{Brown:2001nh}
G.~E. Brown and M.~Rho,
\newblock Phys. Rept. {\bf 363}, 85 (2002), arXiv:hep-ph/0103102.

\bibitem{NA60:2006ymb}
NA60, R.~Arnaldi {\em et~al.},
\newblock Phys. Rev. Lett. {\bf 96}, 162302 (2006), arXiv:nucl-ex/0605007.

\bibitem{NA60:2007lzy}
NA60, R.~Arnaldi {\em et~al.},
\newblock Phys. Rev. Lett. {\bf 100}, 022302 (2008), arXiv:0711.1816.

\bibitem{Salabura:2020tou}
P.~Salabura and J.~Stroth,
\newblock Prog. Part. Nucl. Phys. {\bf 120}, 103869 (2021), arXiv:2005.14589.

\bibitem{STAR:2018xaj}
STAR, J.~Adam {\em et~al.},
\newblock (2018), arXiv:1810.10159.

\bibitem{vanHees:2006ng}
H.~van Hees and R.~Rapp,
\newblock Phys. Rev. Lett. {\bf 97}, 102301 (2006), arXiv:hep-ph/0603084.

\bibitem{Rapp:2013nxa}
R.~Rapp,
\newblock Adv. High Energy Phys. {\bf 2013}, 148253 (2013), arXiv:1304.2309.

\bibitem{Endres:2016tkg}
S.~Endres, H.~van Hees, and M.~Bleicher,
\newblock Phys. Rev. C {\bf 94}, 024912 (2016), arXiv:1604.06415.

\bibitem{Endres:2014zua}
S.~Endres, H.~van Hees, J.~Weil, and M.~Bleicher,
\newblock Phys. Rev. C {\bf 91}, 054911 (2015), arXiv:1412.1965.

\bibitem{Linnyk:2015rco}
O.~Linnyk, E.~L. Bratkovskaya, and W.~Cassing,
\newblock Prog. Part. Nucl. Phys. {\bf 87}, 50 (2016), arXiv:1512.08126.

\bibitem{HADES:2019auv}
HADES, J.~Adamczewski-Musch {\em et~al.},
\newblock Nature Phys. {\bf 15}, 1040 (2019).

\bibitem{Staudenmaier:2017vtq}
J.~Staudenmaier, J.~Weil, V.~Steinberg, S.~Endres, and H.~Petersen,
\newblock Phys. Rev. C {\bf 98}, 054908 (2018), arXiv:1711.10297.

\bibitem{PHENIX:2012jbv}
PHENIX, S.~Afanasiev {\em et~al.},
\newblock Phys. Rev. Lett. {\bf 109}, 152302 (2012), arXiv:1205.5759.

\bibitem{PHENIX:2008uif}
PHENIX, A.~Adare {\em et~al.},
\newblock Phys. Rev. Lett. {\bf 104}, 132301 (2010), arXiv:0804.4168.

\bibitem{PHENIX:2014nkk}
PHENIX, A.~Adare {\em et~al.},
\newblock Phys. Rev. C {\bf 91}, 064904 (2015), arXiv:1405.3940.

\bibitem{PHENIX:2018che}
PHENIX, A.~Adare {\em et~al.},
\newblock Phys. Rev. C {\bf 98}, 054902 (2018), arXiv:1805.04066.

\bibitem{ALICE:2015xmh}
ALICE, J.~Adam {\em et~al.},
\newblock Phys. Lett. B {\bf 754}, 235 (2016), arXiv:1509.07324.

\bibitem{CMOR:1989qzc}
CMOR, A.~L.~S. Angelis {\em et~al.},
\newblock Nucl. Phys. B {\bf 327}, 541 (1989).

\bibitem{AxialFieldSpectrometer:1989nag}
Axial Field Spectrometer, T.~Akesson {\em et~al.},
\newblock Sov. J. Nucl. Phys. {\bf 51}, 836 (1990).

\bibitem{PHENIX:2012krx}
PHENIX, A.~Adare {\em et~al.},
\newblock Phys. Rev. C {\bf 87}, 054907 (2013), arXiv:1208.1234.

\bibitem{PHENIX:2022qfp}
PHENIX, U.~A. Acharya {\em et~al.},
\newblock (2022), arXiv:2203.12354.

\bibitem{WA98:2000vxl}
WA98, M.~M. Aggarwal {\em et~al.},
\newblock Phys. Rev. Lett. {\bf 85}, 3595 (2000), arXiv:nucl-ex/0006008.

\bibitem{Paquet:2015lta}
J.-F. Paquet {\em et~al.},
\newblock Phys. Rev. C {\bf 93}, 044906 (2016), arXiv:1509.06738.

\bibitem{Adam:2015lda}
ALICE, J.~Adam {\em et~al.},
\newblock Phys. Lett. B {\bf 754}, 235 (2016), arXiv:1509.07324.

\bibitem{ALICE:2018dti}
ALICE, S.~Acharya {\em et~al.},
\newblock Phys. Lett. B {\bf 789}, 308 (2019), arXiv:1805.04403.

\bibitem{Gale:2021emg}
C.~Gale, J.-F. Paquet, B.~Schenke, and C.~Shen,
\newblock Phys. Rev. C {\bf 105}, 014909 (2022), arXiv:2106.11216.

\bibitem{Schafer:2021slz}
A.~Sch\"afer, O.~Garcia-Montero, J.-F. Paquet, H.~Elfner, and C.~Gale,
\newblock Phys. Rev. C {\bf 105}, 044910 (2022), arXiv:2111.13603.

\bibitem{Steinheimer:2012tb}
J.~Steinheimer {\em et~al.},
\newblock Phys. Lett. B {\bf 714}, 85 (2012), arXiv:1203.2547.

\bibitem{Csernai:1986qf}
L.~P. Csernai and J.~I. Kapusta,
\newblock Phys. Rept. {\bf 131}, 223 (1986).

\bibitem{Scheibl:1998tk}
R.~Scheibl and U.~W. Heinz,
\newblock Phys. Rev. C {\bf 59}, 1585 (1999), arXiv:nucl-th/9809092.

\bibitem{Vovchenko:2019aoz}
V.~Vovchenko, K.~Gallmeister, J.~Schaffner-Bielich, and C.~Greiner,
\newblock Phys. Lett. B {\bf 800}, 135131 (2020), arXiv:1903.10024.

\bibitem{Neidig:2021bal}
T.~Neidig, K.~Gallmeister, C.~Greiner, M.~Bleicher, and V.~Vovchenko,
\newblock Phys. Lett. B {\bf 827}, 136891 (2022), arXiv:2108.13151.

\bibitem{Danielewicz:1991dh}
P.~Danielewicz and G.~F. Bertsch,
\newblock Nucl. Phys. A {\bf 533}, 712 (1991).

\bibitem{Oliinychenko:2018ugs}
D.~Oliinychenko, L.-G. Pang, H.~Elfner, and V.~Koch,
\newblock Phys. Rev. C {\bf 99}, 044907 (2019), arXiv:1809.03071.

\bibitem{Staudenmaier:2021lrg}
J.~Staudenmaier, D.~Oliinychenko, J.~M. Torres-Rincon, and H.~Elfner,
\newblock Phys. Rev. C {\bf 104}, 034908 (2021), arXiv:2106.14287.

\bibitem{Donigus:2020fon}
B.~D\"onigus,
\newblock Eur. Phys. J. A {\bf 56}, 280 (2020).

\bibitem{Tolos:2020aln}
L.~Tolos and L.~Fabbietti,
\newblock Prog. Part. Nucl. Phys. {\bf 112}, 103770 (2020), arXiv:2002.09223.

\bibitem{Fabbietti:2020bfg}
L.~Fabbietti, V.~Mantovani~Sarti, and O.~Vazquez~Doce,
\newblock Ann. Rev. Nucl. Part. Sci. {\bf 71}, 377 (2021), arXiv:2012.09806.

\bibitem{Becattini:2020ngo}
F.~Becattini and M.~A. Lisa,
\newblock Ann. Rev. Nucl. Part. Sci. {\bf 70}, 395 (2020), arXiv:2003.03640.

\bibitem{ParticleDataGroup:2020ssz}
Particle Data Group, P.~A. Zyla {\em et~al.},
\newblock PTEP {\bf 2020}, 083C01 (2020).

\bibitem{STAR:2017ckg}
STAR, L.~Adamczyk {\em et~al.},
\newblock Nature {\bf 548}, 62 (2017), arXiv:1701.06657.

\bibitem{STAR:2021beb}
STAR, M.~S. Abdallah {\em et~al.},
\newblock Phys. Rev. C {\bf 104}, L061901 (2021), arXiv:2108.00044.

\bibitem{ALICE:2019onw}
ALICE, S.~Acharya {\em et~al.},
\newblock Phys. Rev. C {\bf 101}, 044611 (2020), arXiv:1909.01281.

\bibitem{Okubo:2021dbt}
STAR, K.~Okubo,
\newblock EPJ Web Conf. {\bf 259}, 06003 (2022), arXiv:2108.10012.

\bibitem{STAR:2020xbm}
STAR, J.~Adam {\em et~al.},
\newblock Phys. Rev. Lett. {\bf 126}, 162301 (2021), arXiv:2012.13601.

\bibitem{STAR:2018gyt}
STAR, J.~Adam {\em et~al.},
\newblock Phys. Rev. C {\bf 98}, 014910 (2018), arXiv:1805.04400.

\bibitem{Muller:2018ibh}
B.~M\"uller and A.~Sch\"afer,
\newblock Phys. Rev. D {\bf 98}, 071902 (2018), arXiv:1806.10907.

\bibitem{STAR:2019erd}
STAR, J.~Adam {\em et~al.},
\newblock Phys. Rev. Lett. {\bf 123}, 132301 (2019), arXiv:1905.11917.

\bibitem{Becattini:2022zvf}
F.~Becattini,
\newblock (2022), arXiv:2204.01144.

\bibitem{Singha:2020qns}
STAR, S.~Singha,
\newblock Nucl. Phys. A {\bf 1005}, 121733 (2021), arXiv:2002.07427.

\bibitem{Mohanty:2020bqq}
ALICE, B.~Mohanty,
\newblock PoS {\bf ICHEP2020}, 555 (2021), arXiv:2012.04167.

\bibitem{Florkowski:2018fap}
W.~Florkowski, A.~Kumar, and R.~Ryblewski,
\newblock Prog. Part. Nucl. Phys. {\bf 108}, 103709 (2019), arXiv:1811.04409.

\bibitem{Weickgenannt:2019dks}
N.~Weickgenannt, X.-L. Sheng, E.~Speranza, Q.~Wang, and D.~H. Rischke,
\newblock Phys. Rev. D {\bf 100}, 056018 (2019), arXiv:1902.06513.

\bibitem{Gallegos:2021bzp}
A.~D. Gallegos, U.~G\"ursoy, and A.~Yarom,
\newblock SciPost Phys. {\bf 11}, 041 (2021), arXiv:2101.04759.

\bibitem{Peccei:1977hh}
R.~D. Peccei and H.~R. Quinn,
\newblock Phys. Rev. Lett. {\bf 38}, 1440 (1977).

\bibitem{Kharzeev:2007jp}
D.~E. Kharzeev, L.~D. McLerran, and H.~J. Warringa,
\newblock Nucl. Phys. A {\bf 803}, 227 (2008), arXiv:0711.0950.

\bibitem{Newman:2005hd}
G.~M. Newman,
\newblock JHEP {\bf 01}, 158 (2006), arXiv:hep-ph/0511236.

\bibitem{Son:2009tf}
D.~T. Son and P.~Surowka,
\newblock Phys. Rev. Lett. {\bf 103}, 191601 (2009), arXiv:0906.5044.

\bibitem{Kharzeev:2010gd}
D.~E. Kharzeev and H.-U. Yee,
\newblock Phys. Rev. D {\bf 83}, 085007 (2011), arXiv:1012.6026.

\bibitem{Kharzeev:2015znc}
D.~E. Kharzeev, J.~Liao, S.~A. Voloshin, and G.~Wang,
\newblock Prog. Part. Nucl. Phys. {\bf 88}, 1 (2016), arXiv:1511.04050.

\bibitem{Voloshin:2018qsm}
S.~A. Voloshin,
\newblock Phys. Rev. C {\bf 98}, 054911 (2018), arXiv:1805.05300.

\bibitem{Shi:2017cpu}
S.~Shi, Y.~Jiang, E.~Lilleskov, and J.~Liao,
\newblock Annals Phys. {\bf 394}, 50 (2018), arXiv:1711.02496.

\bibitem{Muller:2010jd}
B.~M{\"u}ller and A.~Sch{\"a}fer,
\newblock Phys. Rev. C {\bf 82}, 057902 (2010), arXiv:1009.1053.

\bibitem{STAR:2021pwb}
STAR, M.~Abdallah {\em et~al.},
\newblock Phys. Rev. Lett. {\bf 128}, 092301 (2022), arXiv:2106.09243.

\bibitem{ALICE:2020siw}
ALICE, S.~Acharya {\em et~al.},
\newblock JHEP {\bf 09}, 160 (2020), arXiv:2005.14640.

\bibitem{CMS:2017lrw}
CMS, A.~M. Sirunyan {\em et~al.},
\newblock Phys. Rev. C {\bf 97}, 044912 (2018), arXiv:1708.01602.

\bibitem{STAR:2019bjg}
STAR, J.~Adam {\em et~al.},
\newblock Nucl. Sci. Tech. {\bf 32}, 48 (2021), arXiv:1911.00596.

\bibitem{STAR:2021mii}
STAR, M.~Abdallah {\em et~al.},
\newblock Phys. Rev. C {\bf 105}, 014901 (2022), arXiv:2109.00131.

\bibitem{PHENIX:2015siv}
PHENIX, A.~Adare {\em et~al.},
\newblock (2015), arXiv:1501.06197.

\bibitem{Adamova:2019vkf}
D.~Adamov\'a {\em et~al.},
\newblock (2019), arXiv:1902.01211.

\bibitem{Accardi:2012qut}
A.~Accardi {\em et~al.},
\newblock Eur. Phys. J. A {\bf 52}, 268 (2016), arXiv:1212.1701.

\bibitem{HADES:2009aat}
HADES, G.~Agakishiev {\em et~al.},
\newblock Eur. Phys. J. A {\bf 41}, 243 (2009), arXiv:0902.3478.

\bibitem{Friman:2011zz}
B.~Friman {\em et~al.}, editors,
\newblock {\em The CBM physics book: Compressed baryonic matter in laboratory
  experiments}, Lecture Notes in Physics Vol. 814 (Springer-Verlag, 2011).

\bibitem{Hachiya:2020bjg}
T.~Hachiya,
\newblock Int. J. Mod. Phys. E {\bf 29}, 2040005 (2020).

\bibitem{Xiaohong:2018weu}
Z.~Xiaohong,
\newblock Nucl. Phys. Rev. {\bf 35}, 339 (2018).

\bibitem{Kekelidze:2017ual}
NICA, V.~D. Kekelidze,
\newblock JINST {\bf 12}, C06012 (2017).

\end{thebibliography}

\end{document}